\documentclass[prb,aps,twocolumn,10pt,floatfix,nofootinbib,superscriptaddress]{revtex4-2}

\usepackage{amsmath}
\usepackage{amssymb}
\usepackage{mathtools}

\usepackage{amsthm}

\usepackage{amsfonts}
\usepackage{txfonts}
\usepackage{dsfont}                 
\usepackage{bbold}                  
\usepackage[normalem]{ulem}         
\usepackage{pifont}                 
\newcommand{\cmark}{\ding{51}}
\newcommand{\xmark}{\ding{55}}

\usepackage{array}
\usepackage{dcolumn}                
\usepackage{makecell}               

\usepackage{enumerate}
\usepackage[shortlabels]{enumitem}

\usepackage[usenames,dvipsnames,table]{xcolor}
\usepackage{graphicx}
\graphicspath{{figs/}} 

\usepackage[version=4]{mhchem}      
\usepackage{physics}                
\usepackage{csquotes}               
\usepackage{multirow}               
\usepackage{booktabs} 
\usepackage[flushleft]{threeparttable}  
\usepackage{placeins}

\usepackage[colorlinks=true,citecolor=cyan,linkcolor=magenta,filecolor=magenta]{hyperref}
\usepackage[capitalize]{cleveref}

\usepackage{orcidlink}

\usepackage[globalcitecopy]{bibunits}
\defaultbibliography{bib}
\defaultbibliographystyle{apsrev4-1}

\newcommand{\e}{\mathrm{e}}         
\renewcommand{\i}{\mathrm{i}}       

\DeclareMathOperator{\sign}{sign}  	
\DeclareMathOperator{\diag}{diag}  	

\newcommand{\argplaceholder}{\overline{\makebox[1.25ex]{$\vphantom{a}\cdot$}}}

\newcommand{\floor}[1]{\left\lfloor #1 \right\rfloor}

\renewcommand{\vec}[1]{{\vb{#1}}}  
\newcommand{\uvec}[1]{\vu{#1}}     
\newcommand{\id}{\mathds{1}}       
\newcommand{\mat}[1]{\begin{pmatrix}#1\end{pmatrix}} 
\newcommand{\adjo}[1]{#1^\dagger}  
\newcommand{\padjo}[1]{#1^{\phantom{\dagger}}}  

\newcommand{\nunit}[1]{\,\mathrm{#1}} 
\newcommand{\angstrom}{\textup{\AA}}

\newcommand{\cconj}[1]{{#1}^*}

\newcommand{\Ham}{\mathcal{H}}      

\newcommand{\sLG}{\mathcal{G}}          
\newcommand{\LG}[1]{\sLG^{\hspace{0.05em}#1}}
\newcommand{\sLcG}{\smash{\overline{\mathcal{G}}}}
\newcommand{\LcG}[1]{\sLcG^{\hspace{0.05em}#1}}
\newcommand{\TG}{\mathfrak{T}}          
\renewcommand{\O}{\mathsf{O}}       
\newcommand{\SO}{\mathsf{SO}}       
\newcommand{\PN}{\mathsf{P}}        
\newcommand{\Spin}{\mathsf{Spin}}   

\newcommand{\Eu}{\chi}              
\newcommand{\SW}[1]{w_{#1}}         
\newcommand{\Qc}[1][]{Q_{\llcorner #1}} 

\newcommand{\Clifford}[2]{\mathcal{C}\hspace{-0.1em}\ell_{#1,#2}} 
\newcommand{\so}{\mathfrak{so}}     
\newcommand{\spin}{\mathfrak{spin}} 
\newcommand{\subg}{<}

\newcommand{\Python}{\textsc{Python}}
\newcommand{\Mathematica}{\textsc{Mathematica}}

\DeclareMathAlphabet{\mathbbold}{U}{bbold}{m}{n}
\def\abs#1{\left|{#1}\right|}      	
\def\bs#1{\boldsymbol{#1}}			

\def\mcH{\mathcal{H}}					
\def\mcT{\mathcal{T}}					
\def\mcP{\mathcal{P}}					
\def\mcK{\mathcal{K}}					

\def\ztwo{\mathbbold{Z}_2}				

\newcommand{\td}{\text{--}}

\newcommand{\frameufinal}{
\mathrlap{\ThisStyle{\raisebox{0.34em}[0pt][0pt]{$\SavedStyle\underline{\phantom{\underline{\mathsf{u}}}}$}}}
\mathrlap{\ThisStyle{\raisebox{0.22em}[0pt][0pt]{$\SavedStyle\underline{\phantom{\underline{\mathsf{u}}}}$}}}
\mathsf{u}\raisebox{-0.3em}[0pt][0pt]{\rule{0pt}{0pt}}_{0,0}
}

\newcolumntype{C}{>{$}c<{$}} 

\AtEndEnvironment{threeparttable}{\vskip10pt}{}{}
\makeatletter
\patchcmd{\TPT@doparanotes}
  {\TPT@hsize}
  {\TPT@hsize\footnotesize}
  {}
  {}
\makeatother

\definecolor{MyOrange}{RGB}{255,180,0}
\definecolor{MyRed}{RGB}{204,0,0}
\definecolor{MyBlue}{RGB}{0,165.75,229.5}
\definecolor{MyIndigo}{RGB}{75,0,130}
\definecolor{MyTan}{RGB}{185,155,126}

\definecolor{MyGreen}{RGB}{0,216.75,0}

\newcolumntype{x}[1]{>{\centering\let\newline\\\arraybackslash\hspace{0pt}}p{#1}}
\newcolumntype{L}{>{$}l<{$}} 
\newcolumntype{R}{>{$}r<{$}} 

\theoremstyle{definition}
\newtheorem{conjecture}{Conjecture}

\newtheorem{corollary}{Corollary}

\theoremstyle{plain}
\newtheorem{lemma}{Lemma}[section]

\crefname{section}{Sec.}{Secs.}
\Crefname{section}{Section}{Sections}

\begin{document}
\begin{bibunit}

\title{Universal higher-order bulk-boundary correspondence of triple nodal points}

\author{Patrick M. Lenggenhager\,\orcidlink{0000-0001-6746-1387}}\email[corresponding author: ]{lenpatri@ethz.ch}
\affiliation{Condensed Matter Theory Group, Paul Scherrer Institute, 5232 Villigen PSI, Switzerland}
\affiliation{Institute for Theoretical Physics, ETH Zurich, 8093 Zurich, Switzerland}
\affiliation{Department of Physics, University of Zurich, Winterthurerstrasse 190, 8057 Zurich, Switzerland}

\author{Xiaoxiong Liu\,\orcidlink{0000-0002-2187-0035}}
\affiliation{Department of Physics, University of Zurich, Winterthurerstrasse 190, 8057 Zurich, Switzerland}

\author{Titus Neupert\,\orcidlink{0000-0003-0604-041X}}
\affiliation{Department of Physics, University of Zurich, Winterthurerstrasse 190, 8057 Zurich, Switzerland}

\author{Tom\'{a}\v{s} Bzdu\v{s}ek\,\orcidlink{0000-0001-6904-5264}}\email[corresponding author: ]{tomas.bzdusek@psi.ch}
\affiliation{Condensed Matter Theory Group, Paul Scherrer Institute, 5232 Villigen PSI, Switzerland}
\affiliation{Department of Physics, University of Zurich, Winterthurerstrasse 190, 8057 Zurich, Switzerland}

\date{\today}

\begin{abstract}
Triple nodal points are degeneracies of energy bands in momentum space at which three Hamiltonian eigenstates coalesce at a single eigenenergy.
For spinless particles, the stability of a triple nodal point requires two ingredients: rotational symmetry of order three, four or six; combined with mirror or space-time-inversion symmetry.
However, despite ample studies of their classification, robust boundary signatures of triple nodal points have until now remained elusive.
In this work, we first show that pairs of triple nodal points in semimetals and metals can be characterized by Stiefel-Whitney and Euler monopole invariants, of which the first one is known to facilitate higher-order topology.
Motivated by this observation, we then combine symmetry indicators for corner charges and for the Stiefel-Whitney invariant in two dimensions with the classification of triple nodal points for spinless systems in three dimensions.
The result is a complete higher-order bulk-boundary correspondence, where pairs of triple nodal points are characterized by fractional jumps of the hinge charge.
We present minimal models of the various species of triple nodal points carrying higher-order topology, and illustrate the derived correspondence on \ce{Sc3AlC} which becomes a higher-order triple-point metal in applied strain.
The generalization to spinful systems, in particular to the \ce{WC}-type triple-point material class, is briefly outlined.
\end{abstract}

\maketitle


\section{Introduction}\label{Sec:Intro}
The hallmark feature of higher-order topological phases in $d$ dimensions is the presence of an anomaly on boundaries with dimensions $d-2$ or lower~\cite{Benalcazar:2017,Benalcazar:2017a,Song:2017,Schindler:2018a,Langbehn:2017,Ezawa:2018,Schindler:2018,Kunst:2018,Calugaru:2019,Park:2019,Serra-Garcia:2018,Imhof:2018,SerraGarcia:2019,Mittal:2019,Xie:2019a,ElHassan:2019,Benalcazar:2017,Benalcazar:2017a,Ezawa:2018,Geier:2018,Khalaf:2018,Ahn:2019,Lee:2020,Song:2017,Schindler:2018a,Schindler:2018,VanMiert:2018,Wang:2019,Trifunovic:2019}.
In two dimensions (2D), \emph{second-order crystalline topological insulators} require additional chiral symmetry or particle-hole symmetry to protect zero-energy corner modes~\cite{Serra-Garcia:2018,Peterson:2018,Hwang:2019,Pahomi:2020}.
If such symmetries are absent and the corner modes merge into the bulk energy bands, the nontrivial second-order topology is instead revealed by a \emph{corner-induced filling anomaly}~\cite{Song:2017,Wieder:2018,Benalcazar:2019,Watanabe:2020}, i.e., an obstruction to simultaneously satisfying charge neutrality and preserving the crystalline symmetries in the presence of corners.
In particular, if the valence bands are filled and the crystalline symmetry preserved, the filling anomaly implies corner-localized charges.
Remarkably, in the presence of rotational symmetry, the corner charges acquire fractional quantized values that can be predicted~\cite{Benalcazar:2019,Takahashi:2021,Fang:2021} using \emph{symmetry indicators}~\cite{Fu:2007,Fang:2012a,Alexandradinata:2014,Kim:2015,Po:2017,Kruthoff:2017,Bradlyn:2017,Song:2018a}, i.e., symmetry eigenvalues of the occupied bands at high-symmetry points (HSPs) in the Brillouin zone (BZ).

The notion of higher-order topology has recently been generalized to 3D nodal phases with two- and four-fold degeneracies, resulting in higher-order Weyl~\cite{Wang:2020b,Ghorashi:2020,Wei:2021,Ezawa:2018d,Ezawa:2019} and Dirac~\cite{Lin:2018,Calugaru:2019,Wieder:2020,Qiu:2021} semimetals, respectively, as well as to a class of nodal-ring semimetals obtained by perturbing Dirac degeneracies~\cite{Wang:2020}.
Such semimetals can be understood as 2D topological insulators which are augmented by a third dimension along which the higher-order topology changes, with the band nodes marking the occurrence of higher-order phase transitions.
Correspondingly, higher-order topological semimetals are characterized by zero-energy hinge states (in the presence of chiral symmetry) or fractional hinge charges (in their absence) for a range of momenta demarcated by the band degeneracies.
In light of these observations, a question of timely interest arises, namely whether the phenomena associated with higher-order topology also extend to more intricate species of band nodes.

Triple nodal points~\cite{Zhu:2016,Heikkila:2015,Hyart:2016,Weng:2016,Lv:2017,Ma:2018,Chang:2017,Zhang:2017,Kim:2018,Chen:2018} -- \emph{triple points} (TPs) for short -- are three-fold degeneracies of energy bands occurring at points in momentum space, and therefore constitute intermediates between Weyl and Dirac points.
They occur on high-symmetry lines (HSL) in the BZ when a two-fold degenerate band [corresponding to a 2D irreducible co-representation (ICR) of the HSL's little co-group and forming what we call the \emph{central} nodal line (NL)] is crossed by a third non-degenerate band [one-dimensional (1D) ICR of the little co-group]~\cite{Hyart:2016}.
In some cases, TPs are accompanied by additional NL arcs that lie off the rotation axis but coalesce with the central NL at a \emph{nexus point}~\cite{Lenggenhager:2022:TPClassif}.
Triple points are thus classified~\cite{Zhu:2016,Chen:2018,Lenggenhager:2021:MBNLs,Lenggenhager:2022:TPClassif} as type~$\mathsf{B}$ if a nexus point coincides with the TP and as type~$\mathsf{A}$ otherwise (where the appearance of a nexus point \emph{near} the TP is optional).
TPs were shown by angle-resolved photoemission spectroscopy to exist in the band structure of \ce{MoP}~\cite{Lv:2017} and \ce{WC}~\cite{Ma:2018}.

In contrast to Weyl and Dirac points, a single TP generically results in a \emph{metallic} state, due to the imbalance between the degeneracy of the crossing bands.
However, \emph{pairs} of TPs, formed when the 2D ICR consecutively crosses two 1D ICRs, can result in a semimetal with small Fermi pockets.
We call such a TP configuration a \emph{triple-point pair} (TPP).
These nodal features can arise both in spinful~\cite{Zhu:2016} and spinless~\cite{Hyart:2016,Lenggenhager:2021:MBNLs} systems; however, although there have recently been several efforts~\cite{Winkler:2019,Xie:2019,Das:2020,Lenggenhager:2021:MBNLs,Lenggenhager:2022:TPClassif} towards their topological description and symmetry classification, it has remained an open question whether TPs or TPPs can be characterized by robust boundary signatures.

\begin{table*}
    \centering
    \caption{
        Symmetry conditions to realize type-$\mathsf{A}$ vs.~type-$\mathsf{B}$ triple points (TPs) along a high-symmetry line (HSL) in the momentum space of spinless systems. 
        The table compactly displays all magnetic point groups (MPGs) that (1) preserve one momentum component (such that the MPG corresponds to a little co-group along some HSL), and that (2) support both 1D and 2D irreducible co-representations (ICRs). 
        The columns and the rows indicate generators of the MPG, where $C_n$ is rotational symmetry of order $n$, $\mcP\mcT$ is space-time inversion symmetry, and $m_v$ is mirror symmetry with respect to a plane containing the rotation axis; furthermore, $C_n \mcP\mcT$ corresponds to a composition of rotoinversion $C_n \mcP$ with time reversal $\mcT$. 
        For brevity, we call $C_n\mcP\mcT$ the \emph{antiunitary rotational symmetry} of order $n$.
        For each entry, the corresponding MPG is first labelled per the notation of Ref.~\onlinecite{Litvin:2013}, and subsequently by the Hermann–Mauguin notation~\cite{Bradley:1972}. 
        The last row in each cell indicates the possible TP types; in certain hexagonal cases both type-$\mathsf{A}$ and type-$\mathsf{B}$ TPs are possible, depending on the specific choice of ICRs as specified in \cref{tab:HOBBC:Qjump}.
        The three MPGs colored in gray violate condition (2) and are therefore dismissed; the remaining 13 MPGs support TPs along HSLs, and are analyzed in the manuscript. 
        The classification of TPs for all cases appears in Ref.~\onlinecite{Lenggenhager:2022:TPClassif}.
        In the derivation of the higher-order bulk-boundary correspondence, we exploit that the cases without $\mcP\mcT$ correspond to subgroups of the cases with $\mcP\mcT$. Similarly, the $C_3$ column corresponds to a subgroup of both the $C_6$ and the $C_6\mcP\mcT$ columns.
    }
    \begin{ruledtabular}
    \begin{tabular}{CCCCCC}
    {} &
    \multicolumn{1}{C}{\text{Trigonal}} & 
    \multicolumn{2}{C}{\text{Tetragonal}} &
    \multicolumn{2}{C}{\text{Hexagonal}}  
    \\  \cmidrule(lr){2-2} \cmidrule(lr){3-4} \cmidrule(lr){5-6}
    \text{Generators: $\downarrow$ and $\rightarrow$} & 
    \multicolumn{1}{C}{C_3} & 
    C_4 & 
    C_4 \mcP\mcT & 
    C_6 & 
    C_6 \mcP\mcT   
    \\  \hline \addlinespace
    \multirow{3}{*}{$\qquad\varnothing$}&
    \cellcolor{gray!25}{\;16.1.60\;} & 
    \cellcolor{gray!25}{\;9.1.29\;} &
    {\;10.3.34\;} &
    \cellcolor{gray!25}{\;21.1.76\;} & 
    {\;22.3.81\;} 
    \\
    {} & 
    \cellcolor{gray!25}(3) & 
    \cellcolor{gray!25}(4) & 
    (\bar{4}') & 
    \cellcolor{gray!25}(6) & 
    (\bar{6}') 
    \\
    {} &
    \cellcolor{gray!25}{\;-\;} &
    \cellcolor{gray!25}{\;-\;} &
    \mathsf{A} &
    \cellcolor{gray!25}{\;-\;} &
    \mathsf{A}
    \\ \addlinespace
    \multirow{3}{*}{$\qquad\mcP\mcT$}&
    \;17.3.64\; &
    \multicolumn{2}{C}{\;11.4.38\;} &
    \multicolumn{2}{C}{\;23.4.85\;}  \\
    {} &
    (\bar{3}') &
    \multicolumn{2}{C}{(4/m')} &
    \multicolumn{2}{C}{(6/m')}     
    \\
    {} &
    \mathsf{B} &
    \multicolumn{2}{C}{\mathsf{A}} &
    \multicolumn{2}{C}{\mathsf{A}\;\textrm{and}\;\mathsf{B}} 
    \\ \addlinespace
    \multirow{3}{*}{$\qquad m_v$}&
    {\;19.1.68\;} &
    {\;13.1.44\;} &
    {\;14.3.50\;} &
    {\;25.1.91\;} &
    {\;26.4.98\;}  \\
    {} &
    (3m) &
    (4mm) &
    (\bar{4}'2'm) &
    (6mm) &
    (\bar{6}'m2')
    \\ 
    {} &
    \mathsf{B} &
    \mathsf{A} &
    \mathsf{A} &
    \mathsf{A}\;\textrm{and}\;\mathsf{B} &
    \mathsf{A}
    \\ \addlinespace
    \multirow{3}{*}{$\qquad\{\mcP\mcT,m_v\}$}&
    \;20.3.73\; &
    \multicolumn{2}{C}{\;15.3.55\;} &
    \multicolumn{2}{C}{\;27.3.102\;}  \\
    {} &
    (\bar{3}'m) &
    \multicolumn{2}{C}{(4/m'mm)} &
    \multicolumn{2}{C}{(6/m'mm)} 
    \\
    {} &
    \mathsf{B} &
    \multicolumn{2}{C}{\mathsf{A}} &
    \multicolumn{2}{C}{\mathsf{A}\;\textrm{and}\;\mathsf{B}} 
    \\
    \end{tabular}
    \end{ruledtabular}
    \label{tab:TP-MPGs}
\end{table*}

In this work, we answer both postulated questions in a single stroke: semimetallic TPPs are \emph{generally} characterized by a higher-order bulk-boundary correspondence; namely, each species of TPP can be assigned a unique value of \emph{fractional jump} of the hinge charge.
We derive the exact correspondence by combining the symmetry classification of TPs~\cite{Lenggenhager:2021:MBNLs,Lenggenhager:2022:TPClassif} with the symmetry indicators of higher order topology.
Note that while we explicitly consider only TPs occurring in spinless systems, our mathematical analysis based on symmetry indicators can be easily generalized to the spinful case too.
Nonetheless, the spinless setting allows us to provide a complementary \emph{geometric} interpretation of the higher-order topology.
Specifically, in the presence of space-time-inversion ($\mcP\mcT$) symmetry, non-Abelian band topology~\cite{Wu:2019} in combination with the known properties of certain monopole and linking invariants~\cite{Fang:2015,Bzdusek:2017,Zhao:2017,Ahn:2018,Ahn:2019,Tiwari:2020}, readily provide the result for the bulk-hinge correspondence for a subclass of TPPs in spinless systems.

Stable TPs can arise along HSLs with $13$ distinct little co-groups.
On the one hand, TPs can be protected if rotational symmetry $C_n$ of order $n\in\{3,4,6\}$ is supplemented with $\mcP\mcT$ \emph{or} with mirror symmetry $m_v$ with respect to a plane containing the rotation axis \emph{or} with both $\mcP\mcT$ and $m_v$.
On the other hand, the combined symmetry $C_n\mcP\mcT$ (which we call \emph{antiunitary rotation}) of order $n\in\{4,6\}$ can stabilize TPs with or without the $\mcP\mcT$ and $m_v$ symmetry.
We provide an easily navigable summary of all admissible symmetry combinations and of their subgroup-supergroup relations in \cref{tab:TP-MPGs}.
Non-symmorphic symmetries do neither affect the possible little co-groups of HSLs along which stable TPs can arise nor the classification of those TPs~\cite{Lenggenhager:2022:TPClassif}, which is reflected in the fact that the symmetries listed in \cref{tab:TP-MPGs} are elements of the little \emph{co-group}, i.e., \emph{point-group} symmetries.

The diverse range of symmetry combinations seemingly complicates the analysis.
However, it turns out that all the cases of interest can be obtained by a proper perturbation of a system with $\mcP\mcT$ symmetry.
Therefore, our approach to analyze the topological invariants and the bulk-hinge correspondence of TPPs is to first deal with the $\mcP\mcT$-symmetric cases, and afterwards consider the effect of perturbations to derive the results also for the $\mcP\mcT$-broken cases.
In particular, in the presence of $\mcP\mcT$ we can characterize the TPP also by a second Stiefel Whitney (2SW) or Euler monopole charge, of which the first one provides further insights into the bulk-hinge correspondence~\cite{Ahn:2019,Wang:2019}.
Our results for the correspondence between TPPs, and their associated fractional jump in the hinge charge, and (if also defined) their 2SW and Euler monopole charge are summarized in \cref{tab:HOBBC:Qjump}.
The results also apply to non-symmorphic space groups (SGs) with an appropriate identification of ICRs~\cite{Lenggenhager:2022:TPClassif}.

The manuscript is organized as follows.
In \cref{Sec:C4Model}, we present a concrete $C_4$-symmetric tight-binding model that illustrates the phenomenology of TPP-induced higher-order topology, including both the hinge-charge jump and the 2SW monopole invariant.
This motivates our study and sets the stage for the subsequent general discussion. 
First, \cref{Sec:Monopole} discusses monopole charges induced by type-$\mathsf{A}$ TPs in $\mcP\mcT$-symmetric spinless systems using simple manipulations with the non-Abelian band topology.
The analysis reveals that pairs of TPs in three-band models can be characterized by the Euler monopole charge, while TPPs in four-band models can carry the 2SW monopole charge.
The discussion is then substantially generalized in \cref{Sec:HOBBC}, where we use symmetry indicators to establish the main result of our work: a general bulk-hinge correspondence principle for TPPs \emph{in spinless systems}.
As intermediate results, we are led to also derive symmetry-indicator formulas for the 2SW and Euler monopole charge in the presence of rotational symmetry and a bulk-corner correspondence principle for the former, and we analyze the effect of symmetry breaking on TPPs.
For simplicity, we first focus on symmorphic SGs and discuss the generalization to non-symmorphic SGs only in \cref{Sec:non-symmorphic}.

After presenting the main results of our theoretical analysis, we apply the theory to concrete models. First, in \cref{Sec:MinimalModels}, we demonstrate the derived bulk-hinge correspondence on minimal tight-binding models for various TPP species.
We first revisit the $C_4$-symmetric model from \cref{Sec:C4Model} and then discuss concrete Hamiltonians for the $C_6$- and $C_3$-symmetric cases.
Next, we consider two material examples in \cref{Sec:Materials}.
On the one hand, we consider \ce{Sc3AlC} in large uniaxial strain to provide a solid-state illustration of TPPs with the predicted fractional hinge-charge jump.
On the other hand, we show that \ce{Li2NaN} under ambient conditions has a pair of TPs carrying a nontrivial Euler monopole charge, and we briefly discuss the consequences of this topological obstruction on the stability of the band nodes.
In \cref{Sec:non-symmorphic}, we first explain how the symmetry-indicator formulas for the corner charge can be applied to compute the fractional hinge charge in a wire geometry even for non-symmorphic SGs and then argue that the bulk-hinge correspondence of TPPs generalizes to non-symmorphic SGs. 
Then, in \cref{sec:spinful} we present a short digression to TPPs in spinful systems, and argue that strongly spin-orbit coupled compounds in the \ce{WC}-type crystal structure studied by Ref.~\onlinecite{Zhu:2016} could host the fractional hing-charge jumps.
Finally, we conclude in \cref{Sec:Conclusion} by summarizing the main results and outlining possible extensions of our work to other nodal configurations.

The main text is accompanied by several appendices which discuss details of our theoretical and numerical analysis: in \cref{App:hingecharges,App:Euler-monopole} we summarize our mathematical derivations that involve the symmetry indicators, in \cref{App:MinimalModels} we include details of the presented tight-binding models and explain how they were systematically constructed, in \cref{App:corner-charge-extraction} we present methods to numerically extract the hinge charges from tight-binding models.
Finally, and in \cref{App:NLBZ} we discuss the Euler monopole charge from the perspective of the non-Abelian band topology (significantly extended in the Supplemental Materials~\cite{SM}, where we proof a conjecture relating the non-Abelian invariant computed on contours shifted by reciprocal lattice vectors) and illustrate this on several material examples in \cref{App:Euler_materials}.
We provide access to all the data and code necessary to reproduce the results presented here in the supplementary data and code~\cite{Lenggenhager:2021:TPHOT:SDC}.

\section{Hinge charges induced by triple points}\label{Sec:C4Model}

\begin{figure}[t]
    \centering
    \includegraphics{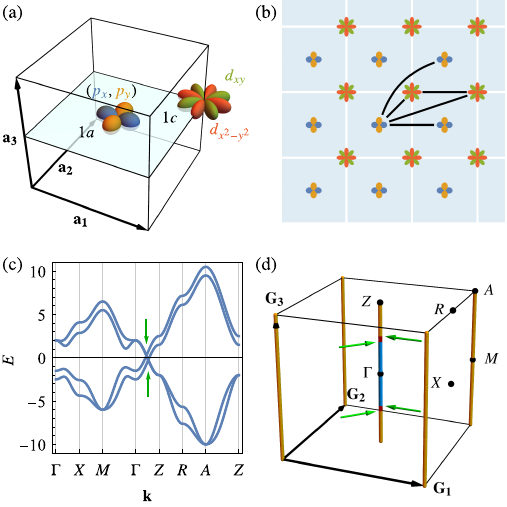}
    \caption{
        Features of the model given in \cref{eq:C4model:def}.
        (a) Real-space unit cell spanned by the lattice vectors $\vec{a}_{1,2,3}$ with the four orbitals: $(p_x,p_y)$ at Wyckoff position $1a$ and $d_{xy}$, $d_{x^2-y^2}$ at $1c$.
        (b) Single layer (for fixed $z$) of the three-dimensional lattice.
        The blue shaded regions indicate the projection of the unit cells with the orbitals indicated in the same colors as in (a).
        Black lines (where further symmetry-related lines are dropped to maintain clarity) indicate the in-plane hopping processes included in the model.
        (c) Band structure along the high-symmetry lines.
        (d) Brillouin zone spanned by the reciprocal lattice vectors $\vec{G}_{1,2,3}$ with high-symmetry points $\{\Gamma,Z,X,R,M,A\}$ shown. Nodal lines in the first, second and third band gap are displayed in orange, red, and blue color, respectively.
        The four triple points of the model are indicated by green arrows in panels (c,d).
    }
    \label{fig:C4model:model}
\end{figure}

To motivate our study of higher-order topology associated with TPs, we introduce a concrete model and investigate its phenomenology.
The model assumes spinless particles on a tetragonal lattice (with lattice constants set to $a=c=1$ for simplicity) and has the (symmorphic) SG $P4/mmm$ (No.~123) with isogonal point group ${D}_{4h}$.
As illustrated in \cref{fig:C4model:model}(a), we place $(p_x,p_y)$ orbitals transforming in the ICR $E_\mathrm{u}$ at Wyckoff position (WP) $1a$, and the orbitals $d_{xy}$ and $d_{x^2-y^2}$, transforming in $B_{1\mathrm{g}}$ and $B_{2\mathrm{g}}$, respectively, at WP $1c$.
The site-symmetry group of both considered WPs corresponds to the complete ${D}_{4h}$ point group.

In the basis ($\i p_x,\i p_y,d_{xy},d_{x^2-y^2}$),
the considered model is expressed by the Bloch Hamiltonian 
\begin{multline}
    \!\!\!\!\Ham(\vec{k}) = {-}\left[t_1{+}2t_2\left(\cos k_x{+}\cos k_y{+}\cos k_z\right)\right]\gamma_3+t_3\left(\gamma_{14}-\gamma_{25}\right)\\
    - t_4\left(\cos k_x{-}\cos k_y\right)\left(\gamma_{14} + \gamma_{25}\right)+2t_5\sin k_x\sin k_y\left(\gamma_{15}-\gamma_{24}\right)\\
    +2\!\sqrt{2}t_6\left(\cos\tfrac{k_x}{2}\sin\tfrac{k_y}{2}\,\gamma_1-\cos\tfrac{k_y}{2}\sin\tfrac{k_x}{2}\,\gamma_2\right),
    \label{eq:C4model:def}
\end{multline}
where $\gamma_1=\sigma_x\otimes \tau_x$, $\gamma_2=\sigma_x\otimes\tau_z$, $\gamma_3=\sigma_z\otimes\id_\tau$, $\gamma_4=\sigma_x\otimes \tau_y$ and $\gamma_5=\sigma_y\otimes \id_\tau$ are Gamma matrices obeying $\{\gamma_a,\gamma_b\}=2\delta_{ab}$, and $\gamma_{ab}=\tfrac{\i}{2}[\gamma_a,\gamma_b]$; Pauli matrices $\sigma_i$ act on the WP degree of freedom, and Pauli matrices $\tau_i$ act on the orbital degree of freedom at fixed WP.
The point group is generated by a $\tfrac{\pi}{4}$-rotation around the $z$-axis $C_{4z}=-\diag(\i\tau_y,\id_\tau)$, a $\pi$-rotation around the $y$-axis $C_{2y}=-\sigma_z\otimes\tau_z$, and inversion $\mcP=-\sigma_z\otimes\id_\tau$. Additionally, the Hamiltonian possesses time-reversal symmetry $\mcT$.
Note that we have intentionally chosen the basis of $p$-orbitals to be \emph{imaginary}, which results in $\mcP\mcT=\mcK$, such that the Bloch Hamiltonian is a real matrix~\cite{Bouhon:2020}.
In real space, \cref{eq:C4model:def} corresponds to the tight-binding model with in-plane hopping indicated by black lines in \cref{fig:C4model:model}(b), whereas the non-vanishing out-of-plane hopping processes (not illustrated) are strictly vertical and intra-orbital.
We choose the model parameters such that the $p$ orbitals have lower energy at all HSPs with the exception of a double band inversion at $\Gamma$: $t_1=4$, $t_2=t_6=-1$, $t_3=t_4=\tfrac{1}{4}$ and $t_5=-\tfrac{1}{4}$.
The resulting band structure is shown in \cref{fig:C4model:model}(c).

The HSL $\Gamma Z$ has little co-group $C_{4v}$ combined with $\mcP\mcT$, corresponding to the magnetic point group (MPG) $4/m'mm$ of \cref{tab:TP-MPGs}.
The eigenstates of the model in \cref{eq:C4model:def} along $\Gamma Z$ transform according to one 2D ICR $E$ ($p$-like) and two 1D ICRs $B_1$ and $B_2$ ($d$-like).
The degeneracy of $p$-orbitals along $\Gamma Z$ can be interpreted as a NL, which due to the double band inversion at $\Gamma$ is crossed by the two remaining $d$-like bands at four places, $k_z=\pm\kappa_1, \pm\kappa_2$, resulting in two TPPs at $(\kappa_1,\kappa_2)$ and $(-\kappa_2,-\kappa_1)$, respectively, visible in \cref{fig:C4model:model}(d).
We observe that the TPs are not attached to additional NLs lying off the rotation axis, consistent with the classification in Refs.~\onlinecite{Lenggenhager:2021:MBNLs,Lenggenhager:2022:TPClassif} that predicts all TPs on $C_{4v}$-symmetric HSLs to be type~$\mathsf{A}$.
The only other band degeneracy exhibited by the model is the 2D ICR along the vertical hinge $MA$ of the BZ, corresponding to a vertical NL formed by the two unoccupied $p$-like bands; however, this NL does not have any effect on the discussed phenomenology.

\begin{figure}[t]
    \centering
    \includegraphics{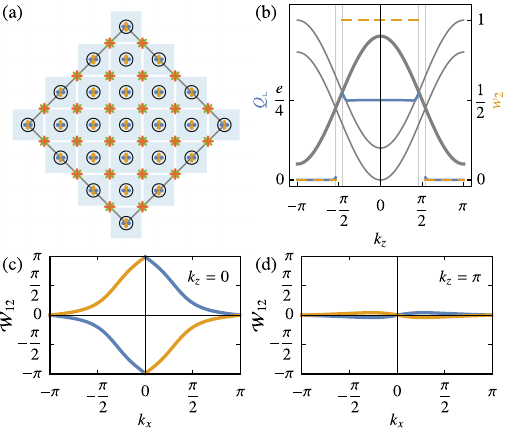}
    \caption{
    (a) In-plane termination for the nanowire geometry considered in the main text. The blue shadowed regions denote unit cells with ionic charge $2\abs{e}$ placed at the center (black circle). The electronic orbitals are illustrated as in \cref{fig:C4model:model}. To establish terminology, we say that the depicted geometry consists of $3.5\times 3.5$ unit cells (cf.~\cref{Sec:MinimalModels:C4}).
    (b) Quantized hinge charges (blue solid line) identified using exact diagonalization (for details, see \cref{Sec:MinimalModels:C4}), and second Stiefel-Whitney (2SW) class (yellow dashed line) obtained from the Wilson-loop winding; both plotted as a function of $k_z$ in the wire geometry. Cuts at fixed $k_z$ can each be interpreted as 2D flakes as shown in (a) that are characterized by a corner charge $\Qc(k_z)$. The edge charge of the model vanishes.
    For reference, the bulk band structure along $k_z$ for $k_x=k_y=0$ is displayed in gray with the 1D (2D) degenerate bands plotted by a thin (thick) lines.
    (c,d) Wilson-loop spectrum of the two occupied bands in the 2D BZ for fixed $k_z=0$ and $k_z=\pi$, respectively. For $k_z=0$ (c) we observe a winding with odd parity, indicating a nontrivial 2SW class.
    }
    \label{fig:C4model:hinge}
\end{figure}

To uncover signatures of higher-order topology, we now consider a nanowire geometry, i.e., a system which is finite in $x$- and $y$- but infinite in $z$-direction, at half-filling. 
Thus, only the momentum $k_z\in [-\pi,\pi]$ that runs along the fourfold rotation axis remains a good quantum number and the BZ is reduced to a \emph{hinge Brillouin zone}.
The considered in-plane termination is displayed in \cref{fig:C4model:hinge}(a) and is chosen to respect all the symmetries of the Hamiltonian.
Furthermore, we choose the ionic charge ($2\abs{e}$ per unit cell to compensate for the half filling of the electron bands; $e<0$ is the elementary electron charge) to be placed at the center of the unit cell.

At fixed $k_z$ in the hinge BZ we can view the bulk Hamilonian as being described by an effective 2D model, $\Ham_{k_z}(k_x,k_y)$ on the 2D lattice shown in \cref{fig:C4model:hinge}(a).
In \cref{Sec:MinimalModels:C4}, we use exact diagonalization to study this family of Hamiltonians $\Ham_{k_z}$.
By computing the charge distribution of all occupied states for various $k_z$, we observe that the model has vanishing edge charge (related to vanishing Berry phase of the occupied bands), and a $k_z$-dependent corner charge (which is interpreted as a hinge charge of the 3D system) $\Qc(k_z)$ plotted as a blue line in \cref{fig:C4model:hinge}(b).
Three regions can be identified: (1) $\Qc=\frac{e}{4}$ for $\abs{k_z}<\kappa_1$, (2) $\Qc=0$ for $\abs{k_z}>\kappa_2$, and (3) the region $\kappa_1\leq \abs{k_z}\leq\kappa_2$ where the bulk is gapless and the corner charge therefore undefined.
These regions reflect exactly the regions defined by which gap the central NL is located in, cf.\ \cref{fig:C4model:model}(d).
In particular, we observe a jump of the hinge charge from region (2) to region (1) by $\Delta \Qc = \tfrac{e}{4}$.

The difference in bulk topology between these regions is reflected in the Wilson-loop spectrum.
\Cref{fig:C4model:hinge}(c,d) shows the Wilson-loop spectrum of the occupied bands for $k_z=0,\pi$, respectively.
While the former winds once around the 2D BZ torus, the latter does not.
Recall~\cite{Bzdusek:2017} that the parity of the winding of the Wilson loop eigenvalues determines the 2SW class.
Note that the $\mcP\mcT$ symmetry of the model allows us to compute the 2SW class for a cut at \emph{each} value of $k_z$ where the central energy gap is open, plotted as a yellow dashed line in \cref{fig:C4model:hinge}(b).
We thus make the observation that the pair of TPs separates a region in the hinge BZ with nontrivial 2SW class and non-vanishing hinge charge $\frac{e}{4}$ from a region with trivial 2SW class and vanishing hinge charge.

Motivated by the fractional jump of the hinge charge observed for the model in \cref{eq:C4model:def} and by the jump's correlation with the 2SW class, we organize the next sections as follows.
First, in \cref{Sec:Monopole} we present a geometric discussion that explains how type-$\mathsf{A}$ TPs give rise to nontrivial 2SW (\cref{Sec:Monopole:SW}) and Euler (\cref{Sec:Monopole:Euler}) monopole charges by appealing to the relation between monopole charges and multi-band nodal links.
These insights will readily explain the 2SW class observed in this section's model. 
In \cref{Sec:HOBBC}, we then adopt the method of symmetry indicators and generalize the observations discussed above to establish a comprehensive bulk-hinge correspondence for TPPs.
Finally, in \cref{Sec:MinimalModels} we first revisit this section's model in the light of those general results, before discussing two $C_6$-symmetric examples: one with type-$\mathsf{A}$ TPs and one with type-$\mathsf{B}$ TPs.

\section{Monopole charges induced by triple points}\label{Sec:Monopole}

In this section we argue that pairs of type-$\mathsf{A}$ TPs in spinless systems can be imbued with monopole charges.
This conclusion is achieved by adopting the geometric interpretation of monopole charges as linking numbers~\cite{Ahn:2018,Tiwari:2020}.
After presenting the general argument, we apply it to two specific configurations of type-$\mathsf{A}$ TPs:
a four-band configuration resulting in a nontrivial second Stiefel-Whitney (2SW) monopole charge in \cref{Sec:Monopole:SW}, and a three-band configuration resulting in a nontrivial Euler (but trivial 2SW) monopole charge in \cref{Sec:Monopole:Euler}. 
In both cases we elaborate on the implications of the monopole charge on the stability of the NL segments and on the bulk topology.
(NL segments demarcated by type-$\mathsf{B}$ TPs are not easily analyzed through the geometric method considered in the present section. We address them using the more general framework of symmetry indicators in \cref{Sec:HOBBC}.)

The 2SW class $\SW{2}$ is a stable $\mathbb{Z}_2$-valued invariant defined on a closed 2D manifold with a spectral gap whenever there is an anti-unitary momentum-preserving symmetry squaring to $+\id$, such as $\mcP\mcT$ or $C_2\mcT$~\cite{Ahn:2019}.
Note that we require only \emph{one} gap, here the \emph{principal} gap, to be open on the 3D manifold.
The 2SW class remains well-defined in the presence of nodes in other band gaps.
On a spherical surface enclosing nodal-ring degeneracies in 3D, the 2SW class defines a monopole charge that enhances the stability of the node~\cite{Fang:2015,Bzdusek:2017,Ahn:2018}.
For 2D insulating systems, on the other hand, the 2SW class on the BZ torus defines a bulk topological invariant of the full system~\cite{Ahn:2018,Wang:2019,Ahn:2019,Lee:2020} similar to the Chern number. 

The $\mathbb{Z}$-valued fragile extension of the 2SW class, which can be defined for a two-band subspace separated from the rest of the spectrum by energy gaps, is the Euler class $\Eu$~\cite{Bzdusek:2017,Bouhon:2020b,Unal:2020}.
Note that only its absolute value is gauge-invariant~\cite{Bouhon:2020} (see also \cref{App:Euler-monopole:charge}); we will therefore restrict $\Eu$ to non-negative values.
Under the addition of trivial bands to the two-band subspace, only the parity of the Euler class is stable and reduces to the 2SW monopole charge discussed above, $\SW{2}=\Eu\mod 2$.

\begin{figure}
    \centering
    \includegraphics{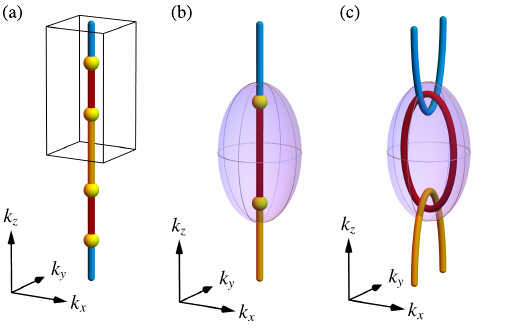}
    \caption{
    (a) Two pairs of type-$\mathsf{A}$ triple points (TPs, yellow dots) formed by consecutive triplets of bands. All TPs are protected by $C_4$ or $C_6$ symmetry with respect to the $z$-axis. In $k_z$-direction the full extent of the Brillouin zone is shown. Nodal lines in the first, second and third gap of a four-band model are shown in orange, red and blue, respectively.
    (b--c) Close up of the boxed region in (a) with the ellipsoid, on which the second Stiefel-Whitney (2SW) class is computed with respect to the principal (red) band gap, shown in purple.
    (c) Multiband nodal links that form after breaking the rotational symmetry.
    The red nodal ring carries a nontrivial value of the 2SW class computed on the purple ellipsoid. 
    By continuity, the red nodal segment in (b) also carries a nontrivial 2SW class.
    }
    \label{fig:Monopole:SW}
\end{figure}

\begin{figure}[t]
    \centering
    \includegraphics{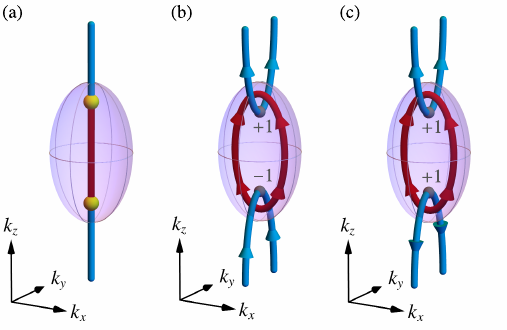}
    \caption{
    (a) Type-$\mathsf{A}$ triple points (yellow dots). The red (blue) line indicates nodal lines formed by the lower (upper) two bands of a three-band model.
    In $k_z$-direction the full extent of the Brillouin zone is shown.
    (b--c) Multiband nodal link formed from (a) by breaking the rotational symmetry protecting the triple points. The flux (gray $\pm1$'s) depends on the orientation of the two blue nodal lines at the intersection points (gray) with the disk bounded by the red nodal ring~\cite{Tiwari:2020}.
    The total flux vanishes in (b) giving rise to $\Eu=0$, but is finite in (c) such that $\abs{\Eu}=2$.
    In the presence of mirror $m_z: k_z\mapsto -k_z$ symmetry, the difference between the situations (b) and (c) is reflected in the mirror eigenvalues of the three bands at $k_z = 0$ vs.~$k_z=\pi$, cf.~the discussion in \cref{Sec:Materials:Euler}.
    }
    \label{fig:Monopole:Euler}
\end{figure}

We now investigate the monopole charges associated with configurations of TPs.
For concreteness, in this section we focus on NL segments that are demarcated by a pair of type-$\mathsf{A}$ TPs (this, in particular, implies the presence of $C_n$ symmetry with $n=4$ or $6$~\cite{Lenggenhager:2021:MBNLs}). 
We will call the energy gap in which that NL segment is formed the \emph{principal gap}~\cite{Bouhon:2020}.
To understand why such NL segments could carry a nontrivial monopole charge, consider one such NL segment and require that there are no additional band nodes in the principal gap for the considered range of $k_z$.
Two examples of such configurations are shown in Figs.~\ref{fig:Monopole:SW}(b) and \ref{fig:Monopole:Euler}(a), where nodes in the principal gap are shown in red.

As has been argued in Ref.~\onlinecite{Lenggenhager:2021:MBNLs}, adding a small $C_n$-breaking but $\mcP\mcT$-preserving perturbation transforms TPs into multiband nodal links as illustrated in Figs.~\ref{fig:Monopole:SW}(c) and \ref{fig:Monopole:Euler}(b,c).
Crucially, the linked nodal rings are in band gaps adjacent to each other, and such linking was shown to be in a one-to-one correspondence~\cite{Ahn:2018,Tiwari:2020} with monopole charges.
Thus, there potentially is a nontrivial monopole charge on a surface enclosing the nodal ring of interest [purple ellipsoid in Figs.~\ref{fig:Monopole:SW}(c) and \ref{fig:Monopole:Euler}(b,c)].
By continuity we can switch this perturbation off without closing the principal energy gap on the enclosing surface, which allows us to assign the same monopole charge also to the original NL segment, i.e., on the ellipsoid in Figs.~\ref{fig:Monopole:SW}(b) and \ref{fig:Monopole:Euler}(a).
In the following two subsections we investigate in detail the implications of the linking for the bulk topology of the two specified type-$\mathsf{A}$ TP configurations.

\subsection{Second Stiefel-Whitney monopole charge from triple points}\label{Sec:Monopole:SW}

We first analyze the four-band configuration with two TPs, in which case the 2D ICR is transferred from occupied to the unoccupied bands along a rotation axis (without loss of generality set to $k_z$) as shown in \cref{fig:Monopole:SW}(b), similar to the case of the model discussed in \cref{Sec:C4Model}.
We call such a configuration of TPs a \emph{triple-point pair} (TPP).
It involves three species of NLs, displayed in the figures in orange, red and blue according to increasing band index.
The individual TPs correspond to locations where the 2D ICR transfers from one energy gap to another.
Due to the periodicity in $k_z$, the minimal model of such a configuration involves \emph{two} such TPPs, cf.~\cref{fig:Monopole:SW}(a).

\begin{figure}
    \centering
    \includegraphics{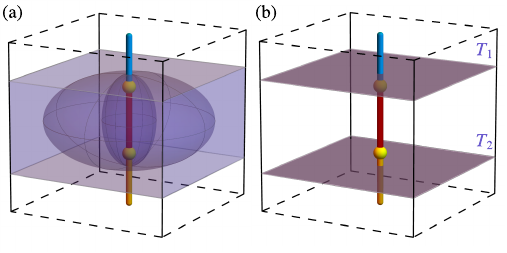}
    \caption{
        Relation between a nodal-line segment carrying a nontrivial second Stiefel-Whitney monopole charge, and a pair of two-dimensional insulators characterized by the $\ztwo$-valued 2SW class.
        The black frame represents the complete momentum-space extent of the Brillouin zone in the two horizontal directions (solid black lines), but not in the vertical direction (dashed black lines).
        (a) The monopole charge of the red nodal-line segment is calculated on a surface enclosing it (innermost purple ellipsoid), which can be continuously inflated (second purple ellipsoid) until it fills the whole BZ in $k_x$- and $k_y$-directions (purple cuboid).
        (b) The purple cuboid in (a) is equivalent to two horizontal planes (actually tori) $T_1$ and $T_2$ at the appropriate $k_z$-values.
    }
    \label{fig:Monopole:SWI}
\end{figure}

Here it is sufficient to consider the upper half of the BZ, \cref{fig:Monopole:SW}(b), with one (red) NL segment in the principal gap.
Applying the argument outlined above for the purple ellipsoid shown in the figure, we consider a small perturbation leading to the multiband nodal link~\cite{Lenggenhager:2021:MBNLs}, cf.~\cref{fig:Monopole:SW}(c).
Since one NL in each of the two adjacent band gaps (first and third) is linked with the red nodal ring, the purple ellipsoid carries nontrivial 2SW monopole charge $\SW{2}=1$~\cite{Ahn:2018}.
This has immediate consequences for the stability of the red NL segment and thus for the two TPs in \cref{fig:Monopole:SW}(b).
In particular, the monopole charge guarantees the persistence of the red NL even when the rotational symmetry is broken (and the 2D ICR split) as long as the $\mathcal{PT}$ symmetry is preserved.

We next discuss implications of the monopole charge for the bulk topology.
Assuming there are no additional NLs in the principal gap in the relevant $k_z$-range, the ellipsoid enclosing the red NL segment can be continuously deformed as illustrated in \cref{fig:Monopole:SWI}(a) until it spans the whole BZ in $k_x$- and $k_y$-directions for a finite range of $k_z$.
Due to the periodicity in the BZ, the contributions from opposite vertical faces of that surface cancel and we are left with the two horizontal planes $T_1$ and $T_2$ shown in \cref{fig:Monopole:SWI}(b).
Note that these are actually 2D tori which can be interpreted as the BZs of certain 2D systems, namely ones with the two-dimensional Hamiltonians $\mathcal{H}_{k_z}(k_x,k_y) := \mathcal{H}(k_x,k_y,k_z)$.

Owing to the continuous deformation of the surface, the 2SW class does not change, such that $\SW{2}(T_1)+\SW{2}(T_2) = 1\mod 2$.
Consequently, one of the two planes will be trivial, $\SW{2}=0$, and the other nontrivial, $\SW{2}=1$.
Viewing them as 2D systems as described above, the latter is a 2D insulator with nontrivial 2SW class, called \emph{Stiefel-Whitney insulator} (SWI)~\cite{Ahn:2018}.
In the simultaneous presence of $C_2$ rotational and of chiral symmetry, SWIs were shown~\cite{Ahn:2019,Wang:2019} to have robust zero-energy corner states accompanied by half-integer corner charges.
Our situation is different: we do not assume chiral symmetry, and the order of rotational symmetry necessary to protect \mbox{type-$\mathsf{A}$} TPs is $n\in\{4,6\}$.
Until now, the (higher-order) bulk-boundary correspondence for (non-chiral) SWIs with rotational symmetry has not been clarified. We derive this piece of information in \cref{Sec:HOBBC:2D}; in particular, we show that the corner charge in the presence of trivial (nontrivial) 2SW class is quantized to even (odd) multiples of $\tfrac{e}{n}$.
We subsequently apply the result to study the bulk-hinge correspondence for TPPs of both type~$\mathsf{A}$ and type~$\mathsf{B}$ in \cref{Sec:HOBBC:3D}. 

\subsection{Euler monopole charge from triple points}\label{Sec:Monopole:Euler}

An even simpler configuration is obtained in a three-band model with two type-$\mathsf{A}$ TPs formed by the \emph{same} triplet of energy bands, as illustrated in \cref{fig:Monopole:Euler}(a).
Since the ordering of the ICRs along the rotation ($k_z$-) axis has to be the same at both $k_z=-\pi$ and $k_z = \pi$ due to the periodicity of the momentum space, it again follows that the number of TPs formed by the three bands is even.
In the minimal model, the NL changes the band gap twice, once from the second (blue) to the first (red) and once back to the second gap, at two TPs (yellow dots).
Here we assume a single occupied and two unoccupied bands, such that the two-band subspace allows us to define the Euler monopole charge $\Eu\in\mathbb{Z}$ on the purple ellipsoid in \cref{fig:Monopole:Euler}(a)~\cite{Bzdusek:2017}.

\begin{figure}
    \centering
    \includegraphics{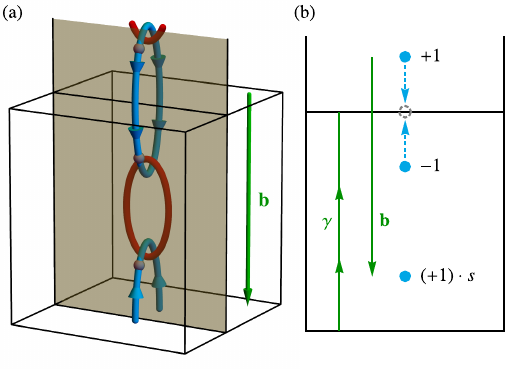}
    \caption{
        (a) Nodal line (NL) configuration (here colored red vs.~blue for NLs formed by the lower vs.~upper two bands of a three-band model) in the first Brillouin zone (BZ, black frame) and part of the second BZ. The two blue displayed NLs are displaced by a reciprocal lattice vector $\vec{b}$ (green arrow). The blue NL is intersected by a 2D plane (pale brown) at two inequivalent points (gray dots). The shown orientation of the blue NLs is compatible with $\phi_2+\phi_3 = \pi$, cf.~\cref{Sec:Monopole:Euler}. 
        (b) Band nodes (blue points) on the 2D plane in panel (a), with winding numbers $\pm 1$ inherited from the orientation of the NLs in 3D. The two upper point nodes have opposite orientation, because they pairwise annihilate (blue arrows pointing to the gray disk) when the vertical plane is shifted away from the NL composition. The point nodes displaced by $\vec{b}$ have winding numbers differing by a factor $s=\e^{\i(\phi_2+\phi_3)}=\pm 1$~\cite{Ahn:2019}.
    }
    \label{fig:Monopole:NLOrientation}
\end{figure}

The Euler monopole charge of a nodal ring is determined by the linking with NLs in the adjacent band gap, similar to the case of the 2SW class, such that we can again apply the argument with the breaking of the rotational symmetry.
However, here the \emph{orientation} of the linked NLs becomes important~\cite{Tiwari:2020}.
Recall that in a generic multi-band setting the orientation of a NL can be formally defined via the non-Abelian generalized quaternion invariant~\cite{Wu:2019}.
In particular, this orientation obeys the same noncommutative rules as the winding number of point nodes in 2D as reported by Ref.~\onlinecite{Ahn:2019}, allowing us to interpret the NLs as braid trajectories.
The orientation [indicated by arrows in \cref{fig:Monopole:Euler}(b,c)] of the adjacent (blue) NL at the point where it crosses the disk bounded by the principal (red) nodal ring determines the flux $\pm 1$, indicated in gray.
We observe that two distinct scenarios can arise after breaking the rotational symmetry: \cref{fig:Monopole:Euler}(b) where the total flux vanishes, and \cref{fig:Monopole:Euler}(c) where it is non-vanishing.
According to Ref.~\onlinecite{Tiwari:2020} the former implies $\Eu=0$ and the latter $\abs{\Eu}=2$.

While we have so far only established the \emph{possibility} of the red NL segment carrying a nontrivial Euler monopole charge, the two cases $\Eu=0$ and $\abs{\Eu}=2$ can be distinguished based on the Zak-Berry phases $\phi_{2,3}\in\{0,\pi\}$ along the $k_z$-axis of the two bands involved in the formation of the blue NL in the adjacent energy gap, i.e., the NL segment on the opposite side of the triple point than the segment under consideration.
Assuming the minimal model with only a single pair of TPs along $k_z$, as shown in \cref{fig:Monopole:Euler}, there are in fact only two nodal rings (the red one in the center, and the blue one on the boundary), since the two blue NL segments belong to the \emph{same} nodal ring, only shifted by a reciprocal lattice vector $\vec{b}$.
Notably, the two \emph{copies} of the blue NL ring do not necessarily exhibit consistent orientations: the orientation of a NL between bands $2$ and $3$ at $\vec{k}+\vec{b}$ is \emph{reversed} compared to the one at $\vec{k}$ if and only if $\phi_2+\phi_3=\pi\mod 2\pi$.
In \cref{App:NLBZ} (with additional information detailed in the Supplemental Material~\cite{SM}), we formalize this statement in the framework of the non-Abelian generalized quaternion invariant~\cite{Wu:2019,Tiwari:2020}.
The result applies to an arbitrary number of bands and provides a general transformation rule for the NL orientation between neighboring BZs depending on the Berry phases.

The orientation reversal can be related to the properties of Dirac points in 2D derived in Ref.~\onlinecite{Ahn:2019} if we consider a 2D cut [brown plane in \cref{fig:Monopole:NLOrientation}(a)] through the BZ that intersects the blue NLs.
Note that continuity implies that the two upper point nodes [i.e. upper two blue dots in \cref{fig:Monopole:NLOrientation}(b)] have opposite winding number because they correspond to cuts through the \emph{same} nodal ring, i.e., they manifestly annihilate when sliding the plane away from the nodal-link composition. 
According to Ref.~\onlinecite{Ahn:2019} the copy of the uppermost node with winding number $+1$ in an adjacent BZ (bottom-most node), i.e., shifted by the reciprocal lattice vector $\vec{b}$, has winding number $(+1)\cdot s$, where $s=\e^{\i(\phi_2+\phi_3)}$ and $\phi_j$ is the Zak-Berry phase of band $j$ on the contour $\gamma$ winding around the 2D BZ in the direction of $\vec{b}$.

Although the even-valued Euler class is a \emph{fragile} topological invariant, it may still imply unusual stability of the corresponding NL under perturbations, which we briefly investigate in \cref{Sec:Materials:Euler}.
The following \cref{Sec:HOBBC,Sec:MinimalModels} investigate in more detail the case of TPs in the four-band configuration, which may exhibit the stable 2SW class, and we derive their associated signatures in the fractional hinge charges.

\section{Higher-order bulk-boundary correspondence}\label{Sec:HOBBC}

In this section we mathematically establish the higher-order bulk-boundary correspondence principle foreshadowed in \cref{Sec:C4Model}: a triple-point pair (TPP) in a SG with (antiunitary) rotational symmetry of order $n\in\{2,3,4,6\}$ is generally associated with a fractional jump of the hinge charge.
While the previous \cref{Sec:C4Model,Sec:Monopole} explicitly considered only type-$\mathsf{A}$ TPs in systems with $\mathcal{PT}$ symmetry, the discussion in the present section is more general and encompasses \emph{all} the TPs reported by the classification in Refs.~\onlinecite{Lenggenhager:2021:MBNLs,Lenggenhager:2022:TPClassif} and reproduced in \cref{tab:TP-MPGs} in \emph{symmorphic SGs}.
(We discuss the generalization to non-symmorphic SGs in \cref{Sec:non-symmorphic}.)
Furthermore, we report a correspondence between the 2SW monopole charge (if it is defined) of the NL segment connecting the two TPs and the value of the hinge-charge jump: a trivial 2SW monopole charge implies a $\pm \tfrac{e}{n}$ jump, while a nontrivial value results in a jump by $\pm \tfrac{2e}{n}$. 
We remark that the hinge-charge jump associated with a TPP always occurs without a change in the bulk polarization.
This implies that if there is no fractional surface charge appearing on one side of the TPP, then there is also no fractional surface charge on the other side of the TPP, thus guaranteeing that the fractional jump in the hinge charge is observable.

We begin in \cref{Sec:HOBBC:2D} by discussing the 2D SWI with $C_n$-rotational symmetry.
Here, we adapt the $C_2$-symmetry-indicator formula~\cite{Ahn:2019} for the 2SW class to $C_n$-rotational symmetry, where $n\in\{4,6\}$, and show that the value of the 2SW class constrains the possible fractional corner charges of the $C_n$-symmetric 2D system with vanishing bulk polarization.
Subsequently, by temporarily adopting a stronger assumption on energy gaps in the band structure, we derive in \cref{Sec:HOBBC:Euler} a symmetry-indicator formula for the Euler monopole charge of TPPs in the presence of $C_n$ rotational symmetry.
In \cref{Sec:HOBBC:3D}, we combine the results obtained in the previous two subsections with the classification of TPPs to derive the general bulk-hinge correspondence principle for all possible TPP configurations in $\mcP\mcT$-symmetric systems.
Finally, in \cref{Sec:HOBBC:SymBreak}, we discuss the effect of breaking various symmetries; in particular, we extend the bulk-hinge correspondence principle to TPPs protected by \emph{all} MPGs listed in \cref{tab:TP-MPGs}.
Our results for the bulk-hinge correspondence are compactly summarized in \cref{tab:HOBBC:Qjump}, while \cref{tab:HOBBC:symbreak} compiles the various topological phase transitions induced by symmetry breaking.

\subsection{Stiefel-Whitney insulator with rotational symmetry}\label{Sec:HOBBC:2D}

We consider an insulating 2D system with $C_2\mcT$ and $C_n$ symmetry, where $C_2$ and $C_n$ are with respect to the same axis perpendicular to the system.
Due to the $C_2\mcT$ symmetry (where $C_2$ acts like inversion in 2D) the insulator is characterized by the 2SW class, while the $C_n$ symmetry implies fractional corner charges~\cite{Benalcazar:2019,Watanabe:2020,Takahashi:2021} if the edge charge vanishes.
For simplicity we assume that all the positive ionic charge, which compensates for the negative charge of the filled electron bands, is located at the maximal Wyckoff position $1a$, i.e., the center of the square unit cell.
In \cref{Sec:HOBBC:3D} we will see that for $n=3$, the 2SW class is not symmetry indicated.
Thus, we restrict the present discussion to $n=4,6$, which implies that $C_2$ and $\mcT$ symmetry are also symmetries of the system.

\begin{figure}[t]
    \centering
    \includegraphics{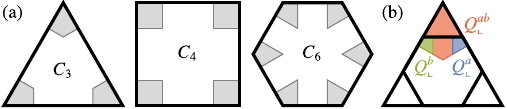}
    \caption{
        (a) The $C_n$-symmetric cross-section geometries for which the hinge charges are defined, and for which the values listed in \cref{tab:HOBBC:Qjump} apply. The corners (gray regions) in the cross sections correspond to the hinges in the 3D system.
        (b) Deformation of the $C_6$-symmetric sample (inner hexagon) to a $C_3$-symmetric sample (outer triangle). Two corners (blue and green) of the hexagon with corner charges $\Qc^a$ and $\Qc^b$ merge into a single corner (red) of the triangle with corner charge $\Qc^{ab}=\Qc^a+\Qc^b$.
    }
    \label{fig:HOBBC:geometry}
\end{figure}

In the following we use symmetry indicators to derive constraints on the corner charges due to a nontrivial 2SW class.
We adopt the notation of Ref.~\onlinecite{Benalcazar:2019}, where the eigenvalues of the $C_n$-rotation operator at the $C_n$-symmetric HSP $\Pi$ for spinless particles are denoted by
\begin{equation}
    \Pi_p^{(n)} = \e^{2\pi\i(p-1)/n},\quad p=1,2,\dotsc,n
    \label{eq:def_eigenvalues}
\end{equation}
and define the quantities
\begin{equation}
    \left[\Pi_p^{(n)}\right] = \#\Pi_p^{(n)} - \# \Gamma_p^{(n)},
    \label{eq:def_SIs}
\end{equation}
where $\#\Pi_p^{(n)}$ and $\#\Gamma_p^{(n)}$ are the number of occupied energy bands with eigenvalue $\Pi_p^{(n)}$ at HSPs $\Pi$ and $\Gamma$, respectively.
For $C_4$ we have HSPs $\Gamma$, $X$ and $M$, while for $C_6$ they are $\Gamma$, $M$ and $K$. 
Then, assuming the specific geometries of the 2D crystals shown in \cref{fig:HOBBC:geometry}(a), the corner charges are given by~\cite{Takahashi:2021}
\begin{subequations}
    \begin{align}
        \Qc^{(4)} &= \frac{e}{4}\left(\mp\left[X_1^{(2)}\right] + 2\left[M_1^{(4)}\right] + 3\left[M_2^{(4)}\right]\right)\mod e,\label{eqn:Q4-sym-ind}\\
        \Qc^{(6)} &= \frac{e}{4}\left[M_1^{(2)}\right] + \frac{e}{6}\left[K_1^{(3)}\right]\mod e,\label{eqn:Q6-sym-ind}
    \end{align}
    \label{eq:HOBBC:Q}%
\end{subequations}
where $e<0$ is the electron charge and the sign in front of $[X_1^{(2)}]$ depends on whether the center of the $C_4$-symmetric crystal is located at Wyckoff position $1a$ (upper sign) or $1b$ (lower sign).
Note that, consistent with the assumptions in the present discussion, \cref{eq:HOBBC:Q} assumes the presence of time-reversal symmetry $\mcT$ with $\mcT^2=+\id$; if $\mcT$ is not a symmetry, \cref{App:eq:Qcorner4,App:eq:Qcorner6} should be used instead.

Since both $C_4$ and $C_6$ symmetry imply the presence of $C_2$ symmetry, we can also consider the symmetry-indicator formula for the 2SW class~\cite{Ahn:2018}, namely
\begin{equation}
    \SW{2} = \sum_{\Pi\in\mathrm{TRIM}}\floor{\frac{1}{2}\#\Pi_2^{(2)}}\mod 2,
    \label{eq:HOBBC:SWc}
\end{equation}
where the sum is over the four time-reversal invariant momenta (TRIM) in the BZ (these are exactly the $C_2$-invariant momenta) and $\floor{\cdot}$ is the floor-function.
In \cref{App:SWICn} we show that in the presence of an enlarged rotational symmetry, the 2SW class can be equivalently written as
\begin{subequations}
    \begin{align}
        C_4: && \SW{2} &= \left[M_2^{(4)}\right] \mod 2\\
        C_6: && \SW{2} &= \frac{1}{2}\left[M_1^{(2)}\right] \mod 2,
    \end{align}
    \label{eq:HOBBC:SWc46}%
\end{subequations}
where on the right-hand side we used the same $C_n$ symmetry indicators that also enter \cref{eq:HOBBC:Q}.

Finally, we observe that for vanishing polarization, the 2SW class constrains the corner charge.
For $n=4$ the polarization vanishes if and only if $[X_1^{(2)}]=0\mod 2$, while for $n=6$ it always vanishes~\cite{Benalcazar:2019}.
Therefore, vanishing polarization also implies that the center-dependent sign ambiguity in \cref{eqn:Q4-sym-ind} drops from our analysis.
Combining the assumption of vanishing polarization with \cref{eq:HOBBC:Q,eq:HOBBC:SWc46}, we show in \cref{App:SWICn} that the nontrivial value $\SW{2}=1$ constrains the corner charge for $C_4$-symmetric SWIs to
\begin{subequations}\label{subeqn:2SW-Q-correspondence}
\begin{equation}
    \Qc^{(4)} \in\left\{\pm\frac{e}{4}\right\}\mod e \label{eqn:c4-2SW-Qcorner}
\end{equation}
and for $C_6$-symmetric SWIs to
\begin{equation}
    \Qc^{(6)} \in\left\{\pm\frac{e}{6}, \frac{e}{2}\right\}\mod e,
\end{equation}
\end{subequations}
i.e., to \emph{odd} multiples of $\frac{e}{n}$, whereas the complementary fractional values correspond to insulators with $\SW{2}=0$.

Before applying the presented formulas to study TPPs in 3D, we briefly comment on some relations of the result in \cref{subeqn:2SW-Q-correspondence} to previous works.
First, note that in the presence of chiral symmetry, the corner charge can only attain value $0$ or $\tfrac{e}{2}\mod e$.
This can be easily understood as follows.
If the occupied states carry localized corner charge $Q \mod e$, then (by completeness of the Hilbert space) the unoccupied states carry localized corner charge $-Q \mod e$.
Chiral symmetry is local in real space and maps occupied onto unoccupied states, therefore guaranteeing that $Q = -Q \mod e$.
That equation has two solutions: $Q = 0,\tfrac{e}{2}\mod e$.
On the other hand, our derivation in \cref{App:SWICn} reveals that the value $\SW{2}=1$ can result in corner charge $\Qc^{(6)}=\frac{e}{2}\mod e$ if and only if $[K_1^{(3)}]=0\mod 6$.
This is compatible with the finding of Ref.~\onlinecite{Noh:2018} which showed that chiral symmetry that commutes with $C_3$-rotation implies $[K_1^{(3)}]=0$.
The value $\tfrac{e}{2}\mod e$ of the corner charge was also reported in the study of a SWI model with chiral and $C_6$ symmetry by Ref.~\onlinecite{Ahn:2019}. 
Second, the result in \cref{eqn:c4-2SW-Qcorner} implies that the three conditions (1) $C_4$ symmetry, (2) $w_2 = 1$, and (3) vanishing polarization are incompatible with (4) the presence of chiral symmetry; equivalently, SWI with chiral and $C_4$ symmetry must necessarily have gapless edges.
This finding is compatible with the observations of Ref.~\onlinecite{Wang:2020} made in the context of second-order NL semimetals.

\subsection{Euler monopole charge in presence of rotational symmetry}\label{Sec:HOBBC:Euler}

If the two occupied bands involved in the TPP formation are separated from lower lying occupied bands by an energy gap, the 2SW can be refined to the Euler monopole charge.
Note that this is a different situation from the one discussed in \cref{Sec:Monopole:Euler}: here we consider a surface enclosing a TPP, i.e., two TPs in adjacent band gaps, as shown in \cref{fig:Monopole:SW}(b).
Monopole charges such as the 2SW and the Euler monopole charge can be inferred from the winding of the Wilson loop spectrum.
Symmetries constrain the latter~\cite{Fang:2012a,Bouhon:2019,Alexandradinata:2020} and therefore lead to symmetry-indicator formulas for the monopole charges.
To our knowledge, no such symmetry-indicator formulas for the Euler monopole charge in the presence of rotational symmetry have been previously derived, therefore we do that here.
Along the way, we identify that TPPs formed by certain combinations of ICRs are necessarily associated with extended NLs in the principal gap, which prevents the system from exhibiting higher-order topological signatures and the monopole charge from being defined.
The details of these derivations are relegated to \cref{App:Euler-monopole}, while here we only outline the main steps and results.

\begin{figure}[t]
    \centering
    \includegraphics{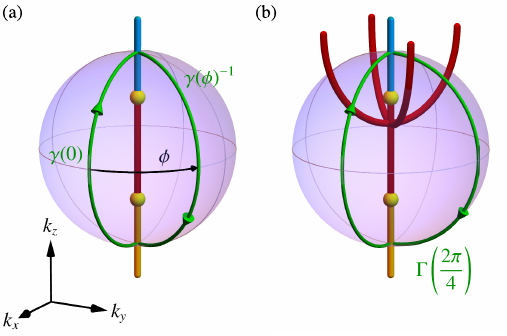}
    \caption{
        (a) Definition of the closed contours $\Gamma(\phi)=\gamma(\phi)^{-1}\circ\gamma(0)$ (green) on which the Wilson loop operator $\mathcal{W}(\phi)$ is computed. 
        The spherical surface (purple) is covered by these contours as the argument is increased in the range $\phi\in[0,2\pi)$.
        If the surface is not crossed by any other nodal lines (NLs) than the orange and blue ones, the Euler monopole charge $\Eu$ can be defined and is determined by the winding of the Wilson loop spectrum. 
        This is the case, for example, if there are no red nexus points enclosed in the surface.
        (b) In other cases, e.g., if there is exactly one red nexus point enclosed, red NLs crossing the surface are \emph{always} present. This is indicated by a $\pi$ Berry phase on the path $\Gamma(2\pi/n)$, where $n$ is the order of the rotational symmetry.
        The non-trivial Berry phase implies that $\Gamma(2\pi/n)$ encircles an odd number of NLs in the principal gap (red NLs), as illustrated.
    }
    \label{fig:Euler_charge}
\end{figure}

To derive the symmetry-indicator formulas, we consider a spherical surface enclosing part of a HSL with $C_n$ rotational symmetry in its little co-group, e.g., containing a TPP, as illustrated in \cref{fig:Euler_charge}(a).
We refer to the two points where the sphere is intersected by the rotation axis as the south and north pole.
Let $N$ and $N_\mathrm{occ}$ be the total number of bands and the number of occupied bands, respectively.
We assume the system to have space-time inversion symmetry $\mcP\mcT$ satisfying $(\mcP\mcT)^2=+\id$, such that there is a basis in which the corepresentation of $\mcP\mcT$ is the identity matrix and the Bloch Hamiltonian a real symmetric matrix.
Let further $D_0$ and $D_1$ label the symmetry representations of the occupied bands in that basis at the south and north pole of the spherical surface, respectively.
In \cref{App:Euler-monopole:Wilson-op} we show that the Wilson loop operator $\mathcal{W}(\phi)$ computed on the path
\begin{equation}
    \Gamma(\phi) = \gamma(\phi)^{-1}\circ\gamma(0)
    \label{eqn:path-composite}
\end{equation}
illustrated in green in \cref{fig:Euler_charge}(a) is constrained by $C_n$-symmetry:
\begin{equation}
    \mathcal{W}\left(\phi+\frac{2\pi}{n}\right) = D_0(C_n)\adjo{P(\phi)}\adjo{D_1(C_n)}P(\phi)\mathcal{W}(\phi),
    \label{eq:Euler_charge:Wilson_loop_constraint}
\end{equation}
where $P(\phi)\in\SO(N_\mathrm{occ})$ is defined by parallel transport [see \cref{App:eq:parallel-transport-P_def} in \cref{App:Euler-monopole:Wilson-op}]. 
Note that in \cref{eqn:path-composite}, one first traverses the path appearing to the right of the composition symbol `$\circ$'.

For arbitrary $N_\mathrm{occ}$, we consider the implication of \cref{eq:Euler_charge:Wilson_loop_constraint} for the Berry phase of the occupied bands on the contours $\Gamma(\phi)$.
The Berry phase is given by $\varphi=\arg\det\mathcal{W}$~\cite{Vanderbilt:2018}, such that
\begin{equation}
    \hspace{-2mm}\varphi\left(\phi+\frac{2\pi}{n}\right) = \varphi(\phi) + \arg\det\left[D_0(C_n)\adjo{D_1(C_n)}\right]\!\!\!\mod 2\pi.\!
    \label{eq:HOBBC:Berry_phase_constraint}
\end{equation}
Due to $\mcP\mcT$ symmetry, $\Delta\varphi = \arg\det\left[D_0(C_n)\adjo{D_1(C_n)}\right]$ is quantized to $0$ vs.\ $\pi$.
If $\Delta\varphi=\pi$, there is an odd number of NLs in the principal gap crossing the surface in each sector $[\phi,\phi+2\pi/n]$, cf.~\cref{fig:Euler_charge}(b). 
This implies that the principal gap is necessarily closed somewhere on the enclosing surface, thus preventing one from assigning a monopole charge to the enclosed TPP.
We will see in Sec.~\ref{Sec:HOBBC:3D} that this conditions also prevents the corresponding TPPs from exhibiting the higher-order signature at the hinges.

In the remainder of this subsection, we call a TPP \emph{admissible} when the principal gap on the enclosing sphere can be open.
This corresponds to the case with $\Delta\varphi=0$, which is equivalent to requiring 
\begin{equation}
    \det\left[D_0(C_n)\adjo{D_1(C_n)}\right] = 1.
    \label{eqn:det-product-unity}
\end{equation}
By our initial assumption, the two occupied bands involved in the TPP are separated from lower lying occupied bands by an energy gap on the whole surface, such that we can ignore the lower lying bands and set $N_\mathrm{occ}=2$.
Then the absence of nodes in the principal gap implies that the Euler monopole charge on the sphere is well-defined and given by the winding number of the Pfaffian of the logarithm of $\mathcal{W}(\phi)$~\cite{Bouhon:2020}.
In \cref{App:Euler-monopole:charge}, we show that if $D_0(C_n), D_1(C_n)\in\SO(2)$, then \cref{eq:Euler_charge:Wilson_loop_constraint} simplifies to
\begin{equation}
    \mathcal{W}\left(\phi+\frac{2\pi}{n}\right) = \padjo{D_0(C_n)}\adjo{D_1(C_n)}\mathcal{W}(\phi).
\end{equation}
Since the $n^\textrm{th}$ power of $D_{0,1}(C_n)$ gives the identity, it follows that for $i\in\{0,1\}$: $D_i(C_n)=\e^{-\frac{2\pi\i}{n}r_is_y}$ with $r_i\in\mathbb{Z}$ and the Pauli matrix $s_y$ acting on the space of the two valence bands.
Then, we extract the following symmetry-indicator formula for the Euler monopole charge $\Eu$
\begin{equation}
    \Eu = r_1-r_0\mod n.
    \label{eq:HOBBC:Euler}
\end{equation}
Note that $\Eu$ is only gauge-invariant up to sign and that the relevant topological invariant therefore is $\abs{\Eu}$.

\begin{figure}[t]
    \centering
    \includegraphics{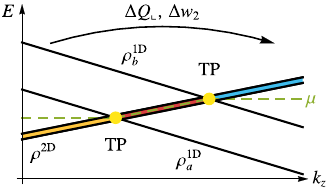}
    \caption{
        Convention for defining the jump of the fractional hinge charge ($\Delta \Qc$) and of the second Stiefel-Whitney monopole ($\Delta\SW{2}$)  associated with a pair of triple points (TP, yellow dots) in a four-band configuration. The black lines indicate the band structure along a high-symmetry line, with the 2D irreducible co-representation (ICR) $\rho^\textrm{2D}$ passing from the bottom (orange) to the central (red) to the upper (blue) energy gap with increasing momentum $k_z$ due to crossing two 1D ICRs $\rho^\textrm{1D}_a$ and $\rho^\textrm{1D}_b$. The green dashed line ($\mu$) indicates half-filling at various $k_z$. The quantities $\Delta \Qc$ and $\Delta\SW{2}$ associated with the TP pair are defined as the characteristics of the 2D cut in the region where $\rho^\textrm{2D}$ is unoccupied (i.e., above $\mu)$ from which we subtract the characteristics of the 2D cut in which $\rho^\textrm{2D}$ is filled (i.e., below $\mu$).
        In practice, the \emph{physical} chemical potential has a value independent of momentum, and will therefore locally deviate from the $\mu$ defined above. 
        Nevertheless, generically, the system will still be at half-filling for extended parts of the orange and blue regions which lie far enough from the TP pair. 
        The differences $\Delta\Qc$ and $\Delta\SW{2}$ are therefore still well-defined and unchanged compared to the situation with variable $\mu(k_z)$.
    }
    \label{fig:HOBBC:BS}
\end{figure}

\subsection{Triple point configurations with hinge charges}\label{Sec:HOBBC:3D}

In this section we apply the symmetry-indicator formulas from \cref{Sec:HOBBC:2D,Sec:HOBBC:Euler} to study four-band TPP configurations in $\mcP\mcT$-symmetric systems.
While in \cref{Sec:Monopole:SW} we have assumed the TPPs to be demarcated by type-$\mathsf{A}$ TPs, we here generalize to the case of both type-$\mathsf{A}$ and type-$\mathsf{B}$ TPs.
Our result, summarized by \cref{tab:HOBBC:Qjump}, provides the complete correspondence between TPPs, their 2SW and Euler monopole charge (if they are defined), and the higher-order signature in the fractional hinge charge.
For simplicity, we first focus on TPs in crystals with a symmorphic SG and generalize the results to non-symmorphic SGs in \cref{Sec:non-symmorphic}.

Our analysis is structured as follows.
First, based on the classification result of Refs.~\onlinecite{Lenggenhager:2021:MBNLs,Lenggenhager:2022:TPClassif}, we know that TPs can be stabilized in spinless systems on HSLs with trigonal, tetragonal or hexagonal symmetry (cf.~\cref{tab:TP-MPGs}).
Such HSLs arise in SGs with the corresponding symmetry.
For crystals with $C_n$ symmetry, where $n\in\{2,3,4,6\}$, some representative HSLs with the specified symmetry (namely ones that are realized in SGs with prismatic BZs), are listed at the top of \cref{tab:HOBBC:Qjump} (however, the derived bulk-hinge correspondence, encoded by $\Delta\Qc$, $\Eu$ and $w_2$, applies equally to \emph{all} HSLs with the prescribed symmetry).
Then, given the little co-group of any such HSL, we consider all possible combinations of ICRs leading to TPPs.
Recall that a TPP requires a 2D ICR ($\rho^\textrm{2D}$) to be crossed consecutively by two 1D ICRs ($\rho^\textrm{1D}_{a}$ and $\rho^\textrm{1D}_{b}$) (cf.~\cref{fig:HOBBC:BS}), and can thus be characterized by the triplet $(\rho^\textrm{2D};\rho^\textrm{1D}_{a},\rho^\textrm{1D}_{b})$ (here, the ordering of the 1D ICRs is unimportant). Finally, we apply the symmetry-indicator formulas from \cref{Sec:HOBBC:2D,Sec:HOBBC:Euler} to derive the higher-order bulk-boundary correspondence of TPPs, listed in the bottom of \cref{tab:HOBBC:Qjump}.

\begin{table*}[t]
    \begin{threeparttable}
    \caption{
        Bulk-hinge correspondence for triple-point pairs (TPPs) in spinless systems.
        Triple points (TPs) can arise only in crystals with rotational symmetry $C_n$ (or screw symmetry) of order $n\in\{2,3,4,6\}$; this fixes the nanowire geometry$^{\,\textrm{a}\,}$ in which we study the hinge charges. 
        In particular, the TPs occur along high-symmetry lines (HSLs) with little co-group (LCG) that has the same rotational symmetry $C_n$ supplemented with space-time inversion ($\mathcal{PT}$) symmetry \emph{or} with mirror ($m_v$) symmetry with respect to plane that contains the HSL \emph{or} with both $\mcP\mcT$ and $m_v$.
        Note that $C_n$ may be generated by an antiunitary rotational symmetry $C_{n/2}\mcP\mcT$ for even $n$, and that the antiunitary rotations $C_4\mcP\mcT$ and $C_6\mcP\mcT$ can stabilize TPs even without additional symmetries. 
        (For an easily navigable summary of all admissible symmetry combinations supporting spinless TPs, see \cref{tab:TP-MPGs}.)
        In the second row we indicate representative HSLs with the required rotational symmetry, which are realized in SGs with a prismatic Brillouin zones (however, the below-listed values of $\Delta\Qc$, $\Eu$, and $w_2$ apply equally to all HSLs with the specified symmetry).
        In the third row we indicate the generator of the LCG rotational symmetry, i.e., the maximal (unitary or antiunitary) rotational symmetry.
        The next row lists the possible pairs of TP types along the corresponding HSL which can exhibit a gapped spectrum on both sides of the NL segment (cf.~\cref{fig:HOBBC:TPconfigs}); for the classification of \emph{individual} TPs in spinless systems into type~$\mathsf{A}$ vs.~type~$\mathsf{B}$ see \cref{tab:TP-MPGs} or Ref.~\onlinecite{Lenggenhager:2022:TPClassif}. The row labelled by \enquote{ICRs} indicates the possible triplets of irreducible co-representations of the LCG which can form a four-band TPP (cf.~\cref{fig:HOBBC:BS}); \enquote{any} means that all combinations of a 2D ICR with two 1D ICRs give the same result.
        The notation for the ICRs follows Ref.~\onlinecite{Bradley:1972}, where we drop the subscripts if they do not affect the result. Note that for the 2D ICRs of $6/m'$ we define: $^2E_2\!\,^1E_2 \mapsto E_1$, and $^2E_1\!\,^1E_1 \mapsto E_2$.
        Finally, we find that \emph{each} TPP is characterized by a fractional hinge-charge jump $\Delta\Qc$.
        If $\mathcal{PT}$ symmetry is present, we also assign the TPPs the Euler $\abs{\Eu}$ and second Stiefel-Whitney $\SW{2}$ monopole charges.
        The hinge charges $\Qc$ are computed for the geometries depicted in \cref{fig:HOBBC:geometry}(a).
    }
    \begin{ruledtabular}
	\begin{tabular}{CCCCCCCC}
	\text{Rotational symmetry} & 
	\multicolumn{1}{C}{C_2{}^\textrm{a}} &
	\multicolumn{2}{C}{C_3} & 
	\multicolumn{1}{C}{C_4} & 
	\multicolumn{3}{C}{C_6} \\\hline\addlinespace
	\text{Example HSLs} & 
	\Gamma Z, MA & 
	\multicolumn{2}{C}{\Gamma A, KH, K'H'} & 
	\multicolumn{1}{C}{\Gamma Z, MA} &
	KH^{\,\textrm{b}} & 
	\multicolumn{2}{C}{\Gamma A} \\ \addlinespace
	\text{LCG rotation generator} & 
	C_{4}\mcP\mcT &
	C_{3} &
	C_{6}\mcP\mcT^{\,\textrm{a}} & 
	\multicolumn{1}{C}{C_{4}} & 
	C_{3} & 
	\multicolumn{2}{C}{C_{6} 
	}\\\addlinespace
	\text{TP types} & 
	\multicolumn{1}{C}{(\mathsf{A}, \mathsf{A})} &
	{(\mathsf{B}, \mathsf{B})} &
	{(\mathsf{A}, \mathsf{A})} &
	\multicolumn{1}{C}{(\mathsf{A}, \mathsf{A})} &
	\multicolumn{1}{C}{(\mathsf{B}, \mathsf{B})} &
	\multicolumn{1}{C}{(\mathsf{A}, \mathsf{A})} &
	\multicolumn{1}{C}{(\mathsf{B}, \mathsf{B})}
	\\ \addlinespace
	\text{ICRs} & 
	\text{any} &
	\multicolumn{2}{c}{any} &
	\makecell{(E;A,A)\\(E;B,B)} &
	\text{any} &
	\makecell{(E_1;A,A)\\(E_2;B,B)} &
	\makecell{(E_1;B,B)\\(E_2;A,A)} \\\hline
	\Delta\Qc\mod e & 
	\displaystyle\frac{e}{2} &
	\multicolumn{2}{C}{\displaystyle+\frac{e}{3}} &
	\rule{0em}{1.5em}\displaystyle+\frac{e}{4} &
	\displaystyle+\frac{e}{3} & 
	\displaystyle+\frac{e}{6} &
	\displaystyle-\frac{e}{3}\\[1em]
	\abs{\Eu\mod n}^{\,\,\textrm{d}} & 
    \td &
	1 & 
	\td & 
	1 & 
	2 & 
	1 & 
	2\\
	\SW{2} & 
    \td &
	0\text{ or }1^{\,\textrm{c}} & 
	\td & 
	1 & 
	0 & 
	1 & 
	0
    \end{tabular}
    \end{ruledtabular}
    \begin{tablenotes}[flushleft]
        \item[$\textrm{a}$] For $C_4\mcP\mcT$ (without additional $\mcP\mcT$), the largest unitary rotational symmetry is $C_2$. Then, we consider a square geometry, and the quantization of the hinge-charge jump to $\tfrac{e}{2}\!\!\mod e$ occurs only if summing over two neighboring hinges. For $C_6\mcP\mcT$ we consider the $C_3$ (triangular) geometry.
        \item[$\textrm{b}$] Due to the double occurrence of the $KH$ line in a $C_6$-symmetric system, the Brillouin zone exhibits two TPPs at the same value of $k_z$. The indicated values of $\Delta\Qc$ and $\SW{2}$ represent the combined contribution of \emph{both} of them.
        \item[$\textrm{c}$] Here, the second Stiefel-Whitney monopole is not symmetry indicated and both options are possible. For an example of each, see \cref{Sec:MinimalModels:C3}.
        \item[$\textrm{d}$] Here we use the convention that $a\mod n\in(-\floor{n/2},\dotsc,\floor{n/2}]$, such that $\abs{a\mod n}\in\{0,1,\dotsc,\floor{n/2}\}$.
        This implies that for $n=6$ the value $4$ ($5$) is equivalent to $2$ ($1$), for $n=4$ the value $3$ is equivalent to $1$ and for $n=3$ only $0,1$ can be distinguished (with $2$ equivalent to $1$ and therefore with undetermined parity).
        For further clarification of the ambiguity involved in defining $\abs{\Eu}\mod n$, see \cref{App:Euler-monopole:TPP}.
        \par
    \end{tablenotes}
    \label{tab:HOBBC:Qjump}
    \end{threeparttable}    
\end{table*}

Choosing $k_z$ along the HSL and taking 2D cuts of the system at constant $k_z$, we obtain a series of 2D systems with Hamiltonian $\Ham_{k_z}(k_x,k_y)$, which we assume to be at half-filling.
This fixes the principal gap to be the second one (such that principal nodes are indicated in red, cf.~\cref{fig:HOBBC:BS}).
Upon increasing $k_z$, the system goes from having $\rho^\mathrm{2D}$ occupied (to the left of the first TP) to instead having $(\rho^\mathrm{1D}_a,\rho^\mathrm{1D}_b)$ occupied (to the right of the second TP).
Let us refer to the three ranges of $k_z$ separated by the two TPs by the corresponding color of the central NL as \emph{orange}, \emph{red} and \emph{blue} $k_z$-range.
The corresponding band inversions lead to changes in the symmetry indicators $\#\Pi_p^{(n)}$, and thus potentially alter both the corner charge [cf.~\cref{eq:HOBBC:Q}] and the 2SW class [cf.~\cref{eq:HOBBC:SWc46}] of the 2D cuts.
Let us remark that, in practice, the \emph{physical} chemical potential of the 3D system will be constant.
However, generally it can be chosen such that we have half-filling with insulating bulk at least for some 2D cuts in the orange and blue $k_z$-ranges; then, the jumps will be observable by comparing only $k_z$ values corresponding to those~cuts.

We briefly explain how to apply the 2D symmetry-indicator formulas to 3D systems with $\mcP\mcT$ and arbitrary $C_n$ symmetry.
For concreteness we assume the rotation axis to be along $k_z$.
In this case, any 2D cut perpendicular to $k_z$ through the 3D BZ inherits both of these symmetries.
In particular, within the 2D plane $\mcP\mcT$ acts as $(C_2\mcT)_\mathrm{2D}$ with $(C_2)_\mathrm{2D} = C_2$ being the rotation around the $k_z$ axis, and $\mcT_\mathrm{2D} = \mathcal{P} C_2 \mathcal{T} = m_z \mathcal{T}$~\cite{Zhu:2016} being the composition of the physical time-reversal symmetry with the horizontal mirror symmetry, $m_z: z\mapsto -z$.
Note that the HSPs entering the symmetry-indicator formulas need to be identified with the intersection of the corresponding HSLs in the 3D BZ with the chosen 2D plane at fixed $k_z$.
For $\Pi^{(n)}$ we thus need to consider the $C_n$-invariant HSLs, while the 2D-TRIM correspond to lines invariant under
$\mathcal{T}_\mathrm{2D}$.
The computed corner charges in the 2D cuts therefore imply the corresponding values of hinge charges in the 1D hinge BZ, while the jump of the 2SW class between two 2D cuts corresponds to the 2SW \emph{monopole charge} as we explained in \cref{Sec:Monopole:SW} and illustrated in \cref{fig:Monopole:SWI}.

\begin{figure}[t]
    \centering
    \includegraphics{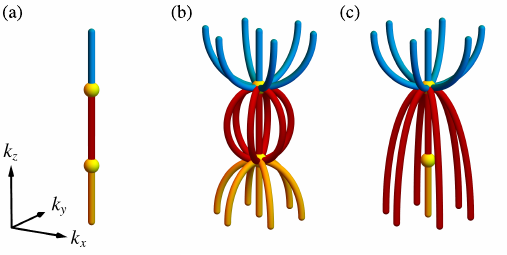}
    \caption{
        Illustration of possible nodal line (NL) compositions for the three considered triple-point pair configurations.
        (a) Two type-$\mathsf{A}$ triple points (TPs) have no attached NL arcs.
        (b) Two type-$\mathsf{B}$ TPs both exhibit attached NL arcs in the principal gap (red), but these NLs \emph{can} compactly tie the two TPs together, leaving the 2D bulk gapped in both the orange and in the blue $k_z$-range.
        (c) If one of the TPs is type~$\mathsf{A}$ and the other type~$\mathsf{B}$, the NL arcs in the principal gap \emph{necessarily} cross into either the orange or into the blue $k_z$-range, making the corresponding 2D cuts gapless.
    }
    \label{fig:HOBBC:TPconfigs}
\end{figure}

\begin{figure}[t]
    \centering
    \includegraphics{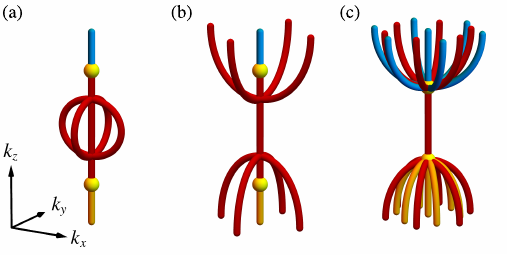}
    \caption{
        Illustration of certain more intricate nodal line (NL) compositions which can also arise near triple-point pairs (TPPs).
        (a)~Type-$(\mathsf{A},\mathsf{A})$ TPP with two red nexus points whose NLs are tied together \emph{inside} the red $k_z$-range, leaving the 2D bulk gapped in both the orange and in the blue $k_z$-range despite the nexus points. 
        Here, both the monopole charges and the hinge-charge jump can be defined.
        (b)~Analogous situation as in the previous panel; however, here the NLs extend into the orange and blue $k_z$-ranges rendering the corresponding 2D cuts gapless. 
        In this case, neither the monopole charges nor the hinge-charge jump can be defined.
        (c)~Type-$(\mathsf{B},\mathsf{B})$ TPP with NL arcs extending \emph{outside} the red $k_z$-range, making the relevant 2D cuts gapless.
        The monopole charges and the hinge-charge jump are undefined.
        This scenario should be contrasted to \cref{fig:HOBBC:TPconfigs}(b), which shows a similar situation with the NL arcs tied together \emph{inside} the red $k_z$-range.
    }
    \label{fig:HOBBC:TPGaplessConfigs}
\end{figure}

The hinge charge is only well-defined for values of $k_z$, where the bulk of $\Ham_{k_z}$ is gapped and the surface charge vanishes~\cite{Benalcazar:2019}.
Similarly, the monopole charges are only well-defined as long as there is a surface enclosing the TPP that is not penetrated by NLs in the principal gap (red).
While these conditions generally depend on the model parameters, we can formulate the following criteria for a gapped bulk, which conversely implies constraints on the admissible TPP configurations (to be precise, we call a combination of a 2D and two 1D ICRs an \emph{admissible TPP configuration} if it is \emph{not} necessarily gapless in the orange or blue $k_z$-range).
\begin{enumerate}
    \item
    In \cref{Sec:HOBBC:Euler} we have discussed a condition [summarized  by \cref{eqn:det-product-unity} and derived in \cref{App:Euler-monopole}], that the representations of $C_n$ in the orange and blue $k_z$-ranges must necessarily fulfill in order for the TPP to be admissible.
    According to \cref{fig:HOBBC:BS}, we have $D_0(C_n)=\rho^\mathrm{2D}(C_n)$ and $D_1(C_n)=\rho_a^\mathrm{1D}(C_n)\oplus\rho_b^\mathrm{1D}(C_n)$, such that the condition for the TPP configuration to be admissible is
    \begin{equation}
        \det \rho^\mathrm{2D}(C_n) = \rho_a^\mathrm{1D}(C_n)\rho_b^\mathrm{1D}(C_n).
        \label{eq:TPP-not-gapless-condition}
    \end{equation}
    Note that this is fully determined by the rotation eigenvalues, since the determinant on the left is just the product of eigenvalues.
    In \cref{App:Euler-monopole:TPP}, we show that this excludes exactly the TPPs with two 1D ICRs that have different rotation eigenvalues, i.e., for the little co-groups $C_{4(v)}$ these are the combinations $(E;A,B)$ and for $C_{6(v)}$ the combinations $(E_i;A,B)$, $i=1,2$ (we have omitted subscripts of the 1D ICR labels, since the argument is insensitive to them).
    In these cases neither the jump of the hinge charge nor the Euler and 2SW monopole charges are defined.
    \item
    For the little co-group $C_{6(v)}$, the gaplessness can be understood from the NL structure implied~\cite{Lenggenhager:2021:MBNLs} by the type of TPs involved, as illustrated in \cref{fig:HOBBC:TPconfigs}.
    A TPP consisting of one \mbox{type-$\mathsf{A}$} and one \mbox{type-$\mathsf{B}$} TP necessarily has NL arcs in the principal gap that extend beyond the red $k_z$-range.
    In contrast, if both involved TPs are of the same type, the orange and the blue $k_z$-ranges can generally be gapped.
    On the one hand, the case of two \mbox{type-$\mathsf{A}$} TPs is very simple since there are no additional NLs and the bulk is gapped in the corresponding regions of $k_z$.
    On the other hand, for two type-$\mathsf{B}$ TPs there must be NL arcs in the principal band gap attached to both TPs.
    However, the NL arcs can tie the two TPs together in the red $k_z$-range, where the gapless 2D cuts are not considered when analyzing the higher-order topology, thus leaving the orange and the blue $k_z$-range gapped.
    Note that in the applicable case, i.e., $C_{6(v)}$ which are the only point groups where both types of TPs are possible, this criterion turns out to be equivalent to the first.
    \item
    While the above two points give necessary criteria for having a gapped bulk, they are not sufficient.
    Two type-$\mathsf{A}$ TPs do not have any NL arcs attached, but they are often accompanied by nexus points, i.e., points on the HSL away from the TPs where NL arcs coalesce~\cite{Lenggenhager:2022:TPClassif}, see \cref{fig:HOBBC:TPGaplessConfigs}(a,b).
    If there is an even number of nexus points in the principal gap, then there are two options: (a)~the NL arcs connect together through the red $k_z$-range or (b)~the NL arcs extend \emph{outside} the red $k_z$-range, making the orange and blue $k_z$-range gapless.
    These two cases are illustrated in \cref{fig:HOBBC:TPGaplessConfigs}(a,b), respectively, for a $C_4$-symmetric example.
    Similarly, the NL arcs of two type-$\mathsf{B}$ TPs might not connect through the red $k_z$-range as indicated in \cref{fig:HOBBC:TPconfigs}(b) but rather extend outside of it as shown in \cref{fig:HOBBC:TPGaplessConfigs}(c).
    These cases are \emph{not} distinguished by symmetry properties but by the model parameters.
    The results of our analysis, e.g., \cref{tab:HOBBC:Qjump}, apply to the cases that do have a gapped bulk in some part of both the orange and the blue $k_z$-range.
\end{enumerate}

Let us remark that the cases disallowed by the criterion 1 for point group $C_{4(v)}$ (which according to \cref{tab:TP-MPGs} involve a pair of type-$\mathsf{A}$ TPs) correspond to situations where there is one (or a larger odd number of) symmetry-imposed nexus points of NLs in the principal gap, i.e., with the NLs attaching to the red region demarcated by the two TPs. 
An example of such a TPP configuration is illustrated in \cref{fig:Euler_charge}(b). 
Since an odd number of nexus points cannot be paired as in \cref{fig:HOBBC:TPGaplessConfigs}(a), there necessarily exist extended NLs which render either the orange or the blue $k_z$-region gapless. 
Therefore, the corresponding TPP is not admissible for the same reason as is visualized for the mixed-type TPP in \cref{fig:HOBBC:TPconfigs}(c).

Having clarified the necessary conditions to realize an admissible TPP, let us tackle their associated hinge-charge jump.
To observe the hinge charge, the system needs to be extended in the $z$-direction and finite in the $x$- and $y$-directions, with a geometry possessing the same $C_n$ symmetry as the SG, see \cref{fig:HOBBC:geometry}(a).
We need to distinguish geometries with the center of the rotational symmetry being placed at different WPs in the unit cell.
In \cref{App:Qjump} we consider all possible cases for each combination of ICRs listed in \cref{tab:HOBBC:Qjump} and compute the hinge-charge jump using the appropriate symmetry-indicator formulas~\cite{Takahashi:2021} [in particular, besides \cref{eqn:Q4-sym-ind,eqn:Q6-sym-ind}, we also need the analogous expression for $\Qc^{(3)}$; see \cref{App:eq:Qcorner3}].
Note that for this analysis the ionic charge distribution is irrelevant, because it does not depend on $k_z$ and therefore it does not contribute to the \emph{jump} of the hinge charge.
\Cref{tab:HOBBC:Qjump} summarizes the results of this analysis.
We observe that the combination of ICRs uniquely determines the jump of the hinge charge.
In particular, there is a non-vanishing jump $\Delta \Qc\neq 0\mod e$ quantized into multiples of $\tfrac{e}{n}$ for $C_n$ symmetry for any~TPP.

Let us briefly comment on the jump of the \emph{surface} charge.
As already stated, the vanishing of the fractional part of the surface charge (which is related to bulk polarization~\cite{Vanderbilt:1993}) is a necessary condition for the hinge charge to be observable~\cite{Benalcazar:2019}.
Therefore, to observe the hinge-charge \emph{jump}, the surface charge must vanish for the $k_z$-ranges on \emph{both} sides of the TPP, which can only happen if the TPP is not associated with a fractional jump $\Delta \vec{P}$ of the 2D bulk polarization.
We present in \cref{App:Qjump} a derivation, based on the symmetry-indicator formulas for the 2D polarization~\cite{Benalcazar:2019,Watanabe:2020,Takahashi:2021}, showing that indeed $\Delta \vec{P} = 0\mod e\vec{R}$ (where the Bravais vectors $\vec{R}$ constitute the usual ambiguity of bulk polarization~\cite{King-Smith:1993}) for all tabulated TPPs.
Therefore, the surface charge does not present an obstacle for the definition of the hinge charges, and we expect the hinge-charge jump to be observable for an appropriate choice of the boundary termination.

To compute the jump of the 2SW class, or equivalently the 2SW monopole charge $\SW{2}=\Delta\SW{2}$ carried by the red NL segment, we proceed as follows.
For systems with $C_4$- or $C_6$ symmetry the 2SW class of the 2D cuts can be computed readily by applying \cref{eq:HOBBC:SWc46} with the appropriate interpretation of the HSPs, as described above.
The full derivation can be found in \cref{App:Qjump} and the results are displayed in the bottom-most row of \cref{tab:HOBBC:Qjump}.
Consistent with the results of \cref{Sec:Monopole:SW}, we find that \mbox{type-($\mathsf{A}, \mathsf{A}$)} TPPs in $\mcP\mcT$-symmetric systems universally carry $\SW{2}=1$.
Furthermore, we recognize the following correspondence principle between the bulk invariant $\SW{2}$ and the hinge signature $\Delta \Qc$ (which is only defined modulo $e$):
a nontrivial 2SW monopole charge $\SW{2}=1$ implies a fractional hinge-charge jump of the minimal non-vanishing magnitude $\abs{\Delta \Qc}=\tfrac{e}{n}$, while $\SW{2}=0$ results in a twice as large a jump $\abs{\Delta \Qc}=\tfrac{2e}{n}$.

Determining the 2SW monopole charge in the case of $C_3$ symmetry is not as straightforward.
However, we note that the 2SW monopole charge is attributed to the red NL segment along the HSL as illustrated in \cref{fig:Monopole:SW}(b) and is therefore a feature local in $k_x,k_y$.
Thus, it cannot depend on the specific choice of HSL in the BZ, but only on the Hamiltonian near the HSL.
Noting that the little co-group of all three HSLs has the same rotational symmetry and there is only one class of combinations of ICRs, the admissible values of the 2SW monopole charge $\SW{2}$ are the same for all three HSLs $\Gamma A$, $KH$ and $K'H'$.
In the next paragraph we argue that, in fact, the 2SW class is not symmetry indicated in the $C_3$-symmetric case, meaning that both $\SW{2}=0$ or $1$ are possible.
We emphasize that this is not in contradiction to the bulk-hinge correspondence principle described above, because the fractional part of the hinge-charge jump satisfies $\pm\tfrac{e}{3}=\mp\tfrac{2e}{3}\mod e$.
Thus, the hinge signatures of nontrivial and trivial 2SW monopole charge become indistinguishable.

The fact that the 2SW class is not symmetry indicated in a SG with $C_3$ symmetry can be easily deduced from the other results in \cref{tab:HOBBC:Qjump}.
In particular, note that a SG with $C_6$ symmetry also has $C_3$ symmetry.
Thus, we can deform any of the possible TPP configurations on the $\Gamma A$ line of a six-fold symmetric system (cf.~last two columns of \cref{tab:HOBBC:Qjump}), having either $\SW{2}=0$ or $1$, into a TPP configuration of a $C_3$-symmetric system by introducing a small $C_2$-breaking perturbation.
However, \cref{tab:HOBBC:Qjump} shows that in the in three-fold symmetric little co-groups, only a single combination of ICRs is possible, such that any initial choice of the TPP in the $C_6$-symmetric case results in that unique TPP type of the $C_3$-symmetric system (in particular, \mbox{type-$\mathsf{A}$} TPs generically transform into \mbox{type-$\mathsf{B}$} TPs).
Since the 2SW class requires only $\mcP\mcT$ symmetry (which is preserved when breaking $C_2$) and the 2D cuts outside the red $k_z$-range remain gapped under the addition of a small perturbation, the 2SW class is unaffected by the perturbation.
This implies that both $\SW{2}=0$ or $1$ are possible for TPPs in $C_3$-symmetric SGs, and that the 2SW class is not symmetry indicated.
We demonstrate this feature for two explicit models in \cref{Sec:MinimalModels:C3}.

The symmetry-indicator formula for the Euler class, \cref{eq:HOBBC:Euler}, is, in constrast to the other formulas, already formulated in 3D.
It involves the corepresentations of $C_n$ in the orange and blue $k_z$-ranges. 
By going through all allowed combinations of ICRs (cf.~\cref{App:Euler-monopole:TPP}), we verify that indeed $\rho^\mathrm{2D}(C_n)$ and $\rho^\mathrm{1D}_a(C_n)\oplus\rho^\mathrm{1D}_b(C_n)$ are $\SO(2)$ matrices; then we extract $r_0=r_\mathrm{2D}$, $r_1=r_\mathrm{1D+1D}$ and compute $\Eu$.
The results are shown in the second to last row of \cref{tab:HOBBC:Qjump} for the cases with $\mcP\mcT$ symmetry.
We observe that in \emph{all} cases where it is defined, the Euler monopole charge is non-trivial and consistent with the independently calculated $\SW{2}$, via the relation $\SW{2}=\Eu\mod 2$.
Note that for $n=3$, $\Eu$ is determined only modulo $3$ and thus has undetermined parity, which is consistent with the 2SW not being symmetry indicated in that case, cf.~\cref{Sec:MinimalModels:C3}.

Finally, we remark on the consistency of the results in \cref{tab:HOBBC:Qjump} for a TPP on the $KH$ line of a $C_6$-symmetric compared to a $C_3$-symmetric system.
Naturally, a system with six-fold rotational symmetry is also three-fold rotation symmetric, such that one expects compatibility of the results for the two cases.
However, there are two caveats: (1) in a $C_6$-symmetric system the $KH$ and $K'H'$ HSLs are equivalent such that there are in fact \emph{two} symmetry-related TPPs per BZ at the same value of $k_z$, and (2) the hinge-charge jumps in the two cases are computed for a \emph{different geometry} of the sample [cf.~\cref{fig:HOBBC:geometry}(a)].
Therefore, if we only consider the $C_3$ symmetry of a $C_6$-symmetric system, we have to (1) count contributions of identical TPPs on the two HSLs $KH$ and $K'H'$, and (2) deform the sample with hexagonal cross section into one with a triangular cross section by combining two corners into a single new corner as depicted in \cref{fig:HOBBC:geometry}(b).

With this insight, it is easily checked that the corresponding entries in \cref{tab:HOBBC:Qjump} are fully
consistent.
First, for the 2SW class the two monopole charge contributions are additive, $\SW{2}^{(6)}=2\SW{2}^{(3)}$, which results in $\SW{2}^{(6)}=0 \mod 2$ for both $\SW{2}^{(3)}=0$ or $1$.
Furthermore, the corner charge of the triangular cross section $\Qc^{ab}$ is given by the sum of two of the corner charges of the hexagonal cross section, $\Qc^{ab}=\Qc^{a}+\Qc^{b}$, cf.~\cref{fig:HOBBC:geometry}(b).
Note that due to the six-fold rotational symmetry $\Qc^{a}=\Qc^{b} = \tfrac{e}{3}\mod e$, such that for the $C_6$-symmetric model with a TPP on $KH$ in the \emph{triangular geometry} we find $\Qc^{ab}=2\tfrac{e}{3}$.
On the other hand, when interpreting the same model as being $C_3$-symmetric, we simply add the corner charges for a TPP on $KH$ and $K'H'$, giving $\Qc^{ab}=2\tfrac{e}{3}$~as~well.

\subsection{Effect of symmetry breaking}\label{Sec:HOBBC:SymBreak}

The discussion in \cref{Sec:HOBBC:3D} has focused on spinless systems with $\mcP\mcT$ symmetry (corresponding to the MPGs in rows two and four of \cref{tab:TP-MPGs}), which has allowed us to characterize the TPPs with the Stiefel-Whitney monopole charge.
However, stable TPPs can exist even when the $\mathcal{PT}$ symmetry is removed~\cite{Lenggenhager:2022:TPClassif} (cf.~\cref{tab:TP-MPGs}); in this case, the monopole charges cease to exist, but the bulk-hinge correspondence of the TPPs persists.
In this subsection we discuss the effect of symmetry breaking on the derived bulk-hinge correspondence of TPPs.
First, in \cref{sec:symmetry-Cn+mirror} we argue that the the hinge-charge jumps derived for the cases with $m_v$ and $\mcP\mcT$ symmetry still apply when the $\mcP\mcT$ symmetry is broken while $m_v$ remains present.
We continue in \cref{sec:symmetry-CnPT} with a discussion how the derived results generalize to the cases where the TPPs are protected by the antiunitary $C_n\mcP\mcT$ rotational symmetry.
Finally, in \cref{sec:symmetry-CnPT} we briefly analyze the effect of breaking other combinations of symmetries, which generally leads to the loss of TPPs and to the formation of other species of band nodes. Our results are summarized in \cref{tab:HOBBC:symbreak}.

\begin{table}
    \begin{threeparttable}
    \caption{
        Effect of symmetry breaking on triple-point pairs (TPPs).
        We initially consider TPPs in the combined presence of $\mcP\mcT$ (space-time inversion), $C_n$ (rotation of order $n\in\{3,4,6\}$) and $m_v$ (vertical mirror) symmetries, corresponding to the first row of the table.
        Such TPPs are characterized by the second Stiefel-Whitney (2SW) monopole, by symmetry indicators ($\# \Pi_p^{(n)}$), and by the fractional hinge-charge jump (integer multiples of $e/n$).
        We then consider the breaking of the various symmetries, as analyzed in \cref{Sec:HOBBC:SymBreak}.
        We find that TPPs can evolve into higher-order Weyl points (HO-Weyl), into nodal lines (NLs) with the 2SW monopole, or into Weyl points or nodal lines without higher-order topology (indicated as \enquote{various}).
    }
    \begin{ruledtabular}
    	\begin{tabular}{ccccccc}
        	$\mcP\mcT$ & $C_n$ & $C_n\mcP\mcT$ & $m_v$ & Nodes & Bulk invariant & Hinge charge\\\hline\addlinespace
        	\cmark & \cmark & \cmark & (\cmark) & TPPs & 2SW, $\# \Pi_p^{(n)}$ & $e/n\,\mathbb{Z}$\\
        	\xmark & \cmark & \xmark & \cmark & TPPs & $\# \Pi_p^{(n)}$ & $e/n\,\mathbb{Z}$\\
        	\xmark & $C_{n/2}{}^\textrm{a}$ & \cmark & (\cmark) & TPPs & $\# \Pi_p^{(n)}$ & $2e/n\,\mathbb{Z}$\\
        	\cmark & \xmark & \xmark & (\cmark) & NLs & 2SW & not quantized \\
        	\xmark & \cmark & \xmark & \xmark & HO-Weyl & $\# \Pi_p^{(n)}$ & $e/n\,\mathbb{Z}$\\
        	\xmark & \xmark & \xmark & (\cmark) & (various) & \-- & not quantized
        \end{tabular}
    \end{ruledtabular}
    \begin{tablenotes}[flushleft]
        \item[$\textrm{a}$] {The requirement of the presence of $C_{n/2}$ restricts $n$ to $4,6$.}
        \par
    \end{tablenotes}
    \label{tab:HOBBC:symbreak}
    \end{threeparttable}    
\end{table}

\subsubsection{Triple points protected by \texorpdfstring{$C_n$}{Cn} and mirror symmetry}\label{sec:symmetry-Cn+mirror}

We first discuss the effect of breaking $\mcP\mcT$ symmetry while keeping the rotational symmetry $C_n$.
According to \cref{tab:TP-MPGs}, mirror symmetry $m_v$ is then required to protect TPs. 
The crucial observation is that if both $\mcP\mcT$ and mirror $m_v$ symmetry are present in the little co-group, then the ICRs are not modified by the removal of $\mcP\mcT$~\cite{Elcoro:2021,Xu:2020}, i.e., the ICRs of the unitary symmetries of the full group are exactly given by the ICRs of the unitary subgroup.
As a consequence, the classification of TPPs in systems with $m_v$ symmetry based on ICRs is \emph{identical} both with and without $\mcP\mcT$ symmetry; in fact, detailed analysis~\cite{Lenggenhager:2022:TPClassif} reveals that the \emph{type} ($\mathsf{A}$ vs.~$\mathsf{B}$) produced by the crossing of specified ICRs is also unaffected by the removal of $\mcP\mcT$.
An analogous statement also applies to the result that the ICR combination $(E;A,B)$ is necessarily gapless, see \cref{App:Euler-monopole:PT-absence}.

To derive the hinge-charge jumps $\Delta \Qc$ associated with TPPs protected by $m_v$ symmetry without $\mcP\mcT$, we should, in principle, repeat the above-described analysis (detailed in \cref{App:Qjump}) with the symmetry-indicator formulas that do not assume time-reversal symmetry, i.e., \cref{App:eq:Qcorner3,App:eq:Qcorner4,App:eq:Qcorner6}, and the ICRs of the point groups without $\mcP\mcT$ symmetry.
However, as described in the previous paragraph, the ICRs (and therefore the rotation eigenvalues) are identical to the ICRs of the corresponding point groups \emph{with}~$\mcP\mcT$ symmetry, such that \cref{App:eq:Qcorner4,App:eq:Qcorner6} still simplify to \cref{App:eq:Qcorner4_TRS,App:eq:Qcorner6_TRS}.
This implies that the jumps in the symmetry indicators are not changed when breaking $\mcP\mcT$ symmetry while keeping the rotational symmetry $C_n$, and the hinge-charge jumps $\Delta \Qc$ remain unaltered as well.
On the other hand, the monopole charges are not defined in the absence of $\mcP\mcT$ symmetry.
Therefore, all results in \cref{tab:HOBBC:Qjump} except for the last two rows apply to systems with mirror $m_v$ symmetry but no $\mcP\mcT$ symmetry.

\subsubsection{Triple points protected by the antiunitary \texorpdfstring{$C_n\mcP\mcT$}{CnPT} symmetry}\label{sec:symmetry-CnPT}

Since TPs can also be protected by the antiunitary rotational symmetry $C_n\mcP\mcT$ (cf.~\cref{tab:TP-MPGs}), $\mcP\mcT$ symmetry can alternatively be broken while keeping $C_n\mcP\mcT$. In such a case, the $C_n$ rotational symmetry is broken, while $C_{n/2} = (C_{n}\mcP\mcT)^2$ is a symmetry of the system.
We have to distinguish two scenarios: the TPs are either protected by $C_6\mcP\mcT$ or by $C_4 \mcP\mcT$.
Note that, as we discuss in \cref{App:Euler-monopole:PT-absence}, \cref{eq:TPP-not-gapless-condition} still applies in these MPGs, yet a careful analysis reveals that this constraint does not imply the inadmissibility of any TPP configurations.

We begin with analyzing the case of $C_6\mcP\mcT$, where the largest remaining rotational symmetry is $C_3$; therefore, a triangular geometry needs to be considered and the relevant rotational symmetry quantizing the hinge charges is $C_3$.
Furthermore, we find~\cite{Elcoro:2021,Xu:2020} that the corresponding MPGs with and without $m_v$ symmetry ($\bar{6}'m2'$ and $\bar{6}'$, respectively) have the same unitary subgroup as the $\mcP\mcT$-symmetric MPGs $\bar{3}'m$ and $\bar{3}'$, respectively, with matching ICRs.
As a consequence, the hinge-charge jump is the same as for the other little co-groups with $C_3$ rotational symmetry, i.e., no change in the bulk-boundary correspondence.
This has motivated us to group $C_3$ and $C_6\mcP\mcT$ within a single column of \cref{tab:HOBBC:Qjump}.
We remark, however, that although the ICRs of the subgroup and the higher-order bulk-boundary correspondence are equivalent to the $C_3$ case, this is not true for the corresponding TP types, which depend on the ICRs of the full point group.
Irrespective of the presence (MPG $\bar{6}'m2'$) vs.~absence (MPG $\bar{6}'$) of $m_v$ symmetry, the TPs protected by $C_6\mcP\mcT$ are \emph{always} type~$\mathsf{A}$~\cite{Lenggenhager:2022:TPClassif}.
In the latter case, this simply follows from the absence of symmetries (no $m_v$ and no $\mcP\mcT$) that can protect NLs lying off the rotation axis, while in the former this requires a more in-depth analysis~\cite{Lenggenhager:2022:TPClassif}.

Similarly, $C_4$ and $\mcP\mcT$ can be broken while keeping $C_4\mcP\mcT$.
However, in that case the largest remaining rotational symmetry is $C_2$, such that in a square geometry only the \emph{sum} over two adjacent hinges gives a quantized charge~\cite{Benalcazar:2019}.
The resulting little co-groups are the MPGs $\bar{4}'2'm$ (with $m_v$ mirrors) and $\bar{4}'$ (without $m_v$ mirrors), which are subgroups of the $\mcP\mcT$-symmetric MPGs $4/m'mm$ and $4/m'$, respectively.
It is easily verified~\cite{Elcoro:2021,Xu:2020} that all four listed MPGs have equivalent $C_2$-rotation eigenvalues for the 1D and 2D ICRs: all 1D ICRs are \emph{even} while all 2D ICRs are \emph{odd} under $C_2$.
Since in the presence of $C_4\mcP\mcT$ the hinge-charge jump is completely determined by $C_2$-rotation eigenvalues, the hinge-charge jump of TPPs protected by $C_4\mcP\mcT$ is, in principle, the same as in the $C_4$ case.
However, since only the sum of charges on two adjacent hinges is quantized, the resulting jump is doubled to $\Delta\Qc=2\times \tfrac{e}{4}\mod e = \tfrac{e}{2}\mod e$.
The TPs protected by $C_4\mcP\mcT$ are always type~$\mathsf{A}$~\cite{Lenggenhager:2022:TPClassif}, i.e., of the same type as before the $\mcP\mcT$ breaking.
With this analysis, we explained the \enquote{$C_2$} column in \cref{tab:HOBBC:Qjump}, and established the higher-order bulk-boundary correspondence for all TPPs listed in \cref{tab:TP-MPGs}.

\subsubsection{Breaking the symmetry protection of triple points}\label{sec:symmetry-other}

Besides breaking the symmetry in ways that keep the TPs robust, here we briefly discuss the effect of breaking symmetries that protect the TPs. 
We find that this generically results in the conversion of the TPs to other species of band nodes~\cite{Sun:2018}.
The case of broken $C_n$ symmetry (with or without $m_v$) in the presence of $\mcP\mcT$ symmetry has already been discussed implicitly in \cref{Sec:Monopole:SW} for type-$(\mathsf{A},\mathsf{A})$ TPPs, where we revealed their conversion~\cite{Lenggenhager:2021:MBNLs} into multiband nodal links carrying a non-trivial 2SW monopole.
Using the results in \cref{tab:HOBBC:Qjump}, we can now extend this discussion to generic TPPs.
If the TPP carries $w_2 = 1$, then the 2SW monopole implies stable multiband nodal links upon breaking the rotational symmetry. On the other hand, note that $w_2 = 0$ only arises for type-($\mathsf{B},\mathsf{B}$) TPPs. 
Since these are characterized by the attached NL arcs carrying a $\pi$-flux of Berry phase~\cite{Burkov:2011}, and this quantization is unaffected by the broken rotational symmetry, a small $C_n$-breaking perturbation leaves behind a NL composition with a trivial 2SW monopole. In both cases, $w_2\in\{0,1\}$, the absence of rotational symmetry implies the \emph{absence} of the hinge-charge quantization for all geometries depicted in \cref{fig:HOBBC:geometry}.

We next analyze the situation where $C_n$ is kept but both $m_v$ and $\mcP\mcT$ are broken.
In such cases, the rotational symmetry ensures that the quantized hinge-charge jump associated with the TPP is maintained.
At the same time, it is impossible to protect NLs (including the 2D ICR along the HSL) without $\mcP\mcT$ and $m_v$~\cite{Fang:2016} using $C_n$ alone, and instead Weyl points~\cite{Wan:2011} become the generic band degeneracy.
Therefore, the breaking of both $m_v$ and $\mathcal{PT}$ symmetry results in a \emph{higher-order Weyl semimetal} with Weyl points along the rotation axis~\cite{Wang:2020b,Ghorashi:2020,Wei:2021}, characterized by quantized hinge-charge jumps.

We remark that in spinless systems higher-order \emph{Dirac} points~\cite{Lin:2018,Calugaru:2019,Wieder:2020,Qiu:2021} cannot be obtained by perturbing a TPP; and vice versa, a TPP cannot be created by perturbing a higher-order Dirac point. 
Nonetheless, the results in \cref{tab:HOBBC:Qjump} readily generalize to include the higher-order Dirac points that are realized in $C_6$-symmetric models by crossing two different 2D ICRs~\cite{Qiu:2021}.
Indicating such a crossing as $(E_1;E_2)$, we formally write $(E_1;E_2) \sim (E_1;\rho_a^\textrm{1D},\rho_b^\textrm{1D}) - (E_2;\rho_a^\textrm{1D},\rho_b^\textrm{1D})$.
This equation represents the fact that the two pairs of TPPs can be evolved into the higher-order Dirac point: to see this, assume that along $k_z$ the two 1D ICRs (for simplicity dispersionless) are first crossed by a 2D ICR $E_1$ with increasing energy and subsequently by 2D ICR $E_2$ with decreasing energy.
By shifting the energy of the dispersionless 1D ICRs to higher values, the nodal feature at half-filling indeed evolves from two TPPs into a single Dirac point.
However, this continuous change preserves the total topological characterization of the band nodes, therefore (cf.~\cref{tab:HOBBC:Qjump}) we deduce that the higher-order Dirac point is characterized by hinge-charge jump $\Delta \Qc^{(6)}=\tfrac{e}{6}-\left(-\tfrac{e}{3}\right) = \tfrac{e}{2}\mod e$ and by a non-trivial Stiefel-Whitney monopole $\SW{2} = 1 - 0 = 1 \mod 2$.

Finally, if $C_n$ and $\mcP\mcT$ are broken and only $m_v$ is kept, then both the 2SW as well as the rotational symmetry indicators become invalid.
As a consequence, no higher-order topology remains.
Depending on the details of the perturbation, the $m_v$ symmetry could leave behind mirror-protected NLs or the system opens an energy gap.
If $m_v$ is also broken, then depending on the details of the perturbation the systems either becomes a Weyl semimetal or opens an energy gap.

\section{Minimal Models}\label{Sec:MinimalModels}
In the previous section we have shown that TPP configurations demarcate a NL segment that can be characterized by an Euler and 2SW monopole charge (if $\mcP\mcT$ is present) and by a fractional hinge-charge jump (for all cases).
Here we verify these theoretical predictions by analyzing several concrete tight-binding models:
First, in \cref{Sec:MinimalModels:C4} we revisit the $C_4$-symmetric model introduced in \cref{Sec:C4Model}, and we relate the phenomenology described there to the theoretical results presented in \cref{Sec:HOBBC}.
Subsequently, in \cref{Sec:MinimalModels:C6AA,Sec:MinimalModels:C6BB} we present two $C_6$-symmetric models, one involving type-$\mathsf{A}$ and the other involving type-$\mathsf{B}$ TPs, and confirm that the jumps in the 2SW class and in the hinge charge as well as the Euler monopole charge follow the predictions of~\cref{tab:HOBBC:Qjump}.

For simplicity, all the discussed models retain both the (spinless) $\mcP\mcT$ symmetry and the mirror $m_v$ symmetry.
It follows from \cref{tab:TP-MPGs} and from the subgroup relations discussed therein, that models for all other species of TPPs can in principle be obtained by applying an appropriate perturbation to one of the $C_4$ or $C_6$ models discussed in the following subsections.
We utilize this feature in \cref{Sec:MinimalModels:C3}, where we break the $C_2$ symmetry in each of the two $C_6$-symmetric models. 
This construction allows us to explicitly show that the 2SW class is not symmetry-indicated in $C_3$ symmetric systems, as previously stated (but not proved) in \cref{Sec:HOBBC:3D}.

\begin{figure}[t!]
    \centering
    \includegraphics{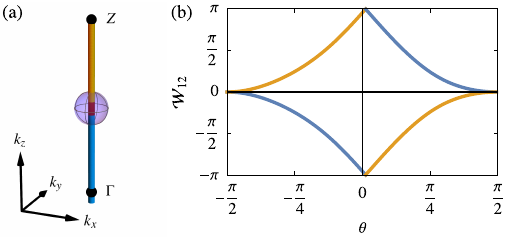}
    \caption{
        Verification of the second Stiefel-Whitney monopole charge $\SW{2}=1$ carried by the red NL segment in the model defined in \cref{eq:C4model:def}.
        (a) Nodal configuration near the high-symmetry line $\Gamma Z$ and the ellipsoid (purple) on which the monopole charge is computed.
        (b) Wilson-loop spectrum $\mathcal{W}_{12}$ of the lower two bands computed on that ellipsoid, parametrized by the latitude $\theta\in[-\pi/2,\pi/2]$. The spectrum shows a single winding (odd parity) and thus implies $\abs{\Eu}=1$ and $\SW{2}=1$.
    }
    \label{fig:MinimalModels:C4:SWonEllipsoid}
\end{figure}

\subsection{\texorpdfstring{$C_4$}{C4}-symmetric model}\label{Sec:MinimalModels:C4}

In \cref{Sec:C4Model} we introduced a tight-binding model with hinge charges induced by a NL configuration with two pairs of TPs in adjacent band gaps, see \cref{eq:C4model:def,fig:C4model:model,fig:C4model:hinge}.
We observed that the red NL segments divide the BZ into two parts: one with nontrivial 2SW class on 2D cuts through the BZ and hinge charge $\tfrac{e}{4}$, and another with trivial 2SW class and a vanishing hinge charge.
We now recast the phenomenology observed for that model in light of the bulk-hinge correspondence derived in \cref{Sec:Monopole,Sec:HOBBC} and summarized in \cref{tab:HOBBC:Qjump}.
In particular, we recognize that the 2SW class and hinge charge are consequences of the nodal configuration of the model with the robust signatures being their jumps as a function of $k_z$.
In addition, we also clarify how the hinge charge $\Qc(k_z)$ plotted in \cref{fig:C4model:hinge}(b) has been extracted.

As we argued in \cref{Sec:Monopole:SW} and proved using symmetry indicators in \cref{Sec:HOBBC:3D}, each four-band TPP along a $C_4$-rotation axis (which are automatically type~$\mathsf{A}$) demarcates a NL segment in the principal gap [red in \cref{fig:MinimalModels:C4:SWonEllipsoid}(a)] which carries 2SW monopole charge $\SW{2}=1$.
This can be diagnosed by computing the Wilson-loop spectrum of the lower two bands on an ellipsoid [purple in \cref{fig:MinimalModels:C4:SWonEllipsoid}(a)] containing one such NL segment.
\Cref{fig:MinimalModels:C4:SWonEllipsoid}(b) shows that there is a single winding, confirming that indeed $\abs{\Eu}=1$ and $\SW{2}=1$.

Furthermore, according to \cref{tab:HOBBC:Qjump}, the TPP should be accompanied with a fractional hinge-charge jump in a nanowire geometry.
To observe the quantized hinge charges and the predicted hinge-charge jump, the termination of the nanowire has to preserve the $C_4$ symmetry. 
Here, we opt to study the model in the rotated square geometry shown in \cref{fig:C4model:hinge}.
To fully specify the model, we also need to choose where to place the ionic charges.
The placement is not fixed by the tight-binding model itself and in a real material would depend on the chemical composition.
While the ionic charge distribution affects the value of the hinge charge and can even lead to fractional hinge or corner charges for systems that are electronically completely trivial~\cite{Watanabe:2021}, it does not affect the hinge-charge \emph{jump}, because of its independence of $k_z$.
For concreteness, we assume the ionic charge to be concentrated at the center of the 2D projection of the unit cell (Wyckoff position $1a$), as indicated with black circles in \cref{fig:C4model:hinge}(a).
However, other distribution of the ionic charge are possible; in particular, moving charge $n\abs{e}$ from $1a$ to $1b$ position changes the corner charge of a 2D system by $-\tfrac{n\abs{e}}{4}$, as is easily verified from Eqs.~(17-19) in Ref.~\onlinecite{Takahashi:2021}.

\begin{figure}[t]
    \centering
    \includegraphics{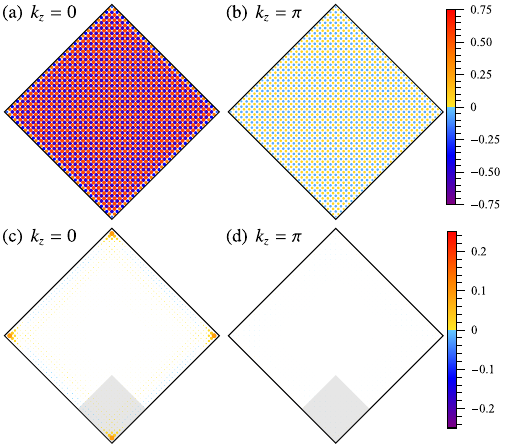}
    \caption{
        Charge distribution of the model defined in \cref{eq:C4model:def} and for the geometry illustrated for fewer unit cells in \cref{fig:C4model:hinge}(b). Assuming the compensating ionic charge to be located at Wyckoff position $1a$, we plot the total charge distribution at $k_z=0$ (a,c) and at $k_z=\pi$ (b,d).
        The total charge at each lattice site is indicated by a disk with its value encoded both in the area of the disk and the color scale (see legend on the right of each row).
        Note that the disks are scaled separately for each panel: compared to panel (a) the disks are enlarged by factors of (b) $2$, (c) $3$ and (d) $6$ to increase visibility.
        To remove sub-unit-cell oscillations of the charge distribution observable in (a,b), we perform coarse-graining of the data as explained in \cref{App:coarse-graining,fig:corner-charge:coarse-graining}(b).
        The results of the coarse-graining are shown in panels (c) and (d), respectively.
    }
    \label{fig:MinimalModels:C4:ChargeDist}
\end{figure}

For fixed $k_z$ of the nanowire geometry model, we consider a 2D system with $20.5\times 20.5$ unit cells [compare \cref{fig:C4model:hinge}(a) for $3.5\times 3.5$ unit cells] and compute the charge distribution using exact diagonalization.
Results for $k_z=0$ and $k_z = \pi$ are shown in \cref{fig:MinimalModels:C4:ChargeDist}(a,b), respectively.
On length scales smaller than a unit cell, we observe strong oscillations, which make it impossible to directly integrate the charge on a corner or even detect any possible localization.
This is a typical problem for the case of ionic crystals and is usually tackled by coarse-graining the charge distribution~\cite{Watanabe:2020,Ren:2021,Takahashi:2021}.
We perform a discrete version of such a coarse-graining by following a scheme explained in \cref{App:coarse-graining}.
Considering each orbital position (i.e., the two per unit cell) as a separate site of a square lattice, we compute the charge on the \emph{dual square lattice} [cf.~\cref{fig:corner-charge:coarse-graining}(b)] by distributing charge on each site of the original lattice with equal weights to the nearest sites of the dual lattice.
Effectively, this coarse-grains the charge distribution to the length scale of one unit cell and significantly reduces the oscillations, as shown in \cref{fig:MinimalModels:C4:ChargeDist}(c,d).

After coarse-graining, we can integrate over various regions to find the bulk, edge and corner charges.
First, we observe that the bulk and edge charge vanish for both $k_z=0,\pi$ as can be seen in \cref{fig:MinimalModels:C4:ChargeDist}(c,d).
Due to the double band-inversion at $\Gamma$, the model has vanishing Berry phases along the path $M\Gamma M$ (perpendicular to the edge), which implies~\cite{Vanderbilt:1993} that the bulk polarization~\cite{King-Smith:1993} and therefore the edge charge vanish.
On the other hand, the corner charge only vanishes for $k_z=\pi$.
The non-vanishing corner-charge at $k_z=0$ is a consequence of the corner-induced filling anomaly~\cite{Benalcazar:2019}.
This excess charge is strongly localized near the four corners.
Integrating over the gray area indicated in the figure, and taking care that the integration region forms $90^\circ$ angles with the edges~\cite{Watanabe:2020}, we find $\Qc=0.2498e$ for $k_z=0$ and $\Qc=8\cdot 10^{-9} e$ for $k_z=\pi$.
The deviations from the expected values $\tfrac{e}{4}$ and $0$ are due to finite-size effects, and can be further reduced by increasing the total system size and the coarse-graining scale while keeping the relative size of the integration region fixed.

We numerically compute the 2SW class and the (coarse-grained) corner-charge of the $C_4$-symmetric model defined in \cref{eq:C4model:def} using the described methods for planes at multiple values of fixed $k_z$.
Our results are summarized in \cref{fig:C4model:hinge}(b) in \cref{Sec:C4Model}.
We find that the model exhibits a jump of the 2SW class by $+1$ and of the hinge charge by $+\tfrac{e}{4}$ as the 2D ICR moves from the occupied to the unoccupied band subspace.
This agrees with the prediction of \cref{tab:HOBBC:Qjump}, since the ICRs of the bands involved in the two TPs on the HSL $\Gamma Z$ are $(E;B_1,B_2)$.
While far away from the TPs the hinge charge is very close to the quantized value, there are finite-size effects.
This is especially observed in the small deviations from the quantized values for $k_z$ near to the closing of the principal gap (i.e., in the vicinity of each TP), where the localization length of the corner charge is expected to grow.

\begin{figure*}
    \centering
    \includegraphics{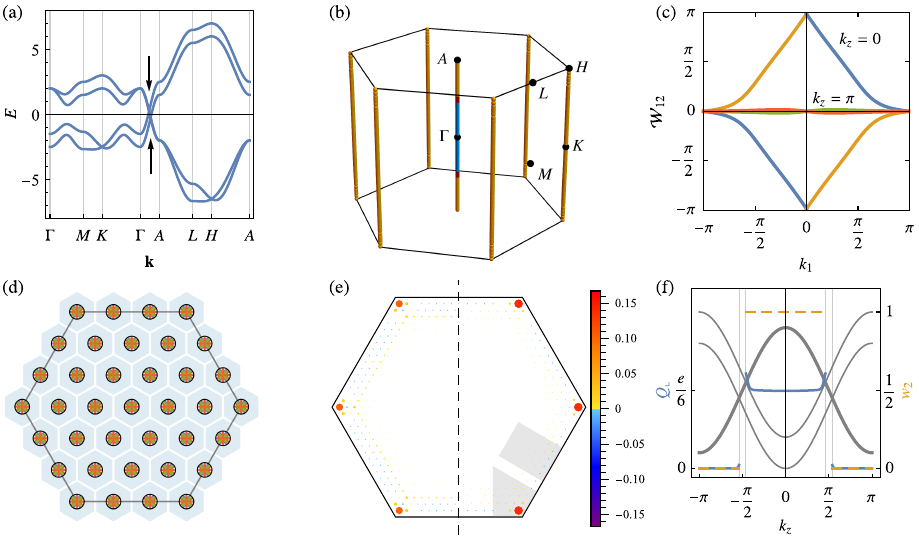}
    \caption{
        Results for the $C_6$-symmetric model with two type-$\mathsf{A}$ triple points (TPs) defined in \cref{eq:MinimalModels:C6AA:def} and discussed in \cref{Sec:MinimalModels:C6AA}.
        (a) Band structure with the two TPs indicated by black arrows.
        (b) Nodal-line structure in the full Brillouin zone with the high-symmetry points indicated. Nodal lines in the first, second and third band gap of the four-band model are shown in orange, red and blue, respectively.
        (c) The second Stiefel-Whitney class can be determined from the Wilson-loop spectrum $\mathcal{W}_{12}$ of the lower two bands on horizontal cuts through the Brillouin zone (parameterized by the projection $k_1$ of the momentum onto the first reciprocal lattice vector). Blue/orange lines show the spectrum for $k_z=0$ (single winding) and green/red lines for $k_z=\pi$ (no winding).
        (d) Cross-section through a hexagonal nanowire three shells wide; light blue hexagons indicate the model's unit cells.
        The orbitals are represented in color and all placed on top of each other in the center of the unit cell. The ionic charge is concentrated at the same position, indicated by the black circles.
        (e) Charge distribution for a cut through a $16$-shell wide nanowire at $k_z=0$. The charge at each site is indicated by a disk with its value encoded both in the area of the disk and the color scale (see legend on the right). The left half shows the charge distribution on the original lattice, and the right half shows the charge after two iterations of coarse-graining [cf.~\cref{App:coarse-graining} and \cref{fig:corner-charge:coarse-graining}(c)].
        A hinge charge of $\Qc=\tfrac{e}{6}$ with a strong localization is observed.
        (f) Hinge Brillouin zone with hinge charge $\Qc$ (blue solid line, left axis), second Stiefel-Whitney class $\SW{2}$ (orange dashed line, right axis) and the projection of the bulk dispersion along $k_x=k_y=0$ (gray).
        The doubly-degenerate band is displayed with a thicker line.
    }
    \label{fig:MinimalModels:C6AA}
\end{figure*}

\subsection{\texorpdfstring{$C_6$}{C6}-symmetric model with type-\texorpdfstring{$\mathsf{A}$}{A} triple points}\label{Sec:MinimalModels:C6AA}

To realize a type-($\mathsf{A},\mathsf{A}$) TPP along a $C_6$-rotational symmetry, we consider the $\Gamma A$ line of the hexagonal SG $P6/mmm$ (No.~191).
More specifically, we consider a model [detailed in \cref{eq:MinimalModels:C6AA:def} of \cref{App:ModelDetails:C6AA}] with $(d_{xy},d_{x^2-y^2})$ orbitals transforming in the 2D ICR $E_{2\mathrm{g}}$ and orbitals $f_{x(x^2-3y^2)}$ and $f_{y(3x^2-y^2)}$ transforming in 1D ICRs $B_{1\mathrm{u}}$ and $B_{2\mathrm{u}}$, respectively, all placed at the Wyckoff position $1a$ with site-symmetry group $D_{6h}$.
We tune the parameters such that the $d$-orbitals have lower energy than the $f$-orbitals at all HSPs with the exception of a double-band-inversion at $\Gamma$.
Consequently, on $\Gamma A$ two 1D ICRs ($B_1$, $B_2$) consecutively cross the 2D ICR ($E_2$) resulting~\cite{Lenggenhager:2021:MBNLs} in two type-$\mathsf{A}$ TPs.
The band structure and the NL configuration of the model are displayed in \cref{fig:MinimalModels:C6AA}(a) and (b), respectively.
We remark that apart from the two-fold degeneracy along the $\Gamma A$ line, the model also exhibits two-fold degeneracies along the $KH$ lines; however, these lie within the occupied band subspace and therefore have no effect on the discussed topological features. The model exhibits no additional NLs beyond these degeneracies along the HSLs.
We remark that for the computation of the hinge charges we set the compensating ionic charge to also reside at site $1a$.

According to \cref{tab:HOBBC:Qjump}, we expect a jump of $+\tfrac{e}{6}$ in the hinge charge and $+1$ both in the Euler and 2SW class when going from the orange $k_z$-range to the blue $k_z$-range.
More precisely, based on a symmetry eigenvalue analysis and \cref{eq:HOBBC:Q,eq:HOBBC:SWc}, we expect that the 2SW class is $1$ in the blue $k_z$-range and $0$ in the orange $k_z$-range, while the hinge charge should be $\tfrac{e}{6}$ and $0$, respectively.
The results for the 2SW class are verified by computing the Wilson-loop spectra of the lower two bands on horizontal cuts through the BZ as a function of $k_z$.
The 2SW class is then given by the parity of the winding of the spectrum.
Two examples, for $k_z=0$ and $k_z=\pi$, are shown in \cref{fig:MinimalModels:C6AA}(c), and demonstrate that indeed $\abs{\Eu}=1$ and $\SW{2}=1$ for $k_z=0$.
\Cref{fig:MinimalModels:C6AA}(f) shows $\SW{2}$ as a function of $k_z$ with the visible jumps at the position of the two TPs along the $\Gamma A$ line.

To verify the results for the hinge charge, we construct a system finite in $x$- and $y$-directions with $16$ hexagonal shells in the cross section [the cross section illustrated in \cref{fig:MinimalModels:C6AA}(d) has three shells].
Using exact diagonalization we find the charge distribution at half-filling for 2D cuts of the BZ that lie outside of the red $k_z$-range.
\Cref{fig:MinimalModels:C6AA}(e) shows the results of such analysis for $k_z=0$.
The charge is strongly localized at the corners of the 2D cross section while it vanishes both in the bulk as well as along the edges.
Note that while the charge vanishes on the edges [integrating over the gray edge area in \Cref{fig:MinimalModels:C6AA}(e) we obtain $8\cdot 10^{-4}\abs{e}$], we also observe small oscillations of the charge in the direction perpendicular to the edges which visually obscure the localization of the corner charge.
To reduce these oscillations, we perform coarse-graining similar to the one described in the previous subsection.
The major technical difference is that here we coarse-grain over a hexagonal supercell with two shells, i.e., seven unit cells, in one iteration [cf.~\cref{fig:corner-charge:coarse-graining}].
After two iterations we end up with a coarse grained lattice that is smaller by two shells.
Integrating over the gray corner area in \Cref{fig:MinimalModels:C6AA}(e) we obtain $0.1652e$, where the deviation from $\tfrac{e}{6}$ can be traced back to finite-size effects.
As can be seen in \cref{fig:MinimalModels:C6AA}(f) those finite-size effects grow stronger with decreasing the central energy gap, i.e., close to the two TPs.

\begin{figure*}
    \centering
    \includegraphics{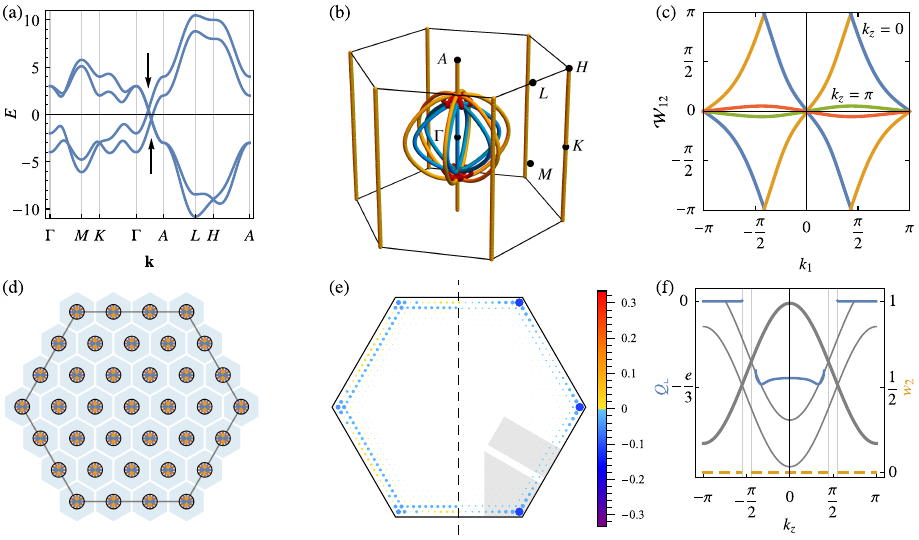}
    \caption{
        Results for the $C_6$-symmetric model with two type-$\mathsf{B}$ triple points (TPs) defined in \cref{eq:MinimalModels:C6BB:def} and discussed in \cref{Sec:MinimalModels:C6BB}.
        The organization of the panels is in one-to-one correspondence with \cref{fig:MinimalModels:C6AA}.
        The key differences to observe are:
        (b) Due to the TPs being type~$\mathsf{B}$, there are nodal line arcs attached to them, visible in the nodal line structure.
        (c) The Wilson loop spectra for both $k_z=0$ and $k_z=\pi$ show double and zero winding, respectively, implying $\abs{\Eu}=2$ and $\SW{2}=0$.
        (e,f) The hinge charge in the region around $k_z=0$ is $\Qc=-\tfrac{e}{3}$; the localization length is larger than for the model shown in \cref{fig:MinimalModels:C6AA}, such that finite-size effects produce a visible deviation from the ideal quantized value of the corner charge in panel (f).
    }
    \label{fig:MinimalModels:C6BB}
\end{figure*}

\subsection{\texorpdfstring{$C_6$}{C6}-symmetric model with type-\texorpdfstring{$\mathsf{B}$}{B} triple points}\label{Sec:MinimalModels:C6BB}

We next consider a $C_6$-symmetric model with type-($\mathsf{B},\mathsf{B}$) TPP realized in the same SG ($P6/mmm$ , No. 191) and on the same hexagonal lattice.
The following four orbitals are placed at Wyckoff position $1a$: $f_{x(x^2-3y^2)}$ and $f_{y(3x^2-y^2)}$ transforming, respectively, in 1D ICRs $B_{1\mathrm{u}}$ and $B_{2\mathrm{u}}$, together with $(p_x,p_y)$ transforming in the ICR $E_{1\mathrm{u}}$.
The model's Bloch Hamiltonian is given in \cref{eq:MinimalModels:C6BB:def} in \cref{App:ModelDetails:C6BB}.
As in the previously discussed tight-binding models, we set the parameters to produce a double-band-inversion at $\Gamma$.
As a consequence, the $p$-orbitals have lower energy than the $f$-orbitals at all HSPs except $\Gamma$.
Then, two 1D ICRs ($B_1$, $B_2$) consecutively cross the 2D ICR ($E_2$) resulting~\cite{Lenggenhager:2021:MBNLs} in two type-$\mathsf{B}$ TPs.
The band structure and the NL configuration are shown in \cref{fig:MinimalModels:C6BB}(a) and (b), respectively.
We observe the NL arcs attached to the type-$\mathsf{B}$ TPs; as desired, the NL arcs in the principal gap (shown in red) are tied together in the red $k_z$-range [cf.~\cref{fig:HOBBC:TPconfigs}(b)].
There are no further degeneracies in the principal band gap of the model.
According to \cref{tab:HOBBC:Qjump}, we expect a jump of $-\tfrac{e}{3}$ in the hinge charge and no jump of the 2SW class when the 2D ICR along $\Gamma A$ moves from the occupied to the unoccupied band subspace but an Euler monopole charge $\abs{\Eu}=2$.
More precisely, based on a symmetry eigenvalue analysis and \cref{eq:HOBBC:Q,eq:HOBBC:SWc}, we anticipate that the 2SW class vanishes in \emph{both} regions, while the hinge charge is expected to be $0$ and $-\tfrac{e}{3}$ in the orange and blue $k_z$-range, respectively, assuming that the compensating ionic charge is placed at position $1a$.

We verify these predictions with numerically.
First, in \cref{fig:MinimalModels:C6BB}(c) we plot the Wilson-loop spectra of the lower two bands on horizontal cuts at both $k_z=0$ and $k_z=\pi$.
Since they both exhibit even winding number, we confirm that $\SW{2}=0$ on both sides of the TPPs.
We observe that the Wilson-loop spectrum for the 2D cut at $k_z = 0$ has winding number $\pm 2$, meaning that the corresponding 2D model is an Euler insulator~\cite{Bouhon:2020b}. 
The double winding also implies that each TPP with its attached (red) NL arcs (the \emph{nodal-line nexus}) carries an Euler class $\abs{\Eu}=2$ on both the occupied and the unoccupied band subspace, as predicted. 
Therefore, as long as $\mcP\mcT$ symmetry is present, the only way to gap out the red NL nexus is to annihilate it with the other NL nexus~\cite{Bzdusek:2017}. 
We remark that this topological obstruction is trivialized in the presence of additional occupied and unoccupied bands~\cite{Tiwari:2020}.

To calculate the hinge charge, we construct a system finite in $x$- and $y$-directions with $20$ hexagonal shells in the cross section [in \cref{fig:MinimalModels:C6BB}(d) we illustrate a system with three hexagonal shells].
By exact diagonalization we find the charge distribution at half-filling for values of $k_z$ where the spectrum is gapped.
In \cref{fig:MinimalModels:C6BB}(e) the charge distribution for $k_z=0$ is shown.
The charge is still localized at the corners of the 2D cross section and vanishes both in the bulk as well as along the edges.
However, due to trivial in-gap states and therefore a reduced energy gap, we observe an oscillation of the charge distribution in the direction perpendicular to the edge, as well as a significantly larger localization length for the corner charge than found in the previous hexagonal model.
In fact, the chosen system is too small to achieve convergence, but it is sufficient to support the theoretical predictions in \cref{tab:HOBBC:Qjump}.
Integrating the coarse-grained charge distribution over the gray edge area in \Cref{fig:MinimalModels:C6BB}(e) we obtain $4\cdot 10^{-2}\abs{e}$; while over the gray corner area we find $-0.2983e$, which is close to the ideal result $-\tfrac{e}{3}$ in the absence of finite-size effects.
We repeat the same analysis for 2D cuts for multiple values of $k_z$, and plot the dependence of the 2SW class and of the hinge charge as a function $k_z$ (where the gap is open) in \cref{fig:MinimalModels:C6AA}(f).

\subsection{\texorpdfstring{$C_3$}{C3}-symmetric models}\label{Sec:MinimalModels:C3}

In this section, we explicitly show that in the presence of $\mcP\mcT$ symmetry there exist $C_3$-symmetric models where the TPPs carry either $\SW{2}=0$ or $\SW{2}=1$.
We therefore confirm that the 2SW monopole charge in $C_3$-symmetric SGs is not symmetry indicated, thus justifying the ambiguous entry in \cref{tab:HOBBC:Qjump}.
This freedom is present despite the fact that (cf.~\cref{tab:TP-MPGs}), $C_3$-symmetric HSLs only support a single species of TPs, namely type~$\mathsf{B}$ with three attached NLs arcs and with the central NL carrying Berry phase $\pi$.

\begin{figure*}
    \centering
    \includegraphics{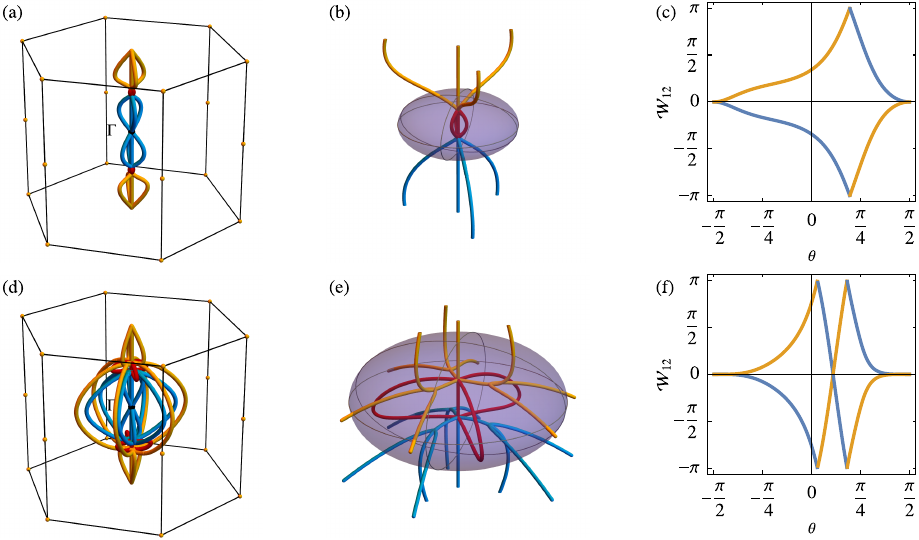}
    \caption{
        Nodal lines and Wilson-loop spectra for the $C_3$-symmetric models obtained by breaking the $C_6$ symmetry in the models discussed in \cref{Sec:MinimalModels:C6AA,Sec:MinimalModels:C6BB} using the trigonal perturbation given in \cref{eq:MinimalModels:C3:pert} with $\delta_{C_3}=8\times 10^{-4}$.
        (a,d) Nodal lines in the full Brillouin zone (black frame). Nodal lines in the first/second/third band gap are shown in orange/blue/red.
        (b,e) Close-ups of the red nodal-line segment. In both cases the triple points are clearly type~$\mathsf{B}$ with three attached nodal line arcs per energy gap.
        (c,f) Wilson-loop spectra on the purple ellipsoids shown in (b,e), respectively. The single winding in (c) implies $\abs{\Eu}=1$ and $\SW{2}=1$ and the double winding in (f) implies $\abs{\Eu}=2$ and $\SW{2}=0$.
    }
    \label{fig:MinimalModels:C3}
\end{figure*}

We can obtain such $C_3$-symmetric models by reducing the $C_6$ symmetry of the models discussed in \cref{Sec:MinimalModels:C6AA,Sec:MinimalModels:C6BB} down to $C_3$ (while keeping $\mcP\mcT$).
A possible perturbation that achieves that is given by \cref{eq:MinimalModels:C3:pert} in \cref{App:ModelDetails:C3}.
The resulting NL structure for each parent model is shown in \cref{fig:MinimalModels:C3}(a,b) and \cref{fig:MinimalModels:C3}(d,e), respectively.
Note that the 2SW monopole charges of the TPPs in the parent hexagonal models are not changed by the trigonal perturbation:
as long as the enclosing ellipsoid [purple in \cref{fig:MinimalModels:C3}(b,e)] is chosen to be sufficiently large to contain the lobes of red NLs, the principal gap does not close on that ellipsoid, keeping the Wilson-loop winding invariant.
We confirm this by explicitly computing the Wilson-loop spectra on the enclosing ellipsoids of the perturbed models, plotted in \cref{fig:MinimalModels:C3}(c,f).

We briefly comment on the change of the NL configurations from \cref{fig:MinimalModels:C6AA}(b) and \cref{fig:MinimalModels:C6BB}(b) to \cref{fig:MinimalModels:C3}(b,e), respectively.
In both cases, the central NL stretched along $\Gamma A$ initially carries quaternion charge $-1$ (corresponding to Berry phase $2\pi$~\cite{Wu:2019}, i.e., it is a quadratic NL) before switching on the trigonal perturbation.
Let us first consider the case of type-$\mathsf{A}$ TPs [$\SW{2}=1$, \cref{fig:MinimalModels:C3}(a--c)].
Here, as soon as the perturbation is switched on, all three segments of the central NL, i.e., in the orange, red, as well as blue $k_z$-range, split into four NLs with Berry phase $\pi$ each.
One NL of the quadruplet is still pinned to the $\Gamma A$ line, while the other three NLs are related by the $C_3$ symmetry and remain attached to the TPs. 
As a result, the hegaonal type-$\mathsf{A}$ TPs with no attached NL arcs have transformed into trigonal type-$\mathsf{B}$ TPs with three attached NL arcs in each gap, as detailed in \cref{fig:MinimalModels:C3}(b).

For the hexagonal model with type-$\mathsf{B}$ TPs [$\SW{2}=0$, \cref{fig:MinimalModels:C3}(d--f)], the process is similar; in particular, one observes the same splitting of the central quadratic NL into four linear NLs around the HSL. However, in this case the hexagonal NL configuration starts out with six attached NLs per gap, because the hexagonal TPs have been type~$\mathsf{B}$.
Here the three new NL arcs either annihilate with three of the original NL arcs (in the principal gap, red) or combine with the original NL arcs to form intersecting nodal chains (other two gaps, orange and blue)~\cite{Bzdusek:2016}. 
As a result, we find that the hexagonal type-$\mathsf{B}$ TPs have been transformed into trigonal type-$\mathsf{B}$ TPs with three attached NLs per gap, as predicted in Ref.~\onlinecite{Lenggenhager:2021:MBNLs}. 
The intersection of the orange and of the blue NLs visible in \cref{fig:MinimalModels:C3}(e) is stablized by the non-Abelian band topology in the presence of $m_v$ and $\mcP\mcT$ symmetry, as discussed in Ref.~\onlinecite{Wu:2019}.

\begin{figure*}
    \centering
    \includegraphics{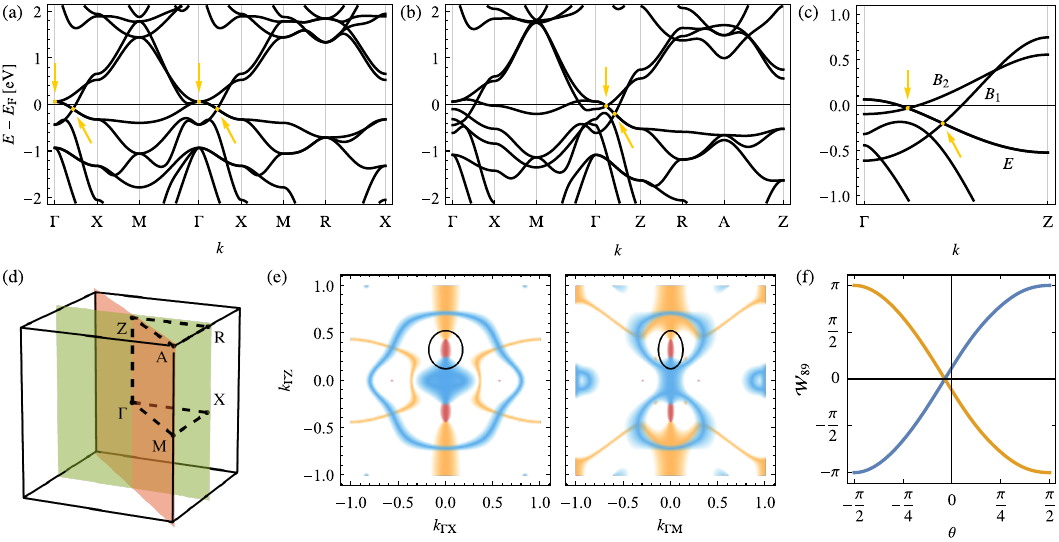}
    \caption{
        Bulk properties of the compound \ce{Sc3AlC}.
        The band structure (a) without any strain and (b) with $6.6\%$ uniaxial compressive strain in $z$-direction, which generates a triple-point pair on the $\Gamma Z$ line. Triple points are indicated by yellow arrows.
        (c) Closeup of the band structure near the triple-point pair with the relevant bands labelled by their irreducible co-representations.
        (d) The Brillouin zone of the tetragonal lattice of strained \ce{Sc3AlC}. The high-symmetry points and mirror planes are indicated.
        (e) Nodal lines in the relevant three gaps (orange, blue, red according to increasing energy) determined from the Wannier-tight-binding model for the mirror planes shown in panel (d). The triple point is clearly type~$\mathsf{A}$. The cross sections of the ellipsoid on which we compute the second Stiefel-Whitney monopole charge with the mirror planes are indicated by black ellipses.
        (f) Wilson-loop spectrum of the relevant bands computed on the ellipsoid indicated in panel (e); the single winding implies $\SW{2}=1$.
    }
    \label{fig:Sc3AlC:bulk}
\end{figure*}

\section{Material Examples}\label{Sec:Materials}

In this section we discuss two concrete material examples that illustrate the introduced phenomenology.
While these examples demonstrate that the higher-order topology of TPs can arise in crystalline solids, the large values of strain required to realize the presented band structures imply that our particular materials predictions are not amenable to experimental studies, and that further research is needed to find realistic material candidates.
Note that in \cref{sec:spinful} we briefly discuss the higher-order bulk-boundary correspondence for TPPs in \emph{spinful systems}.

Our discussion is structured as follows.
In \cref{Sec:Materials:StiefelWhitney} we show that the compound \ce{Sc3AlC} subjected to large uniaxial strain exhibits a TPP with type-$\mathsf{A}$ TPs on the $Z\Gamma Z$ line and no interfering NLs.
Based on first-principles calculations and an \emph{ab-initio} tight-binding model, we compute the 2SW monopole charge and the hinge-charge jump.
We find that the 2SW monopole charge takes value $\SW{2}=1$ and that a fractional hinge-charge jump $\Delta\Qc=\tfrac{e}{4}$ is present, in agreement with the values predicted by \cref{tab:HOBBC:Qjump}.
This demonstrates the higher-order bulk-boundary correspondence introduced in \cref{Sec:HOBBC}.

\begin{figure*}
    \centering
    \includegraphics{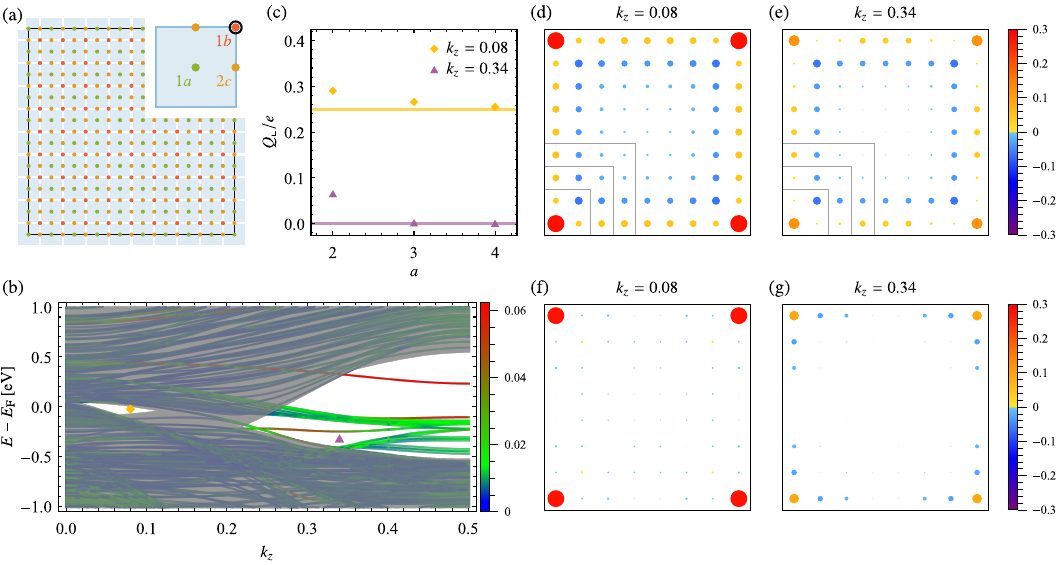}
    \caption{
        Hinge-charge jump observed in the nanowire geometry of \ce{Sc3AlC}.
        (a) Cross section of the nanowire geometry (extended along the $z$-direction) with $9.5\times 9.5$ unit cells. The unit cells are shown as blue shaded squares, and the boundary of the nanowire indicated with a black frame. Two-dimensional Wyckoff positions (WPs) are shown as green, red and orange points with their labels in the inset that shows an enlarged single unit cell. The scandium atoms project onto WPs $1b$ and $2c$, the aluminum atoms onto $1a$ and the carbon atoms onto $1b$. Note the $C_4$ symmetry of the cross section with the center at WP $1b$. The black circle in the enlarged unit cell denotes the location of the ionic charge of $8\abs{e}$ per unit cell.
        (b) Band structure of the nanowire (colored lines) with cross section shown in panel (a) as a function of momentum ($k_z$) in the hinge Brillouin zone. The projected bulk bands (transparent gray) are overlayed on the nanowire spectrum. 
        The coloring of the bands (see legend) encodes the inverse participation ratio, with a larger value indicating localization on fewer sites. Consequently hinge-localized states are colored red, surface-localized states green, and bulk states blue.
        The two energy gaps to the left and to the right of the studied triple-point pair are marked by a yellow square ($k_z=0.08$) and by a purple triangle ($k_z=0.34$), respectively.
        (c) Integrated corner charge (corresponding to the hinge at the selected values of $k_z$) for the two $k_z$-cuts with filling as indicated in panel (b) as a function of the side length $a$ of the square integration region [cf.\ gray squares in panels (d-e)].
        The corner charge for $k_z=0.08$ converges to $\tfrac{e}{4}$ and for $k_z=0.34$ to $0$ (yellow vs.~purple solid lines).
        (d,e) Charge distribution coarse-grained over one unit cell for $k_z=0.08$ and $0.34$, respectively. The magnitude of the charge (in units of $e$) is shown by both the area of the circles as well as the color (see legend on the right).
        (f,g) Charge distribution for the same values of $k_z$ after removing the contributions from the edge based on slab calculations (cf.~\cref{App:edge-signal_removal}). In (f) the localization of a non-zero charge on the corners is clearly visible.
    }
    \label{fig:Sc3AlC:nanowire}
\end{figure*}

In \cref{Sec:Materials:Euler} we study a material example with a NL segment carrying nontrivial Euler monopole charge.
The TP material \ce{Li2NaN}~\cite{Jin:2019,Lenggenhager:2021:MBNLs} has a single pair of inversion-related type-$\mathsf{A}$ TPs along the $A\Gamma A$ line.
Based on an \emph{effective} tight-binding model~\cite{Lenggenhager:2021:MBNLs}, we show that one of the NL segments carries Euler monopole charge $\abs{\Eu}=2$.
Furthermore, we demonstrate that this leads to a topological obstruction preventing the removal of that NL segment when colliding the two TPs at the $A$ point, instead leading to a conversion into a nodal ring. 

\subsection{Nontrivial Stiefel-Whitney class in strained \texorpdfstring{\NoCaseChange{\ce{Sc3AlC}}}{Sc3AlC}}\label{Sec:Materials:StiefelWhitney}

We first consider \ce{Sc3AlC}~\cite{Sc3AlC}, which, on the level of a Wannier tight-binding model, manifests the bulk-hinge correspondence for TPPs derived in \cref{Sec:HOBBC:3D}.
The compound has a cubic crystal structure with SG $Pm\bar{3}m$ (No. 221) and exhibits a three-fold degenerate touching point at $\Gamma$ and TPs on the $\Gamma X$, $\Gamma Y$ and $\Gamma Z$ lines (which are all equivalent), see \cref{fig:Sc3AlC:bulk}(a).
Furthermore, the spin-orbit coupling is expected to be small due to the elements involved being light, therefore, we can neglect it and treat the electrons as spinless.
Then, the system possesses $\mcP\mcT$ symmetry squaring to $+\id$.
Applying uniaxial compressive strain along one of the equivalent crystal axes (which we choose to be the $z$-direction) removes the TPs on the $\Gamma X$ and $\Gamma Y$ lines leaving only those on the $\Gamma Z$ line, and splits the touching point at $\Gamma$ into two additional TPs on the $\Gamma Z$ line, cf.~\cref{fig:Sc3AlC:bulk}(b) for approximately $6.6\%$ strain.
The result is a configuration of two inversion-symmetry-related TPPs on the $\Gamma Z$ line with one TPP shown in \cref{fig:Sc3AlC:bulk}(c).
The strained material has a simple tetragonal crystal structure with SG $P4/mmm$~(No.~123).

We study the material by first obtaining the band structure and wave functions from density functional theory (DFT) calculations with the projected augmented wave (PAW) method implemented in the Vienna ab initio simulation package (VASP)~\cite{Kresse:1996,Kresse:1999} with generalized gradient approximation (GGA) and Perdew-Burke-Ernzerhof (PBE) approximation ~\cite{Perdew:1996}.
We use a $\Gamma$-centered $8 \times 8 \times 8$ $k$-mesh.
Uniaxial compressive strain in $z$-direction is modelled by reducing the lattice constant in that direction by the appropriate amount.
The lattice structure of \ce{Sc3AlC} is obtained from the Materials Project database~\cite{Jain:2013} (material identifier mp-4079~\cite{Sc3AlC}) with lattice vectors $a=4.512\,\angstrom$ ($a=4.512\,\angstrom$ and $c=4.212\,\angstrom$ after strain).
We construct Wannier functions using \textsc{Wannier90}~\cite{Pizzi:2020} resulting in a Wannier tight-binding model with $s,p,d$ orbitals of \ce{Sc} and $s,p$ orbitals of \ce{Al} and \ce{C}.
We process disentanglement with a frozen window from $-20\,\textrm{eV}$ to $4\,\textrm{eV}$ relative to $E_\mathrm{F}$ but do not perform maximum localization~\cite{Marzari:2012}.
The hoppings of the Wannier model are symmetrized~\cite{Liu:Symmetrization,Liu:SymGithub} in real space.

To check for conflicting NLs in the BZ, we perform additional DFT calculations on the two inequivalent mirror planes $X\Gamma Z$ and $M\Gamma Z$.
The results are shown in \cref{fig:Sc3AlC:bulk}(e) and we observe that the only NLs in the principal gap are the two NL segments (red) spanning between the TPs of each TPP.
In particular, there are no additional NLs attached to the TPs off the HSL, confirming that the TPs are type~$\mathsf{A}$, in agreement with \cref{tab:TP-MPGs}.
This allows us to define an ellipsoid [whose intersections with the two mirror planes are indicated by the black ellipses in \cref{fig:Sc3AlC:bulk}(e)] enclosing the TPP on which the relevant 2SW monopole charge is defined.
Using the Wannier tight-binding model and the \Python{} package \textsc{Z2Pack}~\cite{Soluyanov:2011,Gresch:2017} we compute the Wilson-loop spectrum on the ellipsoid, cf.~\cref{fig:Sc3AlC:bulk}(f), and find that it winds once, therefore indicating $\SW{2}=1$.

We further proceed to study the hinge-charge jump in strained \ce{Sc3AlC}.
First, we calculate the traces of matrix representations obtained from VASP to get the ICRs of the energy bands at HSPs in the first BZ with the help of \textsc{Irvsp}~\cite{Gao:2020}.
The co-representations at HSLs are then inferred using compatibility relations obtained from the Bilbao crystallographic server (BCS)~\cite{Aroyo:2011,Aroyo:2006a,Aroyo:2006b,Elcoro:2017}.
The ICRs of the bands forming the TPP are found to be $(E;B_1,B_2)$, cf.~\cref{fig:Sc3AlC:bulk}(c).
Based on \cref{tab:HOBBC:Qjump}, the knowledge of the rotational symmetry $C_4$, the HSL $\Gamma Z$ and the ICRs of the little co-group of that line allows us to predict the fractional part of the hinge-charge jump $\Delta \Qc$ to~be~$+\tfrac{e}{4}$.

\begin{figure*}[t]
    \centering
    \includegraphics{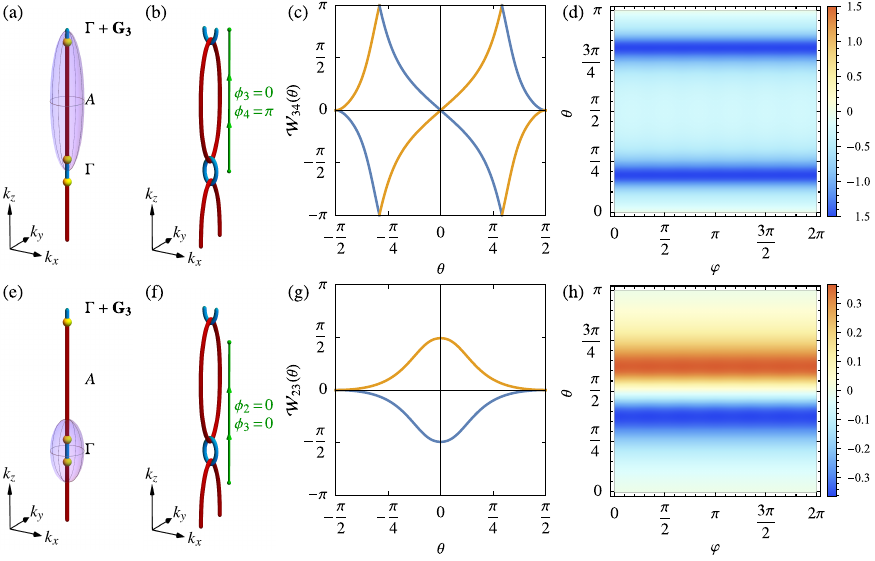}
    \caption{
        Verification of the Euler monopole charge $\abs{\Eu}=2$ ($\Eu=0$) on the red (blue) nodal-line segment in \ce{Li2NaN}, based on the four-band tight-binding model of Ref.~\onlinecite{Lenggenhager:2021:MBNLs}.
        Panels (a--d) show illustrations and data for the red nodal-line segment (centered at $A$) and panels (e--h) for the blue segment (centered at $\Gamma$).
        (a,e) Nodal line and triple point configuration in the first one-and-a-half Brillouin zones $A\Gamma A \Gamma$. Nodal lines in the second and third band gap are shown in red and blue, respectively; triple points are shown in yellow. The purple ellipsoid defines the surface on which we compute the Euler class.
        (b,f) Linking structure of the corresponding nodal chain configuration after applying approximately $5\%$ tensile strain in $y$-direction. The green line shows the closed contour on which the Berry phases are computed.
        (c,g) Spectrum of the Wilson loop operators $\mathcal{W}_{ij}(\theta)$ of bands $i$ and $j$ as indicated in the axis labels of constant latitude contours on the purple ellipsoid in panel (a,e), respectively, parametrized by the latitude angle $\theta$.
        (d,h) Euler curvature $F(\theta,\varphi)$ on the purple ellipsoid in panel (a,e), respectively, calculated using the algorithm from Ref.~\onlinecite{Bouhon:2020}. The curvature integrates to $\abs{\Eu} = 2$ in (d) and to $\Eu = 0$ in (h).
    }
    \label{fig:Li2NaN:EulerClass}
\end{figure*}

To compute the hinge-charge jump explicitly and to verify the above prediction, we use the \Python{} package \textsc{PythTB} to construct a nanowire with a $C_4$-symmetric cross section of $9.5\times 9.5$ unit cells from the bulk Wannier tight-binding model, see \cref{fig:Sc3AlC:nanowire}(a).
Recall that the placement of the ionic charge in the unit cell does not influence the hinge charge \emph{jump}.
Therefore, we choose it such that it simplifies our calculations: we assume all ionic charge of a unit cell to be concentrated at WP $1b$ (at the corner of the 2D projection of the unit cell), which is also the location of the the center of the cross-section, cf.~\cref{fig:Sc3AlC:nanowire}(a).
The band structure of the nanowire is shown in \cref{fig:Sc3AlC:nanowire}(b) as a function of the remaining momentum $k_z$.
Each state $\psi$ is colored according to the inverse participation ratio $\sum_i p_i(\psi)^2$ where $p_i(\psi)$ is the probability of finding an electron in state $\psi$ at site $i$ in the 2D cross section.
Together with the gray overlay of the bulk states, we can easily identify the in-gap surface (green) and hinge (red) states.

We now select two $k_z$-values at which the principal gap is open in the spectrum of the nanowire [yellow diamond and purple triangle in \cref{fig:Sc3AlC:nanowire}(b)].
Assuming that all states below the indicated gap are occupied, we compute the total charge distribution in the nanowire.
The results for $k_z=0.08$ and $k_z=0.34$, after coarse-graining over a unit cell (see \cref{App:coarse-graining}), are shown in \cref{fig:Sc3AlC:nanowire}(d,e), respectively.
Integrating over successively larger square regions at one of the corners [cf.\ gray squares in \cref{fig:Sc3AlC:nanowire}(d,e)], we observe that the corner charge (corresponding to the hinge charge of the 3D model at the selected value of $k_z$) converges to $\tfrac{e}{4}$ and $0$ for $k_z=0.08$ and $k_z=0.34$, respectively.
This verifies that the jump is $\Delta \Qc = \tfrac{e}{4}$, as predicted.

The localization of the charge at the corners is already visible in \cref{fig:Sc3AlC:nanowire}(d,e) and the edge charge (corresponding to the surface charge of the 3D model) is clearly vanishing; however, there are strong oscillations of the charge on the edges.
These oscillations are due to trivial surface states, as is revealed by computing the charge distribution on a slab (data not shown).
Using the information about the charge distribution on the slab, we remove (see \cref{App:edge-signal_removal}) this edge signal and reveal the strong localization of the net charge on the corners, cf.~\cref{fig:Sc3AlC:nanowire}(f,g).
Note that this \emph{removal} of the charge oscillations is performed in a charge neutral way, i.e.\ by changing neither the corner nor the edge charge.

\subsection{Nontrivial Euler monopole charge in \texorpdfstring{\NoCaseChange{\ce{Li2NaN}}}{Li2NaN}}\label{Sec:Materials:Euler}

We now turn our attention to \ce{Li2NaN}.
Reference~\cite{Lenggenhager:2021:MBNLs} identified the compound \ce{Li2NaN} as an ideal candidate to observe the conversion of TPs to multi-band nodal links, and developed its description using an effective four-band model which we assume in the following discussion.
The compound exhibits a single pair of inversion-related TPs and no additional NLs close to the central NL (along the reciprocal lattice vector $\vb{G_3}\parallel k_z$) and TPs, as shown in \cref{fig:Li2NaN:EulerClass}(a).
Therefore, this is also an ideal candidate for a TP-induced Euler monopole charge.
The Berry phases of each band of the tight-binding model in $k_z$-direction can be easily computed using the Wilson-loop method, and we obtain $(0,0,0,\pi)$, with the non-trivial value carried only by the highest-energy band.
Based on the arguments presented in \cref{Sec:Monopole:Euler}, we therefore predict the red NL segment (connecting TPs in two adjacent BZs) to carry Euler monopole charge $\abs{\Eu}=2$ and the blue one $\Eu=0$.
We verify these predictions in the effective four-band tight-binding model by computing the Euler class on ellipsoids enclosing the corresponding NL segment, such as shown in \cref{fig:Li2NaN:EulerClass}(a)] using three independent methods: (i) via the linking numbers obtained when applying strain to get proper linked nodal rings, (ii) via Wilson-loop spectra and (iii) directly via the Euler curvature~\cite{Bouhon:2020}.

We first discuss the red NL segment.
Following the argument of \cref{Sec:Monopole}, we reduce the $C_6$ rotational symmetry of \ce{Li2NaN} down to $C_2$ by applying tensile strain of approximately $5\%$ in $y$-direction.
This turns the NL segment into linked red and blue NL rings, cf.~\cref{fig:Li2NaN:EulerClass}(b).
The Berry phases $\phi_{3,4}$ of the bands forming the blue NL (i.e., bands $3$ and $4$) on the green contour satisfy $\phi_3+\phi_4=\pi\mod 2\pi$.
Thus, according to \cref{Sec:Monopole:Euler}, the red nodal ring and consequently also the red NL segment in \cref{fig:Li2NaN:EulerClass}(a) carry Euler monopole charge $\abs{\Eu}=2$.
We arrive at the same conclusion based on the Wilson-loop spectrum computed on the purple ellipsoid shown in \cref{fig:Li2NaN:EulerClass}(a).
The double winding implies~\cite{Bzdusek:2017} $\abs{\Eu}=2$.

To verify this prediction independently, recall that the Euler class can be brought~\cite{Nakahara:1990,Zhao:2017} to a form analogous to the Chern number, i.e., an integral of a curvature over the base manifold.
By adopting a Hilbert-space basis for which $\mcP\mcT$ is represented by complex conjugation~\cite{Bouhon:2020}, the eigenstates can be gauged to be purely real.
For two real Bloch bands $\ket{u_1(\vec{k})}$ and $\ket{u_2(\vec{k})}$ (which could possibly be degenerate with each other, but which must be separated by energy gaps from all other bands) one then defines the \emph{Euler curvature}
\begin{equation}
    \vec{F}(\vec{k}) = \mel{\grad_{\vec{k}}u_1(\vec{k})}{\times}{\grad_{\vec{k}}u_2(\vec{k})},
\end{equation}
i.e., as the off-diagonal component of the two-band non-Abelian Berry-Wilczek-Zee connection~\cite{Zhao:2017,Bouhon:2020}.
Integrating the curvature over a closed surface $S^2$ gives the Euler class
\begin{equation}
    \Eu = \frac{1}{2\pi}\int_{S^2}\dd{\vec{S}}\cdot\vec{F}(\vec{k}) \in\mathbb{Z}.
\end{equation}
Note that the Euler class is independent of the parametrization of $S^2$, such that we can parametrize the relevant ellipsoid by the spherical coordinates $(\theta,\varphi)\in[0,\pi]\times[0,2\pi)$ and define
\begin{equation}
    F(\theta,\varphi) = \mel{\grad_{(\theta,\varphi)}u_1(\theta,\varphi)}{\times}{\grad_{(\theta,\varphi)}u_2(\theta,\varphi)}.
\end{equation}
We utilize the algorithm of Ref.~\onlinecite{Bouhon:2020} implemented in \Mathematica{} to compute the Euler curvature on the ellipsoid shown in \cref{fig:Li2NaN:EulerClass}(a).
The resulting $F(\theta,\varphi)$ is plotted in \cref{fig:Li2NaN:EulerClass}(d). Integration over the full ellipsoid confirms that $\abs{\Eu}=2$.

\begin{figure}
    \centering
    \includegraphics{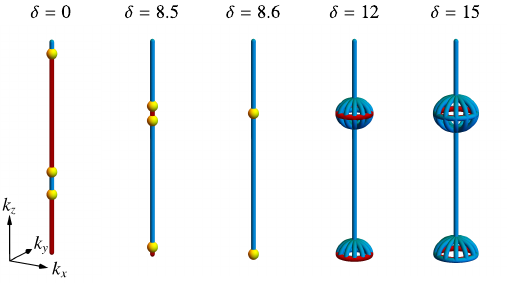}
    \caption{
        Demonstration of the topological obstruction of the red nodal-line segment in \ce{Li2NaN} due to the nontrivial Euler monopole charge.
        The nodal lines (obtained from the tight-binding model in Ref.~\onlinecite{Lenggenhager:2021:MBNLs}) near the rotation axis $A\Gamma A\Gamma$ spanning over one-and-a-half Brillouin zones are shown, with nodes in the second and third band gap shown in red and blue, respectively.
        The two triple points (yellow) can be forced to collide as described in the text by tuning a parameter $\delta$ of the tight-binding model.
        The triple points collide at $\delta=-8.6$, at which point the red NL segment has been reduced to only a single touching point at $A$.
        Increasing $\delta$ even further, a horizontal red nodal ring forms in the $ALH$ plane.
    }
    \label{fig:Li2NaN:TPcollision}
\end{figure}

The nontrivial value of $\Eu$ implies a topological obstruction as discussed in \cref{Sec:Monopole:Euler}:
the red NL segment cannot be removed completely as long as $\mcP\mcT$ symmetry is preserved.
We verify this explicitly in the context of the tight-binding model of Ref.~\onlinecite{Lenggenhager:2021:MBNLs}.
The 1D ICR involved in the TP has most of its weight on the nitrogen $p_z$ orbital, such that we can move the TPs towards the $A$ point by reducing the onsite energy of that orbital.
We denote the change in energy by $\Delta\varepsilon_{\mathrm{N}p_z}=-\delta\,\mathrm{eV}<0$.
As shown in \cref{fig:Li2NaN:TPcollision}, increasing $\delta$ shrinks the red NL segment until the two TPs collide at $A$.
At that stage, the band degeneracy in the red energy gap has reduced to a single touching point at $A$. However, the Euler monopole charge prevents opening of the energy gap; indeed increasing $\delta$ even further results in a conversion of the red TP segment into a nodal ring in the horizontal $k_z = \pi$ plane.
Crucially, the Euler monopole charge persists in $\mcP\mcT$-symmetric systems even if the horizontal $m_z$ symmetry was removed from the model. 

We remark that the critical three-band touching obtained by colliding the two TPs at the $A$ point shares certain similarities with the very recently discussed Euler topology of topological acoustic triple points by Ref.~\onlinecite{Park:2021}.
The notable difference is that in our case the Euler-point degeneracy requires fine-tuning of the model parameters, whereas Ref.~\onlinecite{Park:2021} finds this to be a \emph{generic} feature of the acoustic phonon branches.

Finally, we also comment on the blue NL segment.
We can use the same three arguments to show that $\Eu=0$ in this case.
For the argument via the linking, the relevant change is that now we need to consider bands $2$ and $3$, whose Berry phases along the green contour [cf.~\cref{fig:Li2NaN:EulerClass}(f)] are both vanishing.
Therefore, $\Eu=0$ despite the linking of the rings (see \cref{Sec:Monopole:Euler} and \cref{App:NLBZ}).
We arrive at the same conclusion by observing that the Wilson loop spectrum in \cref{fig:Li2NaN:EulerClass}(g) does not wind.
Finally, in \cref{fig:Li2NaN:EulerClass}(h), we present the Euler curvature for the purple ellipsoid shown in \cref{fig:Li2NaN:EulerClass}(e).
The Euler curvature turns out to be anti-symmetric with respect to reflection at $\theta=\pi/2$, such that integrating over the whole ellipsoid results again in $\Eu=0$.

\clearpage
\section{Extension to non-symmorphic \texorpdfstring{\\}{}space groups}\label{Sec:non-symmorphic}
Up to now, we have simplified the presented analysis by excluding non-symmorphic SGs, i.e., SGs containing symmetries like screw rotations and glide planes.
In this section we discuss these excluded cases, arguing that the inclusion of non-symmorphic elements in the space group does not change the bulk-hinge correspondence.
In particular, we reveal that \cref{tab:HOBBC:Qjump} applies to triple-point pairs in non-symmorphic SGs as well, with the only differences being that (\emph{i}) the first row of the table \enquote{$C_n$} should be interpreted as only the point-group part of the possibly present non-symmorphic screw symmetry (i.e., without the translation), and (\emph{ii}) the mapping of ICRs of the little group to the ICRs of the little \emph{co-group} should be performed as described in Ref.~\onlinecite{Lenggenhager:2022:TPClassif}.

To motivate the advertised result, note that our parallel work~\cite{Lenggenhager:2022:TPClassif} shows that non-symmorphic symmetries do not alter the classification of TPs.
More precisely, as long as the HSL supports TPs (i.e., both 1D and 2D ICRs of the little group exist -- due to non-symmorphicity this property may be lost for certain HSLs on the BZ boundary), the symmetry constraints on the Bloch Hamiltonian $\Ham(\vec{k})$ due to some corepresentation of the little group $\LG{\vec{k}}$ of the HSL supporting TPs, are \emph{equivalent}~\cite{Lenggenhager:2022:TPClassif} to the constraints due to a corresponding corepresentation of the little co-group $\LcG{\vec{k}}$ (which consists of the corresponding point-group symmetries, i.e., without the translation). 
However, although non-symmorphic symmetries have trivial implications for the TP classification, they may non-trivially affect the rotation eigenvalues that enter the symmetry-indicator formulas for the fractional corner/hinge charges (in particular if the rotational symmetry is replaced by a screw rotation).

Here, we argue in two steps that even in the case of a screw rotation, there is no change to the bulk-hinge correspondence derived in \cref{Sec:HOBBC:3D}.
First, we discuss in \cref{Sec:non-symmorphic:screw_hinge_charge} how to apply the symmetry-indicator formulas for corner charges in $C_n$-symmetric 2D systems~\cite{Benalcazar:2019,Takahashi:2021} to compute momentum-resolved hinge-charges in a wire-geometry of 3D crystals with a screw rotation (instead of pure rotation) symmetry.
Along the way, we derive that the symmetry-indicator formulas for corner charges (without assuming time-reversal symmetry, cf.~\cref{App:SIformulas}) are symmetric under cyclic permutation of the rotation eigenvalues.
Next, we consider in \cref{Sec:non-symmorphic:TPP_bulk-hinge} the mapping of ICRs between the little group and the little co-group of the HSL on which the TPs lie.
We show that this mapping is compatible with the application of the symmetry-indicator formulas for the hinge charge, allowing us to establish the generalization of the bulk-hinge correspondence to TPPs protected by a screw symmetry.

\subsection{Fractional hinge charges due to screw rotational symmetry}\label{Sec:non-symmorphic:screw_hinge_charge}

Consider any SG with a non-symmorphic rotational symmetry around the $z$-axis, $\{C_n|\vec{w}\}$, where $n=2,3,4,6$ and $\vec{w}$ is the non-symmorphic translation.
Note that the $x,y$ components of $\vec{w}$, i.e., the ones perpendicular to the rotation axis, can be removed by shifting the rotation axis (this may potentially result in a change of the fractional translations associated with other symmetry elements in the SG, but this is not relevant for our argument).
We end up with the screw symmetry $S_n=\{C_n|\tfrac{m}{n}\vec{e}_z\}$ with $m\in\mathbb{Z}$ that together with the translations $\vec{t}\in\TG$ by lattice vectors generate a subgroup $G$ of the full SG (i.e., we drop the potentially present time-reversal symmetry as well as all point-group operations that are not generated by $C_n$).

Let $D$ be the representation of $G$ in which the eigenfunctions of the Hamiltonian transform and $D_\vec{k}$ its restriction to the $\vec{k}$-sector, i.e., a representation of the little group $G^\vec{k}$.
Then, for any $\{R|\vec{v}\}\in G$,
\begin{equation}
    D_\vec{k}(\{R|\vec{v}\})\Ham(R^{-1}\vec{k})D_\vec{k}(\{R|\vec{v}\})^{-1} = \Ham(\vec{k}).
    \label{eq:non-symmorphic_constraint}
\end{equation}
In analogy with \cref{Sec:HOBBC:3D}, we take a 2D cut at fixed $k_z$ and define the 2D Hamiltonian $\Ham_{k_z}(k_x,k_y)=\Ham(\vec{k})$, which still satisfies \cref{eq:non-symmorphic_constraint} at any $\vec{k}_\mathrm{2D}=(k_x,k_y)$ and for any $\{R|\vec{v}\}\in G$.
In particular, for any 2D HSP $\Pi$ invariant under $S_n^\ell$ for some power $\ell$ of the screw symmetry, where $0<\ell<n$ is a divisor of $n$, $\Ham_{k_z}$ commutes with the operator $D_\vec{k}(S_n^\ell)$ which admits eigenvalues
\begin{equation}
    s^{(n/\ell)}_p = \e^{\i\frac{\ell}{n}\left[-mk_z + 2\pi(p-1)\right]},\quad p = 1,2,\dotsc,\frac{n}{\ell}.
    \label{eq:screw_eigenvalues}
\end{equation}
The above follows because $D_\vec{k}(S_n^\ell)$ is a representation and $(S_n^\ell)^{n/\ell}=S_n^n=\{\id|m\vec{e}_z\}$, where $m\vec{e}_z\in\TG$ is a lattice translation and therefore $D_\vec{k}(\{\id|m\vec{e}_z\})=\e^{-\i mk_z}\id$, which has eigenvalues $\e^{-\i mk_z}$; thus, eigenvalues of $D_\vec{k}(S_n^\ell)$ are $(n/\ell)^\textrm{th}$ roots of $\e^{-\i mk_z}$.

Next, we define the operator
\begin{equation}
    D_{\vec{k}_\mathrm{2D}}'(C_n) = \e^{\i\frac{1}{n}\left(mk_z + 2\pi p'\right)}D_\vec{k}(S_n),
    \label{eq:screw_rotation_mapping}
\end{equation}
which is a symmetry of the 2D Hamiltonian for any $p'\in\mathbb{Z}$:
\begin{equation}
    D_{\vec{k}_\mathrm{2D}}'(C_n)\Ham_{k_z}(C_n^{-1}\vec{k}_\mathrm{2D})D_{\vec{k}_\mathrm{2D}}'(C_n)^{-1} = \Ham_{k_z}(\vec{k}_\mathrm{2D}).
\end{equation}
We observe that $D_{\vec{k}_\mathrm{2D}}'(C_n)$ [together with $D_{\vec{k}_\mathrm{2D}}'(\{\id|\vec{t}_\mathrm{2D}\})=D_\vec{k}(\{\id|\vec{t}_\mathrm{2D}\})$] furnishes a representation of the 2D space group $pn$ (i.e., the 2D symmetry group generated by $C_n$ rotation with respect to a point and by translations).
This implies that the Hamiltonian $\Ham_{k_z}$ describes a 2D system with $C_n$ symmetry, and that the rotation eigenvalues of its energy bands at a $C_{n/\ell}$-invariant HSP $\Pi$ are
\begin{equation}
    r^{(n/\ell)}_{p,p'} = \e^{2\pi\i\frac{p+p'-1}{n/\ell}} 
\end{equation}
where we recognize $\e^{2\pi\i\frac{p+p'-1}{n/\ell}} = \Pi^{(n/\ell)}_{p+p'}$ as defined in \cref{eq:def_eigenvalues}.
With these eigenvalues, the symmetry-indicator formulas for the corner charge in 2D systems with rotational symmetry $C_2$, $C_3$, $C_4$, and $C_6$~\cite{Benalcazar:2019,Takahashi:2021}, e.g., \cref{App:eq:Qcorner3,App:eq:Qcorner4,App:eq:Qcorner6}, can be directly applied to compute the fractional hinge charges even for non-symmorphic 3D~SGs.

Note that a non-trivial symmetry of the formulas for the corner charges can now be easily deduced.
Namely, the non-uniqueness of \cref{eq:screw_rotation_mapping} due to the freedom of choosing $p'\in\mathbb{Z}$ implies that the symmetry-indicator formulas for the corner charges must be symmetric under \emph{cyclic permutations of the rotation eigenvalues} (which correspond to the replacement $p'\mapsto p'+1$).
In particular, this applies irrespective of the non-symmorphicity, as can be seen by setting $m=0$ in the symmetry $S_n$ above.
We are not aware whether this symmetry has been previously pointed out, but it is easily verified explicitly for the symmetry class~A (i.e., without time-reversal symmetry), which we checked using the generating Wannier configurations discussed in Appendix~A of Ref.~\onlinecite{Takahashi:2021}.
Time-reversal symmetry implies that the number of bands with rotation eigenvalues $\Pi_p^{(n/\ell)}$ and $\cconj{(\Pi_p^{(n/\ell)})}$ must match; however, this matching will generically be lost after performing a cyclic permutation of rotation eigenvalues.
For this reason, the symmetry-indicator formulas derived under the assumption that time-reversal symmetry is present, i.e., \cref{App:eq:Qcorner4_TRS,App:eq:Qcorner6_TRS}, do not manifest the symmetry under cyclic permutations of the rotation eigenvalues.
If $n$ is even, a reduced symmetry of the symmetry-indicator formulas under $p'\mapsto p' + \tfrac{n}{2}$ remains.

In summary, we have shown that the fractional hinge charges of non-symmorphic crystals in wire-geometry with screw symmetry $\{C_n|\tfrac{m}{n}\vec{e}_z\}$ are characterized by the same symmetry-indicator formulas that apply for corner charges in 2D systems with rotational symmetry $C_n$, after the $k_z$-dependence of the screw eigenvalues is removed.
It remains to be shown that this removal is compatible with the identification of ICRs of the little group with those of the little co-group described in Ref.~\onlinecite{Lenggenhager:2022:TPClassif}.
This task is left for \cref{Sec:non-symmorphic:TPP_bulk-hinge}.

\subsection{Bulk-hinge correspondence of TPPs}\label{Sec:non-symmorphic:TPP_bulk-hinge}

The characterization of TPs reproduced in \cref{tab:TP-MPGs} is phrased in terms of ICRs of the little \emph{co-group} $\LcG{\vec{k}}$ of the HSL on which the TPs lie, even though the symmetry constraints on the nodal-line structure near the TPs involve the ICRs of the little \emph{group} $\LG{\vec{k}}$.
For symmorphic SGs, this is explained by the fact that the ICRs of $\LG{\vec{k}}$ restricted to elements $\{R|0\}\in\LG{\vec{k}}$ (forming a group isomorphic to $\LcG{\vec{k}}$) are identical to the ones of $\LcG{\vec{k}}$:
the ICR $\varsigma$ of $\LG{\vec{k}}$ is injectively mapped to the ICR $\rho$ of $\LcG{\vec{k}}$ by
\begin{equation}
    \forall \{R|0\}\in\LG{\vec{k}} : \rho(R) = \varsigma(\{R|0\}).
\end{equation}
For non-symmorphic SGs, on the other hand, this is not the case.
Nevertheless, we have shown in Ref.~\onlinecite{Lenggenhager:2022:TPClassif} that for little groups that support TPs a different mapping between ICRs $\varsigma$ of $\LG{\vec{k}}$ restricted to $\LcG{\vec{k}}$ and ICRs $\rho$ of $\LcG{\vec{k}}$ exists (see Appendix A in Ref.~\onlinecite{Lenggenhager:2022:TPClassif}):
Let $\varsigma^\mathrm{1D}$ be any 1D ICR of $\LG{\vec{k}}$ (e.g., the one that facilitates the formation of the discussed TP), then the ICR $\varsigma$ of $\LG{\vec{k}}$ is injectively mapped to the ICR $\rho$ of $\LcG{\vec{k}}$ by
\begin{equation}
    \forall \{R|\vec{v}\}\in\LG{\vec{k}} : \rho(R) = \varsigma^\mathrm{1D}(\{R|\vec{v}\})^{-1}\varsigma(\{R|\vec{v}\}).
    \label{eqn:mapping-of-ICRs}
\end{equation}
Note that, in contrast to the symmorphic case, the restriction of $\varsigma$ to $\LcG{\vec{k}}$ is \emph{not} a representation of $\LcG{\vec{k}}$ but a \emph{projective} representation.

Here, we are interested in the rotation eigenvalues at HSLs, i.e., the eigenvalues of $\rho(C_{n/\ell})$ at momenta invariant under $C_{n/\ell}$-symmetry.
It follows from \cref{eqn:mapping-of-ICRs} that the eigenvalues $s^{(n/\ell)}$ of $\varsigma(S_{n/\ell})$ and $r^{(n/\ell)}$ of $\rho(C_{n/\ell})$ are related as
\begin{equation}
    r^{(n/\ell)} = \rho^\mathrm{1D}(S_{n/\ell})^{-1}s^{(n/\ell)}.
\end{equation}
However, as a 1D ICR, $\rho^\mathrm{1D}(S_{n/\ell})$ must be one of the eigenvalues defined in \cref{eq:screw_eigenvalues}, i.e., $s^{(n/\ell)}_{\tilde{p}}$ for some $\tilde{p}\in\mathbb{Z}$.
Then,
\begin{equation}
    r^{(n/\ell)} = \e^{\i\frac{1}{n}\left(mk_z - 2\pi(\tilde{p}-1)\right)}s^{(n/\ell)},
    \label{eq:screw_rotation_relationship}
\end{equation}
which is fully compatible with \cref{eq:screw_rotation_mapping} for $p'=1-\tilde{p}$ in the sense that if $s^{(n/\ell)}$ is one of the eigenvalues of $D_\vec{k}(S_n)$, then $r^{(n/\ell)}$ is the corresponding eigenvalue of $D_{\vec{k}_\mathrm{2D}}'(C_n)$.
In particular, we observe that the $k_z$-dependence on the right-hand side of \cref{eq:screw_rotation_relationship} cancels, and $r^{(n/\ell)}$ does not depend on $k_z$.

Given a TPP formed by four bands transforming according to certain ICRs of the little group, the relevant rotation eigenvalues used to compute the hinge-charge jump should be obtained by applying the construction described in \cref{Sec:non-symmorphic:screw_hinge_charge}.
However, the preceding paragraph implies that the same results are obtained if we first map the ICRs of the little group to ICRs of the little co-group [cf.~\cref{eqn:mapping-of-ICRs}] and then apply the symmetry-indicator formulas in \cref{App:eq:Qcorner3,App:eq:Qcorner4,App:eq:Qcorner6} to the rotation eigenvalues obtained from the ICRs of the little co-group.
In fact, the non-uniqueness of $\rho^\textrm{1D}$ in \cref{eqn:mapping-of-ICRs} (if 1D ICRs of $\LG{\vec{k}}$ exist) amounts precisely to the non-uniqueness of $p'$ in \cref{eq:screw_rotation_mapping}, further evincing how the two presented descriptions are two facets of the same argument.
We therefore conclude, that both the TP types and the hinge-charge jumps can be determined from the ICRs of the little co-group, which implies that the bulk-hinge correspondence derived in \cref{Sec:HOBBC:3D} directly applies to non-symmorphic SGs as well.

\section{Triple points in spinful band structures}\label{sec:spinful}

The presented analysis of the higher-order bulk-boundary correspondence of TPPs is easy to generalize to the spinful case.
According to the classification of TPs by Ref.~\onlinecite{Zhu:2016}, in spinful systems without magnetic order, TPs can be protected on the HSLs $\Gamma A$, $KH$ and $K'H'$ of SGs with three-fold rotational symmetry.
Only two magnetic point groups (as little co-groups of HSLs) can protect TPs: $C_{3v}$ ($3m$) resulting in type-$\mathsf{B}$ TPs, and $C_{3v}$ supplemented with $m_z\mcT$ ($\bar{6}'m2'$) resulting in type-$\mathsf{A}$ TPs.
The spinful ICRs of the two MPGs are equivalent, such that we do not need to consider them separately for the discussion of the hinge-charge jump and of the bulk-polarization jump (which only depend on the rotation eigenvalues).
The main difference between the two cases is that the additional anti-unitary symmetry $m_z\mcT$ in $\bar{6}'m2'$ forces the NL arcs (characteristic of type-$\mathsf{B}$ TPs) to coalesce on the rotation axis, resulting in the type-$\mathsf{A}$ TPs.
For both MPGs there are two spinful 1D ICRs $\rho_1^\mathrm{1D}$, $\rho_2^\mathrm{1D}$ and only a single spinful 2D ICR $\rho^\mathrm{2D}$, giving rise to three different TPPs $(\rho^\mathrm{2D};\rho_a^\mathrm{1D},\rho_b^\mathrm{1D})$ with $a,b\in\{1,2\}$.
However, $\rho_1^\mathrm{1D}$ and $\rho_2^\mathrm{1D}$ have identical rotation eigenvalues; therefore, they are not distinguished in the corresponding symmetry indicators, and they all result in the \emph{same} value of jumps $\Delta \Qc$ and $\Delta\vec{P}$.

For simplicity, we restrict the explicit analysis to the case when the $C_3$ rotation center of the sample resides at the $1a$ WP (a straightforward analysis using the symmetry-indicator formulas from Ref.~\onlinecite{Takahashi:2021} reveals that our results remain true if the rotation center resides at WP $1b$ or $1c$).
Similar to \cref{Sec:HOBBC}, we set all ionic charge to WP $1a$.
Then, the corner charge on a 2D cut is given by (cf.~class~A in Ref.~\onlinecite{Takahashi:2021})
\begin{equation}
    \Qc^{(3)} = \frac{e}{3}\left(\left[K_1^{(3)}\right]+\left[K_2^{(3)}\right]+\left[{K_1'}^{(3)}\right]+\left[{K_2'}^{(3)}\right]\right)\mod e.
\end{equation}
The square bracket is defined in the exact same way as in the spinless case, $[\Pi_p^{(n)}] = \# \Pi_p^{(n)} - \#\Gamma_p^{(n)}$; however the labelling of rotation eigenvalues in \cref{eq:def_eigenvalues} is replaced by
\begin{equation}
    \Pi_p^{(n)} = \e^{2\pi\i(p-1)/n}\e^{\pi\i/n},\quad p=1,2,\dotsc,n.
\end{equation}
The bulk polarization can similarly be expressed in terms of the symmetry indicators as~\cite{Watanabe:2020,Takahashi:2021}
\begin{multline}
    \vec{P}^{(3)} = \frac{e}{3}\left(2\left[K_1^{(3)}\right]+\left[K_2^{(3)}\right]+\left[{K_1'}^{(3)}\right]+2\left[{K_2'}^{(3)}\right]\right)\left(\vec{a}_1+\vec{a}_2\right)\\\mod e\vec{R}.
\end{multline}%

By performing an analysis (not appended) similar to the one detailed in \cref{App:Qjump} and \cref{App:noPjump} for the spinless case, we derive the bulk-hinge correspondence for spinful TPPs in triangular geometry.
Independent of the HSL on which the TPP lies and regardless of the WP of the rotation center, the hinge-charge jump is universally found to be
\begin{equation}
    \Delta\Qc = \frac{e}{3} \mod e
\end{equation}
and the jump of the bulk-polarization
\begin{equation}
    \Delta\vec{P} = 0\mod e\vec{R}.
\end{equation}
We observe that even in spinful systems, all TPPs carry fractional hinge-charge jump.

The derived results directly apply to TPPs reported in \ce{WC}-type crystalline materials reported by Ref.~\onlinecite{Zhu:2016}.
In particular, \ce{TaN} exhibits a type-$(\mathsf{A},\mathsf{A})$ TPP characterized by a hinge-charge jump $\tfrac{e}{3}$ along the $\Gamma A$ line.
Furthermore, this material is very close to the ideal semimetallic TPP case, with only small additional Fermi pockets around the $K$-point and $A$-point of the BZ. 
We have considered extracting the predicted hinge-charge jump for \ce{TaN} from first-principles calculations (as we have done for \ce{Sc3AlC} under strain), however the absence of a surface gap has prevented us from doing so.
In the future, a systematic check of the materials that host TPPs for open surface gaps could lead to a better material candidate.

\section{Conclusions and Outlooks}\label{Sec:Conclusion}

In this work we exposed higher-order topological fingerprints of triple nodal points (TPs) in three dimensions (3D). Our focus has been on the \emph{spinless case}, where one encounters an interplay of band nodes with a rich topological structure:
nodal lines (NLs) can be protected by mirror ($m_v$) planes, NLs governed by non-Abelian topology and equippied with monopole charges can be stabilized by space-time-inversion ($\mcP\mcT$) symmetry, and either species of NLs can be pinned to high-symmetry lines (HSLs) by rotation ($C_n$) or antiunitary rotation ($C_n\mcP\mcT$) symmetry.
In particular, triple point pairs [TPPs; formed when a 2D irreducible co-representation (ICR) crosses a pair of 1D ICRs] are semimetallic features where a NL along a HSL is transferred across three adjacent energy gaps.

The rich topological structure of spinless systems supporting TPs prompted us to analyze HSLs with 13 distinct magnetic little co-groups, summarized by \cref{tab:TP-MPGs}.
The main finding of our analysis is that triple-point pairs demarcated by a pair of type-$\mathsf{A}$ or by a pair of type-$\mathsf{B}$ TPs are generally characterized by a \emph{fractional jump of the hinge charge}.
These results apply also to non-symmorphic space groups.
On top of the spinless analysis, in the last section we briefly considered the \emph{spinful case}, where TPs arise in the presence of $C_{3v}$ symmetry, and where a fractional hinge-charge jump is also predicted to be a universal higher-order feature of TPPs.
Note that while three-dimensional Weyl nodes and Dirac nodes exist in both the higher-order~\cite{Ezawa:2018d,Ezawa:2019,Wang:2020b,Ghorashi:2020,Wei:2021,Lin:2018,Wieder:2020,Qiu:2021} and first-order~\cite{Wan:2011,Young:2012} varieties, we report that \emph{all} type-$(\mathsf{A},\mathsf{A}$) and type-$(\mathsf{B},\mathsf{B})$ TPPs are \emph{universally} associated with the higher-order bulk-hinge correspondence. 
Our finding thus overcomes the previous unsuccessful search for a general bulk-boundary characterization of TPs in surface states~\cite{Zhu:2016,Weng:2016,Kim:2018,Winkler:2019}, as conventionally associated with first-order topology.

Although our analysis in the presence of $C_4$-rotational symmetry explicitly assumed a primitive tetragonal Bravais lattice, the result in \cref{tab:HOBBC:Qjump} generalizes \emph{with no alterations} to the body-center tetragonal case.
To understand this, note that the derivation of the hinge-charge jump (cf.~\cref{app:TP-C4}) is based on considering a two-dimensional (2D) system with $C_4$ symmetry.
It is found that (1) the ICRs at the $\Gamma$ and at the $M$ point of the 2D square lattice are \emph{identical}, and that (2) the corner-charge jump associated with a band inversion of a 2D ICR with two 1D ICRs is \emph{also identical} for both the $\Gamma$ and at the $M$ point.
The 3D models with  tetragonal (either primitive or body-centered) symmetry are obtained from the 2D systems by interpreting the band-inversion-tuning parameter as a third momentum component, $k_z$.
The difference between the two Bravais lattices merely corresponds to the way the periodically changing Hamiltonians $\Ham_{k_z}(\vec{k}_\textrm{2D}) = \Ham_{k_z+2\pi}(\vec{k}_\textrm{2D})$ are glued together: for the body-centered case, $\Ham_{k_z}(\vec{k}_\textrm{2D}) = \Ham_{k_z+\pi}(\vec{k}_\textrm{2D} + \Gamma - M)$, whereas such a constraint is absent for the primitive case.
It is clear that the mathematical analysis of the hinge-charge jump in \cref{Sec:HOBBC:3D} leading to  \cref{tab:HOBBC:Qjump} applies irrespective of this additional constraint. An analogous argument in the presence of $C_3$ also reveals why in the trigonal case we find identical classification of TPPs along the lines $\Gamma$, $K$ and $K'$.

While the presented spinless material \ce{Sc3AlC}~\cite{Kim:2018} requires unrealistic values of applied strain to realize the quantized hinge-charge jump, our analysis demonstrates that the discussed phenomenology could conceivably be realized and observed in crystalline solids if more appropriate compounds are identified in the future.
Furthermore, meta-material realizations (e.g., in acoustics, similar to those of Refs.~\onlinecite{Wei:2021,Qiu:2021}) are definitely feasible.

Finally, it is interesting to speculate whether a similar universal higher-order bulk-boundary correspondence extends to other previously reported one-dimensional nodal structures; for example, besides the nodal-line segment spanned by a TPP which has been considered by the present work, nodal-line chains~\cite{Bzdusek:2016,Wu:2019,Wang:2017b}, gyroscopes~\cite{Yu:2015,Kim:2015,Weng:2015b}, and starfruits~\cite{Chen:2017b} were proposed.
They all require crossing of bands with particular choices of symmetry eigenvalues which is associated with a change of the symmetry indicators, potentially implying similar phenomenology of higher-order topological hinge-charge jumps.
Given the numerous material candidates for the various nodal-line compositions, it should be of theoretical as well as of experimental interest to consider such generalizations in the future.
\smallskip

\begin{acknowledgments}
We thank A.~Bouhon, A.~Skurativska, L.~Trifunovic, and S.~S.~Tsirkin for valuable discussions.
P.~M.~L. and T.~B. were supported by the Ambizione grant No.~185806 by the Swiss National Science Foundation.
T.~N. acknowledges support from the European Research Council (ERC) under the European Union’s Horizon 2020 research and innovation programm (ERC-StG-Neupert-757867-PARATOP).
T.~B. and T.~N. were supported by the NCCR MARVEL funded by the Swiss National Science Foundation.
X.~L acknowledges the support by the China Scholarship Council (CSC). 
\end{acknowledgments}


\appendix

\section{Corner and hinge charges}\label{App:hingecharges}

In this appendix we derive various statements presented in \cref{Sec:HOBBC} in relation to the second Stiefel-Whitney (2SW) class and hinge charges.
For reference, we include the relevant symmetry-indicator formulas for the corner charge of spinless 2D systems with $C_n$ symmetry~\cite{Benalcazar:2019,Takahashi:2021}, including both the case without resp.~with spinless time-reversal symmetry, in \cref{App:SIformulas}.
Next, in \cref{App:SWICn} we show that in spinless 2D systems with $C_2\mcT$ and $C_n$ symmetry, where $n=4,6$, the 2SW class can be written in terms of $C_4$ and $C_6$ symmetry indicators.
In particular, we prove \cref{eq:HOBBC:SWc46} from \cref{Sec:HOBBC:2D}.
Then, in \cref{App:Qjump} we consider triple-point pairs (TPPs), i.e., pairs of triple points (TPs) formed by consecutive triplets of bands, protected by $C_n$ symmetry, where $n=3,4,6$, in spinless systems with $\mcP\mcT$ symmetry.
We derive \cref{tab:HOBBC:Qjump} presented in \cref{Sec:HOBBC:3D} which gives the jumps in the 2SW class of 2D cuts and in the hinge charge for the different combinations of irreducible co-representations.
Finally, in \cref{App:noPjump} we discuss why TPPs are not associated with a jump in the \emph{surface} charge.

\subsection{Symmetry-indicator formulas for the corner charge}\label{App:SIformulas}
Here we list the symmetry-indicator formulas for the corner charges $\Qc[,mX]^{(n)}$ of 2D crystals with $C_n$ symmetry in the corresponding geometries shown in \cref{fig:HOBBC:geometry}(a) with the center of the sample located at the Wyckoff position (WP) $mX$ (cf.~Ref.~\onlinecite{Takahashi:2021} for the labelling of WPs).
In all cases we assume that the ionic charges are placed at WP $1a$ and we use the notation of Ref.~\onlinecite{Benalcazar:2019}, defined in \cref{eq:def_eigenvalues,eq:def_SIs}, for the labelling of rotation eigenvalues and symmetry indicators.

In the absence of time-reversal symmetry (symmetry class~A), Ref.~\onlinecite{Takahashi:2021} derives the formulas to be the following: for $n=3$
\begin{subequations}
    \begin{align}
    \Qc[,1a]^{(3)} &= \frac{e}{3}\left(\left[K_1^{(3)}\right]+\left[K_2^{(3)}\right]+\left[{K_1'}^{(3)}\right]+\left[{K_2'}^{(3)}\right]\right)\mod e,\label{App:eq:Qcorner31a}\\
    \Qc[,1b]^{(3)} &= -\frac{e}{3}\left(\left[K_1^{(3)}\right]+\left[{K_2'}^{(3)}\right]\right)\mod e,\label{App:eq:Qcorner31b}\\
    \Qc[,1c]^{(3)} &= -\frac{e}{3}\left(\left[K_2^{(3)}\right]+\left[{K_1'}^{(3)}\right]\right)\mod e,\label{App:eq:Qcorner31c}
    \end{align}
    \label{App:eq:Qcorner3}
\end{subequations}
for $n=4$
\begin{subequations}
    \begin{align}
    \Qc[,1a]^{(4)} &= \frac{e}{4}\left(-\left[X_1^{(2)}\right]+\frac{1}{2}\left[M_1^{(4)}\right]-\frac{3}{2}\left[M_3^{(4)}\right]\right)\mod e,\label{App:eq:Qcorner41a}\\
    \Qc[,1b]^{(4)} &= \frac{e}{4}\left(\left[X_1^{(2)}\right]-\frac{3}{2}\left[M_1^{(4)}\right]+\frac{1}{2}\left[M_3^{(4)}\right]\right)\mod e,\label{App:eq:Qcorner41b}
    \end{align}
    \label{App:eq:Qcorner4}
\end{subequations}
and, for $n=6$
\begin{equation}
    \Qc[,1a]^{(6)} = -\frac{e}{6}\left(2\left[K_1^{(3)}\right]+\frac{3}{2}\left[M_1^{(2)}\right]\right).
    \label{App:eq:Qcorner6}
\end{equation}

If spinless time-reversal symmetry $\mcT$ satisfying $\mcT^2=+\id$ is present, i.e., in symmetry class AI, the formulas simplify due to constraints on the topological invariants $[\Pi_p^{(n)}]$~\cite{Takahashi:2021}.
In our work, this case only occurs for $n=4$,
\begin{subequations}
    \begin{align}
        \Qc[,1a]^{(4)} &= \frac{e}{4}\left(-\left[X_1^{(2)}\right] + 2\left[M_1^{(4)}\right] + 3\left[M_2^{(4)}\right]\right)\mod e,\\
        \Qc[,1b]^{(4)} &= \frac{e}{4}\left(\left[X_1^{(2)}\right] + 2\left[M_1^{(4)}\right] + 3\left[M_2^{(4)}\right]\right)\mod e,
    \end{align}
    \label{App:eq:Qcorner4_TRS}%
\end{subequations}
and $n=6$,
\begin{equation}
    \Qc[,1a]^{(6)} = \frac{e}{4}\left[M_1^{(2)}\right] + \frac{e}{6}\left[K_1^{(3)}\right]\mod e.
    \label{App:eq:Qcorner6_TRS}
\end{equation}
Finally, we also need the formula for $n=2$, which is given in Ref.~\onlinecite{Benalcazar:2019}:
\begin{equation}
    \Qc[,1a]^{(2)} = \frac{e}{4}\left(-\left[X_1^{(2)}\right]-\left[Y_1^{(2)}\right]+\left[M_1^{(2)}\right]\right)\mod e.
    \label{App:eq:Qcorner2_TRS}
\end{equation}

\subsection{Stiefel-Whitney insulators with rotational symmetry}\label{App:SWICn}

\subsubsection{\texorpdfstring{$C_4$}{C4}-symmetry}
We first consider $C_4$ symmetry.
The four time-reversal invariant momenta are $\Gamma$, $X$, $X'$ and $M$, where $X$ and $X'$ are equivalent.
According to \cref{eq:HOBBC:SWc} a nontrivial 2SW class is equivalent to
\begin{equation}
    \floor{\frac{1}{2}\#\Gamma_2^{(2)}} + 2\floor{\frac{1}{2}\#X_2^{(2)}} + \floor{\frac{1}{2}\#M_2^{(2)}} = 1\mod 2.
\end{equation}
This does not constrain $\#X_2^{(2)}$ at all, but it is equivalent to either
\begin{equation*}
    \begin{array}{rc}
        & \floor{\frac{1}{2}\#\Gamma_2^{(2)}} = 0\mod 2 \quad\wedge\quad \floor{\frac{1}{2}\#M_2^{(2)}} = 1\mod 2\\[0.5em]
        \Leftrightarrow\quad & \#\Gamma_2^{(2)} \in \{0,1\}\mod 4 \quad\wedge\quad \#M_2^{(2)} \in \{2,3\}\mod 4\\
        \Rightarrow\quad & \# M_2^{(2)} - \# \Gamma_2^{(2)} =:\,
        \left[M_2^{(2)}\right] \in \{1,2,3\}\mod 4
    \end{array}
\end{equation*}
or the same with $\Gamma$ and $M$ exchanged, where the latter implies $[M_2^{(2)}] \in \{1,2,3\}\mod 4$ as well.
Thus, $\SW{2}=1$ implies $[M_2^{(2)}] \in \{1,2,3\}\mod 4$.
On the other hand, $\SW{2}=0$ implies
\begin{equation*}
    \begin{array}{rc}
        & \floor{\frac{1}{2}\#\Gamma_2^{(2)}} = \floor{\frac{1}{2}\#M_2^{(2)}}\mod 2\\[0.5em]
        \Rightarrow\quad & \left[M_2^{(2)}\right] \in \{0,1,3\}\mod 4.
    \end{array}
\end{equation*}
Furthermore, time-reversal combined with $C_4$ symmetry imply~\cite{Benalcazar:2019} that $[M_2^{(4)}] = [M_4^{(4)}]$, such that together with $[M_2^{(2)}]=[M_2^{(4)}]+[M_4^{(4)}]$, (which follows from $C_4^2=C_2$) we find
\begin{equation}
    \left[M_2^{(2)}\right] = 2\left[M_2^{(4)}\right].
\end{equation}
Therefore, $[M_2^{(2)}]$ has to be even.
Since we already restricted $[M_2^{(2)}]$ for $w_2 = 1$ to the set $\{1,2,3\} \mod 4$ (and to the set $\{0,1,3\} \mod 4$ for $w_1 = 0$), selecting the even element of the corresponding set results in
\begin{equation}
    \SW{2} = \frac{1}{2}\left[M_2^{(2)}\right] \mod 2 = \left[M_2^{(4)}\right] \mod 2.
    \label{App:eq:2SW_C4}
\end{equation}
Finally, using \cref{App:eq:2SW_C4} and assuming vanishing bulk polarization, $[X_1^{(2)}]=0$~\cite{Benalcazar:2019}, we can rewrite \cref{eqn:Q4-sym-ind} as
\begin{equation}
    \Qc^{(4)} = \frac{e}{4}\left(2\left[M_1^{(4)}\right] + 3w_2\right) \mod \tfrac{e}{2}.
\end{equation}
We observe that for $\SW{2}=1$ the term in the brackets is an odd integer, while it is an even integer for $\SW{2}=0$.
This restricts the fractional part of $\Qc^{(4)}$ to $\pm\frac{e}{4}$ in the first, and to $0$ or $\frac{e}{2}$ in the second case.

\subsubsection{\texorpdfstring{$C_6$}{C6}-symmetry}\label{sec:2SW-C6-derivation}
Next, we consider $C_6$ symmetry, where the four time-reversal invariant momenta are $\Gamma$, $M$, $M'$ and $M''$ with the latter three being equivalent.
Correspondingly,
\begin{equation}
    \floor{\frac{1}{2}\#\Gamma_2^{(2)}} + 3\floor{\frac{1}{2}\#M_2^{(2)}} = 1\mod 2.
\end{equation}
A nontrivial value $\SW{2} =1$ implies, in analogy with the analysis of the $C_4$-symmetric case, that 
\begin{equation}
    \left[M_2^{(2)}\right] \in \{1,2,3\}\mod 4,
\end{equation}
while $\SW{2}=0$ implies
\begin{equation}
    \left[M_2^{(2)}\right] \in \{0,1,3\}\mod 4.
\end{equation}

\begin{figure}
    \centering
    \includegraphics{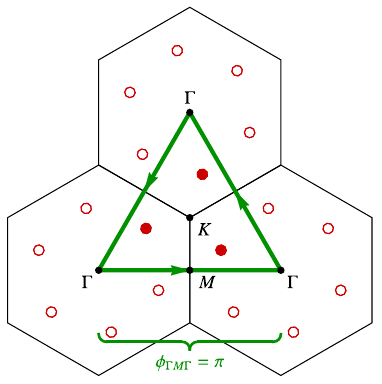}
    \caption{
        Three hexagonal Brillouin zones around a $K$ point of a $C_6$-symmetric model.
        Assuming $[M_1^{(2)}] = 1\mod 2$ the total Berry phase $\phi_{\Gamma M \Gamma}$ of the occupied bands along the closed contour $\Gamma M \Gamma$ equals $\pi$ (cf.~\cref{sec:2SW-C6-derivation}).
        Therefore, the Berry phase along the triangular contour (green) is also $\phi_\triangle = 3\pi = \pi \mod 2\pi$.
        The result implies that the triangle encloses an odd number of band nodes in the principal gap (red disks), i.e., the system is \emph{not an insulator}.
    }
    \label{App:fig:C6ABTPconfig}
\end{figure}

For the next step in our reasoning we need to apply the following property: if $[M_1^{(2)}]=1\mod 2$, then the bulk is necessarily gapless.
This can be seen as follows.
By assumption, the number of occupied bands with $C_2$-rotation eigenvalue $1$ changes by an odd number between $\Gamma$ and $M$
\begin{equation}
    \#M_1^{(2)} - \#\Gamma_1^{(2)} = [M_1^{(2)}] = 1\mod 2. \label{eqn:1D-C2-eigs}
\end{equation}
Since the $C_2$ symmetry acts as inversion on the 1D BZ segment $\Gamma M \Gamma$, \cref{eqn:1D-C2-eigs} implies~\cite{Alexandradinata:2014} that the occupied bands carry a total Berry phase $\pi$ on that segment.
We further consider the triangle formed by the $\Gamma M\Gamma$ paths of the three BZs around a $K$ point, as illustrated in \cref{App:fig:C6ABTPconfig}.
Due to the six-fold rotational symmetry the three sides of the triangle all contribute $\pi\mod 2\pi$ to the Berry phase, such that the total Berry phase along the triangular contour is $\pi\mod 2\pi$, indicating that there is an odd number of band nodes (formed \emph{between} the occupied and the unoccupied bands) inside the triangle, i.e., the bulk is gapless.
Conversely, a gapped bulk implies $[M_1^{(2)}]=0\mod 2$.
Noting that $[M_1^{(2)}]=-[M_2^{(2)}]$ (because $\sum_{p=1}^n[\Pi_p^{(n)}] = 0$~\cite{Benalcazar:2019}), an insulating bulk band structure equivalently requires $[M_2^{(2)}]$ to be even.
Therefore,
\begin{equation}
    \SW{2} = \frac{1}{2}\left[M_1^{(2)}\right]\mod 2.
    \label{App:eq:2SW_C6}
\end{equation}

Using \cref{App:eq:2SW_C6} (and noting that for $C_6$-symmetric systems the bulk polarization \emph{always} vanishes~\cite{Benalcazar:2019}), we can rewrite \cref{eqn:Q6-sym-ind} for the corner charge as
\begin{equation}
    \Qc^{(6)} = \frac{e}{6}\left(\left[K_1^{(3)}\right]+3w_2\right)\mod e.
    \label{App:eq:Q6_2SW}
\end{equation}
It follows from time-reversal symmetry that at HSPs states have either real rotation eigenvalues or come in pairs with complex conjugate rotation eigenvalues.
In particular, this implies that $[K_2^{(3)}]=[K_3^{(3)}]$ and because of the same number of filled bands at points $K$ and $\Gamma$, it follows that
\begin{equation}
    \left[K_1^{(3)}\right] = -\left[K_2^{(3)}\right]-\left[K_3^{(3)}\right] = 0\mod 2,
\end{equation}
i.e., $[K_1^{(3)}]$ is an even integer.
Thus, the term in brackets in \cref{App:eq:Q6_2SW} is an odd integer for $\SW{2}=1$ and an even integer for $\SW{2}=0$.
This restricts $\Qc^{(6)}$ to $\pm\frac{e}{6}$ or $\frac{e}{2}$ in the first case, and to $0$ or $\pm\frac{e}{3}$ in the second case.

\subsection{Hinge-charge jump and 2SW monopole due to triple points}\label{App:Qjump}

We start from the classification of TPs in spinless $\mathcal{PT}$-symmetric systems as presented in Ref.~\onlinecite{Lenggenhager:2021:MBNLs} for little co-groups $\LcG{\vec{k}} = C_{3(v)}, C_{4(v)}, C_{6(v)}$ of some high-symmetry line (HSL) and we consider the situation depicted in \cref{fig:HOBBC:BS}.
We compare quantities in the orange $k_z$-range (to the left of the first TP, where the 2D ICR $\rho^\mathrm{2D}$ is occupied) and the blue $k_z$-range (to the right of the second TP, where the two 1D ICRs $\rho^\mathrm{1D}_a$ and $\rho^\mathrm{1D}_b$ are occupied).
For the ICRs the notation of Ref.~\onlinecite{Bradley:1972} is used.
In particular, when defining the changes $\Delta \Qc$ and $\Delta w_2$, we subtract the first (orange $k_z$-range) from the latter (blue $k_z$-range).
For each 2D or 1D ICR $\rho$ of $\LcG{\vec{k}}\cup (\mcP\mcT)\LcG{\vec{k}}$, we look up the corresponding rotation eigenvalues on the Bilbao crystallographic server (BCS)~\cite{Aroyo:2011,Aroyo:2006a,Aroyo:2006b} using the program \textsc{Corepresentations PG}~\cite{Xu:2020,Elcoro:2021} and, assuming there are no other band inversions in the principal gap, compute the relevant symmetry indicators, see \cref{App:tab:SI:C3,App:tab:SI:C4,App:tab:SI:C6}.
Before proceeding with the mathematical analysis, we make two remarks.
First, note that the HSL on which the TPs lie does not need to go through the center of the BZ (point $\Gamma$ of the 2D cut).
Namely, the little co-group $C_{3(v)}$ is also realized at the $K$ point(s) of space groups (SGs) with three-fold or six-fold rotational symmetry, while $C_{4(v)}$ is also realized at $M$ points of SGs with four-fold rotational symmetry.

In all cases we assume that the SG contains $\mcP\mcT$, acting as $(C_2\mcT)_\mathrm{2D}$ in the 2D cuts, and $C_n$ with order \mbox{$n\in\{3,4,6\}$}.
Furthermore, we place all ionic charge at the center of the unit cell, because ultimately, we are interested in the change of the corner charge as a function of $k_z$ and the ionic charge distribution does not depend on $k_z$.
If the system symmetry is $C_3$, then $\mcT_\mathrm{2D} \,(\, = C_2 \mathcal{PT}=  m_z \mathcal{T})$ is not present, such that the corner charge is given by \cref{App:eq:Qcorner3}.
For SG symmetries with rotation of order $n=4,6$, the $C_2$ combines with $\mcP\mcT$ to give $C_2\mcP\mcT=\mcT_\mathrm{2D}$, such that we can apply the simplified formulas \cref{App:eq:Qcorner4_TRS,App:eq:Qcorner6_TRS}.
For the 2SW class, we use the results of the previous subsection, i.e., \cref{App:eq:2SW_C4,App:eq:2SW_C6}.

\medskip

We next go in detail through all the possible cases.

\subsubsection{Triple points along \texorpdfstring{$C_{3(v)}$}{C3v}-symmetric lines}

For $C_{3(v)}$ there is only one possible configuration of two TPs, namely the one with ICRs $(E;A,A)$. The symmetry indicators (cf.~\cref{App:tab:SI:C3}) imply that $\Delta\#\Pi_1^{(3)}=2$ and $\Delta\#\Pi_2^{(3)}=-1$.

Let us first consider TPs occurring along the HSL $\Pi=\Gamma$ of a $C_{3}$-symmetric system.
If the rotation center of the sample lies at WP $1a$, then $\Delta [K_p^{(3)}]=\Delta [K_p'^{(3)}]=-\Delta\#\Gamma_p^{(3)}$.
Then, according to \cref{App:eq:Qcorner31a}, the hinge-charge jump is
\begin{equation}\label{eqn:Q3-C3-G}
    \Delta \Qc[,1a]^{(3)} = -\frac{2e}{3}\left(\Delta\#\Gamma_1^{(3)}+\Delta\#\Gamma_2^{(3)}\right) \mod e = \frac{e}{3} \mod e
\end{equation}
If the TPs instead occur along HSL $\Pi = K$ (or equivalently $K'$) of a $C_{3}$-symmetric system (with rotation center still set to $1a$, then $\Delta \big[K_p^{(3)}\big]=\Delta\#K_p^{(3)}$ such that
\begin{equation}\label{eqn:Q3-C3-K}
    \Delta \Qc[,1a]^{(3)} = \frac{e}{3}\left(\Delta\#K_1^{(3)}+\Delta\#K_2^{(3)}\right) \mod e = \frac{e}{3} \mod e
\end{equation}
It is easily verified by using \cref{App:eq:Qcorner31b,App:eq:Qcorner31c} that the derived results for $\Delta \Qc$ remain unchanged if the rotation center of the finite system is instead located at WP $1b$ or $1c$.

If $\Pi=K$ of a $C_{6}$-symmetric system, then we need to use \cref{App:eq:Qcorner6_TRS} instead, resulting in the same
\begin{equation}\label{eqn:Q6-C6-K}
    \Delta \Qc^{(6)} = \frac{e}{6}\Delta\#K_1^{(3)} \mod e = \frac{e}{3} \mod e.
\end{equation}
This concludes the derivation of $\Delta \Qc^{(n)} = \tfrac{e}{3} \mod e$ for TPPs along $C_{3(v)}$-symmetric lines.

We briefly analyze the jump $\Delta w_2$ of the 2SW class.
We have argued in \cref{Sec:HOBBC:3D} and demonstrated in \cref{Sec:MinimalModels:C3} that the 2SW class is not symmetry indicated for $C_3$-symmetric systems.
For $C_6$, on the other hand, we use \cref{App:eq:2SW_C6} to find
\begin{equation}
    \Delta\SW{2} = \frac{1}{2}\Delta\left[M_1^{(2)}\right] = 0,
\end{equation}
because, by assumption, there is neither a band inversion at $\Gamma$ nor at $M$.

\begin{table}
    \centering
    \caption{
        Rotation eigenvalues and symmetry indicators for all irreducible co-representations of $C_{3(v)}$ with $\mcP\mcT$ at a high-symmetry point $\Pi=\Gamma,K,K'$ of a $C_3$-symmetric system or at $\Pi=K$ of a $C_6$-symmetric system.
        The list of ICRs is unchanged if $\mathcal{PT}$ symmetry is removed in the case of $C_{3v}$.
    }
    \begin{ruledtabular}
	\begin{tabular}{CCCCCCC}
	\multicolumn{2}{l}{ICRs for} & \multicolumn{1}{l}{Eigenvalue of} & \multicolumn{2}{l}{Symmetry indicators at $\Pi$}\\
	C_{3v} & C_3 & C_3 & \#\Pi_1^{(3)} & \#\Pi_2^{(3)}\\[0.1em]\hline
	A_1, A_2 & A_1 & 1 & 1 & 0\\
	E & ^2E^1E & \e^{2\pi\i/3}, \e^{-2\pi\i/3} & 0 & 1\\
    \end{tabular}
    \end{ruledtabular}
    \label{App:tab:SI:C3}
\end{table}

\subsubsection{Triple points along \texorpdfstring{$C_{4(v)}$}{C4v}-symmetric lines}\label{app:TP-C4}

Next, we consider $C_{4(v)}$-symmetric lines with ICRs and symmetry indicators given in \cref{App:tab:SI:C4}.
Here, we need to distinguish three TP configurations [namely $(E;A,A)$, $(E;B,B)$ and $(E;A,B)$] and two HSLs (namely $\Pi=\Gamma$ and $\Pi=K$).
The differences $\Delta\#\Pi_1^{(2)}$, $\Delta\#\Pi_1^{(4)}$ and $\Delta\#\Pi_2^{(4)}$ for the various TP configurations are listed in \cref{App:tab:TPconfigs:C4}.
By recalling \cref{App:eq:2SW_C4}, we find that for all cases
\begin{equation}
    \Delta \SW{2} = \Delta\left[M_2^{(4)}\right] \mod 2 = 1\mod 2,
\end{equation}
in agreement with the fact that the TPs are type~$\mathsf{A}$ (cf.~\cref{Sec:Monopole:SW}).

To identify the jump in the hinge charge, we have to analyze the two HSLs separately.
First, for $\Pi=\Gamma$ we have $\Delta [\textrm{P}_p^{(n)}] = -\Delta \# \Gamma_p^{(n)}$ for both $\textrm{P}\in\{K,M\}$, and the change in corner charge is given by
\begin{equation}
    \Delta \Qc^{(4)} = -\frac{e}{4}\left(\mp\Delta\#\Gamma_1^{(2)} + 2\Delta\#\Gamma_1^{(4)} + 3\Delta\#\Gamma_2^{(4)}\right)\mod e,
    \label{eqn:C4-charge-Gamma}
\end{equation}
where the negative (positive) sign of $\Delta\#\Gamma_1^{(2)}$ corresponds to setting the center of the system to WP $1a$ ($1b$).
Observe in \cref{App:tab:SI:C4} that $\Delta \Pi_1^{(2)}$ is \emph{even} for all TPPs, which means that the `$\mp$' sign ambiguity is unimportant, and both WPs lead to the same value of the hinge-charge jump.

In contrast, for $\Pi=M$ we obtain contributions from $\Delta [M_p^{(4)}] = \Delta \# M_p^{(4)}$, leading to
\begin{equation}
    \Delta \Qc^{(4)} = \frac{e}{4}\left(2\Delta\#M_1^{(4)} + 3\Delta\#M_2^{(4)}\right)\mod e\label{eqn:C4-charge-K}.
\end{equation}
Careful evaluation of \cref{eqn:C4-charge-Gamma,eqn:C4-P-Gamma,eqn:C4-charge-K} for all combinations of ICRs gives the following results: For TPPs along either $\Gamma$ or $M$, $\Delta \Qc^{(4)}=\tfrac{e}{4}$ for both $(E;A,A)$ and $(E;B,B)$.
The results are summarized in \cref{App:tab:TPconfigs:C4}.

\begin{table}
    \centering
    \caption{
        Rotation eigenvalues and symmetry indicators for all irreducible co-representations of $C_{4(v)}$ with $\mcP\mcT$ at a high-symmetry point $\Pi=\Gamma,M$ of a $C_4$-symmetric system. 
        The list of ICRs is unchanged if $\mathcal{PT}$ symmetry is removed in the case of $C_{4v}$.
    }
    \begin{ruledtabular}
	\begin{tabular}{CCCCCCC}
	\multicolumn{2}{l}{ICRs for} & \multicolumn{2}{l}{Eigenvalues of} & \multicolumn{3}{l}{Symmetry indicators at $\Pi$}\\
	C_{4v} & C_4 & C_2 & C_4 & \#\Pi_1^{(2)} & \#\Pi_1^{(4)} & \#\Pi_2^{(4)}\\[0.1em]\hline
	A_1, A_2 & A & 1 & 1 & 1 & 1 & 0\\
	B_1, B_2 & B & 1 & -1 & 1 & 0 & 0\\
	E & E & -1, -1 & \i, -\i & 0 & 0 & 1
    \end{tabular}
    \end{ruledtabular}
    \label{App:tab:SI:C4}
\end{table}

\begin{table}
    \centering
    \caption{
        Changes in symmetry indicators ($\Delta\#\Pi_p^{(n)}$), hinge charge ($\Delta \Qc^{(4)}$) and second Stiefel-Whitney class ($\Delta w_2$) associated with the different configurations of triple-point pairs (TPPs). Each TPP is specified by the irreducible co-representations (ICRs) $(\rho^\mathrm{2D};\rho^\mathrm{1D}_{a},\rho^\mathrm{1D}_{b})$ on a four-fold rotation axis, corresponding to the high-symmetry point $\Pi=\Gamma,M$ in a 2D cut perpendicular to the rotation axis.
        The changes are defined by subtracting the characteristics of the systems with $\rho^\mathrm{2D}$ occupied from those of the system with $\rho^\mathrm{1D}_{a,b}$ occupied.
        The hinge-charge jump is computed using \cref{eqn:C4-charge-Gamma} for $\Pi=\Gamma$ and \cref{eqn:C4-charge-K} for $\Pi=M$, while $\Delta w_2$ is computed from \cref{App:eq:2SW_C4}).
    }
    \begin{ruledtabular}
	\begin{tabular}{CCC}
	\text{ICRs} & (E;A,A) & (E;B,B) \\\hline\addlinespace
	\Delta\#\Pi_1^{(2)} & 2 & 2 \\\addlinespace
	\Delta\#\Pi_1^{(4)} & 2 & 0 \\\addlinespace
	\Delta\#\Pi_2^{(4)} & -1 & -1 \\\addlinespace
	\Delta \Qc^{(4)}\mod e & +\frac{e}{4} & +\frac{e}{4} \\\addlinespace
	\Delta \SW{2}\mod 2 &  1 & 1
    \end{tabular}
    \end{ruledtabular}
    \label{App:tab:TPconfigs:C4}
\end{table}

\subsubsection{Triple points along \texorpdfstring{$C_{6(v)}$}{C6v}-symmetric lines}

There is only one HSL with little co-group $C_{6(v)}$ (namely $\Gamma$ in $C_{6}$-symmetric systems), and considering the possible crossings of ICRs (listed in \cref{App:tab:SI:C6}) results in six distinct TPP configurations.
We first focus on the cases where the two TPs are of the same type, which gives the four TPPs listed in \cref{App:tab:TPconfigs:C6}.
By combining $\Delta [\mathrm{\Pi}_p^{(n)}] = -\Delta\#\Gamma_p^{(n)}$ (for $\mathrm{\Pi}\in\{M,K\}$) with \cref{App:eq:Qcorner6_TRS,App:eq:2SW_C6}, we calculate the jumps
\begin{align}
    \Delta \Qc^{(6)} &= -\frac{e}{4}\Delta\#\Gamma_1^{(2)} - \frac{e}{6}\Delta\#\Gamma_1^{(3)} \mod e,\label{eqn:C6-DQ-result}  \\
    \Delta \SW{2} &= -\frac{1}{2}\Delta\#\Gamma_1^{(2)} \mod 2.\label{eqn:C6-Dw2-result}
\end{align}
The computed values of $\Delta \Qc^{(6)}$ and of $\Delta w_2$ for all the TPPs where both TPs are of the same type are listed in \cref{App:tab:TPconfigs:C6}.
We find that a nontrivial change of the 2SW class, which occurs for two type-$\mathsf{A}$ TPs, is associated with a hinge-charge jump $\Delta \Qc^{(6)} = +\tfrac{e}{6}$.
In contrast, pairs of type-$\mathsf{B}$ TPs are characterized by trivial $\Delta\SW{2}=0$ and are associated with a hinge-charge jump of $\Delta \Qc^{(6)}=-\tfrac{e}{3}$.

\begin{table}
    \centering
    \caption{
        Rotation eigenvalues and symmetry indicators for all irreducible co-representations of $C_{6(v)}$ with $\mcP\mcT$ at the high-symmetry point $\Gamma$ of a $C_6$-symmetric system.
        The list of ICRs is unchanged if $\mathcal{PT}$ symmetry is removed in the case of $C_{6v}$.
    }
    \begin{ruledtabular}
	\begin{tabular}{CCCCCC}
	\multicolumn{2}{l}{ICRs for} & \multicolumn{2}{l}{Eigenvalues of} & \multicolumn{2}{l}{Symmetry indicators at $\Gamma$}\\
	C_{6v} & C_6 & C_2 & C_3 & \#\Gamma_1^{(2)} & \#\Gamma_1^{(3)}\\[0.1em]\hline
	A_1, A_2 & A & 1 & 1 & 1 & 1\\
	B_1, B_2 & B & -1 & 1 & 0 & 1\\
	E_1 & ^2E_2^1E_2 & -1, -1 & \e^{2\pi\i/3}, \e^{-2\pi\i/3} & 0 & 0\\
	E_2 & ^2E_1^1E_1 & 1, 1 & \e^{2\pi\i/3}, \e^{-2\pi\i/3} & 2 & 0\\
    \end{tabular}
    \end{ruledtabular}
    \label{App:tab:SI:C6}
\end{table}

\begin{table}
    \centering
    \caption{
        Changes in symmetry indicators ($\Delta\#\Gamma_1^{(n)}$), hinge charge ($\Delta \Qc^{(6)}$) and second Stiefel-Whitney class ($\Delta w_2$) associated with the different configurations of triple-point pairs (TPPs) along the $\Gamma$ line of $C_6$-symmetric systems. Each TPP is specified by the irreducible co-representations (ICRs) $(\rho^\mathrm{2D};\rho^\mathrm{1D}_{a},\rho^\mathrm{1D}_{b})$ along the $\Gamma$ line.
        For simplicity we use the labels of ICRs for $C_{6v}$, the corresponding notation for $C_6$ can be extracted from \cref{App:tab:SI:C6}.
        The changes are defined by subtracting the characteristics of the systems with $\rho^\mathrm{2D}$ occupied from those of the system with $\rho^\mathrm{1D}_{a,b}$ occupied.
        The values of $\Delta \Qc^{(6)}$ and $\Delta w_2$ are computed from \cref{eqn:C6-DQ-result} and \cref{eqn:C6-Dw2-result}, respectively.
    }
    \begin{ruledtabular}
	\begin{tabular}{CCCCC}
	\text{TP types} & \multicolumn{2}{C}{{(\mathsf{A},\mathsf{A})}} & \multicolumn{2}{C}{(\mathsf{B}, \mathsf{B})} \\
	\text{ICRs} & (E_1;A_i,A_j) & (E_2;B_i,B_j) & (E_1;B_i,B_j) & (E_2;A_i,A_j) \\\hline\addlinespace
	\Delta\#\Gamma_1^{(2)} & 2 & -2 & 0 & 0 \\\addlinespace
	\Delta\#\Gamma_1^{(3)} & 2 & 2 & 2 & 2 \\\addlinespace
	\Delta \Qc^{(6)}\mod e & +\frac{e}{6} & +\frac{e}{6} & -\frac{e}{3} & -\frac{e}{3} \\\addlinespace
	\Delta \SW{2}\mod 2 &  1 & 1 & 0 & 0
    \end{tabular}
    \end{ruledtabular}
    \label{App:tab:TPconfigs:C6}
\end{table}

Finally, there is the possibility to have two TPs of \emph{different types}, i.e., one type~$\mathsf{A}$ and one type~$\mathsf{B}$, by choosing one of the following ICR combinations: $(E_1;A_i,B_j)$, $(E_2;A_i,B_j)$.
In these two cases, we find that $\Delta[M_1^{(2)}] = -\Delta \# \Gamma_1^{(2)} = 1 \mod 2$ (i.e., it is an \emph{odd number}.
This implies that the value of $[M_1^{(2)}]$ must be odd on one side of the TPP.
However, recall from \cref{App:SWICn} that a gapped bulk requires $[M_1^{(2)}]=0\mod 2$.
It therefore follows that the bulk is necessarily gapless on one side of the TPP, such that the hinge-charge jump cannot be defined in this case.
This finding is consistent with the intuitive explanation given in the main text, where the different type of the two TPs forces the attached NL arcs to cross the orange or blue $k_z$-range.
In \cref{fig:HOBBC:TPconfigs}(c) a 2D cut in the orange $k_z$-range would correspond to the situation shown in \cref{App:fig:C6ABTPconfig} with the red disks corresponding to places where the red attached NL arcs cross the 2D cut.

\subsection{No jump of the surface charge due to triple points}\label{App:noPjump}

For the hinge-charge jump to be observable, it is important that the surface charge (more precisely, the surface charge \emph{density}; this corresponds to the edge charge density of the 2D cuts at fixed $k_z$) is vanishing~\cite{Benalcazar:2019} for the $k_z$-ranges on both sides of the TPP.
This can only be true if the \emph{jump} of the surface charge associated with the TPP is zero.
Here we show that this is indeed true for all the TPPs shown in \cref{tab:HOBBC:Qjump}.

Before deriving the desired fact mathematically from the corresponding symmetry-indicator formulas, let us present a simple argument based on the quantization of Berry phase.
Namely, recall that for a boundary of a high-symmetry orientation with respect to the crystalline axes, the surface charge is in a one-to-one correspondence~\cite{King-Smith:1993} with the Berry phase along a closed momentum-space ($\vec{k}$) path along the direction perpendicular to the considered surface.
Here, first recall~\cite{Lenggenhager:2021:MBNLs,Lenggenhager:2022:TPClassif} that stable TPs require the presence of either $\mcP\mcT$ symmetry (in which case the Berry phase is quantized to $0$ vs.~$\pi$ on \emph{any} closed $\vec{k}$-path due to the reality condition~\cite{Bzdusek:2017}) or of $m_v$ mirror symmetry (in which case the mirror symmetry acts like inversion on the straight $\vec{k}$-paths perpendicular to the boundaries considered in \cref{fig:C4model:hinge,fig:MinimalModels:C6AA,fig:MinimalModels:C6BB}, thus quantizing the Berry phase of interest to $0$ vs.~$\pi$~\cite{Zak:1989}).

The quantization of the Berry phase implies that the surface charge can only change if the $\vec{k}$-paths of the required orientation encounter a bulk nodal line (carrying the $\pi$-quantum of Berry phase).
However, a look at \cref{fig:HOBBC:TPconfigs} reveals that for both type-$(\mathsf{A},\mathsf{A})$ and type-$(\mathsf{B},\mathsf{B})$ TPPs it is possible to continuously shift a horizontal $\vec{k}$-path from the orange $k_z$-range to the blue $k_z$-range without encountering any (red) nodal line in the principal gap.
We therefore anticipate the surface charge to be identical in both $k_z$-ranges (and therefore the hinge-charge jump to be observable) at least when the boundaries are oriented symmetrically with respect to the crystal axes.

More formally, we now derive the same conclusion for an \emph{arbitrarily} oriented $C_n$-symmetric system.
In such a case the surface charge density of a gapped 2D crystal is given~\cite{Vanderbilt:1993} by $\sigma_\textrm{surf} = \vec{P}\cdot\uvec{n}$, where $\vec{P}$ is the bulk polarization and $\uvec{n}$ the surface normal. 
To prove that $\Delta \sigma_\textrm{surf}=0$ across the TPP for arbitrary boundary termination, it needs to be true that $\Delta \vec{P}\cdot\uvec{n} = 0$ when moving across the TPP.
Similar to the corner charge, the bulk polarization $\vec{P}^{(n)}$ of a 2D crystal with $C_n$ rotational symmetry can be expressed~\cite{Benalcazar:2019,Watanabe:2020,Takahashi:2021} using symmetry indicators.
For $C_3$-symmetric crystals, we use the result of Refs.~\onlinecite{Watanabe:2020,Takahashi:2021} which does not assume time-reversal symmetry to be present,
\begin{eqnarray}
        \vec{P}^{(3)} &=& \frac{e}{3}\left(2\left[K_1^{(3)}\right]+\left[{K_1'}^{(3)}\right]+\left[K_2^{(3)}\right]+2\left[{K_2'}^{(3)}\right]\right)\left(\vec{a}_1+\vec{a}_2\right)\nonumber \\
        &\phantom{=}&\hspace{4.8cm}\mod e\vec{R},\label{eqn:polar-charge-C3}
\end{eqnarray}
while for $C_n$ with $n=4,6$, we can use the simplified expressions from Ref.~\onlinecite{Benalcazar:2019}:
\begin{align}
    \vec{P}^{(4)} &= \frac{e}{2}\left[X_1^{(2)}\right]\left(\vec{a}_1+\vec{a}_2\right)\mod e\vec{R},\label{eqn:polar-charge-C4}\\
    \vec{P}^{(6)} &= 0\mod e\vec{R}.\label{eqn:polar-charge-C6}
\end{align}
Here $\vec{a}_1$ and $\vec{a}_2$ are lattice basis vectors and $\vec{R}=m_1\vec{a}_1+m_2\vec{a}_2$ with $m_1,m_2\in\mathbb{Z}$ is a general Bravais lattice vector.

We now briefly apply the symmetry-indicator formulas in \cref{eqn:polar-charge-C3,eqn:polar-charge-C6} to show that the polarization jump $\Delta \vec{P}^{(n)}$ associated with the TPPs in $C_n$-symmetric systems is always vanishing.
First, for a TPP along the $\Gamma$ line of a $C_{3}$-symmetric system [cf.~\cref{eqn:Q3-C3-G}], we obtain from \cref{eqn:polar-charge-C3} the jump of the bulk polarization
\begin{equation}
    \begin{split}
        \Delta\vec{P}^{(3)} &= -e\left(\Delta\#\Gamma_1^{(3)}+\Delta\#\Gamma_2^{(3)}\right)\left(\vec{a}_1+\vec{a}_2\right)\mod e\vec{R}\\
        &= 0\mod e\vec{R}
    \end{split}.
\end{equation}
Similarly, for a TPP along the $K$ line [cf.~\cref{eqn:Q3-C3-K}]
\begin{equation}
    \begin{split}
        \Delta\vec{P}^{(3)} &= \frac{e}{3}\left(2\Delta\# K_1^{(3)}+\Delta\# K_2^{(3)}\right)\left(\vec{a}_1+\vec{a}_2\right)\mod e\vec{R}\\
        &= 0\mod e\vec{R},
    \end{split}
\end{equation}
where we have read from \cref{App:tab:SI:C3} that $\Delta\#K_1^{(3)}= 2$ and $\Delta\#K_2^{(3)}= -1$ for a TPP formed by ICRs $(E;A_i,A_j)$.

Furthermore, note that per \cref{eqn:polar-charge-C3} the polarization always vanishes in $C_6$-symmetric systems.
Therefore $\Delta\vec{P}^{(6)}=0\mod e\vec{R}$ for TPPs along both the $K$ line [cf.~\cref{eqn:Q6-C6-K}] and the $\Gamma$ line [cf.~\cref{eqn:C6-DQ-result}] of $C_{6}$-symmetric systems.
Next, for TPPs along the $\Gamma$ line of $C_{4}$-symmetric systems [cf.~\cref{eqn:C4-charge-Gamma}] we find from \cref{eqn:polar-charge-C4} that the change in bulk polarization is
\begin{equation}
    \begin{split}
    \Delta\vec{P}^{(4)} &= -\frac{e}{2}\Delta\#\Gamma_1^{(2)}\left(\vec{a}_1+\vec{a}_2\right) \mod e\vec{R} \\
    &= 0\mod e\vec{R}.
    \label{eqn:C4-P-Gamma}
    \end{split}
\end{equation}
where we have read from \cref{App:tab:TPconfigs:C4} that $\Delta\#\Gamma_1^{(2)}$ is even for all possible TPPs.
Finally, note that TPPs along the $M$ line [cf.~\cref{eqn:C4-charge-K}] do not have an effect on the symmetry indicator $\big[X_1^{(2)}\big]$ in \cref{eqn:polar-charge-C4}, therefore again $\Delta\vec{P}^{(4)}=0\mod e\vec{R}$.
We conclude that all TPPs listed in \cref{tab:HOBBC:Qjump} are associated with no jump in surface charge, thus making the hinge-charge jump observable for an appropriate choice of the boundary termination.

\section{Euler monopole charge in the presence of rotational symmetry}\label{App:Euler-monopole}

In this Appendix we derive the statements made in \cref{Sec:HOBBC:Euler} and apply them to triple-point pairs (TPPs) to find the combinations of irreducible corepresentations (ICRs) that necessarily lead to a gapless bulk on at least one side of the TPP.
Such TPPs cannot be associated with a higher-order signature on the wire hinges; for brevity, we call them \emph{inadmissible TPP configurations}. 
For the remaining cases, we determine the value of the Euler monopole charge given in \cref{tab:HOBBC:Qjump}. 
The analysis assumes the presence of space-time inversion ($\mcP\mcT$) symmetry unless explicitly stated otherwise (a particular case of the latter is the entire \cref{App:Euler-monopole:PT-absence}).

The presented arguments are based on analyzing the spectrum of Wilson loop operators on certain appropriately chosen paths.
The Wilson loop operators are defined in terms of projectors onto occupied eigenstates $\ket{u_i(\vec{k})}$ of the Bloch Hamiltonian $\Ham(\vec{k})$. 
We organize the occupied eigenstates into an $N\times N_\mathrm{occ}$ matrix
\begin{equation}
    \mathsf{u}(\vec{k}) = \mqty(\ket{u_1(\vec{k})}&\ket{u_2(\vec{k})}&\cdots&\ket{u_{N_\mathrm{occ}}(\vec{k})}),
\end{equation}
where $N$ is the total number of bands and $N_\mathrm{occ}$ is the number of occupied bands.
In the presence of $\mcP\mcT$ symmetry satisfying $(\mcP\mcT)^2=+\id$, we always adopt a basis of the Hilbert space in which $\mcP\mcT$ is represented by complex conjugation $\mcK$.
Then, the Bloch Hamiltonian is a real-symmetric matrix, and the eigenstates (as well as the matrix $\mathsf{u}(\vec{k})$) can be gauged to be real. 
To determine the advertised properties of TPPs, we need to study how the spectrum of the Wilson-operator on certain appropriately chosen paths is constrained by $\mathcal{PT}$, $m_v$, and $C_n$ symmetries.

The discussion in this appendix is structured as follows.
In \cref{App:Euler-monopole:Wilson-op} we study how $C_n$ rotational symmetry constrains the Wilson loop operator on an appropriately chosen path, composed of two $C_n$-related segments.
Next, in \cref{App:Euler-monopole:Berry-phase}, we use the previous result to relate the Berry phase of the occupied bands on a particular choice of path to the $C_n$ eigenvalues of the occupied bands at the high-symmetry line, and in \cref{App:Euler-monopole:charge} we derive a symmetry-indicator formula for the Euler monopole charge.
Then, in \cref{App:Euler-monopole:TPP}, we apply the previous two results to TPPs, identifying which TPPs are inadmissible and giving the Euler monopole charge for each admissible TPP configuration.
Finally, in \cref{App:Euler-monopole:PT-absence} we extend the discussion of inadmissible TPPs to systems without $\mcP\mcT$ symmetry.

\subsection{Symmetry-constraint on Wilson loop operator}\label{App:Euler-monopole:Wilson-op}
We define the following family of closed contours
\begin{equation}
    \Gamma(\phi) = \gamma(\phi)^{-1}\circ\gamma(0)
\end{equation}
consisting of half circles given by $\gamma(\phi)\,:\,\theta\in[0,\pi]\mapsto \gamma(\theta,\phi)$ with
\begin{equation}
    \gamma(\theta,\phi) = R\mat{\sin(\theta)\cos(\phi)\\\sin(\theta)\sin(\phi)\\-\cos(\theta)}.
\end{equation}
One example of such a contour is illustrated in \cref{fig:Euler_charge} in green color.
We abbreviate $\mathsf{u}(\theta,\phi) = \mathsf{u}(\gamma(\theta,\phi))$, $\mathsf{u}_0=\mathsf{u}(0,\phi)$ and $\mathsf{u}_1=\mathsf{u}(\pi,\phi)$ for any $\phi\in[0,2\pi)$.

Given any closed path $\Gamma(t)$ defined on $t\in[0,1]$, the Wilson loop operator is defined by
\begin{equation}
    \mathcal{W}(\Gamma)_{mn} = \lim_{\delta t\to 0}\mel{u_m(\Gamma(1))}{\prod_t^{1\leftarrow 0}\mathbb{P}(\Gamma(t))}{u_n(\Gamma(0))},
\end{equation}
where $\delta t$ is the discretization of $t$ in the product, the arrow over the product symbol indicates the ordering of the factors from small to large values of $t$ (i.e., in this case factors with smaller $t$ are to the right), and
\begin{equation}
    \mathbb{P}(\vec{k}) = \sum_{j=1}^{N_\mathrm{occ}}\op{u_j(\vec{k})}{u_j(\vec{k})}
\end{equation}
is the projector onto the occupied subspace. 
Observing that
\begin{equation}
    \mathsf{u}(\vec{k})\adjo{\mathsf{u}(\vec{k})} = \sum_{j=1}^{N_\mathrm{occ}}\op{u_j(\vec{k})}{u_j(\vec{k})} = \mathbb{P}(\vec{k}),
\end{equation}
we can rewrite the expression for the Wilson loop operator in matrix form ($\mathcal{W}$ is an $N_\mathrm{occ}\times N_\mathrm{occ}$ matrix)
\begin{equation}
    \mathcal{W}(\Gamma) = \lim_{\delta t\to 0}\adjo{\mathsf{u}(\Gamma(1))}\prod_t^{1\leftarrow 0}\mathsf{u}(\Gamma(t))\adjo{\mathsf{u}(\Gamma(t))}\mathsf{u}(\Gamma(0)).
    \label{App:eq:Wilson-operator_def}
\end{equation}

The Wilson loop operator for one of the closed contours $\Gamma(\phi)$ defined above is then found to be
\begin{align}
    \mathcal{W}(\phi) &= \mathcal{W}(\Gamma(\phi))\nonumber\\
    &= \lim_{\delta\theta\to 0}\adjo{\mathsf{u}_0}\left[\prod_\theta^{0\rightarrow \pi}\mathsf{u}(\theta,\phi)\adjo{\mathsf{u}(\theta,\phi)}\right]\left[\prod_\theta^{\pi\leftarrow 0}\mathsf{u}(\theta,0)\adjo{\mathsf{u}(\theta,0)}\right]\padjo{\mathsf{u}_0}.
    \intertext{Reordering the terms in the products, this becomes}
    &= \lim_{\delta\theta\to 0}\adjo{\mathsf{u}_0}\padjo{\mathsf{u}_0}\left[\prod_\theta^{0\rightarrow \pi-\delta\theta}\adjo{\mathsf{u}(\theta,\phi)}\mathsf{u}(\theta+\delta\theta,\phi)\right]\adjo{\mathsf{u}_1}\times\nonumber\\*
    &\quad\times\padjo{\mathsf{u}_1}\left[\prod_\theta^{\pi-\delta\theta\leftarrow 0}\adjo{\mathsf{u}(\theta+\delta\theta,0)}\mathsf{u}(\theta,0)\right]\padjo{\mathsf{u}_0}\nonumber\\
    &= \lim_{\delta\theta\to 0}\left[\prod_\theta^{0\rightarrow \pi-\delta\theta}\adjo{V(\theta,\phi)}\right]\left[\prod_\theta^{\pi-\delta\theta\leftarrow 0}V(\theta,0)\right]\padjo{\mathsf{u}_0}
\end{align}
with
\begin{equation}
    V(\theta,\phi) = \adjo{\mathsf{u}(\theta+\delta\theta,\phi)}\mathsf{u}(\theta,\phi).
\end{equation}
If we choose the real gauge along $\Gamma(\phi)$ by monodromy, $V(\theta,\phi)\in\SO(N_\mathrm{occ})$.

The $C_n$ rotational symmetry relates eigenstates at points in momentum space related by that symmetry:
\begin{equation}
    \mathsf{u}(C_n\vec{k}) = D(C_n)\mathsf{u}(\vec{k})\mathcal{B}(\bs{k}),
\end{equation}
where $D(C_n)$ is the corepresentation matrix of $C_n$ in the chosen basis, i.e., where $D(\mcP\mcT)=\id$, and $\mathcal{B}(\bs{k})\in \mathsf{O}(N_\textrm{occ})$ is a sewing matrix~\cite{Fang:2012a}.
Thus, for any $\theta,\phi$
\begin{equation}
    \mathsf{u}\left(\theta,\phi+\frac{2\pi}{n}\right) = D(C_n)\mathsf{u}(\theta,\phi)\mathcal{B}(\theta,\phi).
\end{equation}
Substituting this into the expression for the Wilson loop operator, we find
\begin{align}
    \mathcal{W}\left(\phi+\frac{2\pi}{n}\right) &= \lim_{\delta\theta\to 0}\adjo{\mathsf{u}_0}\left[\prod_\theta^{0\rightarrow \pi}D(C_n)\mathsf{u}(\theta,\phi)\adjo{\mathsf{u}(\theta,\phi)}\adjo{D(C_n)}\right]\times\nonumber\\*
    &\quad\times\left[\prod_\theta^{\pi\leftarrow 0}\mathsf{u}(\theta,0)\adjo{\mathsf{u}(\theta,0)}\right]\padjo{\mathsf{u}_0},
    \intertext{where the sewing matrices canceled out. Reordering the terms again,}
    &= \lim_{\delta\theta\to 0}\adjo{\mathsf{u}_0}D(C_n)\padjo{\mathsf{u}_0}\left[\prod_\theta^{0\rightarrow \pi-\delta\theta}\adjo{\mathsf{u}(\theta,\phi)}\mathsf{u}(\theta+\delta\theta,\phi)\right]\times\nonumber\\*
    &\quad\times\adjo{\mathsf{u}_1}\adjo{D(C_n)}\padjo{\mathsf{u}_1}\left[\prod_\theta^{\pi-\delta\theta\leftarrow 0}\adjo{\mathsf{u}(\theta+\delta\theta,0)}\mathsf{u}(\theta,0)\right]\padjo{\mathsf{u}_0}.
\end{align}

Note that
\begin{equation}
    \begin{array}{rcl}
        \adjo{\mathsf{u}_0}D(C_n)\padjo{\mathsf{u}_0} &= D_0(C_n),\\
        \adjo{\mathsf{u}_1}D(C_n)\padjo{\mathsf{u}_1} &= D_1(C_n)
    \end{array}
\end{equation}
gives the corepresentation matrices $D_{0,1}(C_n)$ of only the occupied bands at $\Gamma(0)$ and $\Gamma(1/2)$ (i.e., at the south and north pole), respectively.
Thus,
\begin{align}
    \mathcal{W}\left(\phi+\frac{2\pi}{n}\right) &= \lim_{\delta\theta\to 0}D_0(C_n)\left[\prod_\theta^{0\rightarrow \pi-\delta\theta}\adjo{V(\theta,\phi)}\right]\times\nonumber\\*
    &\quad\times\adjo{D_1(C_n)}\left[\prod_\theta^{\pi-\delta\theta\leftarrow 0}V(\theta,0)\right]\padjo{\mathsf{u}_0}.
\end{align}
Defining
\begin{equation}
    P(\phi) = \left[\prod_\theta^{\pi-\delta\theta\leftarrow 0}V(\theta,\phi)\right] \in \SO(N_\mathrm{occ}),
    \label{App:eq:parallel-transport-P_def}
\end{equation}
we arrive at \cref{eq:Euler_charge:Wilson_loop_constraint} of \cref{Sec:HOBBC:Euler},
\begin{equation}
    \mathcal{W}\left(\phi+\frac{2\pi}{n}\right) = D_0(C_n)\adjo{P(\phi)}\adjo{D_1(C_n)}P(\phi)\mathcal{W}(\phi).
    \label{eqn:Wilson-Cn-relation}
\end{equation}

\subsection{Berry phase on a symmetric path}\label{App:Euler-monopole:Berry-phase}
The constraint on the Wilson loop operator derived above has implications for the Berry phase of the occupied bands computed on the contours $\Gamma(\phi)$.
The Berry phase is given by $\varphi=\arg\det\mathcal{W}$, such that
\begin{align}
    \varphi\left(\phi+\frac{2\pi}{n}\right) &= \arg\det\left[D_0(C_n)\adjo{P(\phi)}\adjo{D_1(C_n)}P(\phi)\mathcal{W}(\phi)\right].\nonumber\\
    \intertext{Using that inside the determinant all matrices commute,}
    &= \arg\det\mathcal{W}(\phi) + \arg\det\left[D_0(C_n)\adjo{D_1(C_n)}\right]\nonumber\\
    &= \varphi(\phi) + \arg\det\left[D_0(C_n)\adjo{D_1(C_n)}\right]\mod 2\pi,
    \label{App:eq:Berry-phase_diff}
\end{align}
which is exactly \cref{eq:HOBBC:Berry_phase_constraint}.
The difference in Berry phase on the two contours is found to be
\begin{equation}
    \Delta\varphi = \arg\det\left[D_0(C_n)\adjo{D_1(C_n)}\right].
    \label{eqn:Berry-difference}
\end{equation}

Recall that in the presence of $\mcP\mcT$ symmetry, the Berry phase on any closed loop is (as well as their differences) quantized to $0$ vs.\ $\pi$, and that the Berry curvature vanishes identically away from the nodal lines~\cite{Bzdusek:2017}.
It therefore follows that the trivial value, $\Delta\varphi=0$, indicates that there is an \emph{even} number of nodal lines enclosed by the contour $\gamma(\phi+2\pi/n)^{-1}\circ\gamma(\phi)$, while a non-trivial value, $\Delta\varphi = \pi$, indicates an \emph{odd} number.
This implies that each $\tfrac{2\pi}{n}$ sector of the spherical surface is penetrated by an even (odd) number of nodal lines in the principal gap and thus, if $\Delta\varphi=\pi$, that the surface \emph{must} contain gapless points.

\subsection{Symmetry indicators for Euler monopole charge}\label{App:Euler-monopole:charge}
We now specialize to the case where $N_\mathrm{occ}=2$ and assume that $D_{0,1}(C_n)\in\SO(2)$.
Since the $n^\textrm{th}$ power of $D_{0,1}(C_n)$ gives the identity, it follows that for $i\in\{0,1\}$:
\begin{equation}
    D_i(C_n) = \e^{-\frac{2\pi\i}{n}r_{i}s_y}\quad\textrm{with}\quad r_{i}\in\mathbb{Z}
\end{equation}
and Pauli matrices $s_i$ acting on the space of the two valence bands.
Because $\mathsf{SO}(2)$ is an Abelian group, we find that
\begin{align}
    \mathcal{W}\left(\phi+\frac{2\pi}{n}\right) &= D_0(C_n)\adjo{D_1(C_n)}\times\nonumber\\*
    &\quad\times\lim_{\delta\theta\to 0}\left[\prod_\theta^{0\rightarrow \pi-\delta\theta}\adjo{V(\theta,\phi)}\right]\left[\prod_\theta^{\pi-\delta\theta\leftarrow 0}V(\theta,0)\right]\padjo{\mathsf{u}_0}\nonumber\\*
    &= D_0(C_n)\adjo{D_1(C_n)}\mathcal{W}(\phi).
    \label{eq:Wilsonloop_constraint}
\end{align}
From the family of Wilson loop operators $\mathcal{W}(\phi)$ for $\phi\in[0,2\pi)$, the Euler class can be obtained~\cite{Bouhon:2020}.
Since $\mathcal{W}(\phi)\in\SO(2)$, there is a $\zeta(\phi)\in[0,2\pi)$ such that
\begin{equation}
    \mathcal{W}(\phi) = \e^{\i\zeta(\phi)s_y}
\end{equation}
given by the Pfaffian of the logarithm
\begin{equation}
    \zeta(\phi) = \mathrm{Pf}\left[\log\mathcal{W}(\phi)\right]\mod 2\pi.
\end{equation}
The phase $\zeta(\phi)$ changes continuously in $\phi$ as long as the two-band subspace is separated from the other bands by energy gaps, and its winding number determines the Euler class as
\begin{equation}
    \Eu = \frac{1}{2\pi}\int_0^{2\pi}\dv{\zeta(\phi)}{\phi} = \frac{1}{2\pi}\left[\zeta(2\pi)-\zeta(0)\right].
\end{equation}

We remark that the above expression should be read with caution.
Namely, a gauge transformation $\mathsf{u}_0\mapsto\mathsf{u}_0U$ with $U\in\O(2)$ at the base point of the closed path $\Gamma$ transforms the Wilson loop operator as follows
\begin{equation}
    \mathcal{W}(\phi) \mapsto U^\top\mathcal{W}(\phi)U,
\end{equation}
such that $\zeta(\phi)\mapsto\zeta'(\phi)$ with
\begin{align}
    \zeta'(\phi) &= \mathrm{Pf}\left\{\log\left[U^\top\mathcal{W}(\phi)U\right]\right\}\mod 2\pi\nonumber\\
    &= \mathrm{Pf}\left\{U^\top\log\left[\mathcal{W}(\phi)\right]U\right\}\mod 2\pi.\nonumber\\
    \intertext{Using that $\mathrm{Pf}(BAB^\top)=\det(B)\mathrm{Pf}(A)$, we obtain}
    &= \det(U)\zeta(\phi)\mod 2.
    \label{eqn:Pf-sign-ambiguity}
\end{align}
Since $\det(U)=\pm 1$, the sign of $\zeta(\phi)$ and therefore $\Eu$ is gauge-dependent~\cite{Bouhon:2020}.
Therefore, the well-defined topological invariant that we can extract is the \emph{absolute value} $\abs{\Eu}$ of the Euler monopole charge.

\Cref{eq:Wilsonloop_constraint} implies that
\begin{align}
    \zeta\left(\phi+\frac{2\pi}{n}\right) &= \mathrm{Pf}\left[\log\e^{\frac{2\pi\i}{n}\left(r_1-r_0\right)s_y + \i\zeta(\phi) s_y}\right]\nonumber\\
    &= \frac{2\pi}{n}(r_1-r_0)+\zeta(\phi)\mod 2\pi
\end{align}
and applying that identity $n$ times gives
\begin{equation}
    \zeta\left(\phi+2\pi\right) = 2\pi\left(r_1-r_0\right)+\zeta(\phi)\mod 2\pi.
\end{equation}
Therefore, the Euler class is
\begin{equation}
    \Eu = r_1-r_0\mod n.
    \label{App:eq:Euler-SI}
\end{equation}
Recall that the parity of $\Eu$ is the second Stiefel-Whitney class
\begin{equation}
    \SW{2} = \Eu \mod 2.
    \label{App:eq:2SW_from_Euler}
\end{equation}
Observe that for odd $n$, \cref{App:eq:Euler-SI} leaves the parity of $\Eu$ undetermined.
It follows that for $C_3$, $\SW{2}$ is not constrained by symmetry, which is consistent with what we demonstrated in \cref{Sec:MinimalModels:C3}. 

\subsection{Application to triple-point pairs}\label{App:Euler-monopole:TPP}
We apply the above results to TPPs formed by a 2D ICR $\rho^\mathrm{2D}$ and two 1D ICRs $\rho_a^\mathrm{1D}$, $\rho_b^\mathrm{1D}$.
In that case the representation matrices of $C_n$ in the orange and blue $k_z$-range appearing in \cref{App:eq:Berry-phase_diff,App:eq:Euler-SI} are
\begin{equation}
    \begin{array}{rcl}
        D_0(C_n) &=& \rho^\mathrm{2D}(C_n),\\
        D_1(C_n) &=& \rho_a^\mathrm{1D}(C_n)\oplus\rho_b^\mathrm{1D}(C_n),
    \end{array}
    \label{App:eq:Euler-monopole:reps-at-poles}
\end{equation}
respectively.
For the relevant point groups stabilizing TPs in the presence of $\mcP\mcT$ symmetry, \cref{App:tab:rot_ICRs} lists the representation matrices for all 1D and 2D ICRs.

\begin{table}[t]
    \centering
    \caption{
        Irreducible corepresentations (ICRs) of the $C_n$ rotational symmetry in the point groups $C_{n(v)}$ with space-time inversion symmetry $\mcP\mcT$ for $n\in\{3,4,6\}$.
        The notation for the ICRs follows Ref.~\onlinecite{Bradley:1972}, where we drop the subscripts if they do not affect the result. 
        Note that for the 2D ICRs of the point group $C_{6(v)}$ we define: $^2E_2\!\,^1E_2 \mapsto E_1$, and $^2E_1\!\,^1E_1 \mapsto E_2$.
        The 2D ICRs are all $\SO(2)$ matrices and are therefore given in the form $\e^{-\frac{2\pi\i}{n}rs_y}$ for $r\in\mathbb{Z}$ and the Pauli matrix $s_y$.
    }
    \begin{ruledtabular}
	\begin{tabular}{CCCCCC}
		          \multirow{2}{*}{$C_{6(v)}$}           & \text{ICR} & A & B  &              E_1               &              E_2               \\
		                                                & \rho(C_6)  & 1 & -1 & \e^{-\frac{2\pi\i}{6} s_y} & \e^{-\frac{2\pi\i}{6} 2s_y} \\[0.1em] \hline
		\rule{0pt}{1.25em}
	\multirow{2}{*}{$C_{4(v)}$} & \text{ICR} & A & B  &               E                &                                \\
		                                                & \rho(C_4)  & 1 & -1 & \e^{-\frac{2\pi\i}{4} s_y} &                                \\[0.1em] \hline
		\rule{0pt}{1.25em}
	\multirow{2}{*}{$C_{3(v)}$} & \text{ICR} & A &    &               E                &                                \\
		                                                & \rho(C_3)  & 1 &    & \e^{-\frac{2\pi\i}{3} s_y} &
	\end{tabular}
    \end{ruledtabular}
    \label{App:tab:rot_ICRs}
\end{table}

In \cref{App:Euler-monopole:Berry-phase} we have derived a necessary condition for the enclosing spherical surface (on which the Wilson loop operators are computed) to be gapped.
Substituting \cref{App:eq:Euler-monopole:reps-at-poles} into \cref{App:eq:Berry-phase_diff}, we find that the condition
\begin{equation}
    \arg\det\left[\rho^\mathrm{2D}(C_n)\adjo{\rho_a^\mathrm{1D}(C_n)}\oplus\adjo{\rho_b^\mathrm{1D}(C_n)}\right] = 0
\end{equation}
is equivalent to
\begin{equation}
    \det\rho^\mathrm{2D}(C_n) = \rho_a^\mathrm{1D}(C_n)\rho_b^\mathrm{1D}(C_n).
    \label{App:eq:TPP-gap-condition}
\end{equation}
According to \cref{App:tab:rot_ICRs}, all 2D ICRs have determinant $1$, such that \cref{App:eq:TPP-gap-condition} is violated by all combinations of ICRs that involve two 1D ICRs with \emph{different} rotation eigenvalue.
For $C_4$ symmetries this excludes $(E;A,B)$ and for $C_6$ $(E_i;A,B)$ with $i=1,2$. (Note that we have dropped subscripts at $A$ and $B$ where they do not make a difference; this convention is also adopted in tables \cref{tab:HOBBC:Qjump,App:tab:rot_ICRs,App:tab:TPP-Euler-monopole}.)

Finally, we apply \cref{App:eq:Euler-SI} to all admissible combinations of ICRs from \cref{App:tab:rot_ICRs}, i.e., where the 1D ICRs have \emph{equal} rotation eigenvalues.
In all those cases both $\rho^\mathrm{2D}(C_n)$ and $\rho_a^\mathrm{1D}(C_n)\oplus\rho_b^\mathrm{1D}(C_n)$ are indeed $\SO(2)$ matrices.
The results are shown in \cref{App:tab:TPP-Euler-monopole}.
Note that (\emph{i}) only the absolute value of the Euler class is well-defined, since the sign is gauge-dependent, and (\emph{ii}) the symmetry-indicator formula determines the Euler monopole charge modulo $n$, where $n$ is the order of the rotational symmetry. 
Due to these these ambiguities, $\Big\{0,1,\ldots,\floor{\tfrac{n}{2}}\Big\}$ is the largest set of unique values of the Euler monopole charge that can be distinguished based on the ICRs of the two valence bands. 
More precisely, all values of $\Eu$ such that $\Eu=mn\pm p$ (with $n$ the order of the rotational symmetry, $p$ an element of the just specified set, and $m\in\mathbb{Z}$) are not distinguishable from $\Eu=p$.
With the convention that $a\mod n\in(-\floor{n/2},\dotsc,\floor{n/2}]$, we write $\abs{\Eu\mod n}$ to indicate the representative of the corresponding equivalence class of all such indistinguishable values~of~$\Eu$.

In particular,
for $n=6$, values $\abs{\Eu}=4$ and $\abs{\Eu}=5$ are indistinguishable from $\abs{\Eu}=2$ and $\abs{\Eu}=1$, respectively; while for $n=4$, the value $\abs{\Eu}=3$ is indistinguishable from $\abs{\Eu}=1$.
For these two cases the parity of $\Eu$ is well defined, such that the 2SW monopole charge, given by $\SW{2} = \Eu\mod 2\in\mathbb{Z}_2$, is uniquely determined.
In contrast, for $n=3$, the symmetry-indicator formula for $\Eu$ can only distinguish values $\abs{\Eu}=0$ and $\abs{\Eu}=1$, with $\abs{\Eu}=2$ being indistinguishable from $\abs{\Eu}=1$.
This implies that the parity is not fixed by the symmetry indicators and therefore the 2SW monopole charge in $C_3$-symmetric models is not determined from the symmetry eigenvalues.

\begin{table}[t]
\begin{threeparttable}
    \centering
    \caption{
        Euler class and second Stiefel-Whitney class computed on a surface enclosing a triple point pair for all possible combinations of irreducible corepresentations (ICRs) in systems with space-time inversion symmetry $\mcP\mcT$ squaring to the identity and assuming that on that surface the two-band subspace of occupied bands is separated from the remaining bands by energy gaps.
        Combinations of ICRs leading necessarily to a gapless bulk below or above the TPP are excluded.
        The first column gives the order $n$ of rotational symmetry protecting the triple points.
        In the second column we list possible combinations of one 2D and two         1D ICRs, where the notation follows Ref.~\onlinecite{Bradley:1972}.
        Note that we drop the subscript of the ICRs if the choice of the subscript does not affect the result. 
        Furthermore, for the 2D ICRs of the point group $C_{6(v)}$ we define: $^2E_2\!\,^1E_2 \mapsto E_1$, and $^2E_1\!\,^1E_1 \mapsto E_2$.
        For convenience we repeat the results on triple-point type for the given combinations of ICRs in the third column.
        Finally, we give Euler class $\Eu$ and the second Stiefel-Whitney class $\SW{2}$ in the last two columns.
    }
    \begin{ruledtabular}
	\begin{tabular}{CCCCC}
	    n & \text{ICRs} & \text{Type} & \abs{\Eu\mod n}^{\,\textrm{b}} & \SW{2}\\[0.1em]\hline\addlinespace
	    \multirow{4}{*}{$6$} & (E_1;A,A) & (\mathsf{A},\mathsf{A}) & 1 & 1\\
	    & (E_1;B,B) & (\mathsf{B},\mathsf{B}) & 2 & 0\\
	    & (E_2;A,A) & (\mathsf{B},\mathsf{B}) & 2 & 0\\
	    & (E_2;B,B) & (\mathsf{A},\mathsf{A}) & 1 & 1\\\addlinespace
	    \multirow{2}{*}{$4$} & (E;A,A) & (\mathsf{A},\mathsf{A}) & 1 & 1\\
	    & (E;B,B) & (\mathsf{A},\mathsf{A}) & 1 & 1\\
	    3 & (E;A,A) & (\mathsf{B},\mathsf{B}) & 1\text{ or }2^{\,\textrm{b}} &
	    1\text{ or }0^{\,\textrm{b}}
	\end{tabular}
    \end{ruledtabular}
    \begin{tablenotes}[flushleft]
        \item[$\textrm{a}$] Here we use the convention that $a\mod n\in(-\floor{n/2},\dotsc,\floor{n/2}]$, such that $\abs{a\mod n}\in\{0,1,\dotsc,\floor{n/2}\}$, see text.
        \item[$\textrm{b}$] Explicitly, we find $\Eu=-1$, which cannot be distinguished from $2$. However, because $\Eu$ is only well-defined up to a sign, the Euler class could also take the value $1$.
        This implies that the parity of the Euler class is not uniquely determined by the symmetry indicator and thus the second Stiefel-Whitney class is not constrained by $C_3$ symmetry.
        \par
    \end{tablenotes}
    \label{App:tab:TPP-Euler-monopole}
\end{threeparttable}
\end{table}

\subsection{Gaplessness in absence of \texorpdfstring{$\mcP\mcT$}{PT} symmetry}\label{App:Euler-monopole:PT-absence}

Here, we discuss the implications of \cref{App:eq:Berry-phase_diff} (the derivation of which only assumes the $C_n$ rotational symmetry) in the \emph{absence} of $\mcP\mcT$ symmetry.
By adjusting a few steps leading to \cref{eqn:Berry-difference}, one should recognize that the Berry phase on any closed loop of the form $\Lambda(n, \phi) = \gamma(\phi+2\pi/n)^{-1}\circ\gamma(\phi)$ (and assuming the trajectory does not encounter nodes in the principal gap) is given by
\begin{equation}
    \varphi_n = \arg\det\left[D_0(C_n)\adjo{D_1(C_n)}\right],
\end{equation}
independent of $\phi$.
In the presence of $\mcP\mcT$, we have argued that the Berry phase on any closed path [including the phase $\varphi_n$ on $\Lambda(n,\phi)$] is quantized to $0$ vs.~$\pi$, and that $\varphi_n=\pi$ implies an odd number of nodal lines to pass through the loop.
If $\mcP\mcT$ is absent but vertical mirror symmetries ($m_v$) are present, then an analogous quantization of Berry phase can be established for loops that are symmetric under mirror reflection~\cite{Zak:1989,Hughes:2011,Bzdusek:2017b}, and nodal lines can only be stabilized inside mirror planes.
In this case, a $\pi$ Berry phase implies that the mirror-symmetric loop encloses an odd number of nodal lines inside the mirror-invariant plane.
(The cases with neither $\mcP\mcT$ nor $m_v$ are dealt with trivially towards the end of this section).

In the relevant magnetic point groups with rotational symmetry $C_n$, with vertical mirror symmetry, but without $\mcP\mcT$ symmetry (cf.~\cref{tab:TP-MPGs}), we can therefore choose $\phi$ such that $\gamma(\phi)$ lies within one of the mirror planes $m_1$, $\gamma(\phi+2\pi/n)$ in its copy $m_1'$ related by $C_n$ symmetry, and the two parts of the contour are mapped onto each other by the mirror symmetry with mirror plane $m_2$ lying between $m_1$ and $m_1'$.
However, we need to be more careful, because $\phi$ is now fixed to a mirror plane and therefore $\gamma(\phi)$ might contain band nodes in the principal gap.

\begin{figure}[t]
    \centering
    \includegraphics{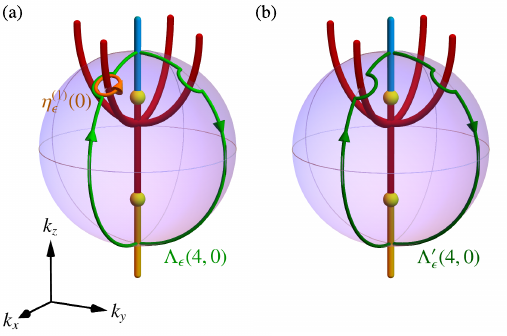}
    \caption{
        Illustration of the closed loops considered in \cref{App:Euler-monopole:PT-absence} for the particular case $n=4$.
        (a) The contour $\Lambda_\epsilon(n,\phi)$ (green) consists of two rotation-related semi-circular arcs inside rotation-related mirror planes (here $m_x$ and $m_y$). 
        If mirror-protected nodal lines occur along the contour, we avoid them with infinitesimal half circles perpendicular to the mirror plane containing the avoided nodal lines.
        The infinitesimal circular contour $\eta_\epsilon^{(1)}(\phi)$ enclosing the nodal line is illustrated as a dark orange ring; it is symmetric under $m_y$.
        (b) Composition of $\Lambda_\epsilon(n,\phi)$ with $\eta_\epsilon^{(1)}(\phi)$ results in the new contour $\Lambda_\epsilon'(n,\phi)$, which is symmetric under the diagonal mirror symmetry $m_{\bar{x}y}$.
    }
    \label{App:fig:Wilson-loop_Berry-phase-path}
\end{figure}

If $\gamma(\phi)$ does not contain any band nodes, then $\varphi_n=\pi$ implies that $m_2$ contains an odd number of nodal lines as illustrated in \cref{fig:Euler_charge}(b) for $\Lambda(4,0)=\Gamma(2\pi/4)$.
On the other hand, if $\gamma(\phi)$ does contain $M_1$ band nodes, we exclude them by modifying $\gamma$ to $\gamma_\epsilon$ such that the band nodes are circumvented (but enclosed by the loop) with infinitesimal half circles, see \cref{App:fig:Wilson-loop_Berry-phase-path}(a).
Now $\Lambda_\epsilon(n,\phi) = \gamma_\epsilon(\phi+2\pi/n)^{-1}\circ\gamma_\epsilon(\phi)$ is not mirror symmetric anymore, but it is still composed of two segments related by rotational symmetry; therefore, the Berry phase on $\Lambda_\epsilon(n,\phi)$ is still $\varphi_n$.
We next compose $\Lambda_\epsilon(n,\phi)$ with an infinitesimal circular contour $\eta_\epsilon^{(i)}(\phi)$ around each excluded band node on $\gamma_\epsilon(\phi)$ such that the resulting loop $\Lambda_\epsilon'(n,\phi)$ does \emph{not} include the band nodes anymore. 
Observe that this new path is mirror symmetric with respect to $m_2$, and its Berry phase is therefore quantized to $\{0,\pi\}$, cf.~\cref{App:fig:Wilson-loop_Berry-phase-path}(b).
Additionally, the Berry phase on $\eta_\epsilon^{(i)}(\phi)$ is quantized as well and is indeed $\pi$, because it is itself mirror-symmetric (albeit with respect to $m_1$ rather than $m_2$) and by assumption it encloses exactly one band node.
Thus, the Berry phase on $\Lambda_\epsilon'(n,\phi)$ satisfies
\begin{equation}
    \varphi_n' = \varphi_n + M_1\pi\mod 2\pi
\end{equation}
and it determines the parity of the number $M_2$ of nodal lines in the mirror plane $m_2$, i.e.,
\begin{equation}
    \varphi_n' = (M_2\mod 2)\pi.
\end{equation}
By combining the previous two equations, we obtain
\begin{equation}
    \varphi_n = (M_1+M_2\mod 2)\pi,
\end{equation}
meaning that $\varphi_n\in\{0,\pi\}$ determines the parity of the number of nodal lines in mirror planes $m_1$ and $m_2$ together, i.e., exactly the ones passing through a $\tfrac{2\pi}{n}$ sector of the enclosing sphere, just as in the presence of $\mcP\mcT$.

Let us finally remark that for TPs stabilized by $C_n\mcP\mcT$ with $n=4,6$, we only need to consider the case \emph{with} mirror symmetry, because if no mirror symmetry is present, then nodal lines cannot be stabilized away from the rotation axis (see also the discussion in \cref{Sec:HOBBC:SymBreak}).
Therefore, there are no stable band nodes bound to the enclosing sphere.
But the case of $C_n\mcP\mcT$ and mirror symmetry, reduces to the case $C_{n'}$ with $n'=n/2\in\{2,3\}$ and mirror symmetry that we have discussed in the previous two paragraphs.

In conclusion, we have shown that even in the absence of $\mcP\mcT$ symmetry,
\begin{equation}
    \det\rho^\mathrm{2D}(C_n) \neq \rho_a^\mathrm{1D}(C_n)\rho_b^\mathrm{1D}(C_n)
\end{equation}
implies that the bulk in the orange or blue $k_z$-ranges is necessarily gapless.
It remains to check this condition for all possible combinations of ICRs.
To that end, recall from Sec.~\ref{Sec:HOBBC:SymBreak} that in the presence of vertical mirror symmetry, addition of $\mcP\mcT$ does not change the ICRs and \cref{App:tab:rot_ICRs} still applies, giving the same result as in the presence of $\mcP\mcT$.
The only remaining cases are the magnetic point groups $\bar{4}'2'm$ and $\bar{6}'m2'$.
For the former we find $\det\rho^\mathrm{2D}(C_2)=1$, $\rho_{a,b}^\mathrm{1D}(C_2)=1$ and for the latter $\det\rho^\mathrm{2D}(C_3)=1$, $\rho_{a,b}^\mathrm{1D}(C_3)=1$, such that in both cases none of the combinations of ICRs are necessarily gapless.

\section{Tight-binding models}\label{App:MinimalModels}

\subsection{Construction of minimal tight-binding models}\label{App:MinimalModelsConstruction}

To construct the three tight-binding models discussed in \cref{Sec:MinimalModels}, we proceed as follows.
We first choose a point group ($D_{4h}$ and $D_{6h}$ for our models) containing the little co-group $\LcG{\vec{k}}$ which we want to protect the triple-points (TPs) as a subgroup.
The choice of orbitals determines the irreducible representations of the point group.
Using the \textsc{3D GenPos} application on the Bilbao crystallographic server (BCS), we determine a set of possible generators of that point group, and then find their matrix representations using the \textsc{Representations PG} application~\cite{Elcoro:2017}.
To find the ICRs corresponding to the irreducible representations of the chosen orbitals after inclusion of $\mcP\mcT$ symmetry with $(\mcP\mcT)^2=+\id$, we follow Ref.~\onlinecite{Bradley:1972}. Given the set of generators, their ICRs and a generating set of hopping vectors, we use the \Python{} package \textsc{Qsymm}~\cite{Varjas:2018} to construct a family of symmetry-allowed Bloch Hamiltonians.

Finally, we tune the parameters in that family of Hamiltonians such that we obtain a four-band triple-point pair (TPP), accompanied by the minimal number of additional nodal lines (NLs).
This is achieved by first introducing a double band inversion at $\Gamma$ by setting selected intra-orbital hopping parameters to be non-vanishing, and then by adding inter-orbital terms until we gap out all band nodes that are not required by the symmetry or topology.
Note that the type of the TPs can be predicted using our previously developed classification~\cite{Lenggenhager:2021:MBNLs}.
Using topological quantum chemistry (TQC) and the \textsc{BandRep} application on BCS~\cite{Bradlyn:2017,Vergniory:2017,Elcoro:2017}, we deduce the ICRs at the high-symmetry points, and using the \textsc{Compatibility Relations}~\cite{Elcoro:2017} infer the ones along the relevant rotation axis.

\subsection{Details of the discussed tight-binding models}\label{App:MinimalModels:Details}

Here, we provide details of the models studied in \cref{Sec:MinimalModels}.
For each model we present the Hamiltonian, the basis in which the Hamiltonian is expressed, the numerical values of parameters considered in the main text, and the matrix representations that we used to construct the models (cf.~\cref{App:MinimalModelsConstruction}).

\subsubsection{\texorpdfstring{$C_6$}{C6}-symmetric model with two type-\texorpdfstring{$\mathsf{A}$}{A} triple points}\label{App:ModelDetails:C6AA}

The model discussed in \cref{Sec:MinimalModels:C6AA} has SG $P6/mmm$ (No.~191) and is defined on a hexagonal lattice with lattice constants $a=c=1$.
As described in the main text, we place $(d_{xy},d_{x^2-y^2})$ and $f_{x(x^2-3y^2)}$, $f_{y(3x^2-y^2)}$ orbitals at Wyckoff position (WP) $1a$ (site-symmetry group $D_{6h}$).
We adopt the basis $(d_{xy},d_{x^2-y^2},\i f_{x(x^2-3y^2)},\i f_{y(3x^2-y^2)})$, such that $\mcP\mcT$ is represented simply by complex conjugation.
According to TQC such a placement of orbitals results in the irreducible representations $(E_{2\mathrm{g}};B_{1\mathrm{u}},B_{2\mathrm{u}})$ of $D_{6h}$ with unchanged ICRs after adding $\mcP\mcT$. In the above basis, the generators of $D_{6h}$ have matrix representations
\begin{subequations}
    \begin{align}
        C_{6z} &= \diag\left(R(-2\pi/3),-\id_\tau\right),\\
        m_v &= \sigma_z\otimes\tau_z,\\
        \mcP &= \sigma_z\otimes\id_\tau,\\
        \mcP\mcT &= \mcK,
    \end{align}
\end{subequations}
where $m_v\,:\,y\to -y$ and $R(\theta)$ is the 2D rotation matrix
\begin{equation}
    R(\theta) = \mat{\cos\theta & -\sin\theta \\ \sin\theta & \cos\theta }.
\end{equation}
The Pauli matrices $\sigma_i$ act on the $d$~vs.~$f$ (i.e., angular momentum) degree of freedom, while $\tau_i$ act on the two-level degrees of freedom with a fixed angular momentum.

Including only the nearest neighbour in-plane as well as the nearest vertical hopping terms, the model's Bloch Hamiltonian can be written as
\begin{multline}
    \Ham_\mathsf{AA}^{(6)}(\vec{k}) = -\left[t_1+2t_2\left(\cos k_x+2\cos\tfrac{k_x}{2}\cos\tfrac{\sqrt{3}k_y}{2}\right)\right.\\
    \left.\vphantom{\tfrac{\sqrt{3}}{2}}+\,t_3\cos k_z\right]\gamma_3 - t_4\left(\gamma_{14}-\gamma_{25}\right)\\
    -\,t_5\left[\left(\cos k_x-\cos\tfrac{k_x}{2}\cos\tfrac{\sqrt{3}k_y}{2}\right)(\gamma_{14}+\gamma_{25})\right.\\
    \left.-\,\sqrt{3}\sin\tfrac{k_x}{2}\sin\tfrac{\sqrt{3} k_y}{2}(\gamma_{15}-\gamma_{24})\right]\\
    -\,2t_6\left[\sin\tfrac{k_x}{2}\left(2\cos\tfrac{k_x}{2}+\cos\tfrac{\sqrt{3} k_y}{2}\right)\gamma_1\right.\\
    \left.-\,\sqrt{3}\cos\tfrac{k_x}{2}\sin\tfrac{\sqrt{3} k_y}{2}\gamma_2\right],
    \label{eq:MinimalModels:C6AA:def}
\end{multline}
with the Gamma matrices defined in \cref{Sec:C4Model}.
We tune the parameters to $t_1=3$, $t_2=-\tfrac{1}{2}$, $t_3=-2$, $t_4=\tfrac{1}{4}$, $t_5=\tfrac{1}{6}$ and $t_6=-\tfrac{1}{3}$ to obtain a double band inversion at $\Gamma$ and no other band inversions.
The resulting band structure is shown in \cref{fig:MinimalModels:C6AA}(a).

\subsubsection{\texorpdfstring{$C_6$}{C6}-symmetric model with two type-\texorpdfstring{$\mathsf{B}$}{A} triple points}\label{App:ModelDetails:C6BB}

The second $C_6$-symmetric model, discussed in \cref{Sec:MinimalModels:C6BB}, has the same space group (SG) $P6/mmm$ (No. 191) and is also defined on a hexagonal lattice with lattice constants $a=c=1$.
However, the $d$ orbitals are replaced by $p$ orbitals, such that we place $(p_x,p_y)$ and $f_{x(x^2-3y^2)}$, $f_{y(3x^2-y^2)}$ at WP $1a$.
We adopt the basis $(\i p_x,\i p_y,\i f_{x(x^2-3y^2)},\i f_{y(3x^2-y^2)})$.
According to TQC, this results in the irreducible representations $(E_{1\mathrm{u}};B_{1\mathrm{u}},B_{2\mathrm{u}})$ of $D_{6h}$, again with unchanged ICRs after adding $\mcP\mcT$.
In the above basis, the generators of $D_{6h}$ have matrix representations
\begin{subequations}
    \begin{align}
        C_{6z} &= \diag\left(R(2\pi/6),-\id_\tau\right),\\
        m_v &= \diag\left(\tau_x,-\tau_z\right),\\
        \mcP &= -\id_\sigma\otimes\id_\tau,\\
        \mcP\mcT &= \mcK,
    \end{align}
\end{subequations}
where Pauli matrices $\sigma_i$ act on the $p$~vs.~$f$ (i.e., angular momentum) degree of freedom, while $\tau_i$ act on the two-level degrees of freedom with a fixed angular momentum.

The model's Bloch Hamiltonian with only the nearest neighbour in-plane and the nearest vertical hopping terms is
\begin{multline}
    \Ham_\mathsf{BB}^{(6)}(\vec{k}) = -\left[t_1+2t_2\left(\cos k_x+2\cos\tfrac{k_x}{2}\cos\tfrac{\sqrt{3}k_y}{2}\right)\right.\\
    \left.\vphantom{\tfrac{\sqrt{3}}{2}}+\,t_3\cos k_z\right]\gamma_3 - t_4\left(\gamma_{14}-\gamma_{25}\right)\\
    -\,t_5\left[\left(\cos k_x-\cos\tfrac{k_x}{2}\cos\tfrac{\sqrt{3}k_y}{2}\right)(\gamma_{15}-\gamma_{24})\right.\\
    \left.+\,\sqrt{3}\sin\tfrac{k_x}{2}\sin\tfrac{\sqrt{3} k_y}{2}(\gamma_{14}+\gamma_{25})\right]\\
    -\,\sqrt{2}t_6\left[\left(\cos k_x - \cos\tfrac{k_x}{2}\cos\tfrac{\sqrt{3} k_y}{2}\right)(\gamma_1-\gamma_2)\right.\\
    \left.-\,\sqrt{3}\sin\tfrac{k_x}{2}\sin\tfrac{\sqrt{3} k_y}{2}(\gamma_1+\gamma_2)\right],
    \label{eq:MinimalModels:C6BB:def}
\end{multline}
with the Gamma matrices defined in \cref{Sec:C4Model}.
We set the model parameters to $t_1=4$, $t_2=-\tfrac{2}{3}$, $t_3=-3$, $t_4=-\tfrac{1}{2}$, $t_5=-\tfrac{1}{3}$ and $t_6=\tfrac{6}{5}$, which results in the band structure plotted in \cref{fig:MinimalModels:C6BB}(a).

\subsubsection{\texorpdfstring{$C_3$}{C3}-symmetric models}\label{App:ModelDetails:C3}

The $C_3$-symmetric models of \cref{Sec:MinimalModels:C3} are obtained by starting from either \cref{eq:MinimalModels:C6AA:def} or \cref{eq:MinimalModels:C6BB:def} and adding the following perturbations to the Hamiltonian,
\begin{multline}
    \Delta\Ham_\mathsf{AA}^{(3)}(\vec{k}) = \\
    \delta_{C_3}\mat{1&0\\0&0}\otimes\left[\mat{-173\sin k_x&0\\0&2\sin\tfrac{k_x}{2}\left(58\cos\tfrac{k_x}{2}+115\cos\tfrac{\sqrt{3}k_y}{2}\right)}\right.\\
    \left.\vphantom{\mat{-173\sin k_x&0\\0&2\sin\tfrac{k_x}{2}\left(58\cos\tfrac{k_x}{2}+115\cos\tfrac{\sqrt{3}k_y}{2}\right)}}+\,200\cos\tfrac{k_x}{2}\,\sin\tfrac{\sqrt{3}k_y}{2}\,\tau_x\right]\sin k_z,
    \label{eq:MinimalModels:C3:pert}
\end{multline}
and
\begin{equation}
    \Delta\Ham_\mathsf{BB}^{(3)}(\vec{k}) = U_\mathrm{BB}\Delta\Ham_\mathrm{AA}^{(3)}(\vec{k})\adjo{U_\mathrm{BB}}
\end{equation}
where
\begin{equation}
    U_\mathsf{BB}=\frac{1}{\sqrt{2}}\mat{1&-1\\1&1}\otimes\mat{1&0\\0&0}+\id\otimes\mat{0&0\\0&1},
\end{equation}
respectively, resulting in the Hamiltonians $\Ham_\mathsf{AA}^{(3)}(\vec{k})$ and $\Ham_\mathsf{BB}^{(3)}(\vec{k})$ whose NL structure is shown in \cref{fig:MinimalModels:C3}(a,d), respectively.
Note that, in constrast to the $C_6$-symmetric models, the subscript of the Hamiltonian does \emph{not} denote the types of TPs, because a $C_3$-symmetric little co-group only allows for type-$\mathsf{B}$ TPs~\cite{Lenggenhager:2021:MBNLs}.
Instead, the subscript indicates from which $C_6$-symmetric model the corresponding $C_3$-symmetric model is derived.

\section{Computation of the corner charges}\label{App:corner-charge-extraction}

In this appendix we provide technical details on some of the methods used in the extraction of the corner charges from the charge distribution obtained from exact diagonalization of the tight-binding models.
First, note that if the electronic or ionic charge distribution leads to finite charge on different Wyckoff positions in the unit cell, we are left with an ionic crystal~\cite{Watanabe:2021}. 
Therefore, a coarse-graining or smoothing via a moving average is necessary to properly define edge and surface charge.
In \cref{App:coarse-graining} we present a discrete method that performs this task.
We used this method for the analysis of the models discussed in \cref{Sec:MinimalModels,Sec:Materials:StiefelWhitney}.
However, while this coarse-graining easily removes strong oscillations of the charge distribution on sub-unit-cell length scales, oscillations due to trivial edge-localized states can remain~\cite{Watanabe:2020} and obscure the corner charge.
Therefore, in \cref{App:edge-signal_removal} we discuss a method that we developed to remove the signal due to such edge states.
The method applies to the case when the \emph{total} edge charge per unit cell (of the ribbon, i.e., the collection of sites periodically repeated in the ribbon geometry) is vanishing, but the presence of edge-localized states induces a non-trivial profile of the charge density as a function of the distance to the boundary, as has been observed for \ce{Sc3AlC} in \cref{Sec:Materials:StiefelWhitney}.

\subsection{Coarse-graining of charge distribution}\label{App:coarse-graining}

\begin{figure}
    \centering
    \includegraphics{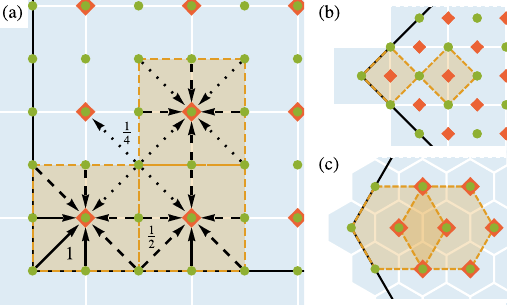}
    \caption{
        Discrete coarse-graining method.
        (a) The charge on the original lattice (defined by the unit cells shaded in blue and sublattice sites shown as green disks) is redistributed to the target lattice (defined by the red diamond sites).
        Here the lattice corresponding to the projection of \ce{Sc3AlC} (see \cref{Sec:Materials:StiefelWhitney}) is shown.
        Each site in the target lattice is assigned a coarse-graining cell (unit cell of that lattice; shaded in orange). The charge on each green site is then redistributed (black arrows) with equal weights (weight $1$ shown by solid, $\tfrac{1}{2}$ by dashed, and $\tfrac{1}{4}$ by dotted lines) to all those red sites whose coarse-graining cell is adjacent to the  green site.
        Only a few arrows are shown to maintain clarity of the illustration.
        (b,c) Definitions of the original (green) and target (red) lattice and the coarse-graining cell (orange) for the $C_4$- and $C_6$-symmetric models discussed in \cref{Sec:MinimalModels}, respectively.
        Note that in (c) the coarse-graining cells overlap; as a consequence, the original and the target lattice coincide except for a missing layer on the boundary.
    }
    \label{fig:corner-charge:coarse-graining}
\end{figure}

To coarse-grain the electric charge distribution defined on the \emph{original lattice}, we first need to choose the \emph{target lattice}.
The choice of target lattice defines the coarse-graining length scale.
To coarse-grain over a single unit cell with multiple sites, it is convenient to consider a single Wyckoff position of maximal symmetry to define the target lattice.
Some examples are shown in \cref{fig:corner-charge:coarse-graining}.
Note that it is also possible to coarse-grain over larger lengths scales, i.e., multiple unit cells.
Such a situation is depicted in \cref{fig:corner-charge:coarse-graining}(c) where 
coarse-graining is performed over seven original hexagonal unit cells.
However, note that due to overlapping of the coarse-graining cells, the Wigner-Seitz tessellation formed around the coarse-graining centers (not illustrated) coincides with the original honeycomb lattice.

The coarse-graining is performed by redistributing the charge from the original to the target lattice.
If a site on the original lattice belongs to $m$ coarse-graining cells, then a fraction $\tfrac{1}{m}$ of the total charge on the original site is distributed to each of the corresponding target sites.
Consequently, charge on corner sites is generally transferred to a single target site.
In \cref{fig:corner-charge:coarse-graining}(a), where we illustrate the coarse-graining as applied to \ce{Sc3AlC}, redistribution weights of magnitude $1$, $\tfrac{1}{2}$ and $\tfrac{1}{4}$ are illustrated with solid, dashed and dotted arrows, respectively.

We have applied the coarse-graining method to achieve a charge redistribution in all the discussed models.
In particular, \cref{fig:MinimalModels:C4:ChargeDist} shows a comparison of the charge distribution before [panels (a,b)] and after [panels (c,d)] coarse-graining for the $C_4$-symmetric model of \cref{Sec:C4Model}.
The definitions of the original and target lattice as well as the coarse-graining unit cell are shown in \cref{fig:corner-charge:coarse-graining}(b).
Panels (a,c) of \cref{fig:corner-charge:coarse-graining} give the same information for the model of \ce{Sc3AlC} (cf.~\cref{Sec:Materials:StiefelWhitney}) and the $C_6$-symmetric models (cf.~\cref{Sec:MinimalModels:C6AA,Sec:MinimalModels:C6BB}), respectively.

\subsection{Removal of edge signal from charge distribution}\label{App:edge-signal_removal}

Even after coarse-graining (as described above), charge oscillations caused by trivial edge-localized states can remain.
Note that in contrast to the oscillations on sub-unit-cell scale, these remaining oscillations are generally perpendicular to the edges, and vanish in the bulk.
Since, the total corner and edge charges are defined via integration over regions with boundaries perpendicular to the edges of the sample~\cite{Watanabe:2020}, the remaining oscillations do not prevent us from computing these total charges, see \cref{fig:Sc3AlC:nanowire}(c) for such a calculation on the \emph{original} coarse-grained data.
However, such oscillations can visually obscure the localization properties of the charge.
Here, we discuss a method to subtract such edge-localized oscillations for the cases where the total edge charge per unit cell (of the ribbon geometry) vanishes.
The method applies to $C_n$-symmetric systems with $n\geq 3$, where the bulk as well as the finite-size sample satisfy the symmetry requirement.
For concreteness we choose $n=4$ for the following discussion.
In particular, the described setup directly applies to the case of \ce{Sc3AlC} as presented in \cref{Sec:Materials:StiefelWhitney}.

\begin{figure}
    \centering
    \includegraphics{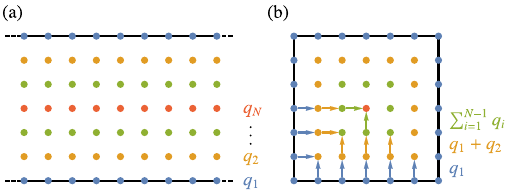}
    \caption{
        Procedure to remove the edge signal from the charge distribution of a $C_4$-symmetric system.
        It is assumed that the total edge charge per unit cell (of the ribbon geometry) vanishes.
        The solid lines in both panels represent the boundary of the lattice.
        (a) Ribbon geometry extended in the horizontal direction with $2N-1=7$ unit cells in the vertical direction (corresponding to $N=4$ inequivalent layers). Unit cells of the lattice are indicated by dots, colored according to their distance to the boundary.
        We denote by $q_i$ the charge per unit cell on the $i^\textrm{th}$ layer (which is constant within the layer).
        (b) Flake geometry for the same system. Here, \emph{shells} of unit cell sites (dots) correspond to the layers from the ribbon geometry (indicated by the corresponding color).
        The charge on the whole flake is redistributed according to the arrows indicated in the bottom-left part of the system. 
        The amount of charge that is transferred from a site that belongs to the $j^\textrm{th}$ shell to the next one is given by $\sum_{i=1}^{j} q_j$, where $q_i$ are the charges determined in the ribbon geometry, cf.~panel~(a). Note that the charge redistribution flows perpendicular to the boundary; therefore, no charge is transferred between the corners.
    }
    \label{fig:corner-charge:edge-signal_removal}
\end{figure}

Two separate calculations need to be performed: one on a ribbon with $2N-1$ layers (of unit cells) [cf.~\cref{fig:corner-charge:edge-signal_removal}(a)] and one on a flake with $N$ \emph{shells} (of unit cells) [cf.~\cref{fig:corner-charge:edge-signal_removal}(b)], such that there is a direct correspondence between the layers and the shells.
Due to the rotational symmetry, the four sides of each shell are equivalent and only a single orientation of the ribbon needs to be computationally modelled.
Note that compatible fillings (in particular, the same choice of chemical potential) need to be considered for the two systems.
The charge distributions obtained for both geometries first need to be coarse-grained to the level of unit-cells (cf.~\cref{Sec:Materials:StiefelWhitney}), such that the coarse-grained lattice has a single site per unit cell.
The resulting lattices are depicted in \cref{fig:corner-charge:edge-signal_removal}(a,b).

From the calculation on the ribbon, the total charge $q_i$ per unit cell in the $i^\textrm{th}$ layer can easily be extracted.
On the flake, we then redistribute the charge as indicated in \cref{fig:corner-charge:edge-signal_removal} with arrows:
from a site in the $j^\textrm{th}$ shell (but not on one of the diagonals) the charge $\sum_{i=1}^j q_i$ is transferred to the $(j+1)^\textrm{th}$ shell.
There is no charge transfer \emph{away} from sites that lie on one of the diagonals.
Therefore, this procedure seemingly leads to a charge accumulation on the diagonals and in the center.
However, recall that, by assumption, the total edge charge vanishes, such that in the limit of large $N$, $\sum_{i=1}^N q_i = 0$.
Consequently, the accumulated charge compensates oscillations on the diagonals and converges to zero at the center of the flake.
Furthermore, note that the charge redistribution flows strictly \emph{perpendicular} to the flake boundaries; therefore, no charge is transferred between the corners of the flake, keeping its value invariant (up to corrections that are exponentially small in the system size).

The effect of the method can be seen in \cref{fig:Sc3AlC:nanowire}, where the original (coarse-grained) distribution is shown in panels (d,e), and the charge distribution after removing the edge signal is displayed in panels (f,g).
We remark that the data in panels (f,g) have undergone one additional coarse-graining step \emph{after} the removal of the edge charge.
Compared to panels (d), the localization of the corner charge in panel (f) is visually much more manifest.

\section{Generalized quaternion charge in different Brillouin zones}\label{App:NLBZ}

\begin{figure}
    \centering
    \includegraphics{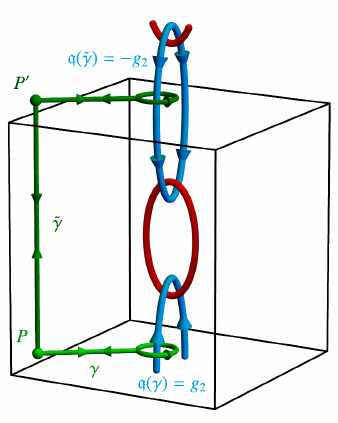}
    \caption{
        Definition of nodal line (NL) orientations via the non-Abelian invariant, illustrated on the example of multiband nodal link~\cite{Lenggenhager:2021:MBNLs} of a three-band model; red (blue) line indicates a NL formed by the lower (upper) two bands.
        The non-Abelian invariant is computed on paths $\gamma$ (light green) and $\tilde{\gamma}$ (dark green), where $P'$ is shifted relative to $P$ by a reciprocal lattice vector.
        In the three-band example shown here, $\mathfrak{q}(\gamma)=g_2$ (corresponding to a nodal line in the second gap, shown in blue) and bands $2$ and $3$ have different Berry phases $\phi_2\neq\phi_3$.
        Therefore, the orientation of the upper blue nodal ring is opposite to the one of the lower ring, $\mathfrak{q}(\gamma)=-g_2$.
    }
    \label{App:fig:NLBZ:NLOrientation}
\end{figure}

In this appendix we formalize the relationship between the orientation of nodal lines (NLs) and their copies in neighbouring Brillouin zones (BZs).
The orientation of a NL as used in \cref{Sec:Monopole:Euler} can be formally defined using the non-Abelian invariant introduced by Ref.~\onlinecite{Wu:2019} (also called \emph{generalized quaternion charge} therein).
While the topological classification of band nodes is given by equivalence classes of the generalized quaternion group (each equivalence class contains either a single element or two elements that differ by a sign), the sign of the invariant becomes well-defined if the contours have a common base point $P$.
In that case the sign of the generalized quaternion invariant computed on a contour enclosing a single NL, \emph{defines}~\cite{Tiwari:2020} the orientation of that NL, cf.~\cref{App:fig:NLBZ:NLOrientation}.

The prior works~\cite{Wu:2019, Tiwari:2020} on the non-Abelian band topology have only considered closed contours which are fully contained within a single BZ.
However, to understand the reversal of NL orientation between neighboring copies of BZ (cf.~\cref{fig:Monopole:NLOrientation}), one must explicitly consider paths that cross the BZ boundary.
Such paths entail additional complications, as is well known from the case of the Zak-Berry phase~\cite{Zak:1989a,Vanderbilt:2018}; namely, in general $\Ham(\vec{k})$ and $\Ham(\vec{k}+\vec{b})$, where $\vec{b}$ is any reciprocal lattice vector, are not identical, but are related by a diagonal unitary rotation.
In the Supplemental Material~\cite{SM} we explain how to compute the non-Abelian invariant in that case and then prove a very general statement relating the values of the invariant computed on contours shifted by a reciprocal lattice vector but connected to a common base point, see e.g., the light and dark green paths $\gamma$ and $\tilde{\gamma}$ in \cref{App:fig:NLBZ:NLOrientation}.
As a corollary of that general statement, we find that the following holds:
\begin{corollary}\label{corollary:NLBZ}
Assume an $N$-band system with $\mcP\mcT$ symmetry squaring to $+\id$ described in the orbital basis by a Hermitian Bloch Hamiltonian $\Ham(\vec{k})$.
Let $\gamma\,:\,t\in[0,1]\mapsto\gamma(t)$ be a closed contour with no band degeneracies located inside the (first) BZ. The path starts at the \emph{base point} $P=\gamma(0)=\gamma(1)$ and we decompose
\begin{equation}
    \mathfrak{q}(\gamma) = s\prod_{j\in J}g_j
    \label{App:eq:q_from_generators}
\end{equation}
with $s\in\{\pm 1\}$, $g_j$ the generators defined in Refs.~\onlinecite{Wu:2019,Lenggenhager:2021:MBNLs} and $J\subseteq\{1,2,\dotsc,N-1\}$ a subset of the energy gaps of the $N$-band Hamiltonian (factors with smaller subscript $j$ appearing to the right).
Then, the generalized quaternion charge on the corresponding contour $\tilde{\gamma}$  with the same base point and enclosing the same band inversions but in the BZ shifted by the reciprocal lattice vector $\vec{b}$ [cf.~\cref{App:fig:NLBZ:NLOrientation}] is
\begin{equation}
    \mathfrak{q}(\tilde{\gamma}) = (-1)^m\mathfrak{q}(\gamma),
    \label{App:eq:q_shifted}
\end{equation}
where $m$ is the number of elements of the set
\begin{equation}
    \left\{j\in J\left|\phi_j\neq\phi_{j+1}\right.\right\}
    \label{App:eq:subset-of-J}
\end{equation}
with $\phi_j\in\{0,\pi\}$ the Berry phase of the $j^\textrm{th}$ band in the direction~$\vec{b}$.
[Note that in the conditioning in \cref{App:eq:subset-of-J} the label $j+1$ may not be in the set $J$.]
\end{corollary}

We briefly discuss the application of \cref{corollary:NLBZ} to the situation discussed in \cref{Sec:Monopole:Euler} and depicted in \cref{fig:Monopole:NLOrientation}.
We are interested in the change of the orientation of the blue nodal ring when comparing two copies displaced by the primitive reciprocal lattice vector in $k_z$-direction.
Recall that the orientation of nodal lines (NLs) is defined via the generalized quaternion invariant of unique paths encircling those NLs with a fixed base point.
This is exactly the situation of \cref{corollary:NLBZ} with $\mathfrak{q}(\gamma) = sg_2$ (in the figure the case $s=+1$ is illustrated).
Then, the corollary implies that the orientation changes if and only if the Berry phases of bands $2$ and $3$ are different, i.e., if $\phi_2+\phi_3=\pi\mod 2\pi$, in agreement with the statements in the main text.
We also remark that if $\mathfrak{q}(\gamma)=\pm 1$ (i.e., if the Berry phase of each band on $\gamma$ is trivial and \emph{no} NLs are enclosed by $\gamma$), we have $J=\varnothing$ and \cref{corollary:NLBZ} therefore immediately implies that $\mathfrak{q}(\gamma) = \mathfrak{q}(\tilde{\gamma})$.

\section{Additional material examples with Euler monopole charge}\label{App:Euler_materials}

\begin{table}[h]
    \centering
    \caption{
        Compounds with space-time inversion symmetry that host type-$\mathsf{A}$ triple points (TPs) on the given high-symmetry line (HSL) and are therefore candidates for having a non-trivial Euler monopole charge.
        Some compounds host multiple TPs.
        For each material and TP the space group, the HSL on which the TP lies, the nodal line (NL) segment carrying the Euler monopole charge, i.e., the segment enclosed by the ellipsoid, the relevant Berry phases and the Euler monopole charge induced from Wilson loop spectra are shown.
    }
    \label{App:tab:Euler_materials}
    \begin{ruledtabular}
    \begin{tabular}{CCCCCC}
    	              \text{Material}                &       \text{SG}        &             \text{HSL}             & \text{NL Segment} & \vec{\phi_B} & \Eu \\ \hline\addlinespace
    	        \multirow{2}{*}{\ce{Na2LiN}}         & \multirow{2}{*}{$129$} & \multirow{2}{*}{$\Gamma-Z-\Gamma$} &   [-0.29, 0.29]   &   (0,\pi)    &  2  \\
    	                                             &                        &                                    &   [ 0.29, 0.71]   &   (\pi,0)    &  2  \\ \addlinespace
    	\multirow{2}{*}{\ce{Li2NaN}~\cite{Jin:2019}} & \multirow{2}{*}{$191$} & \multirow{2}{*}{$\Gamma-A-\Gamma$} &   [-0.08, 0.08]   &    (0,0)     &  0  \\
    	                                             &                        &                                    &   [ 0.08, 0.92]   &   (0,\pi)    &  2  \\ \addlinespace
    	    \text{\ce{TiB2}~\cite{Zhang:2017}}       &          191           &          \Gamma-A-\Gamma           &   [-0.26, 0.26]   &   (0,\pi)    &  2  
    \end{tabular}
    \end{ruledtabular}
\end{table}

\begin{figure*}
    \centering
    \includegraphics{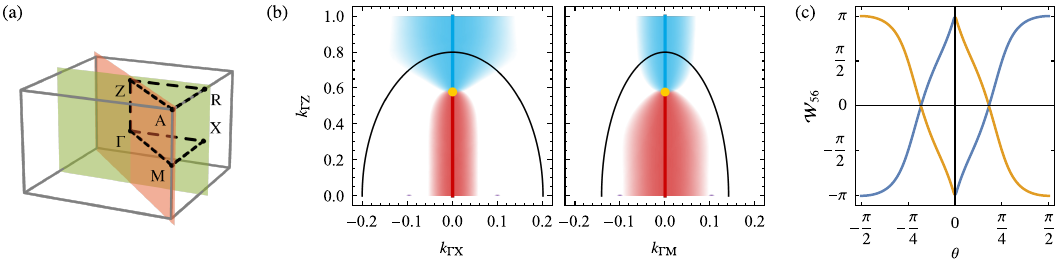}
    \caption{
    Computation of Euler monopole charges of a pair of single triple points in \ce{Na2LiN} with the surface centred at $\Gamma$.
    Panel (a) shows the Brillouin zone, the high-symmetry points and the planes in which the nodal lines are shown in panel (b).
    (b) Size of the four relevant gaps (orange, red, blue, and purple, according to increasing energy; at the triple point both the red and blue gaps are closed) in the two mirror planes shown in panel (a) encoded by the intensity of the color (and with a cutoff at a gap of $0.02\nunit{eV}$). The triple point (yellow) and the central nodal line are emphasized by appropriately colored overlays.
    (c) Wilson loop spectrum computed for the relevant bands on the ellipsoid indicated in panel (b).
    We observe that the Wilson loop spectrum winds twice, which implies $\abs{\Eu}=2$.
    }
    \label{fig:TPmat:Na2LiN_1}
\end{figure*}

\begin{figure*}[t]
    \centering
    \includegraphics{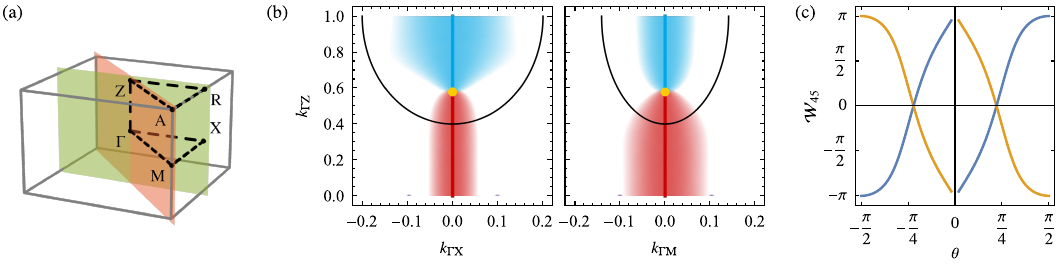}
    \caption{
    Computation of Euler monopole charges of a pair of single triple points in \ce{Na2LiN} with the surface centred at $Z$.
    The organization of the panels is in one-to-one correspondence with \cref{fig:TPmat:Na2LiN_1} (the same cutoff of $0.02\nunit{eV}$ is used).
    In panel (c) we observe that the Wilson loop spectrum winds twice, which implies $\abs{\Eu}=2$.
    }
    \label{fig:TPmat:Na2LiN_2}
\end{figure*}

\begin{figure*}
    \centering
    \includegraphics{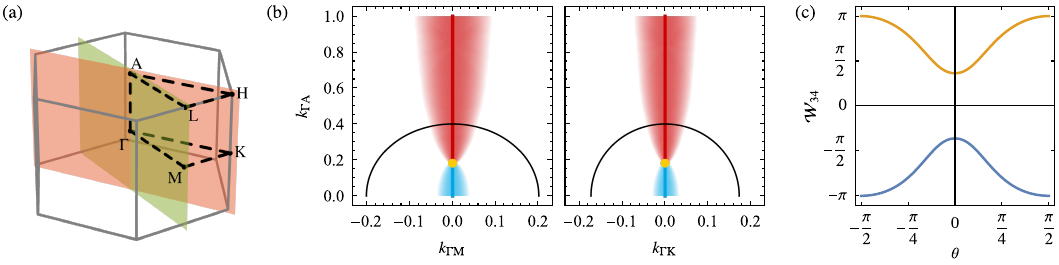}
    \caption{
    Computation of Euler monopole charges of a pair of single triple points in \ce{Li2NaN} with the surface centred at $\Gamma$.
    The organization of the panels is in one-to-one correspondence with \cref{fig:TPmat:Na2LiN_1} (cutoff $0.02\nunit{eV}$).
    In panel (c) we observe that the Wilson loop spectrum winds twice, which implies $\Eu=0$.
    }
    \label{fig:TPmat:Li2NaN_1}
\end{figure*}

\begin{figure*}
    \centering
    \includegraphics{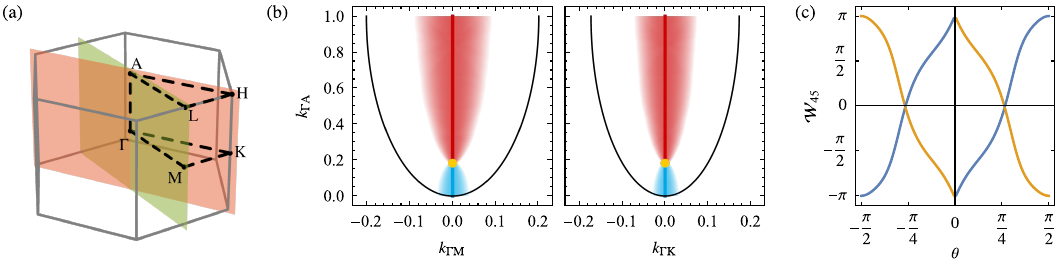}
    \caption{
    Computation of Euler monopole charges of a pair of single triple points in \ce{Li2NaN} with the surface centred at $A$.
    The organization of the panels is in one-to-one correspondence with \cref{fig:TPmat:Na2LiN_1} (cutoff $0.02\nunit{eV}$).
    In panel (c) we observe that the Wilson loop spectrum winds twice, which implies $\abs{\Eu}=2$.
    }
    \label{fig:TPmat:Li2NaN_2}
\end{figure*}

\begin{figure*}
    \centering
    \includegraphics{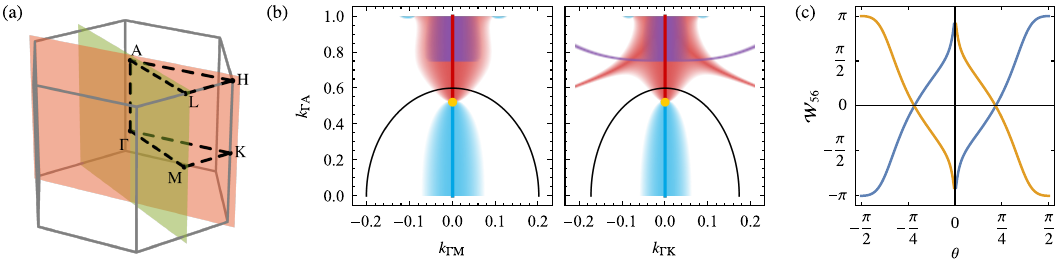}
    \caption{
    Computation of Euler monopole charges of a pair of single triple points in \ce{TiB2} with the surface centred at $\Gamma$.
    The organization of the panels is in one-to-one correspondence with \cref{fig:TPmat:Na2LiN_1} (cutoff $0.05\nunit{eV}$).
    In panel (c) we observe that the Wilson loop spectrum winds twice, which implies $\abs{\Eu}=2$.
    }
    \label{fig:TPmat:TiB2}
\end{figure*}

From the list of triple-point (TP) materials in Ref.~\onlinecite{Lenggenhager:2022:TPClassif}, we select additional candidates (besides \ce{Li2NaN}) that potentially host TP-induced Euler monopole charges, namely \ce{Na2LiN}, \ce{Li2NaN} and \ce{TiB2}.
Based on the first-principles calculations performed in Ref.~\onlinecite{Lenggenhager:2022:TPClassif} and whose results are given in the corresponding supplemental data and code~\cite{Lenggenhager:2022:TPClassif:SDC}, we construct maximally-localized Wannier functions using \textsc{Wannier90}~\cite{Pizzi:2020} with $s$ orbitals of alkali metal elements, $d$ orbitals of transition metal elements and $p$ orbitals of non-metal elements (for \ce{Li2NaN} no orbitals of \ce{Li} are included).
For the resulting Wannier tight-binding models we use the \Python{} package \textsc{Z2Pack}~\cite{Soluyanov:2011,Gresch:2017} to compute the Wilson loop spectra on ellipsoids enclosing the appropriate NL segments as well as the relevant Berry phases (cf.~\cref{fig:Monopole:NLOrientation}).
The results are summarized in \cref{App:tab:Euler_materials} and in \cref{fig:TPmat:Na2LiN_1,fig:TPmat:Na2LiN_2,fig:TPmat:Li2NaN_1,fig:TPmat:Li2NaN_2,fig:TPmat:TiB2} we show for each TP material and NL segment the resulting Wilson loop spectrum from the value $\Eu$ of the Euler monopole charge is deduced.
By comparing the Berry phases and the values of the Euler monopole charge, we verify the condition for non-trivial Euler monopole charge discussed in \cref{Sec:Monopole:Euler}.
We provide access to the Wannier tight-binding models as well as the Wilson loop spectra in the supplementary data and code~\cite{Lenggenhager:2021:TPHOT:SDC}.

\FloatBarrier
\clearpage
\makeatletter\onecolumngrid@push\makeatother

\clearpage
\makeatletter\onecolumngrid@pop\makeatother



\end{bibunit}
\let\oldaddcontentsline\addcontentsline     
\renewcommand{\addcontentsline}[3]{}        
\bibliography{bib}
\bibliographystyle{apsrev4-1}
\let\addcontentsline\oldaddcontentsline     
\clearpage


\begin{bibunit}
\onecolumngrid
\renewcommand\thesection{S\Roman{section}}
\renewcommand\thesubsection{\Alph{subsection}}
\setcounter{page}{1}

\setcounter{section}{0}
\setcounter{figure}{0}
\setcounter{equation}{0}
\setcounter{table}{0}
\setcounter{corollary}{0}

\renewcommand{\theequation}{S\arabic{equation}}
\renewcommand{\thefigure}{S\arabic{figure}}
\renewcommand{\thetable}{S\arabic{table}}
\renewcommand*{\theHsection}{S.\the\value{section}}

\title{Supplemental Material to:\texorpdfstring{\\}{ }Universal higher-order bulk-boundary correspondence of triple nodal points}

\author{Patrick M. Lenggenhager\,\orcidlink{0000-0001-6746-1387}}\email[corresponding author: ]{lenpatri@ethz.ch}
\affiliation{Condensed Matter Theory Group, Paul Scherrer Institute, 5232 Villigen PSI, Switzerland}
\affiliation{Institute for Theoretical Physics, ETH Zurich, 8093 Zurich, Switzerland}
\affiliation{Department of Physics, University of Zurich, Winterthurerstrasse 190, 8057 Zurich, Switzerland}

\author{Xiaoxiong Liu\,\orcidlink{0000-0002-2187-0035}}
\affiliation{Department of Physics, University of Zurich, Winterthurerstrasse 190, 8057 Zurich, Switzerland}

\author{Titus Neupert\,\orcidlink{0000-0003-0604-041X}}
\affiliation{Department of Physics, University of Zurich, Winterthurerstrasse 190, 8057 Zurich, Switzerland}

\author{Tom\'{a}\v{s} Bzdu\v{s}ek\,\orcidlink{0000-0001-6904-5264}}\email[corresponding author: ]{tomas.bzdusek@psi.ch}
\affiliation{Condensed Matter Theory Group, Paul Scherrer Institute, 5232 Villigen PSI, Switzerland}
\affiliation{Department of Physics, University of Zurich, Winterthurerstrasse 190, 8057 Zurich, Switzerland}

\date{\today}

\maketitle

\section{Quaternion charge in different Brillouin zones}

In this supplement we prove the following conjecture:
\begin{conjecture}\label{SM:conjecture:NLBZ}
We assume an $N$-band system with $\mcP\mcT$-symmetry squaring to $+\id$, described in the orbital basis by a Hermitian Bloch Hamiltonian $\Ham(\vec{k})$.
Let $\gamma\,:\,t\in[0,1]\mapsto\gamma(t)$ be a closed contour without any band degeneracies on it that lies in the (first) BZ starting at the \emph{base point} $P=\gamma(0)=\gamma(1)$, and let be $\vec{b}$ a reciprocal lattice vector.
We define the following three paths:
\begin{itemize}
\item[(\emph{i})] the shifted contour $\gamma'(t)\equiv \gamma(t)+\vec{b}$, 
\item[(\emph{i})] the path $\gamma_{P,\vec{b}}$ along $\vec{b}$ connecting the base point $P$ to $P'\equiv\gamma'(0)$ (assuming once more that there are no band degeneracies along it), and
\item[(\emph{iii})] their concatenation (read from left to right)
\begin{equation}
    \tilde{\gamma} = \gamma_{P,\vec{b}}\circ\gamma'\circ\gamma_{P,\vec{b}}^{-1},
    \label{SM:eq:gammatilde_path}
\end{equation}
which is a closed contour with the same base point $P$ as~$\gamma$.
\end{itemize}
Then, the quaternion invariants on $\gamma$ and $\tilde{\gamma}$ (computed with the same gauge choice for the real eigenstates at $P$, cf.~\cref{SM:Subsec:quaternion}) are related by
\begin{equation}
    \mathfrak{q}(\tilde{\gamma}) = \overline{F_{P,\vec{b}}}\mathfrak{q}(\gamma)\overline{F_{P,\vec{b}}}^{\,-1},
    \label{SM:eq:q_conjugation}
\end{equation}
where
\begin{equation}
    \overline{F_{P,\vec{b}}} \equiv \prod_{i\,:\,\e^{\i\phi_i}=-1}\epsilon_i
    \label{SM:eq:defFPmubar}
\end{equation}
with $\phi_i$ the Berry phase of band $i$ along $\vec{b}$ and $\{\epsilon_i\}_{i=1}^{N}$ the generators ($\epsilon_i\epsilon_j+\epsilon_j\epsilon_i=-2\delta_{ij}$) of the Clifford algebra $\Clifford{0}{N}$ as used in the construction of  $\Spin(N)$ [cf.~Sec.~\ref{sec:construct-spin-N}]. 
(Let us emphasize that this is \emph{different} from the Clifford algebra $\Clifford{0}{N-1}$, the particular subset $\overline{P}_N$ of which corresponds to the generalized quaternion charge~\cite{Wu:2019}; cf.~Sec.~\ref{sec:PN-and-double-cover}.)
Note that the ordering of the factors in \cref{SM:eq:defFPmubar} does not affect \cref{SM:eq:q_conjugation} as long as the same ordering is used in both occurrences of $\overline{F_{P,\vec{b}}}$.
Nevertheless, for concreteness, we fix the ordering such that factors with smaller subscript $i$ appear to the right inside the product.
\end{conjecture}

Before proving the above conjecture, let us briefly restate the corollary from \cref{App:NLBZ} and referred to in \cref{Sec:Monopole:Euler}:
\begin{corollary}\label{SM:corollary:NLBZ}
Assume an $N$-band system with $\mcP\mcT$-symmetry squaring to $+\id$ described in the orbital basis by a Hermitian Bloch Hamiltonian $\Ham(\vec{k})$.
Let $\gamma\,:\,t\in[0,1]\mapsto\gamma(t)$ be a closed contour with no band degeneracies located inside the (first) BZ. The path starts at the \emph{base point} $P=\gamma(0)=\gamma(1)$ and we decompose
\begin{equation}
    \mathfrak{q}(\gamma) = s\prod_{j\in J}g_j
    \label{SM:eq:q_from_generators}
\end{equation}
with $s\in\{\pm 1\}$, $g_j$ the generators defined in Refs.~\cite{Wu:2019,Lenggenhager:2021:MBNLs} and $J\subseteq\{1,2,\dotsc,N-1\}$ a subset of the energy gaps of the $N$-band Hamiltonian (factors with smaller subscript $j$ appearing to the right).
    Then, the quaternion charge on the corresponding contour $\tilde{\gamma}$  with the same base point and enclosing the same band inversions but in the BZ shifted by the reciprocal lattice vector $\vec{b}$ [cf.~\cref{App:fig:NLBZ:NLOrientation}] is  
\begin{equation}
    \mathfrak{q}(\tilde{\gamma}) = (-1)^m\mathfrak{q}(\gamma),
    \label{SM:eq:q_shifted}
\end{equation}
where $m$ is the number of elements of the set
\begin{equation}
    \left\{j\in J\left|\phi_j\neq\phi_{j+1}\right.\right\}\label{eqn:subset-of-J}
\end{equation}
with $\phi_j\in\{0,\pi\}$ the Berry phase of the $j^\textrm{th}$ band in the direction~$\vec{b}$.
[Note that in the conditioning in \cref{eqn:subset-of-J} the label $j+1$ may not be in the set $J$.]
\end{corollary}
\begin{proof}
Given \cref{SM:conjecture:NLBZ}, the only thing to show is that \cref{SM:eq:q_conjugation} reduces to \cref{SM:eq:q_shifted} given \cref{SM:eq:q_from_generators}.
The generators $g_j$ of the generalized quaternion group can be defined in terms of the generators $\epsilon_i$ of $\Clifford{0}{N}$ [see \cref{SM:eq:g_generators}] as
\begin{equation}
    g_1 = -\epsilon_1\epsilon_2,\quad g_{j\geq 2} = \epsilon_j\epsilon_{j+1}.
\end{equation}
The above two expressions can be jointly encoded as $g_j = (-1)^{\delta_{j1}} \epsilon_j\epsilon_{j+1}$. 
By combining \cref{SM:eq:q_conjugation,SM:eq:q_from_generators}, we first find (through a repeated insertion of the identity $\overline{F_{P,\vec{b}}}^{\,-1}\overline{F_{P,\vec{b}}}=1$ into the product over $j\in J$) that
\begin{eqnarray}
    \mathfrak{q}(\tilde{\gamma})
    &=& s\left(\prod_{\,i\,:\,\e^{\i\phi_i}=-1}\epsilon_i\right)\left(\prod_{j\in J} g_j\right) \left(\prod_{\,i\,:\,\e^{\i\phi_i}=-1}\epsilon_i\right)^{-1}
    \nonumber \\
    &=& s\prod_{j\in J}\left[\left(\prod_{\,i\,:\,\e^{\i\phi_i}=-1}\epsilon_i\right)(-1)^{\delta_{j1}}\epsilon_j\epsilon_{j+1}\left(\prod_{\,i\,:\,\e^{\i\phi_i}=-1}\epsilon_i\right)^{-1}\right].
\end{eqnarray}
Observe that for every $j\in J$, the conjugation with $\overline{F_{P,\vec{b}}} = \prod_{\,i\,:\,\e^{\i\phi_i}=-1}\epsilon_i$ results in an overall minus sign if and only if exactly one of $\epsilon_j$ and $\epsilon_{j+1}$ is present in $\overline{F_{P,\vec{b}}}$; otherwise, the conjugation does not affect the factor $\epsilon_j\epsilon_{j+1}$.
Therefore, the number of $-1$ factors picked up through the conjugation is equal to the number of $j\in J$ where $\e^{\i\phi_j}\neq \e^{\i\phi_{j+1}}$.
This is equal to the order $m$ of the set in \cref{eqn:subset-of-J}, implying \cref{SM:eq:q_shifted}.
\end{proof}

In preparation for the proof of \cref{SM:conjecture:NLBZ}, we prove several lemmas.
For that we proceed as follows.
In \cref{SM:Subsec:Hamiltonian} we discuss the form and the \mbox{(non-)periodicity} of a $\mathcal{PT}$-symmetric Bloch Hamiltonian in the extended momentum space (i.e., beyond the first Brillouin zone).
In \cref{SM:Subsec:doublecovers,SM:Subsec:conjlifts} we first revisit the construction of the double covers of $\SO(N)$ and of its subgroup $\PN_N$, and then show how to lift elements of $\SO(N)$ close to the identity after being conjugated by elements in $\PN_{Nh}\subg\O(N)$.
Armed with the derived lemmas, we continue by discussing (\emph{i}) the quaternion charge on the various paths involved in the conjecture in \cref{SM:Subsec:quaternion} and (\emph{ii}) the Berry phases of the contour winding around the Brillouin zone torus in \cref{SM:Subsec:Berryphases}.
Finally, we use the derived lemmas and results to prove the conjecture \ref{SM:conjecture:NLBZ} in \cref{SM:Subsec:proof}.

\subsection{Bloch Hamiltonian}\label{SM:Subsec:Hamiltonian}

Before analyzing the quaternion charges, we review in the present section several properties of Bloch Hamiltonians in $\mathcal{PT}$-symmetric systems.
Note that we adopt the Bloch convention which takes into account the positions of the orbitals within the unit cell when forming the Bloch basis, see \cref{eqn:Bloch-basis}.
This is the convention in which the Berry curvature respects the symmetries of the lattice~\cite{Fruchart:2014,Dobardzic:2015} and the Zak phase of energy bands is in one-to-one correspondence with their electric polarization~\cite{Nelson:2021b}. 
The prize to pay for this physical interpretability is that the resulting Bloch Hamiltonian may be non-periodic in reciprocal-lattice vectors.
\medskip

Recall~\cite{Bouhon:2020} that, due to $(\mcP\mcT)^2=+\id$, for each $\vec{k}$, there is a change of basis given by a unitary matrix $V_\vec{k}$ such that $\mcP\mcT$ is represented by complex conjugation $\mcK$: 
\begin{equation}
V_\vec{k}\overline{D}_\vec{k}(\mcP\mcT)\mcK\adjo{V_\vec{k}}=\mcK \label{eqn:PT=K}
\end{equation} 
(here the unitary matrix $\overline{D}_\vec{k}(\mcP\mcT)$ is the \emph{corepresentation} $\mcP\mcT$).
It follows that 
\begin{equation}
V_\vec{k}\Ham(\vec{k})\adjo{V_\vec{k}}\equiv\Ham_\mathrm{R}(\vec{k})
\end{equation} 
is a real symmetric matrix.
Note that $V_\vec{k}$ is not unique, but any $W_\vec{k}V_\vec{k}$ for $W_\vec{k}\in\O(N)$ defines another such basis.

Given a real symmetric Hamiltonian $\Ham_\mathrm{R}(\vec{k})$, it is natural to write its eigenstates in a real gauge; if we additionaly order the eigenstates as columns from left to right according to increasing energy, we obtain the \emph{eigenframe} $\mathsf{u}\in\O(N)$, and
\begin{equation}
    \Ham_\mathrm{R}(\vec{k}) = \mathsf{u}(\vec{k})\mathcal{E}(\vec{k})\mathsf{u}(\vec{k})^\top.
    \label{SM:eq:eigenframe}
\end{equation}
However, the orthogonal eigenframe is not unique, but exhibits a gauge degree of freedom, $\mathsf{u}\mapsto \mathsf{u}F$, where $F\in \mathsf{P}_{Nh} \cong \ztwo^N$ is a diagonal matrix of $\pm 1$'s.
In particular, note that $F^2 = F F^\top = \id$.

It is important to note that while the eigenenergies are periodic in momentum space, the Bloch Hamiltonian $\Ham(\vec{k})$ in general is only periodic up to a unitary transformation,
\begin{equation}
    \Ham(\vec{k}+\vec{b}) = U_{\vec{b}}\Ham(\vec{k})\adjo{U_{\vec{b}}},
    \label{SM:eq:shift_unitary_transform}
\end{equation}
where $\vec{b}$ is a reciprocal lattice vector, and the unitary matrix is defined up to an overall multiplication by a phase factor.

In the context of a tight-binding model with orbital $\alpha$ placed at position $\vec{r}_\alpha$ relative to the unit cell with origin at $\vec{R}$ and Hamiltonian $\hat{\Ham}$, we define the Bloch basis
\begin{equation}
    \ket{\vec{k},\alpha} = \sum_{\vec{R}}\e^{\i\vec{k}\cdot(\vec{R}+\vec{r}_\alpha)}\ket{\vec{R},\alpha},
    \label{eqn:Bloch-basis}
\end{equation}
where $\ket{\vec{R},\alpha}$ forms the tight-binding basis, and the sum is over Bravais lattice vectors $\vec{R}$.
Then, the Bloch Hamiltonian $\Ham(\vec{k})$ is the matrix with components $\Ham^{\alpha\beta}(\vec{k})$ defined by
\begin{equation}
    \mel{\vec{k},\alpha}{\hat{\Ham}}{\vec{k}',\beta} = \Ham^{\alpha\beta}(\vec{k})\delta_{\vec{k},\vec{k}'}.
\end{equation}
and the unitary transformation in Eq.~(\ref{SM:eq:shift_unitary_transform}) is given by 
\begin{equation}
    U_\vec{b}^{\alpha\beta}=\e^{-\i\vec{b}\cdot\vec{r}_\alpha}\delta_{\alpha\beta}
    \label{SM:eq:U_b}
\end{equation}
in the orbital basis $\ket{\vec{R},\alpha}$.
\medskip

In the basis, where the Hamiltonian is real, we have the following lemma:
\begin{lemma}
In the context of an $N$-orbital tight-binding model with $\mcP\mcT$ symmetry, $U_{\vec{b},\mathrm{R}}\equiv V_\vec{k}U_{\vec{b}}\adjo{V_\vec{k}}\in\O(N)$ and $U_{\vec{b},\mathrm{R}}$ is independent of $\vec{k}$.
\end{lemma}

\begin{proof}
In real space $\mcP\mcT$ acts like inversion, mapping a position vector $\vec{r}$ (relative to the center of the unit cell) to $-\vec{r}$.
If $\mcP\mcT$ is a symmetry of the system, then there are two options for each tight-binding orbital $\alpha$: (1) it is mapped to itself (potentially to a Bravais-translation-related copy of itself in another unit cell) or (2) it is exchanged with another orbital.
This implies that, in the orbital basis $\ket{\vec{R},\alpha}$, the corepresentation matrix $\overline{D}_\vec{k}(\mcP\mcT)$ is block-diagonal with one- and two-dimensional blocks.
Below we consider both cases~(1) and~(2) to explicitly construct $V_\vec{k}$ such that
\begin{equation}
    V_\vec{k}^{\phantom{\top}}\overline{D}_\vec{k}(\mcP\mcT)V_\vec{k}^\top = \id,
    \label{SM:eq:def_V}
\end{equation}
i.e., the unitary rotation from the orbital basis to the basis in which $\mcP\mcT$ is represented by complex conjugation [cf.~\cref{eqn:PT=K}].

Let us first consider option (1).
For the orbital $\alpha$ to be mapped to itself under $\mcP\mcT$, its position $\vec{r}_\alpha$ must be an inversion-symmetric point, i.e., $-\vec{r}_\alpha=\vec{r}_\alpha+\vec{R}_\alpha$ for some Bravais lattice vector $\vec{R}_\alpha$.
Then, $\vec{r}_\alpha=-\tfrac{1}{2}\vec{R}_\alpha$, and the corresponding block of the corepresentation matrix must take the form
\begin{equation}
    \overline{D}_\vec{k}(\mcP\mcT) = \e^{-\i\vec{k}\cdot\vec{R}_\alpha+\i\varphi_\alpha}
\end{equation}
with some phase $\varphi_\alpha\in\mathbb{R}$. 
One trivially verifies that $\overline{D}_\vec{k}(\mcP\mcT)\cconj{\overline{D}_\vec{k}(\mcP\mcT)}=\id$, as expected for spinless bands.
We next observe that the corresponding block of $V_\vec{k}$ satisfying \cref{SM:eq:def_V}~is
\begin{equation}
    V_\vec{k} = \e^{-\frac{1}{2}\left(-\i\vec{k}\cdot\vec{R_\alpha}+\i\varphi_\alpha\right)}.
\end{equation}
Therefore $U_\vec{b}$ defined in \cref{SM:eq:U_b} is transformed to
\begin{equation}
    U_{\vec{b},R} = V_\vec{k}\e^{-\i\vec{b}\cdot\vec{r}_\alpha}\adjo{V_\vec{k}} = \e^{\i\frac{1}{2}\vec{b}\cdot\vec{R}}.
\end{equation}
Note that $\vec{b}$ is a reciprocal lattice vector and $\vec{R}$ a Bravais lattice vector, such that $\frac{1}{2}\vec{b}\cdot\vec{R}$ is $0$ or $\pi$ (modulo $2\pi$) and $U_{\vec{b},R}=\pm1\in\O(1)$ is independent of $\vec{k}$.

Next, we consider option (2).
For concreteness let us assume that the orbitals $\alpha$ and $\beta$ are mapped to each other, then the corresponding block of the (unitary) corepresentation matrix must be off-diagonal,
\begin{equation}
    \overline{D}_\vec{k}(\mcP\mcT) = \mqty(0&\e^{\i\varphi_{\alpha\beta}}\\\e^{\i\varphi_{\alpha\beta}}&0),
\end{equation}
where the phases of the off-diagonal matrix elements must be equal in order to satisfy $\overline{D}_\vec{k}(\mcP\mcT)\cconj{\overline{D}_\vec{k}(\mcP\mcT)}=\id$.
Again we construct the corresponding block of $V_\vec{k}$ satisfying \cref{SM:eq:def_V}:
\begin{equation}
    V_\vec{k} = \frac{\e^{-\i\frac{1}{2}\varphi_{\alpha\beta}}}{\sqrt{2}}\mqty(1&1\\\i&-\i).
\end{equation}
The fact that orbitals $\alpha$ and $\beta$ are mapped to each other under $\mcP\mcT$ implies that $-\vec{r}_\alpha=\vec{r}_\beta$ and thus
\begin{equation}
    U_\vec{b} = \mqty(\e^{-\i\vec{b}\cdot\vec{r}_\alpha}&0\\0&\e^{\i\vec{b}\cdot\vec{r}_\alpha}).
\end{equation}
In the transformed basis, we therefore end up with
\begin{equation}
    \!\! U_{\vec{b},R} = V_\vec{k}U_\vec{b}\adjo{V_\vec{k}} = \mqty(\cos(-\vec{b}\cdot\vec{r}_\alpha)&\!\sin(-\vec{b}\cdot\vec{r}_\alpha)\\-\sin(-\vec{b}\cdot\vec{r}_\alpha)&\!\cos(-\vec{b}\cdot\vec{r}_\alpha))\in\O(2),
\end{equation}
which is independent of $\vec{k}$.

We have shown that the full $V_\vec{k}$ is block-diagonal and because $U_\vec{b}$ is diagonal [cf.~\cref{SM:eq:U_b}] this implies that $U_{\vec{b},R}$ is also block diagonal with $\vec{k}$-independent blocks either in $\O(1)$ or $\O(2)$.
Therefore, $U_{\vec{b},R}\in\O(N)$ is $\vec{k}$-independent as well.
\end{proof}

\subsection{Double covers of \texorpdfstring{$\SO(N)$}{SO(N)} and \texorpdfstring{$\PN_N$}{PN}}\label{SM:Subsec:doublecovers}

\subsubsection{Lie algebra $\so(N)$} 
We consider the special orthogonal group $\SO(N)$, whose Lie algebra is denoted $\so(N)$.
A basis for $\so(N)$ is given by the $N\times N$ matrices
\begin{equation}
	L_{ij} = -E_{ij}+E_{ji},\quad i<j,
	\label{SM:eq:Lij}
\end{equation}
where $(E_{ij})_{ab}=\delta_{ai}\delta_{bj}$ is the matrix with a single non-zero element $1$ at position $(i,j)$.
The matrices $E_{ij}$ satisfy 
\begin{equation}
    E_{ij}E_{k\ell} = \delta_{jk}E_{i\ell},
\end{equation}
such that
\begin{equation}
	\comm{L_{ij}}{L_{k\ell}} = -\delta_{i\ell}L_{kj}+\delta_{ik}L_{\ell j}-\delta_{j\ell}L_{ik}+\delta_{jk}L_{i\ell},
	\label{SM:eq:soNcomm}
\end{equation}
which defines the Lie algebra.

\subsubsection{Construction of \texorpdfstring{$\Spin(N)$}{Spin(N)}}\label{sec:construct-spin-N}
The simply-connected double cover of $\SO(N)$, called $\Spin(N)$, can be constructed~\cite{Woit:2012} via the Clifford algebra $\Clifford{0}{N}$ with generators $\epsilon_i$ satisfying
\begin{equation}
	\epsilon_i\epsilon_j+\epsilon_j\epsilon_i = -2\delta_{ij}.\label{eqn:eps-ij}
\end{equation}
We note that the quadratic elements
\begin{equation}
	t_{ij}\equiv -\frac{1}{2}\epsilon_i\epsilon_j,\quad i<j,\label{eqn:t-ij}
\end{equation}
satisfy the same commutation relation as \cref{SM:eq:soNcomm}:
\begin{equation}
	\comm{t_{ij}}{t_{k\ell}} = -\delta_{i\ell}t_{kj}+\delta_{ik}t_{\ell j}-\delta_{j\ell}t_{ik}+\delta_{jk}t_{i\ell},
\end{equation}
such that we can construct the isomorphism
\begin{equation}\label{eqn:algebra-iso}
	\begin{array}{rcl}
		\so(N) &\to& \spin(N)\\
		L_{ij} &\mapsto& \overline{L_{ij}} \equiv t_{ij}=-\frac{1}{2}\epsilon_i\epsilon_j
	\end{array}.
\end{equation}
of the two Lie algebras.
\medskip

Using the exponential map of elements in the Lie algebras, we obtain the corresponding Lie groups:
\begin{equation}
	\e^{\sum_{i<j}\alpha_{ij}L_{ij}}\in\SO(N),\qquad \e^{\sum_{i<j}\theta_{ij}t_{ij}}\in\Spin(N).
\end{equation}
The isomorphism of Lie algebras in \cref{eqn:algebra-iso} induces a homomorphism of the corresponding Lie groups.
In particular, for elements of $\SO(N)$ close to the identity one can construct the \emph{lift map}:
\begin{equation}
	\argplaceholder:\e^{\sum_{i<j}\alpha_{ij}L_{ij}} \mapsto \e^{\sum_{i<j}\alpha_{ij}t_{ij}}.
	\label{SM:eq:liftSON}
\end{equation}
For $A\in \SO(N)$ close to the identity, we denote the unique lift to $\Spin(N)$ close to the identity, as defined in \cref{SM:eq:liftSON}, by $\overline{A}$.\medskip

We proceed to develop an elementary intuition about the lift map~$\argplaceholder: \SO(N)\to\Spin(N)$ for elements close to the identity.\medskip

Because $\Spin(N)$ is the double cover of $\SO(N)$, we have the following short exact sequence of group homomorphisms:
\begin{equation}
    1\longrightarrow\mathbb{Z}_2=\{\pm1\}\stackrel{\varphi}{\longrightarrow}\Spin(N)\stackrel{\sigma}{\longrightarrow}\SO(N)\longrightarrow 1\label{eqn:SES}
\end{equation}
where $\varphi(\pm1)$ is in the center of $\Spin(N)$, and $\sigma$ is a two-to-one projection map. Let as clarify in some detail what the sequence in \cref{eqn:SES} entails.
Since $\varphi$ is a group homomorphism, we have
\begin{equation}
    1 = \varphi(1) = \varphi((-1)(-1)) = \varphi(-1)\varphi(-1),
\end{equation}
thus we identify $\varphi(-1)=-1\in\Spin(N)$, i.e., as the non-trivial element of $\Spin(N)$ that commutes with all other elements of $\Spin(N)$.
The exactness of the above sequence implies that
\begin{equation}
    1 = (\sigma\circ \varphi)(\pm 1) = \sigma(\pm 1),
\end{equation}
i.e., the center $\pm 1\in \Spin(N)$ projects onto the identity element $1\in\SO(N)$.

Now consider $M_1,M_2\in\Spin(N)$ with $\sigma(M_1)=\sigma(M_2)$, then
\begin{equation}
    \sigma(M_1M_2^{-1}) = \sigma(M_1)\sigma(M_2)^{-1} = 1,
\end{equation}
i.e., $M_1M_2^{-1}$ is in the kernel of $\sigma$, which due to the exactness of the sequence is the image of $\varphi$; therefore
\begin{equation}
    M_1M_2^{-1} \in \{\pm 1\}\Rightarrow M_1 = \pm M_2.
    \label{SM:eq:cover_product}
\end{equation}
We also have that 
\begin{equation}
    \sigma\left(\overline{A}\right)=A \label{eqn:project-of-lift}
\end{equation} (i.e., the lift map followed by the projection map is equivalent to the identity operation) for all elements $A\in\SO(N)$ close to identity.
\medskip

One should bear in mind that we use the same symbol `$\,\overline{\phantom{a}}\,$' for two different (but closely related) maps: (1) the isomorphism of the Lie algebras [\cref{eqn:algebra-iso}], and the lift map of Lie-group elements close to the identity [\cref{SM:eq:liftSON}].
These maps are canonically related, because the Lie algebra $\mathfrak{g}$ corresponds to the tangent space of the Lie group $\mathsf{G}$ at the identity.

\begin{lemma}\label{SM:lemma:lift_of_inverse}
Let $A\in \SO(N)$ be close to the identity, $A^{-1}$ its inverse (which is also close to the identity). Let us denote their (unique) lifts to $\Spin(N)$, according to \cref{SM:eq:liftSON}, as $\overline{A}$ and $\overline{A^{-1}}$. Then:
\begin{equation}
    \overline{A}^{\,-1} = \overline{A^{-1}}.
\end{equation}
\end{lemma}
\begin{proof}
Set $M_1:=\overline{A}^{\,-1}$, $M_2:=\overline{A^{-1}}$, then
\begin{equation}
    \sigma(M_1) = \sigma\left(\overline{A}^{\,-1}\right)=\sigma\left(\overline{A}\right)^{-1} = A^{-1} = \sigma(M_2),
\end{equation}
where we used that $\sigma$ is a group homomorpshim. It follows, 
according to \cref{SM:eq:cover_product}, that
\begin{equation}
    \overline{A}\cdot\overline{A^{-1}} = \pm 1.
\end{equation}
However, because the left hand side is a product of elements close to the identity, the right hand side has to be close to the identity as well, such that it can only be $1$ (and not $-1$).
Thus, $\overline{A}^{\,-1} = \overline{A^{-1}}$, i.e., for elements $A\in\SO(N)$ close to the identity, the lift map commutes with the matrix inverse map.
\end{proof}

\subsubsection{The group \texorpdfstring{$\PN_N$}{PN} and its double cover}\label{sec:PN-and-double-cover}
We proceed to discuss the generalized quaternion charges, as introduced in the supplementary material of Ref.~\onlinecite{Wu:2019}.
Consider elements of the discrete subgroup,
\begin{equation}
	\PN_N = \left\langle\left\{\e^{\pi L_{ij}}\right\}_{i<j}\right\rangle\subg\SO(N),
\end{equation}
where the angular brackets denote the group generated by taking products of the elements of the set inside the brackets.
Note that
\begin{equation}
	L_{ij}^n = 
	\begin{cases}
		\id,&n=0\\
		(-1)^kL_{ij},&n=2k+1,k\in\mathbb{N}_0\\
		(-1)^k(E_{ii}+E_{jj}),&n=2k,k\in\mathbb{N}_{>0}
	\end{cases},
\end{equation}
resulting in
\begin{equation}
    \e^{\alpha L_{ij}} = \id + \sin(\alpha) L_{ij} + \left(\cos(\alpha)-1\right)(E_{ii}+E_{jj}).
\end{equation}

For the elements of $\PN_N$ we thus have
\begin{equation}
	\e^{\pi L_{ij}} = \id - 2(E_{ii}+E_{jj}),
	\label{SM:eq:PNgen}
\end{equation}
i.e., diagonal matrices with $+1$ on the diagonal except for the $i^\textrm{th}$ and $j^\textrm{th}$ element, which are $-1$.
Note that $\e^{-\pi L_{ij}} = \e^{\pi L_{ij}} = (\e^{-\pi L_{ij}})^{-1}$.
(Let us also remark that throughout this supplemental material we do \emph{not} use the Einstein summation convention, i.e., there is no implicit summation over repeated indices.)

The double cover of $\PN_N$, denoted $\overline{\PN}_N\subg\Spin(N)$, can be constructed starting from the generators of $\PN_N$: $\e^{\alpha_{ij}L_{ij}}$ and applying the algebra isomorphism in the exponent, then
\begin{equation}
	\overline{\PN}_N = \left\langle\left\{\e^{\pi t_{ij}}\right\}_{i<j}\right\rangle.
\end{equation}
Since for $i\neq j$, $t_{ij}^2 = -\frac{1}{4}$, we find
\begin{equation}
    \e^{\theta t_{ij}}  = \cos\left(\frac{\theta}{2}\right)+2t_{ij}\sin\left(\frac{\theta}{2}\right)
\end{equation}
such that the generators of $\overline{\PN}_N$ are $\e^{\pi t_{ij}} = 2t_{ij}$, and  
\begin{equation}
    \sigma\left(2t_{ij}\right) = \e^{\pi L_{ij}}.
    \label{SM:eq:cover_of_PNbar}
\end{equation}
The above results allow us to relate the generators $\{e_j\}_{j=1}^{N-1}$ of $\overline{\PN}_N$ and of the Clifford algebra $\Clifford{0}{N-1}$ introduced in Ref.~\onlinecite{Wu:2019} to the generators $\{\epsilon_i\}_{i=1}^N$ of $\Clifford{0}{N}$ adopted in \cref{sec:construct-spin-N}, namely"
\begin{equation}
    e_j \equiv 2t_{1,j+1} = -\epsilon_1\epsilon_{j+1}.
\end{equation}
The same reference introduces an alternative, physically motivated set of generators
\begin{equation}
    g_j \equiv
	\begin{cases}
		e_1,&j=1\\
		e_{j-1}e_j,& 2 \leq j \leq N-1
	\end{cases}.
\end{equation}
It follows from combining the preceding equations that
\begin{equation}
    g_j =
	\begin{cases}
		-\epsilon_1\epsilon_2,&j=1\\
		\epsilon_j\epsilon_{j+1},& 2 \leq j \leq N-1
	\end{cases}.
	\label{SM:eq:g_generators}
\end{equation}

\subsubsection{The group \texorpdfstring{$\PN_{Nh}$}{PNh}}
The group $\PN_{Nh}$ of diagonal matrices with $\pm1$ on the diagonal is not a subgroup of $\SO(N)$, but a subgroup of $\O(N)$.
It can be written as
\begin{equation}
    \PN_{Nh} = \PN_N\cup(\id-2E_{ii})\PN_N
\end{equation}
for any $1\leq i \leq N$.
In the following, we will consider conjugation of elements of $\SO(N)$ with elements of $\PN_{Nh}$.
Since the determinant of a product equals the product of the determinants of its factors, conjugation of an element in $\SO(N)$ with an element of $\PN_{Nh}$ does not leave $\SO(N)$.

\
Similarly, the double cover $\overline{\PN}_{Nh}$ of $\PN_{Nh}$ can be constructed as
\begin{equation}
    \overline{\PN}_{Nh} = \overline{\PN}_{N}\cup\epsilon_i\overline{\PN}_{N}
\end{equation}
for any $1\leq i \leq N$.
Based on this and the construction of $\overline{\PN}_N$, any $\mathfrak{p}\in\overline{\PN}_{Nh}$ can, for example, be written as
\begin{equation}
    \mathfrak{p} = \epsilon_1^{p_0}\prod_{i<j}(2t_{ij})^{p_{ij}}
\end{equation}
for $p_0,p_{ij}\in\{0,1\}$.
In this decomposition, $p_0$ distinguishes whether $\mathfrak{p}$ lies in the proper subgroup $\PN_N$ ($p_0=0$) or not ($p_0=1$).
Finally, we remark that
\begin{equation}
    \sigma(\mathfrak{p}) = (\id-2E_{ii})^{p_0}\e^{\pi \sum_{i<j} p_{ij}L_{ij}}.
\end{equation}
for the projection of any element $\mathfrak{p}\in\overline{P}_{Nh}$.

\subsection{Lift of conjugated elements}\label{SM:Subsec:conjlifts}
\begin{lemma}\label{SM:lemma:liftsoconjPN}
    We consider two basis elements of $\so(N)$: $L_{ij}$ ($i<j$) and $L_{k\ell}$ ($k<\ell$). Then $\e^{\pi L_{ij}}L_{k\ell}\e^{-\pi L_{ij}}\in\so(N)$ and
    \begin{equation}
    	\overline{\e^{\pi L_{ij}}L_{k\ell}\e^{-\pi L_{ij}}} = \e^{\pi t_{ij}}t_{k\ell}\e^{-\pi t_{ij}} = (2t_{ij})t_{k\ell}(2t_{ij})^{-1},
    \end{equation}
    which is an element of $\spin(N)$.
\end{lemma}

\begin{proof}
Using \cref{SM:eq:Lij,SM:eq:PNgen}, and that
\begin{subequations}
\begin{eqnarray}
	\forall i {\neq} j, k {\neq} \ell &:& E_{ii}L_{kl}E_{jj} = \left(-\delta_{ik}\delta_{j\ell} + \delta_{i\ell}\delta_{jk}\right)E_{ij},\\
	\forall k {\neq} \ell &:& E_{ii} L_{kl} E_{ii} = 0
\end{eqnarray}
\end{subequations}
we find after some algebra that
\begin{multline}
    \e^{\pi L_{ij}}L_{k\ell}\e^{-\pi L_{ij}}\\
    \quad\begin{aligned}
        =& \left[\id - 2(E_{ii}+E_{jj})\right]L_{k\ell} \left[\id - 2(E_{ii}+E_{jj})\right] \\
        =& \left(1-2\left(\delta_{ik}+\delta_{i\ell}+\delta_{jk}+\delta_{j\ell}\right)+4\left(\delta_{ik}\delta_{j\ell} + \delta_{i\ell}\delta_{jk}\right)\right)L_{k\ell} \\
        \stackrel{i\neq j}{=}& \left(1-2\delta_{ik}-2\delta_{i\ell}\right)\left(1-2\delta_{jk}-2\delta_{j\ell}\right) L_{k\ell}
        .
    \end{aligned}
    \label{SM:eq:PNconj}
\end{multline}
We observe that, since $i<j$ and $k<\ell$, the prefactors in front of $L_{\ell k}$ are $\pm 1$, and thus $\e^{\pi L_{ij}}L_{k\ell}\e^{-\pi L_{ij}} = \pm L_{ij}$ is, up to a sign, one of the basis elements of $\so(N)$.
Applying the algebra isomorphism, this gives
\begin{equation}
	\overline{\e^{\pi L_{ij}}L_{k\ell}\e^{-\pi L_{ij}}} = \left(1-2\delta_{ik}-2\delta_{i\ell}\right)\left(1-2\delta_{jk}-2\delta_{j\ell}\right)t_{k\ell},
    \label{SM:eq:PNconj_lift}
\end{equation}
which is an element of $\spin(N)$.

On the other hand, an explicit calculation [e.g., via a repeated application of \cref{eqn:eps-ij,eqn:t-ij}] reveals that for $i<j$ and $k<\ell$
\begin{equation}
    (2t_{ij})t_{k\ell}(-2t_{ij}) = \left(1-2\delta_{ik}-2\delta_{i\ell}\right)\left(1-2\delta_{jk}-2\delta_{j\ell}\right)t_{k\ell},
    \label{SM:eq:PNbarconj}
\end{equation}
where $t_{k\ell}\in\spin(N)$ and $\e^{\pi L_{ij}}=2t_{ij}\in\overline{\PN_N}$
Noting that $(2t_{ij})^{-1}=-2t_{ij}$, we observe that the right-hand sides of \cref{SM:eq:PNconj_lift,SM:eq:PNbarconj} are equal, such the left-hand sides must be equal as well.
\end{proof}

\begin{lemma}\label{SM:lemma:liftSOconjPN}
    Let $A\in\SO(N)$ be close to the identity and $D\in\PN_N$, then $DAD^{-1}\in\SO(N)$ is close to the identity as well, and
    \begin{equation}
    	\overline{DAD^{-1}} = \overline{D}\,\overline{A}\,\overline{D}^{\,-1},
    	\label{SM:eq:groupPNconj}
    \end{equation}
    with $\overline{D}$ defined as follows:
    Any $D\in\PN_N$ can be written as 
    \begin{equation}
        D=\prod_{i<j}(\e^{\pi L_{ij}})^{d_{ij}}\label{eqn:D-as-product}
    \end{equation}
        for a (non-unique) set of $\{d_{ij}\}$, $d_{ij}\in\{0,1\}$, then
    \begin{equation}
        \overline{D} \equiv \prod_{i<j}(2t_{ij})^{d_{ij}},
        \label{SM:eq:groupPNconj:b}
    \end{equation}
    where the ordering of the factors in \cref{SM:eq:groupPNconj:b} matches the ordering in \cref{eqn:D-as-product}.
\end{lemma}
Let us remark that while $\overline{A}$ is given by \cref{SM:eq:liftSON}, $\overline{D}$ needs to be defined explicitly, since $D$ is not an element close to the identity.

\begin{proof}
Obviously, any $D\in\PN_N$ can be written as a (non-unique) product of generators:
\begin{equation}
    D = \prod_{i<j}(\e^{\pi L_{ij}})^{d_{ij}},
\end{equation}
where $d_{ij}\in\{0,1\}$.
First we consider $L_{k\ell}\in\so(N)$, then
\begin{equation}
    DL_{k\ell}D^{-1} = \e^{\pi L_{i'j'}}\cdots\e^{\pi L_{ij}}L_{k\ell}\e^{-\pi L_{ij}}\cdots\e^{-\pi L_{i'j'}},
\end{equation}
where only $\e^{\pi L_{ij}}$ with $i<j$ and $d_{ij}=1$ appear in the product.
This is just consecutive conjugation by generators of $\PN_N$ and, by \cref{SM:lemma:liftsoconjPN}, the result of each conjugation is an element of $\so(N)$ again, such that $DL_{k\ell}D^{-1}\in\so(N)$ and
\begin{equation}
    \overline{DL_{k\ell}D^{-1}} = \overline{D}t_{k\ell}\overline{D}^{\,-1},
\end{equation}
where
\begin{equation}
    \overline{D} \equiv \prod_{i<j}(\e^{\pi t_{ij}})^{d_{ij}} = \prod_{i<j}(2t_{ij})^{d_{ij}}.
\end{equation}

Now, any $A\in\SO(N)$ close to the identity can be expanded in $\so(N)$:
\begin{equation}
    A \approx \id + \sum_{k<\ell}\alpha_{k\ell}L_{k\ell}.
\end{equation}
Therefore,
\begin{equation}
    DAD^{-1} = \id + \sum_{k<\ell}\alpha_{k\ell}DL_{k\ell}D^{-1}
\end{equation}
is an element of $\SO(N)$ close to the identity and we can lift it to $\Spin(N)$ following \cref{SM:eq:liftSON}, resulting in \begin{eqnarray}
    \overline{DAD^{-1}} &=& 1 + \sum_{k<\ell}\alpha_{k\ell}\overline{D}t_{k\ell}\overline{D}^{\,-1} \\
    &=& \overline{D}\left(1 + \sum_{k<\ell}\alpha_{k\ell}t_{k\ell}\right)\overline{D}^{\,-1} = \overline{D} \,\overline{A}\, \overline{D}^{\,-1}.
\end{eqnarray}

It remains to be shown that conjugation with $\overline{D}$ is independent of the choice of $\{d_{ij}\}$ (among the ones that give the same $D$) and their order.
But this follows immediately, because the effect of conjugation with one factor in $D$ amounts to just a single prefactor $\pm1$ [cf.~the text following \cref{SM:eq:PNconj}] which is identical to the prefactor resulting from the conjugation with the corresponding factor in $\overline{D}$ [cf.\ \cref{SM:eq:PNbarconj}].
If two choices of $\{d_{ij}\}$ represent the same $D$, then in particular the overall prefactor is the same and thus the results of conjugating with the two versions of $\overline{D}$ match.
\end{proof}
Note that $\overline{D}\in\overline{\PN}_N\subg\Spin(N)$ and it covers $D$, i.e.,
\begin{equation}
    \sigma\left(\overline{D}\right) = D,
\end{equation}
which follows directly from the definition of $\overline{D}$ and \cref{SM:eq:cover_of_PNbar}.\medskip

We now extend the above results to conjugation with an element of $\PN_{Nh}$.
\begin{lemma}\label{SM:lemma:liftsoconjPNh}
    Let $L_{k\ell}$ ($k<\ell$) be a basis element of $\so(N)$, then $(\id-2E_{ii})L_{k\ell}(\id-2E_{ii})\in\so(N)$ and
    \begin{equation}
        \overline{(\id-2E_{ii})L_{k\ell}(\id-2E_{ii})} = \epsilon_it_{k\ell}\epsilon_i^{-1}\in\spin(N),
    \end{equation}
    where $\epsilon_i$ is the corresponding generator of $\Clifford{0}{N}$.
\end{lemma}

\begin{proof}
First observe that
\begin{equation}
    (\id-2E_{ii})L_{k\ell}(\id-2E_{ii}) = \left[1-2\left(\delta_{ik}+\delta_{i\ell}\right]\right)L_{k\ell}\in\so(N).
\end{equation}
On the other hand, we find that
\begin{equation}
    \epsilon_it_{k\ell}(-\epsilon_i) = \left[1-2\left(\delta_{ik}+\delta_{i\ell}\right)\right]t_{k\ell},
    \label{SM:eq:ete_conj}
\end{equation}
which implies that
\begin{equation}
    \overline{(\id-2E_{ii})L_{k\ell}(\id-2E_{ii})} = \epsilon_it_{k\ell}(-\epsilon_i)\in\spin(N).
\end{equation}
Note that $\epsilon_i^{-1}=-\epsilon_i$, because $\epsilon_i^2=-1$.
\end{proof}

\begin{lemma}\label{SM:lemma:liftSOconjPNh}
    Let $A\in\SO(N)$ be close to the identity and $P\in\PN_{Nh}$, then $PAP^{-1}\in\SO(N)$ is close to the identity as well, and
    \begin{equation}
    	\overline{PAP^{-1}} = \overline{P}\,\overline{A}\,\overline{P}^{\,-1},
	    \label{SM:eq:groupPNhconj}
    \end{equation}
    where $\overline{P}$ is defined as
    \begin{equation}
        \overline{P} \equiv \epsilon_1^{(1-\det(P))/2}\overline{D}
        \label{SM:eq:groupPNhconj:b}
    \end{equation}
    for $D\equiv(\id-2E_{11})^{(1-\det(P))/2}P\in\PN_N$ and $\overline{D}$ is defined according to \cref{SM:eq:groupPNconj:b}.
\end{lemma}

\noindent Note that $\overline{P}$ must be explicitly defined because $P$ is not close to the identity, implying that $\overline{P}$ has a sign ambiguity.
This ambiguity is irrelevant for \cref{SM:eq:groupPNhconj}, because the $\pm$ sign appears twice and therefore cancels. Nevertheless, for concreteness we opt to work with the particular choice of sign fixed by \cref{SM:eq:groupPNhconj:b}.

\begin{proof}
Obviously, for any $P\in \PN_{Nh}$,
\begin{equation}
    D \equiv (\id-2E_{11})^{(1-\det(P))/2}P \in \PN_N,
\end{equation}
because 
\begin{equation}
    \det(D) = (-1)^{(1-\det(P))/2}\det(P) = \det(P)^2 = +1.
\end{equation}
If $\det(P)=+1$, then $D=P$ and the statement of the Lemma~\ref{SM:lemma:liftSOconjPNh} reduces to the already proved \cref{SM:lemma:liftSOconjPN}.
On the other hand, if $\det(P)=-1$, we have $D = (\id-2E_{11})P$ and
\begin{equation}
    PAP^{-1} = (\id-2E_{11})DAD^{-1}(\id-2E_{11}).
\end{equation}
\cref{SM:lemma:liftSOconjPN} implies that $B\equiv DAD^{-1}\in\SO(N)$ is close to the identity and $\overline{B}=\overline{D}\,\overline{A}\,\overline{D}^{\,-1}$, such that we only need to prove that for any $B\in\SO(N)$ close to the identity $(\id-2E_{11})B(\id-2E_{11})\in\SO(N)$ is close to the identity and
\begin{equation}
    \overline{(\id-2E_{11})B(\id-2E_{11})} = \epsilon_1\overline{B}\epsilon_1^{-1}.
\end{equation}
But this follows from \cref{SM:lemma:liftsoconjPNh} with analogous arguments as those used in the proof of \cref{SM:lemma:liftSOconjPN}
\end{proof}

Note that given some $P\in\PN_{Nh}$ as a matrix
\begin{equation}
    P = \diag\left(p_1,p_2,\dotsc,p_N\right) = \prod_{i\,:\,p_i=-1}(\id-2E_{ii})
    \label{SM:eq:PNh_canonical}
\end{equation}
with $p_i\in\{\pm 1\}$ we can use \cref{SM:lemma:liftsoconjPNh} to conclude that the $\overline{P}$ defined above has the canonical parametrization
\begin{equation}
    \overline{P} = \prod_{i\,:\,p_i=-1}\epsilon_i,
    \label{SM:eq:PNhbar_canonical}
\end{equation}
which is obtained by replacing each factor of $(\id-2E_{ii})$ with~$\epsilon_i$.
The choice of sign mentioned above now corresponds to fixing the ordering of the factors in \cref{SM:eq:PNhbar_canonical}.
Here we choose the convention that in \cref{SM:eq:PNh_canonical,SM:eq:PNhbar_canonical} factors with smaller indices appear to the right.
\medskip

Finally, we consider the conjugation of an element of $\overline{\PN}_N$ with an element of $\overline{\PN}_{Nh}$.
\begin{lemma}\label{SM:lemma:PN_PNh_conj}
Let $\mathfrak{d}\in\overline{\PN}_N$ and $\mathfrak{p}\in\overline{\PN}_{Nh}$, then
\begin{equation}
    \mathfrak{p}\mathfrak{d}\mathfrak{p}^{-1} = s(\mathfrak{d},\mathfrak{p})\mathfrak{d}
\end{equation}
with $s(\mathfrak{d},\mathfrak{p})\in\{-1,+1\}$.
\end{lemma}
\begin{proof}
First we recall from \cref{SM:Subsec:doublecovers} that any $\mathfrak{d}\in\overline{\PN}_N$ can be written as
\begin{equation}
    \mathfrak{d} = \prod_{k<\ell}(-\epsilon_k\epsilon_\ell)^{d_{k\ell}}
\end{equation}
for some (non-unique) $d_{k\ell}\in\{0,1\}$ and any $\mathfrak{p}\in\overline{\PN}_{Nh}$ can be written as
\begin{equation}
    \mathfrak{p} = \epsilon_1^{p_0}\prod_{i<j}(-\epsilon_i\epsilon_j)^{p_{ij}}
\end{equation}
for $p_0\in\{0,1\}$ and some (non-unique) $p_{ij}\in\{0,1\}$, where $\{\epsilon_i\}_{i=1}^N$ are the generators of $\Clifford{0}{N}$.

\Cref{SM:eq:ete_conj,eqn:t-ij} imply that for $k<\ell$
\begin{equation}
    \epsilon_i\epsilon_k\epsilon_\ell\epsilon_i^{-1} =  \left[1-2\left(\delta_{ik}+\delta_{i\ell}\right)\right]\epsilon_k\epsilon_\ell,
\end{equation}
such that
\begin{equation}
    \epsilon_i\mathfrak{d}\epsilon_i^{-1} = \underbrace{\left(\prod_{k<\ell}\left[1-2\left(\delta_{ik}+\delta_{i\ell}\right)\right]^{d_{k\ell}}\right)}_{=:s(\mathfrak{d},\epsilon_i)\in\{-1,+1\}}\mathfrak{d}.
\end{equation}
Therefore,
\begin{equation}
    \begin{split}
    	\mathfrak{p}\mathfrak{d}\mathfrak{p}^{-1} &= \epsilon_1^{p_0}\left(\prod_{i<j}\epsilon_i^{p_{ij}}\epsilon_j^{p_{ij}}\right)\mathfrak{d}\left({\prod_{i<j}}'\epsilon_j^{-p_{ij}}\epsilon_i^{-p_{ij}}\right)\epsilon_1^{-p_0}\\
        &= \underbrace{\left(s(\mathfrak{d},\epsilon_1)^{p_0}\prod_{i<j}\left(s(\mathfrak{d},\epsilon_i)s(\mathfrak{d},\epsilon_j)\right)^{p_{ij}}\right)}_{=:s(\mathfrak{d},\mathfrak{p})}\mathfrak{d},
    \end{split}
\end{equation}
where the dashed product indicates reversed order of the terms.
The prefactor $s(\mathfrak{d},\mathfrak{p})$ is obviously just a sign, because all the factors $s(\mathfrak{d},\epsilon_i)$ are just signs.
The independence of this conclusion and in particular of the prefactor $s(\mathfrak{d},\mathfrak{p})\in\{\pm 1\}$ from the parametrization of $\mathfrak{d}$ and of $\mathfrak{p}$ follows trivially, because the left hand side of the above equation is obviously independent of those parameterizations, and the same is true for the remainder $\mathfrak{d}$ on the right-hand side of the equation.
\end{proof}

\subsection{Quaternion invariant}\label{SM:Subsec:quaternion}
Recall the definition~\cite{Wu:2019,Bouhon:2020} of the quaternion invariant on a closed path $\gamma$ (here assumed to be completely contained in the first Brillouin zone) based at point $P$. 
We will in the following label the same point as $(0,0)$ (the numbers do not correspond to the $\vec{k}$-space coordinates).
It is assumed that there is no band degeneracy of the $N$-band real-symmetric Hamiltonian $\Ham_\mathrm{R}(\vec{k})$ along $\gamma$.
We partition the path into infinitesimally spaced points, which we label $(0,1), (0,2), \dotsc, (0,n-1)$. 
The next point in the sequence is $(0,n)\equiv P$, i.e., the initial point again.
(The motivation for the additional $0$'s in the label for each listed point will become clear in later paragraphs.)

At each point $\vec{k}$ on $\gamma$ we can find the eigenframe [\cref{SM:eq:eigenframe}] $\mathsf{u}(\vec{k})$.
However, as discussed above, it is not unique and has the gauge freedom $\mathsf{u}\mapsto \mathsf{u}F$, where $F\in \mathsf{P}_{Nh}$.
Starting with an initial right-handed eigenframe $\mathsf{u}_{0,0}\in\SO(N)$ of $\Ham_\mathrm{R}(P)$, we define the eigenframes $\mathsf{u}_{0,j}$ of $\Ham_\mathrm{R}$ at the subsequent points such that the rotation $\mathsf{u}_{0,j}^\top \mathsf{u}_{0,j+1}^{\phantom{\top}}$ is close to identity $\id\in\mathsf{SO}(N)$.
We will refer to this continuous choice of frame either as a \emph{parallel transport} or \emph{monodromy} of $\mathsf{u}_{0,0}$.
In the last step of the closed path $\gamma$, we define $\underline{\mathsf{u}}_{0,0}$ such that $u_{0,n-1}^\top \underline{u}_{0,0}^{\phantom{\top}}$ is close to identity; the underline indicates that the final $\underline{\mathsf{u}}_{0,0}$ is in general different from the initial $\mathsf{u}_{0,0}$ due to the possible presence of Berry phases on $\gamma$.
We denote the gauge transformation that relates the two eigenframes as $F_{\gamma} \in \PN_N$, i.e.
\begin{equation}
    \underline{\mathsf{u}}_{0,0} = \mathsf{u}_{0,0} F_\gamma.
    \label{SM:eq:F_gamma}
\end{equation}
The quanternion charge is then defined as
\begin{equation}
    \mathfrak{q}(\gamma) = \overline{\mathsf{u}_{0,0}^\top \mathsf{u}_{0,1}^{\phantom{\top}}} \cdot \overline{\mathsf{u}_{0,1}^\top \mathsf{u}_{0,2}^{\phantom{\top}}} \cdot  \ldots \cdot \overline{\mathsf{u}_{0,n-1}^\top \underline{\mathsf{u}}_{0,0}^{\phantom{\top}}},
    \label{SM:eq:def_q_charge}
\end{equation}
where $\overline{\mathsf{u}_{0,j}^\top \mathsf{u}_{0,j+1}^{\phantom{\top}}}$ is defined according to \cref{SM:eq:liftSON}, since $\mathsf{u}_{0,j}^\top \mathsf{u}_{0,j+1}^{\phantom{\top}}\in\SO(N)$ is close to the identity by construction.
Note that $\mathfrak{q}(\gamma)$ implicitly depends on both the base point $P$ of the closed path $\gamma$ as well as on the choice of gauge of the initial eigenframe $\mathsf{u}_{0,0}$.

Let us also establish the notation for the other two paths that we defined in \cref{SM:conjecture:NLBZ}.
We partition $\gamma_{P,\vec{b}}$ that connects $P\equiv(0,0)$ to $P+\vec{b} \equiv P' \equiv (m,0)$ into infinitesimally spaced points labelled sequentially as $(0,0), (1,0), (2,0),\dotsc,(m-1,0),(m,0)$.
Note that this path is not closed, $P'\neq P$; in particular, the Hamiltonians $\mcH(P)$ and $\mcH(P')$ may differ by the unitary transformation in \cref{SM:eq:shift_unitary_transform}.

Furthermore, we partition the shifted (closed) contour $\gamma'$ based at $P'$ into the points $(m,j) = (0,j) + \vec{b}$ for $1 \leq j \leq n$.
Note that the energy spectra on $\gamma$ and $\gamma'$ are identical, i.e., by our previous assumption there is no band degeneracy on $\gamma'$.

Analogously to the case of $\gamma$, we define the eigenframes $\mathsf{u}_{i,0}$, $1\leq i\leq m$, on $\gamma_{P,\vec{b}}$ such that the rotation $\mathsf{u}_{i,0}^\top\mathsf{u}_{i+1,0}^{\phantom{\top}}$ is close to identity $\id\in\mathsf{SO}(N)$ (i.e., through parallel transport), and similarly for $\mathsf{u}_{m,j}$, $1 \leq j \leq n$.

The final frame on $\gamma'$, i.e., at $P'$ \emph{after} traversing $\gamma'$, is related to the initial frame on $\gamma'$ via a gauge transformation $F_{\gamma'}\in\PN_N$ [cf.~\cref{SM:eq:F_gamma}]
\begin{equation}
    \underline{\mathsf{u}}_{m,0} = \mathsf{u}_{m,0} F_{\gamma'}.
    \label{SM:eq:F_gamma_prime}
\end{equation}
Therefore, the eigenframes on $\gamma_{P,\vec{b}}^{-1}$ are not $\mathsf{u}_{i,0}$ and we instead define new eigenframes $\underline{\mathsf{u}}_{i,0}$, $1\leq i\leq m$, through parallel transport, i.e., such that $\underline{\mathsf{u}}_{i,0}^\top\underline{\mathsf{u}}_{i-1,0}^{\phantom{\top}}$ is close to identity $\id\in\mathsf{SO}(N)$ (note the reversed order in the subscript $i$ because the path $\gamma_{P,\vec{b}}$ is traversed in reverse).
The final frame on $\gamma_{P,\vec{b}}^{-1}$, defined via parallel transport as well, is instead denoted by $\frameufinal$ to avoid confusion with the already defined $\underline{\mathsf{u}}_{0,0}$ [however, we will later show that $\frameufinal=\underline{\mathsf{u}}_{0,0}$, cf.~\cref{eqn:u-under-0-0-gamma}].
All the frames on $\gamma_{P,\vec{b}}^{-1}$ are related to the corresponding frames on $\gamma_{P,\vec{b}}$ by some gauge transformation $F_{(i,0)}\in\PN_N$:
\begin{subequations}\label{SM:eq:F_i0}
    for $1\leq i\leq m$
    \begin{equation}
        \underline{\mathsf{u}}_{i,0} = \mathsf{u}_{i,0} F_{(i,0)},
    \end{equation}
    where $F_{(m,0)}=F_{\gamma'}$; for the last point on $\gamma_{P,\vec{b}}^{-1}$ we have 
    \begin{equation}
        \frameufinal = \mathsf{u}_{0,0} F_{(0,0)}.
    \end{equation}
\end{subequations}

\begin{lemma}\label{SM:lemma:F_i0}
Under the assumptions of \cref{SM:conjecture:NLBZ}, $F_{(i,0)}$ defined in \cref{SM:eq:F_i0} and $F_{\gamma'}$ defined in \cref{SM:eq:F_gamma_prime} are equal,
\begin{equation}
    \forall 0\leq i\leq m : F_{(i,0)} = F_{\gamma'}.
\end{equation}
\end{lemma}
\begin{proof}
    The statement follows from \cref{SM:eq:F_gamma_prime} due to the monodromy (flatness) of the parallel transport.
    We construct an explicit proof via recursion starting from $i=m$ [cf.~\cref{SM:eq:F_gamma_prime}].
    Assuming $F_{(i+1,0)} = F_\gamma$ for some $0\leq i<m$, the frame $\underline{\mathsf{u}}_{i,0}$ is defined via parallel transport such that
    \begin{equation}
        \underline{\mathsf{u}}_{i+1,0}^\top\underline{\mathsf{u}}_{i,0}^{\phantom{\top}} = F_{\gamma'}^\top\mathsf{u}_{i+1,0}^\top\mathsf{u}_{i,0}^{\phantom{\top}}F_{(i,0)}
    \end{equation}
    is close to the identity.
    On the other hand the frames $\mathsf{u}_{i,0}$, $0\leq i\leq m$, are defined such that $\mathsf{u}_{i,0}^\top\mathsf{u}_{i+1,0}^{\phantom{\top}}$ is close to the identity, which is equivalent to the condition that $\mathsf{u}_{i+1,0}^\top\mathsf{u}_{i,0}^{\phantom{\top}}$ is close to the identity.
    Thus, $F_{(i,0)}$ and $F_{\gamma'}$ have to be close, but because the gauge transformations $F_{\gamma'}$ and $F_{(i,0)}$ are diagonal with $\pm 1$'s on the diagonal, it follows that $F_{(i,0)}=F_{\gamma'}$.
\end{proof}

\begin{lemma}\label{SM:lemma:F_gamma_prime}
Under the assumptions of \cref{SM:conjecture:NLBZ}, $F_\gamma$ defined in \cref{SM:eq:F_gamma} and $F_{\gamma'}$ defined in \cref{SM:eq:F_gamma_prime} are equal
\begin{equation}
    F_{\gamma'} = F_\gamma.
\end{equation}
\end{lemma}
\noindent The physical interpretation of this result is that the Berry phases of the individual bands on closed paths $\gamma$ and $\gamma'$ match.

\begin{proof}
The gauge transformations $F_\gamma$ and $F_{\gamma'}$ are defined as gauge transformations at $P$ and $P'$, respectively.
We can obtain an eigenframe of $\Ham(P+\vec{b})$ from the initial eigenframe $\mathsf{u}_{0,0}$ of $\mcH(P)$ in two distinct but canonical ways.
On the one hand, through parallel transport we have defined an eigenframe $\mathsf{u}_{m,0}$ at $P'$.
On the other hand, the unitary relation through $U_{\vec{b},\mathrm{R}}$ defines the eigenframe $U_{\vec{b},\mathrm{R}}\mathsf{u}_{0,0}$ at $P'$.
These two eigenframes in general differ by a gauge transformation $F_{P,\vec{b}}\in\PN_{Nh}$, so we write
\begin{equation}
    U_{\vec{b},\mathrm{R}} \mathsf{u}_{0,0}^{\phantom{\top}} = \mathsf{u}_{m,0}^{\phantom{\top}}F_{P,\vec{b}}.\label{eqn:FPb-def}
\end{equation}
Thus, we have
\begin{equation}
    F_{P,\vec{b}} = \mathsf{u}_{m,0}^\top U_{\vec{b},\mathrm{R}}\mathsf{u}_{0,0}.
\end{equation}
We emphasize that it is possible for $F_{P,\vec{b}}$ to have a negative determinant.
This happens when an odd number of bands carry non-trivial Berry phase $\pi$ in the $\vec{b}$ direction.

It also follows from the monodromy (flatness) of the parallel transport that
\begin{equation}
    U_{\vec{b},\mathrm{R}} \mathsf{u}_{0,j}^{\phantom{\top}} = \mathsf{u}_{m,j}^{\phantom{\top}} {F_{P,\vec{b}}}^{\phantom{\top}}
\end{equation}
for all $j$, such that we can rewrite $\mathsf{u}_{m,j}^{\phantom{\top}}$ in terms of $\mathsf{u}_{0,j}$:
\begin{equation}
    \mathsf{u}_{m,j} = U_{\vec{b},\mathrm{R}} \mathsf{u}_{0,j}^{\phantom{\top}} {F_{P,\vec{b}}}^{\top}.
    \label{SM:eq:umj_u0j}
\end{equation}
For the same reason, i.e., monodromy, an analogous equation holds for $\underline{\mathsf{u}}_{m,0}$ and $\underline{\mathsf{u}}_{0,0}$, which are defined via parallel transport from $\mathsf{u}_{m,n-1}$ and $\mathsf{u}_{0,n-1}$, respectively:
\begin{equation}
    \underline{\mathsf{u}}_{m,0} = U_{\vec{b},\mathrm{R}} \underline{\mathsf{u}}_{0,0}^{\phantom{\top}}F_{P,\vec{b}}^\top
\end{equation}
Together with \cref{SM:eq:F_gamma,SM:eq:umj_u0j}, this gives
\begin{align}
    \underline{\mathsf{u}}_{m,0} &= U_{\vec{b},\mathrm{R}} \mathsf{u}_{0,0}F_\gamma F_{P,\vec{b}}^\top\nonumber\\
    &= \mathsf{u}_{m,0} F_{P,\vec{b}}^{\phantom{\top}} F_\gamma F_{P,\vec{b}}^\top.
    \label{eqn:u-under-m-0}
\end{align}
By comparing \cref{eqn:u-under-m-0} to \cref{SM:eq:F_gamma_prime}, we see that $F_{\gamma'} = F_{P,\vec{b}}^{\phantom{\top}} F_\gamma F_{P,\vec{b}}^\top$.
However, since $F_\gamma$, $F_{\gamma'}$ and $F_{P,\vec{b}}$ are all elements of $\PN_{Nh}$ (i.e., diagonal matrices with $\pm 1$ on the diagonal) and therefore commute with each other, we immediately find $F_{\gamma'}=F_\gamma$.
\end{proof}

Besides the quaternion charge $\mathfrak{q}(\gamma')$, which is defined analogously to \cref{SM:eq:def_q_charge}, we now define the total eigenframe rotations along $\gamma_{P,\vec{b}}$ and $\gamma_{P,\vec{b}}^{-1}$
\begin{align}
    \mathfrak{b}\left(\gamma_{P,\vec{b}}\right) &= \overline{\mathsf{u}_{0,0}^\top\mathsf{u}_{1,0}^{\phantom{\top}}}\cdot \overline{\mathsf{u}_{1,0}^\top\mathsf{u}_{2,0}^{\phantom{\top}}} \cdot \ldots \cdot \overline{\mathsf{u}_{m-1,0}^\top\mathsf{u}_{m,0}^{\phantom{\top}}},\label{SM:eq:b_gammaPmu}\\
    \mathfrak{b}\left(\gamma_{P,\vec{b}}^{-1}\right) &= \overline{\underline{\mathsf{u}}_{m,0}^\top \underline{\mathsf{u}}_{m-1,0}^{\phantom{\top}}}\cdot\ldots\cdot \overline{\underline{\mathsf{u}}_{2,0}^\top \underline{\mathsf{u}}_{1,0}^{\phantom{\top}}}\cdot \overline{\underline{\mathsf{u}}_{1,0}^\top \frameufinal^{\phantom{\top}}}.\label{SM:eq:b_gammaPmu_inv}
\end{align}
With these definitions, we can proceed to prove several relations between the introduced elements in $\Spin(N)$.
One should bear in mind that $\mathfrak{b}\left(\gamma_{P,\vec{b}}\right)$ depends not only on the base point $P$ but also implicitly on the initial frame $\mathsf{u}_{0,0}$; the quaternion charge $\mathfrak{q}(\gamma')$ depends implicitly on $P$, $\mathsf{u}_{0,0}$ and the path $\gamma_{P,\vec{b}}$, while $\mathfrak{b}\left(\gamma_{P,\vec{b}}^{-1}\right)$ additionally depends on the whole path $\gamma$.
Therefore, the following statement only make sense in the context of \cref{SM:conjecture:NLBZ}, i.e., when we consider a path $\tilde{\gamma} = \gamma_{P,\vec{b}}\circ\gamma'\circ\gamma_{P,\vec{b}}^{-1}$ with base point $P$, fixed initial frame $\mathsf{u}_{0,0}$.

\begin{lemma}\label{SM:lemma:q_gamma_prime}
Under the assumptions of \cref{SM:conjecture:NLBZ}, we have
\begin{equation}
    \mathfrak{q}(\gamma') = \overline{F_{P,\vec{b}}}\mathfrak{q}(\gamma)\overline{F_{P,\vec{b}}}^{\,-1},
\end{equation}
where
\begin{equation}
    F_{P,\vec{b}} = \mathsf{u}_{m,0}^\top U_{\vec{b},\mathrm{R}}\mathsf{u}_{0,0}\in\PN_{Nh}
\end{equation}
and given the above $F_{P,\vec{b}}$, $\overline{F_{P,\vec{b}}}$ is defined according to \cref{SM:eq:groupPNhconj:b}.
\end{lemma}

\begin{proof}
\Cref{SM:eq:umj_u0j} allows us to rewrite the factors in the definition of $\mathfrak{q}(\gamma')$ [\cref{SM:eq:def_q_charge}]: for $1\leq j<n-1$
\begin{equation}
    \begin{split}
    	\overline{\mathsf{u}_{m,j}^\top \mathsf{u}_{m,j+1}^{\phantom{\top}}} &= \overline{F_{P,\vec{b}}^{\phantom{\top}}\mathsf{u}_{0,j}^\top\left(U_{\vec{b},\mathrm{R}}\right)^\top U_{\vec{b},\mathrm{R}} \mathsf{u}_{0,j+1}^{\phantom{\top}}F_{P,\vec{b}}^\top}\\
    	&= \overline{F_{P,\vec{b}}^{\phantom{-1}}\mathsf{u}_{0,j}^\top \mathsf{u}_{0,j+1}^{\phantom{\top}}F_{P,\vec{b}}^{-1}}\\
        &= \overline{F_{P,\vec{b}}^{\phantom{\top}}}\cdot\overline{\mathsf{u}_{0,j}^\top\mathsf{u}_{0,j+1}^{\phantom{\top}}}\cdot\overline{F_{P,\vec{b}}^{\phantom{\top}}}^{\,-1},
    \end{split}
\end{equation}
where we used that $U_{\vec{b},\mathrm{R}}$ and $F_{P,\vec{b}}$ are orthogonal and applied \cref{SM:lemma:liftSOconjPNh} with $\overline{F_{P,\vec{b}}}$ as defined therein.
Thus, in $\mathfrak{q}(\gamma')$ all the $\overline{F_{P,\vec{b}}}$ between the factors cancel and we are left with
\begin{equation}
    \begin{split}
    	\mathfrak{q}(\gamma') &= \overline{\mathsf{u}_{m,0}^\top \mathsf{u}_{m,1}^{\phantom{\top}}} \cdot \overline{\mathsf{u}_{m,1}^\top \mathsf{u}_{m,2}^{\phantom{\top}}} \cdot  \ldots \cdot \overline{\mathsf{u}_{m,n-1}^\top \underline{\mathsf{u}}_{m,0}^{\phantom{\top}}}\\
        &= \overline{F_{P,\vec{b}}}\mathfrak{q}(\gamma)\overline{F_{P,\vec{b}}}^{\,-1},
    \end{split}
\end{equation}
as desired.
\end{proof}

\begin{lemma}\label{SM:lemma:b_gammaP_inv}
Under the assumptions of \cref{SM:conjecture:NLBZ}, we have
\begin{equation}
    \mathfrak{b}\left(\gamma_{P,\vec{b}}^{-1}\right) = \mathfrak{q}(\gamma)^{-1}\mathfrak{b}\left(\gamma_{P,\vec{b}}\right)^{-1}\mathfrak{q}(\gamma).
\end{equation}
\end{lemma}

\begin{proof}
Applying \cref{SM:lemma:F_i0,SM:lemma:F_gamma_prime} to \cref{SM:eq:F_i0} gives for all $1 \leq i \leq m$ that
\begin{equation} 
    \underline{\mathsf{u}}_{i,0} = \mathsf{u}_{i,0}F_{\gamma}
    \label{eqn:u-under-i-0-gamma}
\end{equation}
and
\begin{equation}
    \frameufinal = \mathsf{u}_{0,0}F_{\gamma} = \underline{\mathsf{u}}_{0,0}.
    \label{eqn:u-under-0-0-gamma}
\end{equation}
Substituting \cref{eqn:u-under-i-0-gamma,eqn:u-under-0-0-gamma} into the expression for $\mathfrak{b}\left(\gamma_{P,\vec{b}}^{-1}\right)$ in \cref{SM:eq:b_gammaPmu_inv}, and applying \cref{SM:lemma:liftSOconjPN}, we find for $0\leq i\leq m$
\begin{equation}
    \begin{split}
    	\overline{\underline{\mathsf{u}}_{i,0}^\top \underline{\mathsf{u}}_{i-1,0}} &= \overline{F_{\gamma}^{-1}\mathsf{u}_{i,0}^\top \mathsf{u}_{i-1,0}^{\phantom{\top}}F_{\gamma}^{\phantom{-1}}}\\
        &= \overline{F_\gamma}^{\,-1}\cdot\overline{\mathsf{u}_{i,0}^\top \mathsf{u}_{i-1,0}^{\phantom{\top}}}\cdot\overline{F_\gamma},
    \end{split}
\end{equation}
such that
\begin{equation}
    \mathfrak{b}\left(\gamma_{P,\vec{b}}^{-1}\right) = \overline{F_\gamma}^{\,-1}\cdot\overline{\mathsf{u}_{m,0}^\top \mathsf{u}_{m-1,0}^{\phantom{\top}}}\cdot\ldots\cdot\overline{\mathsf{u}_{1,0}^\top\mathsf{u}_{0,0}^{\phantom{\top}}}\cdot\overline{F_\gamma}.\label{eqn:b+decorations}
\end{equation}
Since $\mathsf{u}_{i-1,0}^\top\mathsf{u}_{i,0}^{\phantom{\top}}\in\SO(N)$ is close to the identity, we can apply \cref{SM:lemma:lift_of_inverse} and find
\begin{equation}
    \overline{\mathsf{u}_{i-1,0}^\top\mathsf{u}_{i,0}^{\phantom{\top}}}^{\,-1} = \overline{\left(\mathsf{u}_{i-1,0}^\top\mathsf{u}_{i,0}^{\phantom{\top}}\right)^\top} = \overline{\mathsf{u}_{i,0}^\top \mathsf{u}_{i-1,0}^{\phantom{\top}}}.\label{eqn:overline-inverse}
\end{equation}

On the other hand, by inverting \cref{SM:eq:b_gammaPmu} we obtain
\begin{eqnarray}
    \mathfrak{b}\left(\gamma_{P,\vec{b}}\right)^{-1} &=&  \overline{\mathsf{u}_{m-1,0}^\top\mathsf{u}_{m,0}^{\phantom{\top}}}^{\,-1} \cdot \ldots \cdot \overline{\mathsf{u}_{1,0}^\top\mathsf{u}_{2,0}^{\phantom{\top}}}^{\,-1} \cdot \overline{\mathsf{u}_{0,0}^\top\mathsf{u}_{1,0}^{\phantom{\top}}}^{\,-1} \nonumber \\
    &\stackrel{\textrm{Eq.~(\ref{eqn:overline-inverse})}}{=}&
    \overline{{\mathsf{u}}_{m,0}^\top {\mathsf{u}}_{m-1,0}^{\phantom{\top}}}\cdot\ldots\cdot \overline{{\mathsf{u}}_{2,0}^\top {\mathsf{u}}_{1,0}^{\phantom{\top}}}\cdot \overline{{\mathsf{u}}_{1,0}^\top {\mathsf{u}}_{0,0}^{\phantom{\top}}}\label{eqn:b+stuffs}
\end{eqnarray}
and by comparing to \cref{eqn:b+decorations} find that
\begin{equation}
    \mathfrak{b}\left(\gamma_{P,\vec{b}}^{-1}\right) = \overline{F_\gamma}^{\,-1}\mathfrak{b}\left(\gamma_{P,\vec{b}}\right)^{-1}\overline{F_\gamma}.
\end{equation}
Additionally, recall that $\overline{F_\gamma}$ satisfies
\begin{equation}
    \sigma\left(\overline{F_\gamma}\right) = F_\gamma
\end{equation}
and by the definition of the quaternion invariant in \cref{SM:eq:def_q_charge},
\begin{equation}
    \sigma(\mathfrak{q}(\gamma)) = \mathsf{u}_{0,0}^\top \mathsf{u}_{0,1}^{\phantom{\top}}\mathsf{u}_{0,1}^\top \mathsf{u}_{0,2}^{\phantom{\top}}\ldots\mathsf{u}_{0,n-1}^\top \underline{\mathsf{u}}_{0,0}^{\phantom{\top}} = \mathsf{u}_{0,0}^\top\underline{\mathsf{u}}_{0,0}^{\phantom{\top}} = F_\gamma,
\end{equation}
such that, according to \cref{SM:eq:cover_product},
\begin{equation}
    \mathfrak{q}(\gamma) = \pm\overline{F_\gamma}.
\end{equation}
Thus,
\begin{equation}
    \mathfrak{b}\left(\gamma_{P,\vec{b}}^{-1}\right) = \mathfrak{q}(\gamma)^{-1}\mathfrak{b}\gamma_{P,\vec{b}}^{-1}\mathfrak{q}(\gamma),
\end{equation}
because the inverse $\mathfrak{q}(\gamma)^{-1}$ comes with the same sign as $\mathfrak{q}(\gamma)$ and consequently the overall sign ambiguity cancels.
\end{proof}

\subsection{Berry Phases}\label{SM:Subsec:Berryphases}
\begin{lemma}\label{SM:lemma:FPmu_Berry_phases}
The gauge transformation relating the eigenframes at $P$ and $P+\vec{b}$ with the reciprocal lattice vector $\vec{b}$,
\begin{equation}
    F_{P,\vec{b}} = \mathsf{u}_{m,0}^\top U_{\vec{b},\mathrm{R}}\mathsf{u}_{0,0}^{\phantom{\top}}\in\PN_{Nh},
\end{equation}
is given by the Berry phases $\phi_i$ of the bands $1\leq i\leq N$ in the direction $\vec{b}$:
\begin{equation}
    F_{P,\vec{b}} = \diag\left(\e^{\i\phi_1},\e^{\i\phi_2},\dotsc,\e^{\i\phi_N}\right).\label{eqn:diag-Berry-phases}
\end{equation}
\end{lemma}
\begin{proof}
Recall the construction of the eigenframes following the path $\gamma_{P,\vec{b}}$.
At each of the infinitesimally spaced points $(0,0),(1,0),\dotsc,(m,0)$, we find a right-handed eigenframe $\hat{\mathsf{u}}_{i+1,0}^{\phantom{\top}}\in\SO(N)$.
At $P$ we fix an arbitrary gauge 
\begin{equation}
    \mathsf{u}_{0,0}^{\phantom{\top}}:=\hat{\mathsf{u}}_{0,0}^{\phantom{\top}}\in\SO(N) \label{eqn:initial}
\end{equation} 
and then recursively choose the right gauge as follows:
given $\mathsf{u}_{i-1,0}^{\phantom{\top}}$ the next eigenframe (in the right gauge) is
\begin{equation}
    \mathsf{u}_{i,0}^{\phantom{\top}} = \hat{\mathsf{u}}_{i,0}^{\phantom{\top}}F_i
    \label{eqn:point-wise-GT}
\end{equation}
with $F_i\in\PN_N$ such that
\begin{equation}
    \mathsf{u}_{i-1,0}^{\top}\mathsf{u}_{i,0}^{\phantom{\top}} = \mathsf{u}_{i-1,0}^{\top}\hat{\mathsf{u}}_{i,0}^{\phantom{\top}}F_i
\end{equation}
is close to the identity.
The latter is equivalent to
\begin{equation}
    \sign\diag\left(\mathsf{u}_{i-1,0}^{\top}\hat{\mathsf{u}}_{i,0}^{\phantom{\top}}F_i\right) = \id, \label{eqn:id-ego-superego}
\end{equation}
where, given a matrix $M$, $\sign\diag(M)$ is the matrix with entries
\begin{equation}
    \left(\sign\diag(M)\right)_{ij} = \sign(M_{ii})\delta_{ij}.
\end{equation}
Obviously, for any matrix $D\in\PN_{Nh}$,
\begin{subequations}
    \begin{align}
        [\sign\diag(M)]D &= \sign\diag(MD), \label{eqn:sign-diag-right}\\
        D[\sign\diag(M)] &= \sign\diag(DM),\label{eqn:sign-diag-left}
    \end{align}
\end{subequations}
such that from \cref{eqn:id-ego-superego,eqn:sign-diag-right} we get
\begin{equation}
    F_i^\top = \sign\diag\left(\mathsf{u}_{i-1,0}^{\top}\hat{\mathsf{u}}_{i,0}^{\phantom{\top}}\right).
\end{equation}
Expressing $\mathsf{u}_{i,0}^{\top}$ from \cref{eqn:point-wise-GT} and using \cref{eqn:sign-diag-left}, we further obtain
\begin{equation}
    F_i^\top = F_{i-1}^\top \sign\diag \left(\hat{\mathsf{u}}_{i-1,0}^{\top}\hat{\mathsf{u}}_{i,0}^{\phantom{\top}}\right),
\end{equation}
which constitutes a recursion relation for the gauge transformations $F_i$.

With $F_0=\id$ [cf.~\cref{eqn:initial}], the above recursion relation has solution
\begin{equation}
    F_j = \prod_{i=1}^{j}\sign\diag\left(\hat{\mathsf{u}}_{i,0}^{\top}\hat{\mathsf{u}}_{i-1,0}^{\phantom{\top}}\right), \label{eqn:recurrence-solution}
\end{equation}
where factors with smaller $i$ appear to the \emph{right} (note also the reversed ordering of the $\hat{\mathsf{u}}_{i,0}$ in each factor due to the transposition).
Then, according to \cref{eqn:point-wise-GT},
\begin{equation}
    \mathsf{u}_{m,0}^{\phantom{\top}} = \hat{\mathsf{u}}_{m,0}^{\phantom{\top}} \left(\prod_{i=1}^{m}\sign\diag\left(\hat{\mathsf{u}}_{i,0}^{\top}\hat{\mathsf{u}}_{i-1,0}^{\phantom{\top}}\right)\right),
\end{equation}
such that
\begin{equation}
    F_{P,\vec{b}} = \left(\prod_{i=1}^{m}\sign\diag\left(\hat{\mathsf{u}}_{i-1,0}^{\top}\hat{\mathsf{u}}_{i,0}^{\phantom{\top}}\right)\right)\hat{\mathsf{u}}_{m,0}^{\top}U_{\vec{b},\mathrm{R}}\hat{\mathsf{u}}_{0,0}^{\phantom{\top}}.\label{eqn:FPb-expression}
\end{equation}
where factors with smaller $i$ appear to the \emph{left}.
Note that $F_{P,\vec{b}}\in\PN_{Nh}$, such that we can apply \cref{eqn:sign-diag-right} with $M=\hat{\mathsf{u}}_{m-1,0}^{\top}\hat{\mathsf{u}}_{m,0}^{\phantom{\top}}$ and $D=F_{P,\vec{b}}$ and obtain
\begin{multline}
    F_{P,\vec{b}} = \left(\prod_{i=1}^{m-1}\sign\diag\left(\hat{\mathsf{u}}_{i-1,0}^{\top}\hat{\mathsf{u}}_{i,0}^{\phantom{\top}}\right)\right)\times\\
    \times\sign\diag\left(\hat{\mathsf{u}}_{m-1,0}^{\top}\hat{\mathsf{u}}_{m,0}^{\phantom{\top}}\hat{\mathsf{u}}_{m,0}^{\top}U_{\vec{b},\mathrm{R}}\hat{\mathsf{u}}_{0,0}^{\phantom{\top}}\right).
\end{multline}
Orthogonality of $\hat{\mathsf{u}}_{m,0}$ finally gives
\begin{equation}
    F_{P,\vec{b}} = \left(\prod_{i=1}^{m-1}\sign\diag\left(\hat{\mathsf{u}}_{i-1,0}^{\top}\hat{\mathsf{u}}_{i,0}^{\phantom{\top}}\right)\right)\sign\diag\left(\hat{\mathsf{u}}_{m-1,0}^{\top}U_{\vec{b},\mathrm{R}}\hat{\mathsf{u}}_{0,0}^{\phantom{\top}}\right),
\end{equation}
where again factors with smaller $i$ appear to the \emph{left}.

Next, we consider the Berry phase $\phi_j$ of the $j^\textrm{th}$ band along $\gamma_{P,\vec{b}}$.
It can be defined in terms of the Wilson operator $\mathcal{W}_j$ as follows.
Let $\vec{u}_j^{(i)}$ be the $j^\textrm{th}$ column of the eigenframe $\hat{\mathsf{u}}_{i,0}$, i.e., the $j^\textrm{th}$ eigenvector of the real Bloch Hamiltonian at the point $(i,0)$ (in an arbitrary gauge), then
\begin{equation}
    \mathcal{W}_j\left(\gamma_{P,\vec{b}}\right) = \frac{\left(U_{\vec{b},\mathrm{R}}\vec{u}_{j}^{(0)}\right)^\top}{\abs{\left(U_{\vec{b},\mathrm{R}}\vec{u}_{j}^{(0)}\right)^\top\vec{u}_{j}^{(m-1)}}}\left(\prod_{i=1}^{m-1}\frac{\vec{u}_{j}^{(i)}\left(\vec{u}_{j}^{(i)}\right)^\top}{\abs{\left(\vec{u}_{j}^{(i)}\right)^\top\vec{u}_{j}^{(i-1)}}}\right)\vec{u}_{j}^{(0)}
\end{equation}
where factors with smaller $i$ appear to the right, and we continue to assume the limit of a partitioning into infinitesimal steps.
Rearranging the terms, we arrive at (factors with smaller $i$ still appear to the right)
\begin{equation}
    \mathcal{W}_j\left(\gamma_{P,\vec{b}}\right) = \frac{\left(U_{\vec{b},\mathrm{R}}\vec{u}_{j}^{(0)}\right)^\top \vec{u}_{j}^{(m-1)} }{\abs{\left(U_{\vec{b},\mathrm{R}}\vec{u}_{j}^{(0)}\right)^\top\vec{u}_{j}^{(m-1)}}} \left(\prod_{i=1}^{m-1}\frac{\left(\vec{u}_{j}^{(i)}\right)^\top\vec{u}_{j}^{(i-1)}}{\abs{\left(\vec{u}_{j}^{(i)}\right)^\top\vec{u}_{j}^{(i-1)}}}\right).
\end{equation}
Since $\mathcal{W}$ is a number, it is its own transpose and we find
\begin{equation}
    \mathcal{W}_j\left(\gamma_{P,\vec{b}}\right) = \left(\prod_{i=1}^{m-1}
    \frac{\left(\vec{u}_{j}^{(i-1)}\right)^\top \vec{u}_{j}^{(i)}}
    {\abs{\left(\vec{u}_{j}^{(i-1)}\right)^\top \vec{u}_{j}^{(i)}}}\right) \frac{\left(\vec{u}_{j}^{(m-1)}\right)^\top U_{\vec{b},\mathrm{R}}\vec{u}_{j}^{(0)}}
    {\abs{\left(\vec{u}_{j}^{(m-1)}\right)^\top U_{\vec{b},\mathrm{R}}\vec{u}_{j}^{(0)}}},
\end{equation}
where factors with smaller $i$ now appear to the \emph{left} because of the transposition.
Each factor is just a phase, and because the eigenvectors are real (by assumption) this implies that it is a just a sign.
Assuming that the points along the path are sufficiently close, the denominators are all close to $1$ and we can rewrite the Wilson operator~as
\begin{multline}
    \mathcal{W}_j\left(\gamma_{P,\vec{b}}\right) = \left(\prod_{i=1}^{m-1}\sign\left(\left(\vec{u}_{j}^{(i-1)}\right)^\top\vec{u}_{j}^{(i)}\right)\right)\times\\
    \times\sign\left(\left(\vec{u}_{j}^{(m-1)}\right)^\top U_{\vec{b},\mathrm{R}}\vec{u}_{j}^{(0)}\right).
\end{multline}

We observe that
\begin{equation}
    \left(\hat{\mathsf{u}}_{i-1,0}^\top\hat{\mathsf{u}}_{i,0}^{\phantom{\top}}\right)_{jj} = \left(\vec{u}_{j}^{(i-1)}\right)^\top\vec{u}_{j}^{(i)},
\end{equation}
such that
\begin{equation}
    \left(F_{P,\vec{b}}\right)_{ij} = \mathcal{W}_i\delta_{ij}.
\end{equation}
With the definition of the Berry phase, $\mathcal{W}_i=\e^{\i\phi_i}$, the desired \cref{eqn:diag-Berry-phases} follows.
\end{proof}

\subsection{Proof of \texorpdfstring{\cref*{SM:conjecture:NLBZ}}{Conjecture 1}}\label{SM:Subsec:proof}
We can finally prove \cref{SM:conjecture:NLBZ}.
\begin{proof}
By definition, the quaternion charge of $\tilde{\gamma}$ is
\begin{equation}
    \mathfrak{q}(\tilde{\gamma}) = \mathfrak{b}\left(\gamma_{P,\vec{b}}\right)\mathfrak{q}(\gamma')\mathfrak{b}\left(\gamma_{P,\vec{b}}^{-1}\right)
\end{equation}
and using \cref{SM:lemma:q_gamma_prime,SM:lemma:b_gammaP_inv,SM:lemma:PN_PNh_conj}, this gives
\begin{align}\allowdisplaybreaks
    	\mathfrak{q}(\tilde{\gamma}) &= \mathfrak{b}\left(\gamma_{P,\vec{b}}\right)\underbrace{\overline{F_{P,\vec{b}}}\mathfrak{q}(\gamma)\overline{F_{P,\vec{b}}}^{\,-1}}_{=s\left(\mathfrak{q}(\gamma),\overline{F_{P,\vec{b}}}\right)\mathfrak{q}(\gamma)}\mathfrak{q}(\gamma)^{-1}\mathfrak{b}(\gamma_{P,\vec{b}})^{-1}\mathfrak{q}(\gamma)\nonumber\\
        &= \mathfrak{b}\left(\gamma_{P,\vec{b}}\right)s\left(\mathfrak{q}(\gamma),\overline{F_{P,\vec{b}}}\right)\underbrace{\mathfrak{q}(\gamma)\mathfrak{q}(\gamma)^{-1}}_{=1}\mathfrak{b}(\gamma_{P,\vec{b}})^{-1}\mathfrak{q}(\gamma)\nonumber\\
        &= \underbrace{\mathfrak{b}\left(\gamma_{P,\vec{b}}\right)\mathfrak{b}(\gamma_{P,\vec{b}})^{-1}}\underbrace{s\left(\mathfrak{q}(\gamma),\overline{F_{P,\vec{b}}}\right)\mathfrak{q}(\gamma)}_{=\overline{F_{P,\vec{b}}}\mathfrak{q}(\gamma)\overline{F_{P,\vec{b}}}^{\,-1}}\nonumber\\
        &= \overline{F_{P,\vec{b}}}\mathfrak{q}(\gamma)\overline{F_{P,\vec{b}}}^{\,-1}
        \label{SM:eq:qgamma_qgammatilde_proof}
\end{align}
with $\overline{F_{P,\vec{b}}}$ defined according to \cref{SM:eq:groupPNhconj:b}, given
\begin{equation}
    F_{P,\vec{b}} = \mathsf{u}_{m,0}^\top U_{\vec{b},\mathrm{R}}\mathsf{u}_{0,0}^{\phantom{\top}}\in\PN_{Nh}.
\end{equation}
But according to \cref{SM:lemma:FPmu_Berry_phases}
\begin{equation}
    F_{P,\vec{b}} = \diag\left(\e^{\i\phi_1},\e^{\i\phi_2},\dotsc,\e^{\i\phi_N}\right).
\end{equation}
where $\phi_i\in\{0,\pi\}$ is the Berry phase of the $i^\textrm{th}$ band in the direction $\vec{b}$ and, according to \cref{SM:eq:PNhbar_canonical},
\begin{equation}
    \overline{F_{P,\vec{b}}} = \overline{\prod_{i\,:\,\e^{\i\phi_i}=-1}(\id-2E_{ii})} = \prod_{i\,:\,\e^{\i\phi_i}=-1}\epsilon_i,
    \label{SM:eq:FPb_proof}
\end{equation}
as desired.
The ordering of the product in \cref{SM:eq:FPb_proof} is fixed by the convention in \cref{SM:lemma:FPmu_Berry_phases} such that $\epsilon_i$ with smaller $i$ appear to the right.
Without fixing the ordering in the product, there would be a sign ambiguity in $\overline{F_{P,\vec{b}}}$, because $F_{P,\vec{b}}$ is not close to the identity.
However, note that in \cref{SM:eq:qgamma_qgammatilde_proof} $\overline{F_{P,\vec{b}}}$ appears twice, such that result holds for both choices of the sign.
\end{proof}

\end{bibunit}

\end{document}